\documentclass[11pt,twoside,a4paper,pdf]{article}  

\usepackage[T1]{fontenc}
\usepackage{relsize}
\usepackage{xspace}
\usepackage{booktabs}
\usepackage{feynmp}
\usepackage{subcaption}
\usepackage{afterpage}
\usepackage{times}
\usepackage{booktabs}
\usepackage[english]{babel}
\usepackage{amssymb}
\usepackage{cite}
\usepackage{tabularx}
\usepackage{xspace}
\usepackage{a4wide}
\usepackage{authblk,setspace,caption3,babel}
\usepackage{amsmath,multicol,graphicx,color}
\usepackage[colorlinks=true, pdftex, citecolor=red, urlcolor=blue]{hyperref}

\usepackage[normalem]{ulem}
\usepackage{ifthen}
\usepackage{marginnote}

\renewcommand{\d}{\text{d}}
\renewcommand{\t}[1]{\text{#1}}
\newcommand{\mc}[1]{\mathcal{#1}}
\newcommand{\mrm}[1]{\ensuremath{\mathrm{#1}}\xspace}

\newcommand{\graphNoSpace}[2]{\includegraphics[#1]{#2}\vspace*{-0.5mm}}

\newcommand{\saj}{s_{aj}}
\newcommand{\sjb}{s_{jb}}
\newcommand{\sab}{s_{ab}}
\newcommand{\sAB}{s_{AB}}
\newcommand{\sjk}{s_{jk}}
\newcommand{\sak}{s_{ak}}
\newcommand{\sAK}{s_{AK}}
\newcommand{\sij}{s_{ij}}
\newcommand{\sik}{s_{ik}}
\newcommand{\sIK}{s_{IK}}

\newcommand{\tsc}[1]{\textsc{#1}}

\newcommand{\vc}{\tsc{Vincia}}
\newcommand{\py}{\tsc{Pythia}}
\newcommand{\mg}{\tsc{MadGraph}}
\newcommand{\ari}{\tsc{Ariadne}}
\newcommand{\riv}{\tsc{Rivet}}
\newcommand{\hw}{\tsc{Herwig}}
\newcommand{\sh}{\tsc{Sherpa}}

\newcommand{\eqRef}[1]{eq.~\eqref{#1}\xspace}
\newcommand{\eqsRef}[1]{eqs.~\eqref{#1}\xspace}
\newcommand{\appRef}[1]{app.~\ref{#1}\xspace}
\newcommand{\AppRef}[1]{App.~\ref{#1}\xspace}
\newcommand{\figRef}[1]{fig.~\ref{#1}\xspace}
\newcommand{\figRefCite}[1]{fig.~{#1}}
\newcommand{\figsRef}[1]{figs.~\ref{#1}\xspace}
\newcommand{\FigRef}[1]{Fig.~\ref{#1}\xspace}
\newcommand{\FigsRef}[1]{Figs.~\ref{#1}\xspace}
\newcommand{\secRef}[1]{sec.~\ref{#1}\xspace}
\newcommand{\SecRef}[1]{Sec.~\ref{#1}\xspace}
\newcommand{\secsRef}[1]{secs.~\ref{#1}\xspace}

\newcommand{\tabRef}[1]{tab.~\ref{#1}\xspace}

\newcommand{\inst}[1]{$^{#1}$}
\renewcommand{\and}{, }

\newenvironment{emergency}[1]{
  \par
  \setlength{\emergencystretch}{#1}
}{
  \par
}

\graphicspath{{images/}}

\begin{document}

\vspace*{-1.8cm}
\begin{minipage}{\textwidth}
\flushright
CoEPP-MN-16-11\\
MCNET-16-13\\
SLAC-PUB-16529\\
NIKHEF-2016-020
\end{minipage}
\vskip1.25cm

{\Large\bf
\begin{center}
\vc\footnote{Virtual Numerical Collider with Interleaved Antennae: 
\url{http://vincia.hepforge.org}}
for Hadron Colliders
\end{center}}
\vskip5mm

\begin{center}{\large
N.~Fischer\inst{1}\and 
S.~Prestel\inst{2}\and
M.~Ritzmann\inst{3,4}\and
P.~Skands\inst{1}
}\end{center}
\vskip3mm
\begin{center}
\parbox{0.9\textwidth}{
\inst{1}: School of Physics and Astronomy, Monash University, Clayton VIC-3800, Australia\\
\inst{2}: SLAC National Accelerator Laboratory, Menlo Park, CA 94025, USA\\
\inst{3}: Nikhef, Theory Group, Science Park 105, 1098 XG Amsterdam, The Netherlands\\
\inst{4}: Institut de Physique Th\'eorique, CEA Saclay, F--91191 Gif--sur--Yvette cedex, France}
\end{center}
\vskip5mm

\begin{center}
\parbox{0.85\textwidth}{
\begin{center}
\textbf{Abstract}
\end{center}
We present the first public implementation of antenna-based QCD initial- and final-state 
showers. The shower kernels are $2\to 3$ antenna functions, which capture not only the 
collinear dynamics but also the leading soft (coherent) singularities of QCD matrix elements. 
We define the evolution measure to be inversely proportional to the leading poles, hence 
gluon emissions are evolved in a $p_\perp$ measure inversely proportional to the eikonal, 
while processes that only contain a single pole (e.g., $g\to q\bar{q}$) are evolved in 
virtuality. Non-ordered emissions are allowed, suppressed by an additional power of $1/Q^2$. 
Recoils and kinematics are governed by exact on-shell $2\to 3$ phase-space factorisations. 
This first implementation is limited to massless QCD partons and colourless resonances. 
Tree-level matrix-element corrections are included for QCD up to $\mathcal{O}(\alpha_s^4)$ 
(4 jets), and for Drell-Yan and Higgs production up to $\mathcal{O}(\alpha_s^3)$ ($V/H$ + 3 jets). 
The resulting algorithm has been made publicly available in \vc~2.0.
}
\end{center}

\newpage
\tableofcontents
\newpage

\clearpage\section{Introduction}

The basic differential equations governing
renormalization-group-improved (resummed) perturbation theory for initial-state
partons were derived in the 1970s~\cite{Gribov:1972ri,Dokshitzer:1977sg,Altarelli:1977zs}. 
The resulting DGLAP\footnote{Dokshitzer-Gribov-Lipatov-Altarelli-Parisi. We mourn 
the recent passing of Guido Altarelli (1941-2015), a founder of this field and a 
great inspirer.} equations 
remain a cornerstone of high-energy phenomenology, underpinning our understanding 
of perturbative corrections and scaling in many contexts, in particular the structure 
of QCD jets, parton distribution functions, and fragmentation functions. 

In the context of event generators~\cite{Buckley:2011ms},  DGLAP splitting kernels 
are still at the heart of several present-day parton showers~(including,
e.g.,~\cite{Bengtsson:1986hr,Marchesini:1987cf,Gieseke:2003rz,Sjostrand:2004ef,Krauss:2005re}). 
Although the DGLAP kernels themselves are derived in the collinear (small-angle) limit 
of QCD, which is dominated by radiation off a single hard parton, the destructive-interference
effects~\cite{Bassetto:1984ik} which dominate for wide-angle soft-gluon emission can 
also be approximately accounted for in this formalism; either by choosing the shower evolution 
variable to be a measure of energy times angle~\cite{Marchesini:1983bm} or by imposing 
a veto on non-angular-ordered  emissions~\cite{Bengtsson:1986et}. The resulting 
parton-shower algorithms are called \emph{coherent}. A third alternative, increasingly 
popular and also adopted in this work, is to replace the parton-based DGLAP picture 
by so-called colour dipoles~\cite{Gustafson:1987rq} (known as antennae in the context 
of fixed-order subtraction schemes~\cite{Azimov:1986sf,Kosower:1997zr,GehrmannDeRidder:2005cm,
GehrmannDeRidder:2005hi,GehrmannDeRidder:2005aw}\footnote{Note that a ``Catani-Seymour''
dipole~\cite{Catani:1996jh,Catani:1996vz} corresponds roughly speaking to a 
partial-fractioning of an antenna or Lund colour dipole into two pieces.}),
which incorporate all single-unresolved (i.e., both soft and collinear) limits 
explicitly. In the context of shower algorithms, this approach was originally 
pioneered by the \ari\ program~\cite{Gustafson:1987rq,Lonnblad:1992tz} and is now 
widely used~\cite{Nagy:2005aa,Giele:2007di,Dinsdale:2007mf,Schumann:2007mg,
Winter:2007ye,Platzer:2009jq,LopezVillarejo:2011ap,Ritzmann:2012ca,Hoche:2015sya}. 
We note that, the word ``coherence'' is used in different contexts, such
as angular ordering. When we use coherence in the context of antenna functions, 
we define it at the lowest level, as follows: antenna functions sum up the radiation
from two sides of the leading-$N_C$ dipole coherently, at the amplitude level;
see also Ref.~\cite{Platzer:2009jq}.

In addition, shower algorithms rely on several further improvements that go beyond the 
LO DGLAP picture, including: exact momentum conservation (related to the choice of recoil 
strategy), colour-flow tracing (in the leading-$N_C$ limit, related to coherence at both 
the perturbative and non-perturbative levels), and higher-order-improved scale choices 
(including the use of $\mu_R=p_\perp$ for gluon emissions and the so-called CMW scheme 
translation which applies in the soft limit~\cite{Amati:1980ch,Catani:1990rr}). Each 
of these are associated with ambiguities, with \secRef{sec:antShowers} containing the 
details of our choices and motivations. 

Finally, in the context of initial-state parton showers, the evolution from a high 
factorisation scale to a low one corresponds to an evolution in spacelike (negative) 
virtualities, ``backwards'' towards lower resolution. The correct equations for 
backwards parton-shower evolution were first derived by 
Sj\"ostrand~\cite{Sjostrand:1985xi}; in particular it is essential to multiply the 
evolution kernels by ratios of parton distribution functions (PDFs), to recover the 
correct low-scale structure of the incoming beam hadrons.
We shall use a generalisation of backwards evolution to the case of simultaneous 
evolution of the two incoming-hadron PDFs, similar to that presented 
in~\cite{Winter:2007ye}.

The merits of different shower algorithms is a frequent topic of debate, with individual approaches differing by which compromises are made and by the effective higher-order terms that are generated. We emphasise the  following three attractive properties of antenna showers: 
\begin{itemize}
\item They are intrinsically coherent, in the sense that the correct eikonal structure is generated for each single-unresolved soft gluon, up to corrections suppressed by at least $1/N_C^2$.  Especially for initial-final antennae, where gluon emission off initial- and final-state legs interfere, has some challenges\footnote{Older parton shower models often treat initial-state (ISR) and final-state (FSR) evolution in disjoint sequences. In this case, it is challenging to ensure that FSR evolution from the enlarged and changed parton ensemble after ISR evolution recovers the coherent features. Implementations of a combined simultaneous evolution chain for ISR and FSR may also be challenging. The current $p_\perp$-ordered showers in \py~8~\cite{Sjostrand:2014zea,Corke:2010yf,Sjostrand:2004ef} do, for example, not account for the coherence structure of the hardest gluon emission in $t\bar{t}$ events~\cite{Skands:2012mm}. In contrast, we suspect that due to the fact that physical output states (parsed through hadronization) are only constructed at the end of the evolution, the angular-ordered algorithms of \hw~\cite{Marchesini:1987cf} and \hw++~\cite{Gieseke:2003rz} produce coherent sequences of emission angles in IF configurations correctly. This assessment relies on the assumption that the algorithms ensure that the angular constraints on final-state emission variables are unchanged by the ISR shower evolution, and vice versa.}. In the final-final case which was already testable with previous \vc\ versions, a recent OPAL study of 4-jet events~\cite{Fischer:2015pqa} found good agreement between \vc\ and several recently proposed coherence-sensitive observables~\cite{Fischer:2014bja}.
\item They are extremely simple, relying on local and universal $2\to 3$ phase space maps which represent an exact factorization of the $n$-particle phase spaces not only in the soft and collinear regions but over all of phase space. This makes for highly tractable analytical expansions on which our accompanying matrix-element correction formalism is based~\cite{Giele:2011cb}. The pure shower is in some sense merely a skeleton for generating the leading singularities, with corrections for both hard and soft emissions regarded as an intrinsic part of the formalism, restoring the emission patterns to at least LO accuracy up to the matched orders.
\item There is a close correspondence with the antenna-subtraction formalism used in fixed-order calculations~\cite{GehrmannDeRidder:2005cm,GehrmannDeRidder:2005hi,GehrmannDeRidder:2005aw}, which is based on the same subtraction terms and phase-space maps. 
This property was already utilised in \cite{Hartgring:2013jma} to implement a simple and highly efficient procedure for NLO corrections to gluon emission off a $q\bar{q}$ antenna.  Highly non-trivial fixed-order results which have recently been obtained within the antenna formalism include NNLO calculations for $Z+\mrm{jet}$~\cite{Ridder:2016rzm}, $H+\mrm{jet}$~\cite{Chen:2016vqn} (for $m_t\to\infty$), $gg\to gg$~\cite{Currie:2013dwa}, and leading-colour $q\bar{q}\to t\bar{t}$~\cite{Abelof:2015lna} production at hadron colliders. While it is (far) beyond the scope of the present work to connect directly with these calculations, their feasibility is encouraging to us, and provides a strong motivation for future developments of the antenna-shower formalism.
\end{itemize}

The aim with this work is to present the first full-fledged and publicly available antenna shower for hadron colliders, extending from previous work on final-state antenna showers developed in \cite{Giele:2007di,Giele:2011cb} and building on the proof-of-concept studies for hadronic initial states reported in \cite{Ritzmann:2012ca,Fischer:2016ryl}. The model is implemented in -- and defines -- version 2.0 of the \vc\ plug-in to the \py~8 event generator~\cite{Sjostrand:2014zea}. This article is also intended to serve as the first physics manual for \vc~2.0. It is accompanied by a more technical HTML User Reference documenting each of the user-modifiable parameters and switches at the technical level~\cite{VinciaUserReference} and an author's compendium documenting more detailed algorithmic aspects~\cite{VinciaCompendium}; both of these auxiliary documents are included with the code package, which is publicly available via the HepForge repository at \url{vincia.hepforge.org}.

In \secRef{sec:antShowers} we introduce the basic antenna-shower formalism, including 
our notation and conventions. We mainly focus on initial-initial and initial-final
configurations and summarize finial-final configurations only briefly, as a more
extensive description is available in \cite{Giele:2007di,Giele:2011cb}.
Our conventions for colour flow are specified in \secRef{sec:colour}. These are intended to maximise information on coherence while simultaneously generating a state in which all colour tags obey the index-based treatment of subleading colour correlations proposed 
in~\cite{Bierlich:2014xba,Christiansen:2015yqa}. By assigning these indices after each branching and tracing them through the shower evolution, rather than statistically assigning them at the end of the evolution as was done in~\cite{Christiansen:2015yqa}, we remove the risk of accidentally generating unphysical colour flows\footnote{E.g., in our treatment the case illustrated by~\cite[fig.~19b]{Christiansen:2015yqa} cannot occur: $Z\to qgg\bar{q}$ with the two gluons collinear to each other, non-collinear to any of the quarks, and in a singlet state.}. We therefore believe the procedure proposed here represents an improvement on the one in~\cite{Christiansen:2015yqa}.
The extension of \vc's automated treatment of perturbative shower uncertainties to hadron collisions 
is documented in \secRef{sec:uncertainties}.

In \secRef{sec:MECs}, we present the extension of the GKS\footnote{Giele-Kosower-Skands~\cite{Giele:2011cb}.} matrix-element-correction (MEC)
formalism~\cite{Giele:2011cb} to initial-state partons, starting with the case of a basic process accompanied by one or more jets whose scales are
nominally harder than that of the basic process in \secRef{sec:hardJets}.
In \secRef{sec:validations}, we present some basic numerical comparisons between tree-level 
matrix elements and our shower formalism expanded to the equivalent level (i.e., 
setting all Sudakov factors and coupling constants to unity), to validate that 
combinations of $2\to3$ antenna branchings do produce a reasonable agreement with the 
full $n$-parton matrix elements. We discuss our extension of 
``smooth ordering''~\cite{Giele:2011cb} to reach non-ordered parts of phase space
in \secRef{sec:Impr24Branch}, again focusing on the initial-state context.  \SecRef{sec:hardJetsQCD} summarises the application of smooth ordering to the specific case of hard jets in QCD processes.
In \secRef{sec:MG4MECs} we extend and document \vc's existing use of 
\mg~4~\cite{Alwall:2007st} matrix elements.

The set of numerical parameters which define the default ``tune'' of \vc~2.0 is documented in \secRef{sec:results}, including our preferred convention choice for $\alpha_s$, the most important parameter of any shower algorithm. A set of comparisons to a selection of salient experimentally measured distributions for hadronic $Z$ decays, Drell-Yan, and QCD jet production are included to document and validate the performance of the shower algorithm with these parameters.

Finally, in \secRef{sec:conclusions}, we summarise and give an outlook. Additional material, as referred to in the text, is collected in the Appendices.

\section{\vc's Antenna Showers \label{sec:antShowers}}
A QCD antenna represents a colour-connected parton pair which undergoes a (coherent) $2\to3$ 
branching process~\cite{Azimov:1986sf,Gustafson:1987rq,Kosower:1997zr,GehrmannDeRidder:2005cm,
Dokshitzer:2008ia}. 
In contrast to conventional shower models (including both DGLAP and Catani-Seymour dipole ones) 
which single out one parton as the ``emitter'' with one (or more) other partons acting as
``recoiler(s)'', the antenna formalism treats the two pre-branching ``parent'' partons as a 
single entity, with a single radiation kernel (an antenna function) driving the amount of 
radiation and a single ``kinematics map'' governing the exact relation between the pre-branching 
and post-branching momenta. Formally, the antenna function represents the approximate 
(to leading order in the vanishing invariant(s)) factorisation 
between the pre- and post-branching squared amplitudes, while the kinematics 
map encapsulates the exact on-shell factorisation of the $(n+1)$-parton phase space into 
the $n$-parton one and the $(2\to 3)$ antenna phase space. 

Note that for branching processes involving flavour changes of the parent partons,
such as $g\to q\bar q$, 
a distinction between ``emitter'' and ``recoiler'' and thus a treatment independent
of the above description is possible. However, this is not compulsory and we are 
therefore still using the same $(2\to 3)$ antenna phase space and kinematics
map as in the case of gluon emission. Moreover, applying a $2\to3$ branching amounts 
to using the lowest number of involved partons which admit an on-shell to on-shell mapping.

In this section we briefly review the notation
and conventions that will be used throughout this paper (\secRef{sec:conventions}), 
followed by definitions for all of the phase-space convolutions or factorisations
respectively, antenna functions, and evolution variables on which \vc's treatment 
of initial-initial, initial-final, and final-final configurations are based 
(\secsRef{sec:antDefsII}, \ref{sec:antDefsIF}, and \ref{sec:antDefsFF}). The expressions for final-final configurations are unchanged relative to those
in~\cite{Giele:2007di,Giele:2011cb}, with the default antenna functions chosen to be those of~\cite{Larkoski:2013yi} averaged over helicities. 
Some further details on the explicit kinematics constructions
are collected in \AppRef{app:showerDetails}. The explicit form of the shower-generation 
algorithm is presented in \secRef{sec:generator}. Finally, we round off in 
\secRef{sec:limitations} with comments on some features of earlier incarnations 
of \vc\ which have not (yet) been made available in \vc~2.0. 

\subsection{Notation and Conventions \label{sec:conventions}}
We use the following notation for labelling partons: capital letters for pre-branching 
(parent) and lower-case letters for post-branching (daughter) partons. We label incoming
partons with the first letters of the alphabet, $a$, $b$, and outgoing ones with
$i$, $j$, $k$. Thus, for example, a branching occurring in an initial-final antenna (a colour antenna spanned between an initial-state parton and a final-state one)  would be labeled $ AK \to a jk$. This is consistent with the conventions used in the most recent \vc\ papers~\cite{Giele:2011cb,Ritzmann:2012ca}\footnote{The earliest \vc\ paper on final-final antennae~\cite{Giele:2007di} used an alternative convention: $\hat{a} + \hat{b} \to a + r + b$.}. The recoiler or recoiling system will be denoted by $R$ and $r$ respectively (compared with $R'$ and $R$ in \cite{Ritzmann:2012ca}).

We restrict our discussion to massless partons and denote the Lorentz-invariant momentum four-product between two  partons 1 and 2 by 
\begin{equation}
s_{12}~\equiv~2p_1^{\mu}p_{2\mu}~=~(p_1+p_2)^2~,
\end{equation}
which is always positive regardless of whether the partons involved are in the initial or final state. 
Momentum conservation then yields:
\begin{eqnarray}
\mrm{FF} & : & s_{IK}~=~s_{ij} + s_{jk} + s_{ik}~,\\
\mrm{IF} & : & s_{AK}~=~s_{ak} + s_{aj} - s_{jk}~, \\
\mrm{II} & : & s_{AB}~=~s_{ab} - s_{aj} - s_{jb}~,
\end{eqnarray}
for final-final (FF), initial-final (IF), and initial-initial (II) branchings respectively. 

The evolution variable, which we denote $t$, is evaluated on the post-branching partons, 
hence, e.g., $t_{\mrm{FF}}=t(s_{ij},s_{jk})$. It serves as a dynamic factorisation scale for the shower, separating resolved from unresolved regions. As 
such, it must vanish for singular configurations. Generally, we define the evolution variable for each branching type to vanish 
with the same power of the momentum invariants as the leading poles of the corresponding antenna functions, see below. The complementary phase-space variable will be denoted $\zeta$.

\paragraph{Colour Factors $\mc C$}
We use the following convention: for gluon emission the colour factors are $\mc C=C_A=3$ 
for gluon-only antennae, $\mc C=2\,C_F=8/3$ for quark-only antennae, and the mean, 
$\mc C=(C_A+2C_F)/2$, for quark-gluon antennae. For gluon splitting the colour factor
is $\mc C=2\,T_R=1$. Note that symmetry factors, taking into account that gluons
contribute to two antennae, are included in the antenna functions.

\paragraph{Shower Basics}
A shower algorithm is based on the probability that no branching occurs between two 
scales $t_{n}$ and $t_{n+1}$, with $t_{n}>t_{n+1}$. (For an introduction to conventional 
showers, see, e.g., \cite[Chp.40]{Agashe:2014kda} or \cite{Buckley:2011ms}. For antenna 
showers more specifically, see~\cite{Giele:2011cb,Hartgring:2013jma}.) 
In the case of initial-state radiation in the 
antenna picture the no-emission probability is
\begin{equation}
  \Pi_n(t_{n},t_{n+1}) 
  ~=~\exp\left( -\sum_{i \in \{n\to n+1\}}\int_{t_{n+1}}^{t_{n}} \!\!\!
  \d\Phi_\t{ant}\, 4\pi\alpha_s(t)\,
  \mc C\,\bar a_i(t,\zeta)\,R_{\t{pdf}\,i} \right) ~, 
\label{eq:NoEmiProb}
\end{equation}
with the colour- and coupling-stripped antenna function $\bar a$ and
the (double) ratio of PDFs,
\begin{equation}
  R_\t{pdf}~=~\frac{f_a(x_a,t)}{f_A(x_A,t)}\frac{f_b(x_b,t)}{f_B(x_B,t)}~.
\end{equation}
Note that although the integral over $\d\Phi_\t{ant}$ in \eqRef{eq:NoEmiProb} is 3-dimensional, 
we only explicitly wrote down the boundaries in the evolution variable $t$, with 
integration over the complementary invariant, $\zeta$, and over the azimuth angle, $\phi$, 
implied. Given specific choices for $t$ and $\zeta$ as functions of the phase-space invariants, the boundaries of the $\zeta$ integral are derived from energy-momentum conservation, as usual for shower algorithms 
(see, \cite{Gieseke:2004tc,Giele:2007di,Buckley:2011ms,VinciaCompendium}). This generates modifications to the LL structure 
which --- since $(E,p)$ conservation is a genuine physical effect --- is expected to 
improve the shower approximation at the subleading level. (We are not aware of a rigorous proof of this statement, however.) 

The sum in \eqRef{eq:NoEmiProb} runs over all possible $(n+1)$-parton states that can be created 
from the $n$-parton state, and will be implicit from here on. 
$\d\Phi_\t{ant}$ is the antenna phase space, providing a mapping from two 
to three on-shell partons while preserving energy and momentum. The 
specific form for the two configurations, initial-initial and initial-final 
are defined below, along with the specific forms of the evolution variable.

We define the Sudakov factor as
\begin{equation}
  \Delta_n(t_{n},t_{n+1}) ~=~ \exp\left( -\sum_{i \in \{n\to n+1\}}
  \int_{t_{n+1}}^{t_{n}} \!\!\! \d\Phi_\t{ant}\,\frac{x_A\,x_B}{x_a\,x_b}\,
  4\pi\alpha_s(t)\,\mc C\,\bar a_i(t,\zeta)\right) ~.
\end{equation}
This object does not depend on parton distribution functions or other non-perturbative input, and may thus be regarded as a purely perturbative object. Following the arguments of \cite{Hoche:2015sya}, we \emph{define} the no-emission probability in terms of the Sudakov factor, as follows (generalised from~\cite{Ellis:1991qj}):
\begin{equation}
  \Pi_n(t_{n},t_{n+1})=\frac{f_A(x_A,t_{n+1})}{f_A(x_A,t_{n})}
  \frac{f_B(x_B,t_{n+1})}{f_B(x_B,t_{n})}\Delta_n(t_{n},t_{n+1})~.
  \label{eq:PiDelta}
\end{equation}
This in turn implicitly defines the evolution equation for the antenna shower, which, as shown in \cite{Hoche:2015sya}, is consistent with the DGLAP equation, provided the antenna functions used in \vc\ have the correct (AP-kernel) behaviour in close to $z=1$, where $z$ is an energy-sharing variable\footnote{The resulting evolution equation will contain objects that are very close to the unintegrated parton densities of \cite{Kimber:1999xc,Kimber:2001sc}.}. This is shown in appendix \ref{app:collLimit} in which the collinear limits of all antenna functions used in this work are given. Note that a similar strategy of using \eqRef{eq:PiDelta} as a definition was also used when defining perturbative states in \cite{Nagy:2009vg}. For final-final configurations, \eqRef{eq:PiDelta} simplifies to $\Pi_n(t_{n},t_{n+1})=\Delta_n(t_{n},t_{n+1})$.

\subsection{Initial-Initial Configurations \label{sec:antDefsII}}
\begin{figure}[t]
\centering
Initial-Initial Antenna Branching\\
\begin{tabular}{lcl}
\raisebox{0.85cm}{\includegraphics*[width=.4\textwidth]{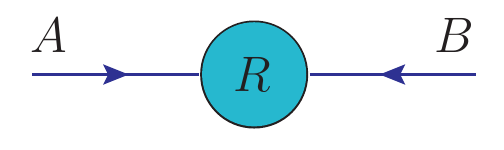}} & \raisebox{1.6cm}{$\huge\to$}& 
\includegraphics*[width=.4\textwidth]{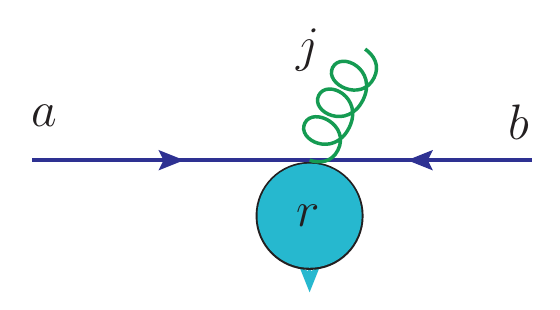}\\[-5mm]
\emph{a)} Before & & \emph{b)} After 
\end{tabular}
\caption{Illustration of pre-branching (left) and post-branching (right) on-shell momenta, for an initial-initial antenna branching, emphasising the transverse kick imparted to the hard system, $R$, which consists of all particles produced in the collision $A+B \to R$. The hard system is treated as a rigid body (i.e., any internal invariants are not modified) by the branching. It is subjected to a single overall Lorentz transformation, $R\to r$, equivalent to a frame reinterpretation required to orient the new incoming partons along the $z$ axis. Note that we define our kinematics maps to preserve not only the invariant mass but also the rapidity of the recoiling system: $m_r^2 = m_R^2$ and $y_r = y_R$, cf.~\appRef{app:IIkinematics}. 
\label{fig:IIkinematics}}
\end{figure}

We denote the pre- and post-branching partons participating in an initial-initial branching by $AB\to abj$ and the (system of) particles produced by the collision by $R\to r$, cf.~the illustrations in \figRef{fig:IIkinematics}.
In the following, we specify the phase-space convolution, antenna functions, evolution
variables and the resulting no-emission probability.

\paragraph{Phase Space}
The phase-space convolution reads
\begin{multline}
  \int \frac{\d x_a}{x_a}\,\Theta(1-x_a)\,\frac{\d x_b}{x_b}\,\Theta(1-x_b)\,
  \d\Phi_2(p_a,p_b\to p_j,p_r) = \\
  \int \frac{\d x_A}{x_A}\,\Theta(1-x_A)\,\frac{\d x_B}{x_B}\,\Theta(1-x_B)
  \,\d\Phi_1(p_A,p_B\to p_R)\,\d\Phi_\t{ant}^\t{II}
\end{multline}
with the antenna phase space
\begin{align}
  \d\Phi_\t{ant}^\t{II} = \frac1{16\pi^2}\,\frac{\sAB}{\sab^2}\,
  \Theta(x_a-x_A)\,\Theta(x_b-x_B)\,
  \d\saj\,\d\sjb\,\frac{\d\phi}{2\pi}~.
\end{align}
See \appRef{app:IIkinematics} for the explicit construction of the post-branching momenta. 

\paragraph{Antenna functions}
The gluon emission antenna functions are
\begin{align}
  \bar a_{q\bar q\,g}^\t{II}=\bar a(a_q,b_{\bar q},j_g) =&~
  \frac1\sAB\left(2\,\frac{\sab\sAB}{\saj\sjb} + \frac\sjb\saj + \frac\saj\sjb\right)~, \label{eq:IIemit1}\\
  \bar a_{gg\,g}^\t{II}=\bar a(a_g,b_g,j_g) =&~
  \frac1\sAB\left(2\,\frac{\sab\sAB}{\saj\sjb} + 
  2\,\frac\sjb\saj\frac\sab\sAB+2\,\frac\sjb\saj\frac\sAB{\sab+\saj}\right.\nonumber \\
  &~\phantom{\frac1\sAB\left(2\,\frac{\sab\sAB}{\saj\sjb}\right.} +
  \left.2\,\frac\saj\sjb\frac\sab\sAB+2\,\frac\saj\sjb\frac\sAB{\sab+\sjb}\right)~, \\
  \bar a_{qg\,g}^\t{II}=\bar a(a_q,b_g,j_g) =&~ \frac1\sAB\left(2\,\frac{\sab\sAB}{\saj\sjb} +
  \frac\sjb\saj + 2\,\frac\saj\sjb\frac\sab\sAB+2\,\frac\saj\sjb\frac\sAB{\sab+\sjb}\right)~. \label{eq:IIemit3}\\
  \intertext{The antenna function for a gluon evolving backwards to a quark
  (and similar to an antiquark) is}
  \bar a_{qx\,q}^\t{II}=\bar a(a_q,b_x,j_q) =&~ 
  \frac1{2\saj}\frac{\sjb^2+\sab^2}{\sAB^2}~, \label{eq:IIconv1}\\
  \intertext{and for a quark evolving backwards into a gluon}
  \bar a_{gx\,\bar q}^\t{II}=\bar a(a_g,b_x,j_{\bar q}) =&~ 
  \frac1{\sAB}\left(-2\,\frac{\sjb\sAB}{\saj(\sab-\saj)}+\frac\sab\saj\right)~.\label{eq:IIconv2}
\end{align}
In \appRef{app:IIdglapLimit} we show that the antenna functions correctly
reproduce the DGLAP splitting kernels in the collinear limit.

\paragraph{Evolution Variables}
We evolve gluon emission in the physical transverse momentum of the emission 
(relative to the $p_a$--$p_b$--axis),
\begin{equation}
  t_\t{II}^\t{emit} = p_{\perp\,\t{II}}^2=\frac{\saj\sjb}\sab~,
  \label{eq:tIIemit}
\end{equation}
which exhibits the same ``antenna-like'' $a\leftrightarrow b$ symmetry as the leading 
(double) poles of the corresponding antenna functions, \eqsRef{eq:IIemit1}--\eqref{eq:IIemit3} 
above. The upper phase-space limit for this variable is 
$p_{\perp\,\t{II}}^2\le (s-\sAB)^2/(4\,s)$, where $s$ denotes the hadronic 
centre-of-mass energy squared.
\begin{figure}[tp]
\centering
\begin{tabular}{ll}
{\it a)} Phase Space for Initial-Initial Antennae &
{\it b)} Phase Space for Initial-Final Antennae \\
\includegraphics[width=.45\textwidth]{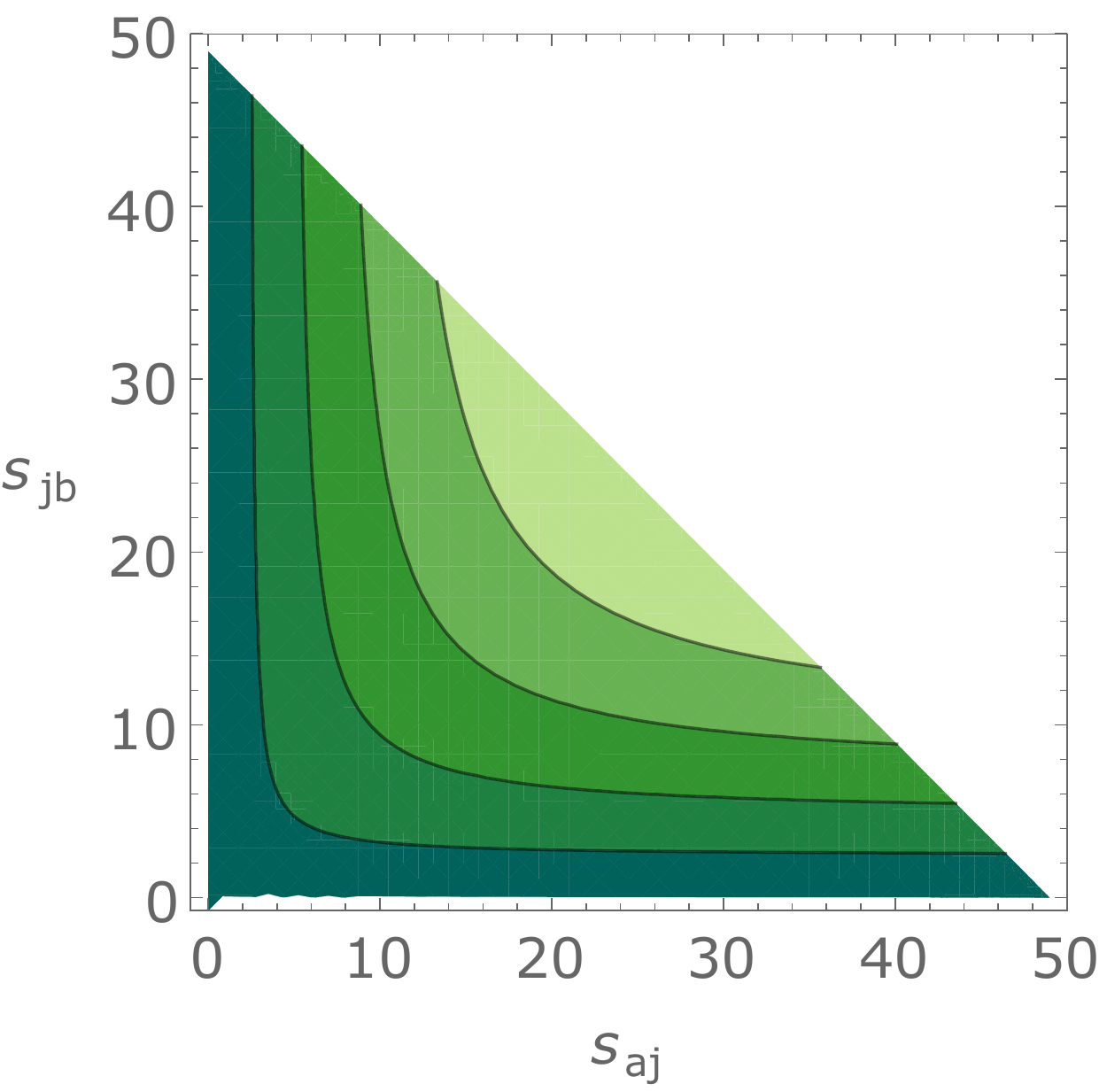} &
\includegraphics[width=.45\textwidth]{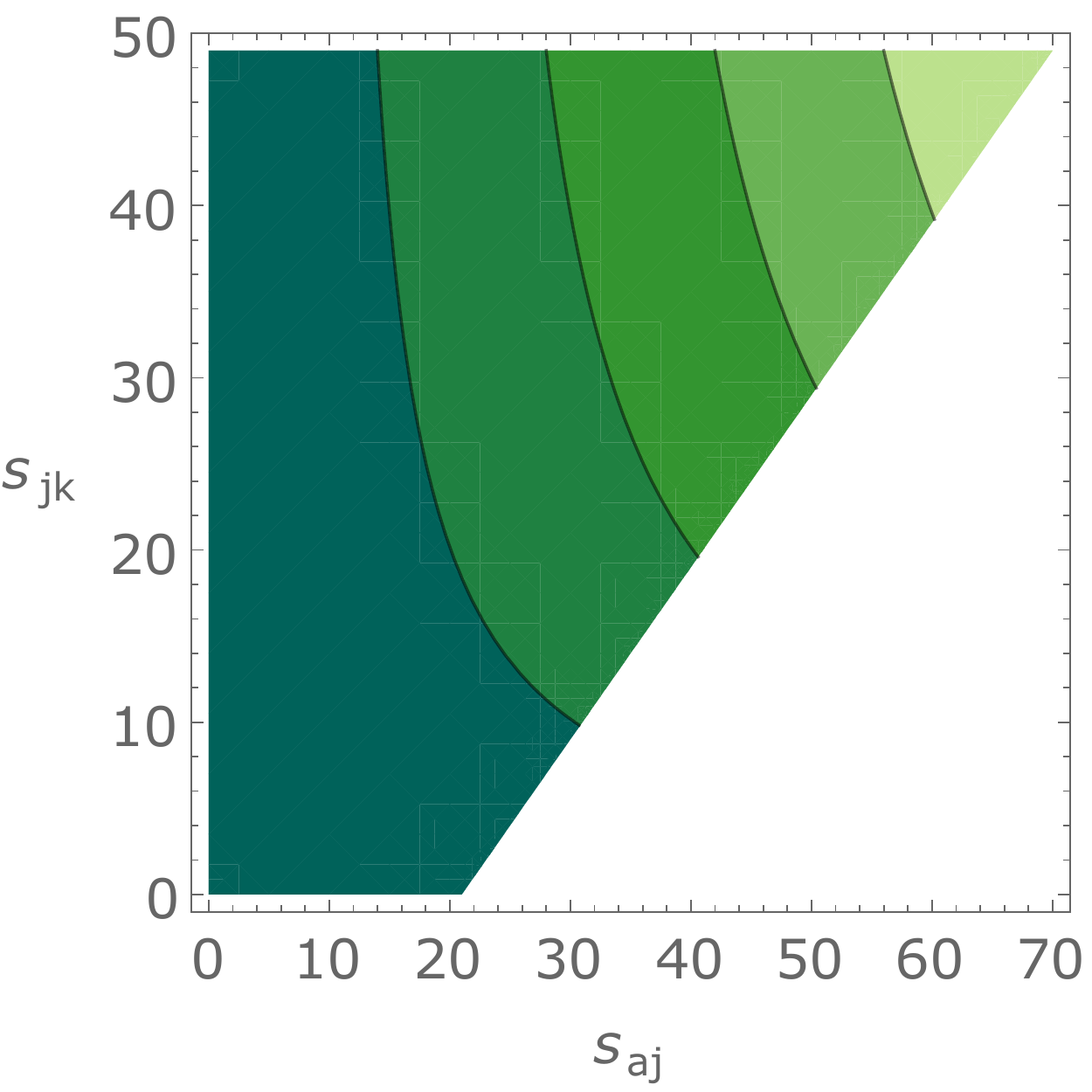}
\end{tabular}
\caption{\label{fig:phasespace} Contours of constant gluon-emission evolution
variable for {\it a)} initial-initial and {\it b)} initial-final configurations.
For {\it a)} the recoiler is chosen to be a Higgs boson, $\sAB=m_H^2$, and for
{\it b)} $\sAK=10500\,\t{GeV}^2$ and $x_A=0.3$. For both cases, the total hadronic 
$\sqrt s=7\,\t{TeV}$.}
\end{figure}
\FigRef{fig:phasespace}\,{\it a)} shows constant contours of $p_{\perp\,\t{II}}^2$, 
as a function of the two branching invariants $\saj$ and $\sjb$. As the phase
space is symmetric in $\saj$ and $\sjb$ it has a triangular shape whose hypotenuse is 
defined by the upper phase-space bound $\sAB+\saj+\sjb\le s$. 
For branchings with flavour changes in the initial state (gluon evolving backwards to a quark or vice versa) for which the antenna functions only contain single poles, cf.~\eqsRef{eq:IIconv1}--\eqref{eq:IIconv2} above, 
we use the corresponding invariant, $\saj$ or $\sjb$ respectively, 
\begin{equation}
  t_\t{II}^\t{conv} = Q^2_\t{II} = \left\{ \begin{array}{cl}
  \saj & \mbox{for $a$ converting to/from a gluon}\\
  \sjb & \mbox{for $b$ converting to/from a gluon}
  \end{array}
  \right.~,
\end{equation}
where the phase-space limit is $s_{xj}\le s-\sAB$. Note that the conversion measure is
equivalent to the Mandelstam $|t|$ variable for the relevant diagrams. Since only 
one parton can convert at a time --- either $A$ \emph{or} $B$ --- these diagrams are 
unique, with no interferences, as is also reflected by the corresponding antenna 
functions containing only single (collinear) poles.

\paragraph{No-emission Probability} With the definitions given
above \eqref{eq:NoEmiProb} for initial-initial configurations reads
\begin{multline}
  \Pi_n(t_{n},t_{n+1})~=~\exp\left(-
  \int_{{\saj}_{n+1}}^{{\saj}_{n}}\!\!\!\!\!\!\!\!\!\d\saj\,\Theta(x_a-x_A)\,
  \int_{{\sjb}_{n+1}}^{{\sjb}_{n}}\!\!\!\!\!\!\!\!\!\d\sjb\,\Theta(x_b-x_B)\,
  \int_0^{2\pi}\!\frac{\d\phi}{2\pi}\,
  \right. \\ \left.
  \frac{\alpha_s(t)}{4\pi}\,\frac{\sAB}{\sab^2}\,\mc C\,
  \bar a(\saj,\sjb,\sAB)\,
  \frac{f_a(x_a,t)}{f_A(x_A,t)}\frac{f_b(x_b,t)}{f_B(x_B,t)}
  \right) ~,
\end{multline}
with $t=t_\t{II}(\saj,\sjb,\sab)$. The subscript of the $\saj$ and $\sjb$ integration
limits indicates the association with the branching scales $t_n$ and $t_{n+1}$
respectively.

\subsection{Initial-Final Configurations\label{sec:antDefsIF}}
In traditional (DGLAP-based) parton-shower formulations, the radiation emitted by a colour line flowing from the initial to the final state is handled by two separate algorithms, one for ISR and one for FSR. Coherence can still be imposed by letting these algorithms share information on the angles between colour-connected partons and limiting radiation to the corresponding coherent radiation cones. But even so, several subtleties can arise in the context of specific processes or corners of phase space. Examples of problems encountered in the literature involving \py's $p_\perp$-ordered showers include how radiation in dipoles stretched to the beam remnant is treated~\cite{Banfi:2010xy}, whether the combined ISR+FSR evolution is interleaved or not~\cite{Dasgupta:2013ihk} and whether/how coherence is imposed on the first emission~\cite{Skands:2012mm}. 

\begin{figure}[t]
\centering
Initial-Final Antenna Branching\\
\begin{tabular}{lcl}
{\includegraphics*[width=.4\textwidth]{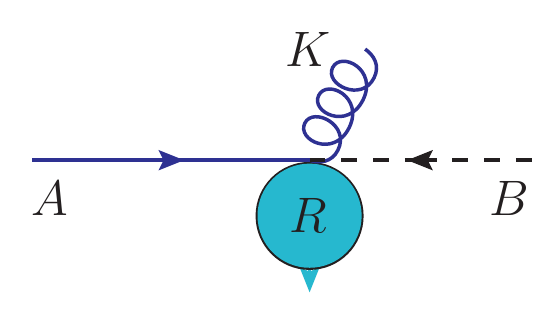}} & \raisebox{1.6cm}{$\huge\to$}& 
\raisebox{2mm}{\includegraphics*[width=.4\textwidth]{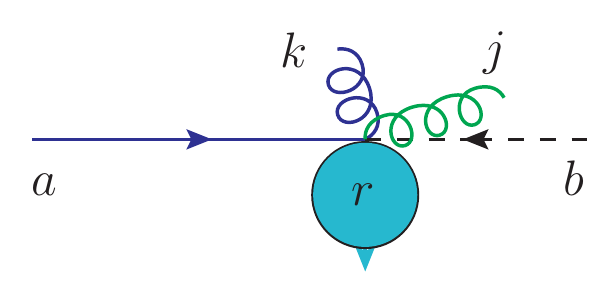}}\\[-5mm]
\emph{a)} Before & & \emph{b)} After 
\end{tabular}
\caption{Illustration of pre-branching (left) and post-branching (right) on-shell momenta, for an initial-final (IF) antenna branching, emphasising that the momenta of the spectators $B$ and $R$ are unchanged: $p_b=p_B$ and $p_r = p_R$, cf.~\appRef{app:IFkinematics}.
\label{fig:IFkinematics}}
\end{figure}

In the context of antenna showers, the radiation off initial-final (IF) colour flows is 
generated by IF antennae, which are coherent ab initio. We therefore expect the treatment 
of wide-angle radiation to be more reliable and plagued by fewer subtleties. The main 
issue one faces instead is technical. Denoting the pre- and post-branching 
partons participating in an IF branching by $AK\to akj$, the choice of kinematics map 
specifying the global orientation of the $akj$ system with respect to the $AK$ one is 
equivalent to specifying the Lorentz transformation that connects the pre-branching frame, 
in which $A$ is incoming along the $z$ axis with momentum fraction $x_A$, to the 
post-branching one, in which $a$ is incoming along the $z$ axis with momentum fraction 
$x_a$. For a general choice of kinematics map, this can result in boosted angles entering 
in the relation between $x_a$ and the branching invariants, producing highly nontrivial 
expressions, and the phase-space boundaries can likewise become very complicated. To retain 
a simple structure for this first implementation, and since we anyway intend our shower 
as a baseline to be improved upon with matrix-element corrections, the algorithm we present 
in this paper is based on the simplest possible kinematics map, in which momentum is 
conserved locally within the antenna, $p_a - p_j - p_k = p_A - p_K$. This implies that 
the momentum of the hard system, $R$, is left unchanged, meaning IF branchings
doe not produce a transverse recoil in the hard system. This is indicated by the unchanged 
momenta of the other incoming parton $B$ and the final-state $R$, cf.~the illustrations in
\figRef{fig:IFkinematics}. 
Though we do perceive of this as artificial (e.g., a parton emitting near-collinear radiation will only generate recoil to the hard system if its colour partner happens to be in the initial state) and presumably a weak point of the physics generated by the IF algorithm~\cite{Parisi:1979se}, it is nevertheless worth pointing out that:
\begin{itemize}
\item Even in cases where there is only one original II antenna (as e.g., in Drell-Yan), it is not true that recoil can only be generated by the first emission. In particular, if the first branching is a (sea) quark  evolving backwards to a gluon, that gluon will participate in a new II antenna, which will generate added recoil according to the above prescription for the II case. For cases with more than one II antenna (e.g., $gg\to H$), the number of possible $p_\perp$ kicks of course increases accordingly. 
\item In \vc\, matrix-element corrections (MECs) are regarded as an integral component of the evolution. Up to the first several orders (typically three powers of $\alpha_s$) we therefore expect to be able to apply MECs which will change the relative weighting of branching events in phase space, emphasising those regions which would have benefited most from large recoils and de-emphasising complementary ones. Matrix-element corrections will ensure that the emission pattern is correctly described with fixed-order precision. The all-orders resummation of non-LL configurations (e.g., configurations
with balancing soft emissions), is however not formally improved, meaning a residual effect of the recoil strategy remains. Note that the MECs will nonetheless attribute a sensible lowest-order weight to hard configurations that are usually out of reach of strongly-ordered parton showers.
\item As already pointed out above and illustrated by~\cite[figs.~3 \& 4]{Ritzmann:2012ca}, the IF radiation patterns remain \emph{coherent}, in the sense that large colour opening angles are a prerequisite for wide-angle radiation. This is a nontrivial and important property of the antenna-shower formalism, which is preserved independently of the recoil strategy.
\end{itemize}
Given these arguments, we regard the maintained simplicity of the resulting formalism as the primary goal at this stage, which has the added benefit of producing faster, more efficient algorithms. For completeness, we note that the strategy adopted in \cite{Platzer:2009jq} for ``finite recoils'' would not be applicable to \vc\ since it does not cover all of phase space and hence could not be used as the starting point for our matrix-element correction strategy. 

In the following, we describe the phase-space convolution, antenna functions and 
resulting no-emission probability used for initial-final evolution.

\paragraph{Phase Space}
The phase-space convolution reads
\begin{multline}
  \int \frac{\d x_a}{x_a}\Theta(1-x_a)\,\frac{\d x_B}{x_B}\,\Theta(1-x_B)\,
  \d\Phi_3(p_a,p_B\to p_R,p_j,p_k) = \\
  \int \frac{\d x_A}{x_A}\Theta(1-x_A)\,\frac{\d x_B}{x_B}\,\Theta(1-x_B)\,
  \d\Phi_2(p_A,p_B\to p_R,p_K)\,\d\Phi_\t{ant}^\t{IF}
\end{multline}
with the antenna phase space
\begin{align}
  \d\Phi_\t{ant}^\t{IF} = \frac1{16\pi^2}\,\frac{\sAK}{(\sAK+\sjk)^2}\,
  \Theta(x_a-x_A)\,
  \d\saj\,\d\sjk\,\frac{\d\phi}{2\pi}~.
\end{align}
See \appRef{app:IFkinematics} for the explicit construction of the post-branching momenta.
\paragraph{Antenna functions}
The gluon emission antenna functions are
\begin{align}
  \bar a_{qq\,g}^\t{IF}=\bar a(a_q,k_q,j_g) &= \frac1\sAK\left(2\,\frac{\sak\sAK}{\saj\sjk} +
  \frac\sjk\saj + \frac\saj\sjk\right)~, \\
  \bar a_{gg\,g}^\t{IF}=\bar a(a_g,k_g,j_g) &= \frac1\sAK\left(2\,\frac{\sak\sAK}{\saj\sjk} +
  2\,\frac\sjk\saj\frac\sak\sAK + 2\,\frac{\sjk\sAK}{\saj(\sAK+\sjk)} + 
  \frac\saj\sjk\frac\sak\sAK\right)~, \\
  \bar a_{qg\,g}^\t{IF}=\bar a(a_q,k_g,j_g) &= \frac1\sAK\left(2\,\frac{\sak\sAK}{\saj\sjk} +
  \frac\sjk\saj + \frac\saj\sjk\frac\sak\sAK\right)~, \\
  \bar a_{gq\,g}^\t{IF}=\bar a(a_g,k_q,j_g) &= \frac1\sAK\left(2\,\frac{\sak\sAK}{\saj\sjk} +
  2\,\frac\sjk\saj\frac\sak\sAK + 2\,\frac{\sjk\sAK}{\saj(\sAK+\sjk)} + 
  \frac\saj\sjk\right)~. \\
  \intertext{The antenna function for a gluon evolving backwards to a quark
  (and similar to an antiquark) is}
  \bar a_{qx\,q}^\t{IF}=\bar a(a_q,k_x,j_q) &= 
  \frac1{2\saj}\frac{\sjk^2+\sak^2}{\sAK^2}~, \\
  \intertext{for a quark evolving backwards to a gluon}
  \bar a_{gx\,\bar q}^\t{IF}=\bar a(a_g,k_x,j_{\bar q}) &= 
  \frac1{\sAK}\left(-2\,\frac{\sjk(\sAK-\saj)}{\saj(\sAK+\sjk)}+\frac\sak\saj\right)~, \\
  \intertext{and for a final-state gluon splitting}
  \bar a_{xq\,\bar q}^\t{IF}=\bar a(a_x,k_q,j_{\bar q}) &= 
  \frac1{2\sjk}\frac{\saj^2+\sak^2}{\sAK^2}~.
\end{align}
In \appRef{app:IFdglapLimit} we show that the antenna functions correctly
reproduce the DGLAP splitting kernels in the collinear limit.

\paragraph{Evolution Variables}
We evolve gluon emission in the transverse momentum of the emission, defined as
\begin{align}
  t_\t{IF}^\t{emit} = p_{\perp\,\t{IF}}^2=\frac{\saj\sjk}{\sAK+\sjk}
  =\frac{\saj\sjk}{\saj+\sak}~,
\label{eq:IFpTmax}
\end{align}
with the phase-space limit $ p_{\perp\,\t{IF}}^2\le\sAK(1-x_A)/x_A$.
\FigRef{fig:phasespace}\,{\it b)}
shows constant contours of $p_{\perp\,\t{IF}}^2$, as a function of the two
branching invariants $\saj$ and $\sjk$. Note that the phase space is limited by
$\sjk\le\sAK(1-x_A)/x_A$ and $\saj\le\sAK+\sjk$. \\
For branchings with flavour changes in the initial or final state we use the corresponding
invariant, $\saj$ or $\sjk$ respectively, 
\begin{equation}
  t_\t{IF}^\t{conv} = Q^2_\t{IF} = \left\{ \begin{array}{cl}
  \saj & \mbox{for $a$ converting to/from a gluon}\\
  \sjk & \mbox{for $K\to q\bar{q}$} 
  \end{array} 
  \right.~, 
\end{equation}
with the phase-space limits $\saj\le\sAK/x_A$ and $\sjk\le\sAK(1-x_A)/x_A$.

\paragraph{No-emission Probability} With the definitions given
above \eqref{eq:NoEmiProb} for initial-final configurations reads
\begin{multline}
  \Pi_n(t_{n},t_{n+1})~=~\exp\left(-
  \int_{{\saj}_{n+1}}^{{\saj}_{n}}\!\!\!\!\!\!\!\!\!\!\!\!\d\saj\,
  \int_{{\sjk}_{n+1}}^{{\sjk}_{n}}\!\!\!\!\!\!\!\!\!\!\!\!\d\sjk\,
  \int_0^{2\pi}\!\frac{\d\phi}{2\pi}\,
  \right. \\ \left.
  \frac{\alpha_s(t)}{4\pi}\,\frac{\sAK}{(\sAK+\sjk)^2}\,
  \mc C\, \bar a(\saj,\sjk,\sAK)\,
  \frac{f_a(x_a,t)}{f_A(x_A,t)}
  \right) ~,
\end{multline}
with $t=t_\t{IF}(\saj,\sjk,\sak)$. The subscript of the $\saj$ and $\sjk$ integration
limits indicates the association with the branching scales $t_n$ and $t_{n+1}$
respectively.

\subsection{Final-Final Configurations \label{sec:antDefsFF}}
We denote the pre- and post-branching partons participating in a final-final branching 
by $IK\to ijk$, with no recoils outside the antenna.
In the following, we specify the phase-space factorisation, antenna functions, evolution
variables and the resulting no-emission probability. More extensive descriptions of \vc's final-state antenna-shower formalism can be found in \cite{Giele:2007di,Giele:2011cb}.

\paragraph{Phase Space}
The phase-space factorisation reads
\begin{align}
  \d\Phi_3(P\to p_i,p_j,p_k) = \d\Phi_2(P\to p_I,p_K)\,\d\Phi_\t{ant}^\t{FF}
\end{align}
with the antenna phase space
\begin{align}
  \d\Phi_\t{ant}^\t{FF} = \frac1{16\pi^2}\,\frac1{\sIK^2}\,
  \d\sij\,\d\sjk\,\frac{\d\phi}{2\pi}~.
\end{align}
See \appRef{app:FFkinematics} for the explicit construction of the post-branching momenta. 

\paragraph{Antenna functions}
The default final-final antenna functions are chosen to be the ones of~\cite{Larkoski:2013yi} averaged over helicities.
For gluon-emission antennae, these are 
\begin{align}
  \bar a_{q\bar q\,g}^\t{FF}=\bar a(i_q,k_{\bar q},j_g) =&~
  \frac1\sIK\left(2\,\frac{\sik\sIK}{\sij\sjk} + \frac\sjk\sij + \frac\sij\sjk+1\right)~,\\
  \bar a_{gg\,g}^\t{FF}=\bar a(i_g,k_g,j_g) =&~
  \frac1\sIK\left(2\,\frac{\sik\sIK}{\saj\sjb} + \frac\sjk\sij + \frac\sij\sjk
  - \frac{\sjk^2}{\sij\sIK} - \frac{\sij^2}{\sjk\sIK} + 
  \frac32 + \frac{\sij+\sjk}{2\,\sIK} \right)~, \\
  \bar a_{qg\,g}^\t{FF}=\bar a(i_q,k_g,j_g) =&~ 
  \frac1\sIK\left(2\,\frac{\sik\sIK}{\saj\sjb} + \frac\sjk\sij + \frac\sij\sjk
  - \frac{\sij^2}{\sjk\sIK} + \frac32\right)~.\\
  \intertext{For a final-state gluon splitting, the default is}
  \bar a_{xq\,\bar q}^\t{FF}=\bar a(i_x,k_q,j_{\bar q}) &= 
  \frac1{2\sjk}\frac{\sij^2+\sik^2}{\sIK^2}+\frac12\frac\sjk{\sIK^2}+\frac\sik{\sIK^2}~.
\end{align}
In \appRef{app:FFdglapLimit} we show that the antenna functions correctly
reproduce the DGLAP splitting kernels in the collinear limit.

\paragraph{Evolution Variables}
We evolve gluon emission either in transverse momentum, which is the
default choice, or in the antenna mass,
\begin{equation}
  t_\t{FF}^\t{emit} = \left\{
  \begin{array}{l} p_{\perp\,\t{FF}}^2=4\,\dfrac{\sij\sjk}\sIK \\
  m_{\t{A}\,\t{FF}}^2=2\,\t{min}(\sij,\sjk)
  \end{array}\right.~.
\end{equation}
The upper phase-space limit is the parent antenna mass, $t_\t{FF}^\t{emit}\le \sIK$.
Gluon splittings are evolved in the invariant mass of the quark-antiquark
pair,
\begin{equation}
  t_\t{FF}^\t{conv} = Q^2_\t{FF} = \left\{ \begin{array}{cl}
  \sij & \mbox{for $i$ being the gluon}\\
  \sjk & \mbox{for $k$ being the gluon}
  \end{array}
  \right.~,
\end{equation}
with the same phase-space limit as before.

\paragraph{No-emission Probability} With the definitions given
above \eqref{eq:NoEmiProb} for final-final configurations reads
\begin{align}
  \Pi_n(t_{n},t_{n+1})~&=~\Delta_n(t_{n},t_{n+1}) \\
  &=~\exp\left(-
  \int_{{\sij}_{n+1}}^{{\sij}_{n}}\!\!\!\!\!\d\sij 
  \int_{{\sjk}_{n+1}}^{{\sij}_{n}}\!\!\!\!\!\d\sjk 
  \int_0^{2\pi}\!\frac{\d\phi}{2\pi}\,
  \frac{\alpha_s(t)}{4\pi}\,\frac1{\sIK^2}\,\mc C\,
  \bar a(\sij,\sjk,\sIK)\right) ~,
\end{align}
with $t=t_\t{FF}(\sij,\sjk,\sIK)$. The subscript of the $\sij$ and $\sjk$ integration
limits indicates the association with the branching scales $t_n$ and $t_{n+1}$
respectively.

\subsection{The Shower Generator \label{sec:generator}}
We now illustrate how the shower algorithm generates branchings, starting from trial 
branchings generated according to a simplified version of the no-emission
probability in eq.~\eqref{eq:NoEmiProb}. For definiteness we consider the specific 
example of initial-initial antennae, 
initial-final ones being handled in much the same way, with a PDF ratio that only 
involves one of the beams, and final-final ones not involving any PDF ratios at all. 
The full antenna-shower evolution (II+IF+FF) is combined with \py's $p_\perp$-ordered
multiple-parton-interactions (MPI) model, in a common interleaved sequence of evolution
steps~\cite{Sjostrand:2004ef}.

With the explicit form of the antenna phase space the no-emission probability reads
\begin{align}
  \Pi_n(t_\t{start},t_{n+1}) &= \exp\left( -\int_{t_{n+1}}^{t_\t{start}}\!
  \d\saj\,\d\sjb\,\frac{\alpha_s(t)\,\mc C}{4\pi}\,\frac{\sAB}{\sab^2}\,
  \bar a(\saj,\sjb,\sAB)\,R_\t{pdf} \right) \nonumber \\
  &= \exp\left( -\int_{t_{n+1}}^{t_\t{start}}\!\d\saj\,\d\sjb\,
  a(\saj,\sjb,\sAB)\,R_\t{pdf} \right) ~,
  \label{eq:PiII1}
\end{align}
where the integral is written in terms of the invariants $\saj$ and $\sjb$ and we have
suppressed the trivial integration over $\phi$. In the second line, colour and coupling 
factors, as well
as leftover factors coming from the antenna phase space are absorbed into a redefined 
antenna function, $a(\saj,\sjb,\sAB)$. To impose the evolution measure, we first change 
the integration variables from $\saj$ and $\sjb$ to $t$ and $\zeta$, where $t$
has dimension $\mrm{GeV}^2$ and $\zeta$ is dimensionless. The definition of 
$\zeta$ is in somewhat arbitrary, as long as it is linearly independent of $t$ and 
there exists a one-to-one map back and forth 
between $(\saj,\sjb)$ and $(t,\zeta)$. Generally, the freedom to choose $\zeta$ can be 
utilised to make the $(t,\zeta)$ integrands and phase-space boundaries as simple and 
efficient as possible. Transformed to arbitrary $(t,\zeta)$, eq.~\eqref{eq:PiII1} now reads
\begin{equation}
  \Pi_n(t_\t{start},t_{n+1}) = \exp\left( -\int_{t_{n+1}}^{t_\t{start}}\!\d t\,\d\zeta\,
  |J|\,a(\saj,\sjb,\sAB)\,R_\t{pdf} \right)~,
  \label{eq:PiII2}
\end{equation}
with the Jacobian $|J|$ associated with the transformation from $(\saj,\sjb)$ to
$(t,\zeta)$. Rather than solving the exact expression, we make three simplifications, 
the effects of which we will later cancel by use of the veto algorithm:
\begin{itemize}
\item Instead of the physical antenna functions, $a$, we use simpler (trial) overestimates, 
$\hat a(\saj,\sjb,\sAB)$. For instance the trial antenna function for gluon emission off an
initial-state quark-antiquark pair is chosen to be
\begin{align}
  \hat a_{q\bar q\,g}^\t{II}=2\,\frac{\sab^2}{\sAB\saj\sjb}~.
\end{align}
\item Instead of the PDF ratio, $R_\t{pdf}$, we use the overestimate
\begin{align}
  \hat R_\t{pdf} = \left(\frac{x_A}{x_a}\frac{x_B}{x_b}\right)^\alpha
  \frac{f_a(x_A,t_\t{min})}{f_A(x_A,t_\t{min})}
  \frac{f_b(x_B,t_\t{min})}{f_B(x_B,t_\t{min})}~,
\end{align}
where $t_\t{min}$ is the lower limit of the range of evolution variable
under consideration and $\alpha$ a parameter, whose value is, wherever possible,
chosen differently, depending on the type of branching, to give a good performance.
\item In cases where the physical $\zeta$ boundaries depend on the evolution
variable $t$, we allow trial branchings to be generated in a larger hull
encompassing the physical phase space, with $\zeta$ boundaries that only depend on 
the $t$ integration limits.
\end{itemize}
Having the trial no-emission probability, $\hat\Pi_n(t_\t{start},t_{n+1})$, at hand we solve 
\begin{align}
  \hat\Pi_n(t_\t{start},t_{n+1}) = \mc R\qquad\t{with}\quad \mc R\in[0,1]
\end{align}
for $t_{n+1}$ to obtain the scale of the next branching. Due to the simplifications 
discussed above, this can be done analytically.
We then generate another uniformly distributed random number, $\mc R_{\zeta}$, 
from which we obtain a trial $\zeta$ value by solving (again analytically),
\begin{equation}
\mc R_\zeta = \frac{\hat{I}_\zeta(\hat{\zeta}_\t{min},\zeta)}
{\hat{I}_\zeta(\hat{\zeta}_\t{min},\hat{\zeta_\t{max}})}\\
\end{equation}
where $\hat{I}$ is the integral over all $\zeta$ dependence in 
$\hat\Pi_n(t_\t{start},t_{n+1})$.

Finally, a uniformly distributed trial $\phi = 2\pi \mc R_\phi$ can be generated, furnishing the last branching variable.  
We now make use of the veto algorithm to
recover the exact integral in eq.~\eqref{eq:PiII2}, as shown in \cite{Sjostrand:2006za}.
First, any trial branching outside the physical phase space is rejected. Each physical trial branching is then accepted with the probability
\begin{align}
  \mc O(\hat t,t_{n+1})\,P^\t{shower}=\mc O(\hat t,t_{n+1})\,\frac{a(\saj,\sjb,\sAB)}
  {\hat a(\saj,\sjb,\sAB)}\,\frac{R_\t{pdf}}{\hat R_\t{pdf}}~,
  \label{eq:acceptProbShower}
\end{align}
where $\mc O(\hat t,t_{n+1})$ represents the ordering condition with respect to some scale
$\hat t$. In traditional, strongly ordered showers this scale is equal to the scale of the 
last branching $t_{n}$ and the ordering condition therefore is
\begin{align}
  \mc O(\hat t,t_{n+1}) = \mc O(t_{n},t_{n+1}) = \Theta(t_{n}-t_{n+1})~.
\end{align}
For more details on the algorithm see \appRef{app:PSdetails} and the 
\vc\ compendium distributed alongside with the code.

\subsection{Colour Coherence and Colour Indices \label{sec:colour}}

When assigning colour indices to represent colour flow after a branching, we 
adopt a set of conventions that are designed to approximately capture 
correlations between partons that are not LC-connected, based on the arguments 
presented in~\cite{Christiansen:2015yqa}. Specifically, we let the last digit 
of the "Les Houches (LH) colour tag"~\cite{Boos:2001cv,Alwall:2006yp} run 
between 1 and 9, and refer to this digit as the "colour index". LC-connected 
partons have matching LH colour tags and therefore also matching colour/anticolour 
indices, while colours that are in a relative octet state are assigned non-identical 
colour/anticolour indices. Hence the last digit of a gluon colour tag will never 
have the same value as that of its anticolour tag. This does not change the LC 
structure of the cascade; if using only the LH tags themselves to decide between 
which partons string pieces should be formed, the extra information is effectively 
just ignored. It does however open for the possibility of allowing strings to form 
between non-LC-connected partons that "accidentally" end up with matching indices, 
in a way that at least statistically gives a more faithful representation of the 
full SU(3) group weights than the strict-LC one~\cite{Christiansen:2015yqa}. 

The new aspect we introduce here is to assign colour indices after each branching, 
whereas the model in~\cite{Christiansen:2015yqa} operated at the purely non-perturbative 
stage just before hadronisation. 
Furthermore, for gluon emissions, we choose to let the colour tag of the parent 
antenna be inherited by the daughter antenna with the largest invariant mass, while 
the one with the smaller invariant mass is assigned a new colour tag (subject to the 
rules described above). This is intended to preserve the coherence structure as seen 
by the rest of the event, so that, for instance, the new colour created in a 
near-collinear branching is attributed to the new small antenna, while the colour 
tag of the parent antenna continues on as the tag of the larger of the daughters.  
An advantage of this approach is that the octet nature of intermediate gluons, e.g.\ 
in collinear $g\to gg$ branchings, is preserved by our treatment, which is not the 
case in the implementation of~\cite{Christiansen:2015yqa}. 
\begin{figure}[t]
\centering
\begin{fmffile}{diagram}
\vspace*{2ex}
\!\begin{tabular}{cc}
\begin{fmfgraph*}(145,90)
\fmfleft{l1,l2}
\fmfright{r1,r2}
\fmftop{t000,t00,t0,t1,t2,t22,t3,t4,t44,t444}
\fmf{plain}{v,l1}
\fmf{plain}{r1,v}
\fmf{gluon,tension=2,label=~\small$g^*_{12}$}{v,vg}
\fmf{gluon,tension=1}{vg,t1}
\fmf{gluon,tension=1}{vg,t3}
\fmffreeze
\fmfv{d.sh=circ,d.siz=5}{v}
\fmfv{label=\!$q_{1}$}{l1}
\fmfv{label=\!$\bar{q}_{2}$}{r1}
\fmfv{label=$g_{13}$}{t1}
\fmfv{label=$g_{32}$}{t3}
\fmfdot{v}
\fmf{dashes,fore=red}{t1,t3}
\fmf{dashes,fore=red,label=\small$\text{string}$,l.side=left}{l1,t1}
\fmf{dashes,fore=red,label=\small$\text{string}$,l.side=left}{t3,r1}
\end{fmfgraph*}~&~
\begin{fmfgraph*}(145,90)
\fmfleft{l1,l2}
\fmfright{r1,r2}
\fmftop{t000,t00,t0,t1,t2,t22,t3,t4,t44,t444}
\fmf{plain}{l1,qgl}
\fmf{plain,tension=0.7,label=$q_3$--$\bar{q}_3$}{qgl,qgr}
\fmf{plain}{qgr,r1}
\fmf{gluon,tension=0.5}{qgl,t1}
\fmf{gluon,tension=0.5}{qgr,t3}
\fmffreeze
\fmf{phantom}{qgl,z,qgr}
\fmfdot{z}
\fmfv{label=\!$q_{1}$}{l1}
\fmfv{label=\!$\bar{q}_{2}$}{r1}
\fmfv{label=$g_{13}$}{t1}
\fmfv{label=$g_{32}$}{t3}
\fmf{dashes,fore=red}{t1,t3}
\fmf{dashes,fore=red}{l1,t1}
\fmf{dashes,fore=red}{t3,r1}
\end{fmfgraph*}\\[1ex]
(a) & (b) 
\end{tabular}\!
\caption{Illustration of colour flow in $Z\to qgg\bar{q}$, using subscripts to denote colour indices. Note that both $x$ and $y$ axes illustrate spatial dimensions, with time indicated roughly by the distance from the location of the original $Z$, denoted by $\bullet$. Two Feynman diagrams contribute to the same leading-colour string topology. \label{fig:z4LC}}
\vspace*{5ex}
\!\begin{tabular}{cc}
\begin{fmfgraph*}(145,90)
\fmfleft{l1,l2}
\fmfright{r1,r2}
\fmftop{t000,t00,t0,t1,t2,t22,t3,t4,t44,t444}
\fmf{plain}{l1,qgl}
\fmf{plain,tension=0.7,label=$q_3$--$\bar{q}_3$}{qgl,qgr}
\fmf{plain}{qgr,r1}
\fmf{gluon,tension=0.5}{qgl,t1}
\fmf{gluon,tension=0.5}{qgr,t3}
\fmffreeze
\fmf{phantom}{qgl,z,qgr}
\fmfdot{z}
\fmfv{label=\!$q_{1}$}{l1}
\fmfv{label=\!$\bar{q}_{1}$}{r1}
\fmfv{label=$g_{13}$}{t1}
\fmfv{label=$g_{31}$}{t3}
\fmf{dashes,fore=red}{t1,t3}
\fmf{dashes,fore=red}{l1,t1}
\fmf{dashes,fore=red}{t3,r1}
\end{fmfgraph*}~&~
\begin{fmfgraph*}(145,90)
\fmfleft{l1,l2}
\fmfright{r1,r2}
\fmftop{t000,t00,t0,t1,t2,t22,t3,t4,t44,t444}
\fmf{plain}{l1,qgl}
\fmf{plain,tension=0.7,label=$q_3$--$\bar{q}_3$}{qgl,qgr}
\fmf{plain}{qgr,r1}
\fmf{gluon,tension=0.5}{qgl,t1}
\fmf{gluon,tension=0.5}{qgr,t3}
\fmffreeze
\fmf{phantom}{qgl,z,qgr}
\fmfdot{z}
\fmfv{label=\!$q_{1}$}{l1}
\fmfv{label=\!$\bar{q}_{1}$}{r1}
\fmfv{label=$g_{13}$}{t1}
\fmfv{label=$g_{31}$}{t3}
\fmf{dashes,fore=red,right=0.2}{t1,t3}
\fmf{dashes,fore=red,left=0.2}{t1,t3}
\fmf{dashes,fore=red}{l1,r1}
\end{fmfgraph*}\\[1ex]
 (c) & (d) 
\end{tabular}\!
\end{fmffile}
\caption{With a probability suppressed by $1/N_C^2$, the same colour index may occur twice in the diagram shown in \figRef{fig:z4LC}.b, illustrated here in the left-hand pane. When this occurs, the string topology shown in the right-hand pane is also possible. (The model of \cite{Christiansen:2015yqa} invokes a string-length minimisation argument to decide which is realised.) 
\label{fig:colRec}}
\end{figure}

In \figsRef{fig:z4LC} \& \ref{fig:colRec} we illustrate our approach, and the 
ambiguity it addresses. For definiteness, and for simplicity, we consider the 
specific case of $Z\to q gg\bar{q}$, but the arguments are general. The two 
diagrams in \figRef{fig:z4LC} show the outgoing partons, produced by a $Z$ 
boson decaying at the point denoted by $\bullet$. Both axes correspond to 
spatial dimensions, hence time is indicated roughly by the radial distance from 
the $Z$ decay point. Examples of the colour indices defined above are indicated 
by subscripts, hence e.g., $g_{13}$ denotes a gluon carrying anticolour index 1 
and colour index 3. Due to our selection rule, the type of assignment represented 
by \figRef{fig:z4LC}.a is always selected when $m_{gg}$ is small, $s_{gg}< s_{qg}$, 
while the one represented by  \figRef{fig:z4LC}.b is selected when $s_{qg}<s_{gg}$ 
(when the second emission occurs in the $q-g$ antenna, and completely analogously 
when it occurs in the $g-\bar{q}$ one). The subleading-colour ambiguity 
illustrated by \figRef{fig:colRec} can only occur for the latter type of 
assignment, hence will be absent in our treatment for collinear $g\to gg$ 
branchings (where the flow represented by \figRef{fig:z4LC}.a dominates), in 
agreement with the collinearly branching gluon having to be an octet. We regard 
this as an improvement on the treatment in \cite{Christiansen:2015yqa}, in which 
there was no mechanism to prevent collinear gluons from ending up in an overall 
singlet state; see also the remarks accompanying~\cite[Fig.~15]{Christiansen:2015yqa}.

As a last point, we remark that this new assignment of colour tags is currently 
left without impact, but is implemented in order to enable future studies, such as colour 
reconnection within \vc.

\subsection{Uncertainty Estimations \label{sec:uncertainties}}

Traditionally, shower uncertainties are evaluated by systematic up/down variations 
of each model parameter, which mandates the generation of multiple event sets, one 
for each variation. 
To avoid this time-consuming procedure, \vc\ instead generates a vector 
of variation weights for each event~\cite{Giele:2011cb}, where each of the weights corresponds to varying 
a different parameter. A separate publication 
 details the formal proof of the validity of the method~\cite{Mrenna:2016sih}, which we 
have here extended to cover both the initial- and final-state showers in \vc. 
(\textbf{Note added in proof:} during the publication of this manuscript, two further papers appeared reporting similar implementations in \hw\ and \sh, see  \cite{Bellm:2016voq,Bothmann:2016nao}.)
In this section, we only give 
a brief overview of the implementation, referring to \cite{Giele:2011cb,Mrenna:2016sih} for details and illustrations. Technical specifications for how to switch the uncertainty bands on and off in the code, and how to access them, are provided in \vc's HTML User Reference~\cite{VinciaUserReference}.

During the shower step, in which a trial branching gets accepted with the 
probability $P_\t{def}$ given in \eqRef{eq:acceptProbShower}, the probability 
of the same branching to occur with
a variation in e.g. the choice of renormalization scale or antenna
function is calculated,
\begin{align}
  P_\t{var} = \frac{\t{VAR}}{\t{DEF}}\,\,P_\t{def}
\end{align}
where $\t{DEF}$ and $\t{VAR}$ are symbols representing the default and
variation choice respectively. In case of an accepted branching the variation
weight of the event gets simply multiplied with $P_\t{var}/P_\t{def}$, 
and for rejected branchings with
\begin{align}
  \frac{1-P_\t{var}}{1-P_\t{def}}
\end{align}
to correctly take the no-emission probability into account.

The variations currently implemented in \vc\ are the following:
\begin{itemize}
  \item \vc's default settings, with default antenna functions, scale choices
  and colour factors.
  \item Variation of the renormalization scale. Using $\alpha_s(t/k_\mu)$ and
  $\alpha_s(t\,k_\mu)$, with a user-specifiable value of the
  additional scaling factor $k_\mu$.
  \item Variation of the antenna functions. Using antenna sets with large and small non\-singular terms, representing unknown (but finite) process-dependent LO matrix-element terms. Note that these are cancelled by LO MECs (up to the matched orders). 
  \item $\alpha_s$-suppressed counterparts of the finite-term variations above\footnote{Up to and including \vc\ 2.001, these variations were erroneously applied by multiplying or dividing the antenna functions by $(1+\alpha_s)$, which is degenerate with the renormalisation-scale variations.} 
  which are not cancelled by (LO) MECs. 
  \item Variation of the colour factors. All gluon emissions use colour factor 
  of either $C_A=3$ or $2C_F=8/3$.
  \item Modified $P_\t{imp}$ factor, 
  \begin{align} P_\t{imp}' = \frac{\hat t^2}{\hat t^2+t^2}~. \end{align}
\end{itemize}
Note that, except for the first one, the variations are taken with
respect to the user-defined settings.
All of these variations are applied in the shower and the MECs,
and are limited to branchings in the hard system, i.e. they are
for instance not applied in the showering of multi-parton interactions.

\subsection{Limitations \label{sec:limitations}}

For completeness, we note that a few options and extensions of the existing \vc\ final-state shower have not yet been implemented in \vc~2.0. These will remain available in earlier versions of the code (limited to pure final-state radiation hence mostly of interest for $e^+e^-$ studies) and may reappear in future versions, subject to interest and available manpower. Briefly summarised, this concerns the following features:
\begin{itemize}
\item Sector Showers~\cite{LopezVillarejo:2011ap}: a variant of the antenna-shower formalism in which a single term is responsible for generating all contributions to each phase-space point. It has some interesting and unique properties including being one-to-one invertible and producing fewer (one) term at each order of GKS matrix-element corrections leading to the numerically fastest matching algorithm we are aware of (see~\cite{LopezVillarejo:2011ap}), at the price of requiring more complicated antenna functions with more complicated phase-space boundaries. For the initial-state extension of \vc\, we have so far focused on the technically simpler case of ``global'' (as opposed to sector) antennae. 
\item One-loop matrix-element corrections. The specific case of one-loop corrections for hadronic $Z$ decays up to and including 3 jets was studied in detail by HLS~\cite{Hartgring:2013jma}. The extension of this method to hadronic initial states, and a more systematic approach to one-loop corrections in \vc\ in general, will be a major goal of future efforts. 
\item Helicity dependence~\cite{Larkoski:2013yi}. The shower and matrix-element-correction algorithms described in this paper pertain to unpolarised partons. Although this is fully consistent with the unpolarised nature of the initial-state partons obtained from conventional parton distribution functions (PDFs), we note that an extension to a helicity-dependent formalism could nonetheless be a relatively simple future development. Moreover, we expect this would provide useful speed gains for the GKS matrix-element correction algorithm equivalent to those observed for the final-state algorithm~\cite{Larkoski:2013yi} . 
\item
Full-fledged fermion mass effects~\cite{GehrmannDeRidder:2011dm}. Our treatment of mass effects for initial-state partons is so far limited to one parallelling the simplest treatment in conventional PDFs,  the ``zero-mass-variable-flavour-number (ZMVF) scheme''. In this scheme, heavy-quark PDFs are set to zero below the corresponding mass threshold(s) and are radiatively generated above them by $g\to Q\bar{Q}$ splittings, with $m_Q$ formally set to zero in those splittings and for the subsequent heavy-quark evolution. Thus, in \vc~2.0, all partons are assigned massless kinematics, but $g\to Q\bar{Q}$ splittings are switched off (also in the final state) below the physical mass thresholds. This only gives a very rough approximation of mass effects~\cite{Norrbin:2000uu,Thorne:2012az} but at least avoids generating unphysical singularities. Beyond the strict ZMVF scheme, optionally and for final-state branchings only, we allow for a set of universal antenna mass corrections to be applied and/or for tighter phase-space constraints to be imposed, with the latter obtained from the would-be massive phase-space boundaries. We note that a mixed treatment similar to the one currently employed by \py, with massive/massless kinematics for outgoing/incoming partons respectively, would not be straightforward to adopt in \vc\, as it would be inconsistent with the application of on-shell matrix-element corrections.
\item
The so-called ``Ariadne factor''~\cite{Lonnblad:1992tz} for gluon splitting antennae
\begin{align}
  P_\t{Ari} = \frac{2s_N}{s_N+s_P}~,
\end{align}
with $S_N$ the invariant mass squared of the colour neighbour on the other side 
of the splitting gluon and $s_P$ the invariant mass squared of the parent (splitting) 
antenna is limited to its original purpose, that of improving the description of 4-jet observables in $Z$ decay, and is not applied outside that context. 
\end{itemize}

\section{Matrix-Element Corrections \label{sec:MECs}}

In this section we focus on the MEC formalism in \vc\ 
and discuss our strategy for reaching the non-ordered parts of phase space,
both with respect to the factorization scale in the case of the first 
branching and with respect to previous branching scales.

Note that in this paper all matrix elements are generated with
\mg~\cite{Alwall:2007st,Murayama:1992gi}. The output is suitably 
modified to extract the leading colour matrix element, i.e. to not 
sum over colour permutations, but pick the (diagonal) entry in \mg's colour 
matrix that corresponds to the colour order of interest. All plots shown
in this paper are based on leading colour matrix elements.

\subsection{Hard Jets in non-QCD processes \label{sec:hardJets}}

In this section we describe our formalism to combine events which are 
accompanied by at least one very hard jet, with the ones which are not. We emphasise 
that the considerations are general and apply to any processes that do not exhibit 
QCD jets at the Born level. 

We first consider the Born inclusive cross section, differential in the Born phase 
space,
\begin{align}
  \d\sigma_B^\t{incl}(t_\t{fac}) = f_0(x_0,t_\t{fac})\,|\mc M_B|^2\,\d\Phi_B~,
  \label{eq:BornIncl}
\end{align}
where $t_\t{fac}$ is the factorisation scale, subscript zero emphasises 
that flavour and energy fraction correspond to the state $\Phi_B$
(subscript one will then correspond to the state $\Phi_{B+1}$ and so on),
and the second PDF factor has been dropped for the sake of readability. 

Since the ISR shower formally corresponds to a ``backwards'' evolution of the PDFs~\cite{Sjostrand:1985xi}, the factorisation scale represents the natural upper bound (starting scale) for the initial-state shower evolution. This implies that any phase-space points with $t > t_\t{fac}$ will not be populated by the shower, potentially leaving a ``dead zone'' for high-$t$ emissions. In principle, the freedom in choosing the evolution variable can be exploited to define $t$ in such a way that the entire physical phase space becomes associated with scales $t<t_\t{fac}$~\cite{Hoche:2015sya}, including points with physical $p_\perp^2\gg t_\t{fac}$. Here, however, we wish to maintain a close correspondence between the evolution variable and the physical (kinematic) $p_\perp$, requiring the development of a different strategy.

The approach used internally in \py\ is that of ``power showers'' (with~\cite{Miu:1998ju} 
or without~\cite{Plehn:2005cq} matrix-element corrections): starting the shower from a 
scale $t_\t{start}$ that is higher than the factorization scale. 
This method has been criticised for producing too hard jet emission spectra and violating 
the factorisation ansatz. Though the improved power showers defined 
in~\cite{Skands:2010ak,Corke:2010zj} are better behaved (dampening the LL $1/p_\perp^2$ 
kernels to explicitly subleading $Q^2/p_\perp^4$ ones for emissions above the $Q$ scale 
of the basic process), shortcomings are still present. 
Consider, for example, the Born exclusive cross section at an arbitrary shower cutoff, 
differential in the Born phase space, 
scale $t_\t{cut}$,
\begin{align}
  \d\sigma_B^\t{excl}(t_\t{cut}) &= \Pi_0(t_\t{start},t_\t{cut})\,f_0(x_0,t_\t{fac})\,
  |\mc M_B|^2\,\d\Phi_B \\
  &= \frac{f_0(x_0,t_\t{fac})}{f_0(x_0,t_\t{start})}\,f_0(x_0,t_\t{cut})\,
  \Delta_0(t_\t{start},t_\t{cut})\,|\mc M_B|^2\,\d\Phi_B~.
  \label{eq:BornExclOrdPower}
\end{align}
Unless $t_\t{start}=t_\t{fac}$, there appears an undesired PDF ratio, which reflects the
difference in the factorization and shower starting scale. To avoid this problem, we
introduce two separate event samples, both initiated by the same matrix element 
with the same factorization scale, as in \eqRef{eq:BornIncl}. They are generated simultaneously,
producing a single stream of ordinary randomly mixed, weighted events, with no need for external
recipes to combine them.
The first sample creates events that do not have a hard jet, by starting
the shower at the factorization scale (hence leaving the region $t>t_\t{fac}$ unpopulated). The second event sample is responsible for all events 
with at least one jet with scale $t>t_\t{fac}$. This sample is initialised by first reweighting 
the Born-level events such that the (temporary) factorization scale is set to the phase-space
maximum, $t_\t{max}$, and the shower algorithm is started from that scale. Events that do not
produce at least one branching before the original (Born-level) factorisation scale is reached 
are vetoed, resulting in a total contribution to the inclusive cross section in 
\eqRef{eq:BornIncl} of  
\begin{align}
  \frac{f_0(x_0,t_\t{max})}{f_0(x_0,t_\t{fac})}\,f_0(x_0,t_\t{fac})\,
  \left(1-\Pi_0(t_\t{max},t_\t{fac})\right)\,|\mc M_B|^2\,\d\Phi_B~.
  \label{eq:Sample2}
\end{align}
Adding the two event samples together yields the new inclusive cross section,
\begin{align}
  \d\sigma_B^\t{incl}(t_\t{fac}) =&~ f_0(x_0,t_\t{fac})\,|\mc M_B|^2\,\d\Phi_B
  ~+\, f_0(x_0,t_\t{start})\, \left(1-\Pi_0(t_\t{start},t_\t{fac})\right)\,
  |\mc M_B|^2\,\d\Phi_B \nonumber\\
  =&~ f_0(x_0,t_\t{fac})\,|\mc M_B|^2\,\d\Phi_B
   ~+\, \int\limits_{t_\t{fac}}^{t_\t{start}}\!\!\d t\,
     f_1(x_1,t)\,\mc A(t)\,
     \Delta_0(t_\t{start},t_\t{})\,|\mc M_B|^2\,\d\Phi_B
  ~,
  \label{eq:BornInclNew}
\end{align}
where $\mc A(t)$ contains all antenna functions, coupling and colour factors.
By virtue of adding and subtracting $f_0(x_0,t_\t{fac})\,|\mc M_B|^2\,\d\Phi_B
\Delta_0(t_\t{start},t_\t{fac})$ and using the DGLAP equation
\begin{align}
  f_0(x_0,t_\t{start}) = f_0(x_0,t_\t{fac})\,\Delta_0(t_\t{start},t_\t{fac})
  + \int\limits_{t_\t{fac}}^{t_\t{start}}\!\!\d t\,f_1(x_1,t)\,\mc A(t)\,
  \Delta_0(t_\t{start},t_\t{})
\end{align}
this becomes 
\begin{align}
  \d\sigma_B^\t{incl}(t_\t{fac}) =&~ f_0(x_0,t_\t{start})\,|\mc M_B|^2\,\d\Phi_B
  \nonumber \\
  &~+~ f_0(x_0,t_\t{fac})\, \left(1-\Delta_0(t_\t{start},t_\t{fac})\right)\,
  |\mc M_B|^2\,\d\Phi_B~.
  \label{eq:BornInclNew2}
\end{align}
Expanding \eqref{eq:BornInclNew} to $\mathcal{O}(\alpha_s)$ yields
\begin{align}
  \d\sigma_B^\t{incl}(t_\t{fac}) = &~ f_0(x_0,t_\t{fac})\,|\mc M_B|^2\,\d\Phi_B
  ~+\, \int_{t_\t{fac}}^{t_\t{start}}\!\!\d t\,f_1(x_1,t)\,\mc A(t)\,
  |\mc M_B|^2\,\d\Phi_B~.
  \label{eq:BornInclNewOrderAlpha}
\end{align}
Expanding \eqref{eq:BornInclNew2} instead yields
\begin{align}
  \d\sigma_B^\t{incl}(t_\t{fac}) = &~ f_0(x_0,t_\t{start})\,|\mc M_B|^2\,\d\Phi_B
  ~+\, f_0(x_0,t_\t{fac})
   \,\int_{t_\t{fac}}^{t_\t{start}}\!\!\d t\,\mc A(t)\,
  |\mc M_B|^2\,\d\Phi_B~,
  \label{eq:BornInclNew2OrderAlpha}
\end{align}
which is seemingly at odds with \eqref{eq:BornInclNewOrderAlpha}. The problem
 is that both \eqref{eq:BornInclNewOrderAlpha} and 
\eqref{eq:BornInclNew2OrderAlpha} have been derived by expanding, so that their
relation through the DGLAP equation is lost. The crucial point --
which is obscured after expanding -- is already contained in 
\eqref{eq:BornInclNew}: The inclusive cross section is calculated
with a sensible factorisation scale $t_\t{fac}$, while all branchings with 
scales $t>t_\t{fac}$ contribute, in a controlled way, at higher orders. \SecRef{sec:results} contains some illustrations of the effects of these corrections for physical observables such as the dilepton rapidity and $p_\perp$ spectra in Drell-Yan processes.

The inclusive cross section obtained from \eqRef{eq:BornInclNew} does not reduce to the zero-parton Born cross section, the changes being only due to hard emissions which have not been incorporated in the first term in \eqref{eq:BornInclNew}. This differs from cross section changes in CKKW-inspired merging prescriptions \cite{Catani:2001cc,Alwall:2007fs}, which arise from real-virtual 
mismatches at the merging scale\footnote{The value of merging scales is typically well below $t_\t{fac}$.}, or from the definition of the inclusive cross section in unitarised merging schemes \cite{Lonnblad:2012ng,Platzer:2012bs}. In the latter, the inclusive cross section is almost entirely given by the first term in \eqref{eq:BornInclNew}, and only changed by "incomplete" 
states which cannot be associated with valid parton shower histories. The definition of what is deemed an "incomplete state" is not conventional and thus may depend on the details of a particular implementation. Note however that \cite{Lonnblad:2012ng,Platzer:2012bs,Bellm:2015epm} do not advocate including the factors "$\Delta_0$" when reweighting "incomplete" states. This could lead to interesting differences in observables relying on very boosted $Z$-boson momenta.

We note that, although the described method of adding hard jets in non-QCD processes is the default choice in \vc, we include the possibility to perform an ordinary shower, starting off the factorization scale $t_\t{fac}$. This is the recommended option when combining \vc's shower with external matching and merging schemes.

\begin{figure}[tbp]
\centering
\includegraphics[width=.6\textwidth]{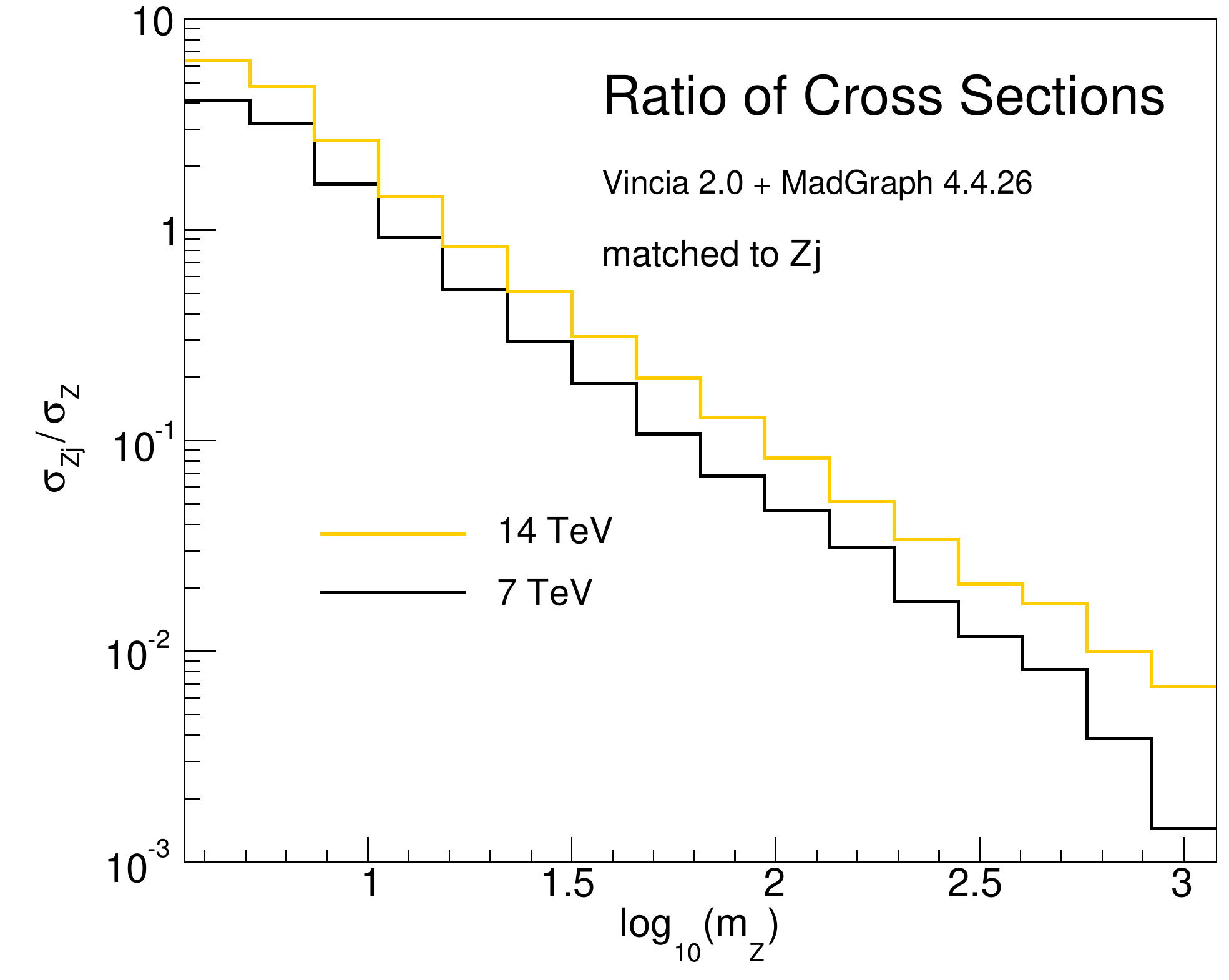}
\caption{\label{fig:XSEC} Ratio of cross sections in $Z$ production as a function
of the $Z$ mass for $\sqrt s=7~\t{GeV}$ (black) and $\sqrt s=14~\t{GeV}$ (orange).}
\end{figure}

In \figRef{fig:XSEC} we show the relative contribution of the two event samples
in $Z$ production, as a function of the $Z$ mass,
\begin{align}
  \frac{\sigma_{Zj}}{\sigma_{Z}}\,\,(m_Z) = \frac
  {f_0(x_0,s)\,\left(1-\Pi_0(s,m_Z^2)\right)\,\left|\mc M_{Zj}\right|^2\,\Phi_{Zj}}
  {f_0(x_0,m_Z^2)\,\left|\mc M_Z\right|^2\,\Phi_Z}~,
\end{align}
with $\sqrt s=7~\t{GeV}$ (black) and $\sqrt s=14~\t{GeV}$ (orange). 
As expected, the contribution of events with at least one hard jets is larger
for decreasing $Z$ masses and increasing centre-of-mass energies.
For both values of $\sqrt{s}$ the Born event sample eventually
dominates for $Z$ masses above $\mc O\,(10~\t{GeV})$.

\subsection{Strong Ordering compared with Tree-Level Matrix Elements \label{sec:validations}}

To validate the quality of the antenna shower, we use large samples of $pp\to Zjj$ phase space
points, generated with \tsc{Rambo}~\cite{Kleiss:1985gy} (an implementation of which is
included in \vc). We cluster all of the phase space points back to the corresponding
$pp\to Z$ phase space point, using the exact inverse of the $2\to3$ recoil prescription 
used in the shower as a clustering algorithm; see \appRef{app:showerDetails} for the 
kinematics map used here. This allows to reconstruct all possible ways in which the shower 
could have populated a certain phase space point, analogously to the study carried out for
final-state radiation in~\cite{Giele:2011cb}~(see also~\cite{Andersson:1991he}). Comparing the shower approximation with the 
LO matrix element for $q_1\bar q_2\to Zg_3g_4$ yields the tree-level PS-to-ME ratio
\begin{align}
  \label{eq:R4long}
  R_4 \equiv&~ \frac{\Theta(t_{\,\widehat{43}}-t_3)\,\mc C_{qg\,g}\,
  \bar a_{qg\,g}^\t{IF}(1,4,3)\,
  \mc C_{q\bar q\,g}\,\bar a_{q\bar q\,g}^\t{II}(\widehat{13},2,\widehat{43})\,
  \left|\mc M_Z(Z)\right|^2}
  {\left|\mc M_{Zgg}(1,2;Z,3,4)\right|^2} \nonumber \\
  &~+~ \frac{\Theta(t_{\,\widehat{34}}-t_4)\,\mc C_{\bar qg\,g}\,
  \bar a_{\bar qg\,g}^\t{IF}(2,3,4)\,
  \mc C_{q\bar q\,g}\,\bar a_{q\bar q\,g}^\t{II}(1,\widehat{24},\widehat{34})\,
  \left|\mc M_Z(Z)\right|^2}
  {\left|\mc M_{Zgg}(1,2;Z,3,4)\right|^2} \vphantom{\frac{\dfrac12}{1}}
\end{align}
where the strong-ordering condition is incorporated by the $\Theta$ step functions and the hatted
variables $\widehat{aj}$ denote clustered momenta. The two terms correspond to the
the two possible shower histories --- obtained from starting by clustering either 
gluon 3 or 4 respectively --- with the sequential clustering scales
\begin{align}
  t_3 = p_{\perp\,\t{IF}}^2(g_3)\quad\t{and}\quad 
  t_{\,\widehat{43}} = p_{\perp\,\t{II}}^2(g_{\,\widehat{43}})~, \\
  t_4 = p_{\perp\,\t{IF}}^2(g_4)\quad\t{and}\quad 
  t_{\,\widehat{34}} = p_{\perp\,\t{II}}^2(g_{\,\widehat{34}})~.
\end{align}
$R_4$ therefore gives a measure of how much the shower under- or overcounts
the tree-level matrix element.
With the first emission already corrected\footnote{This is trivial for
$q\bar q\to Zgg$ as the corresponding antenna function already is the
ratio of the LO matrix elements.} \eqRef{eq:R4long} reduces to
\begin{align}
  \label{eq:R4}
  R_4 =&~ \frac{\Theta(t_{\,\widehat{43}}-t_3)\,\mc C_{qg\,g}\,
  \bar a_{qg\,g}^\t{IF}(1,4,3)\,
  \left|\mc M_{Zg} (\widehat{13},2,\widehat{43}) \right|^2}
  {\left|\mc M_{Zgg}(1,2;Z,3,4)\right|^2} \nonumber \\
  &~+~ \frac{\Theta(t_{\,\widehat{34}}-t_4)\,\mc C_{\bar qg\,g}\,
  \bar a_{\bar qg\,g}^\t{IF}(2,3,4)\,
  \left|\mc M_{Zg} (1,\widehat{24},\widehat{34}) \right|^2}
  {\left|\mc M_{Zgg}(1,2;Z,3,4)\right|^2}~. \vphantom{\frac{\dfrac12}{1}}
\end{align}
Higher-order PS-to-ME ratios are constructed in a similar way.

\begin{figure}[tbp]
\centering
\includegraphics[width=.9\textwidth]{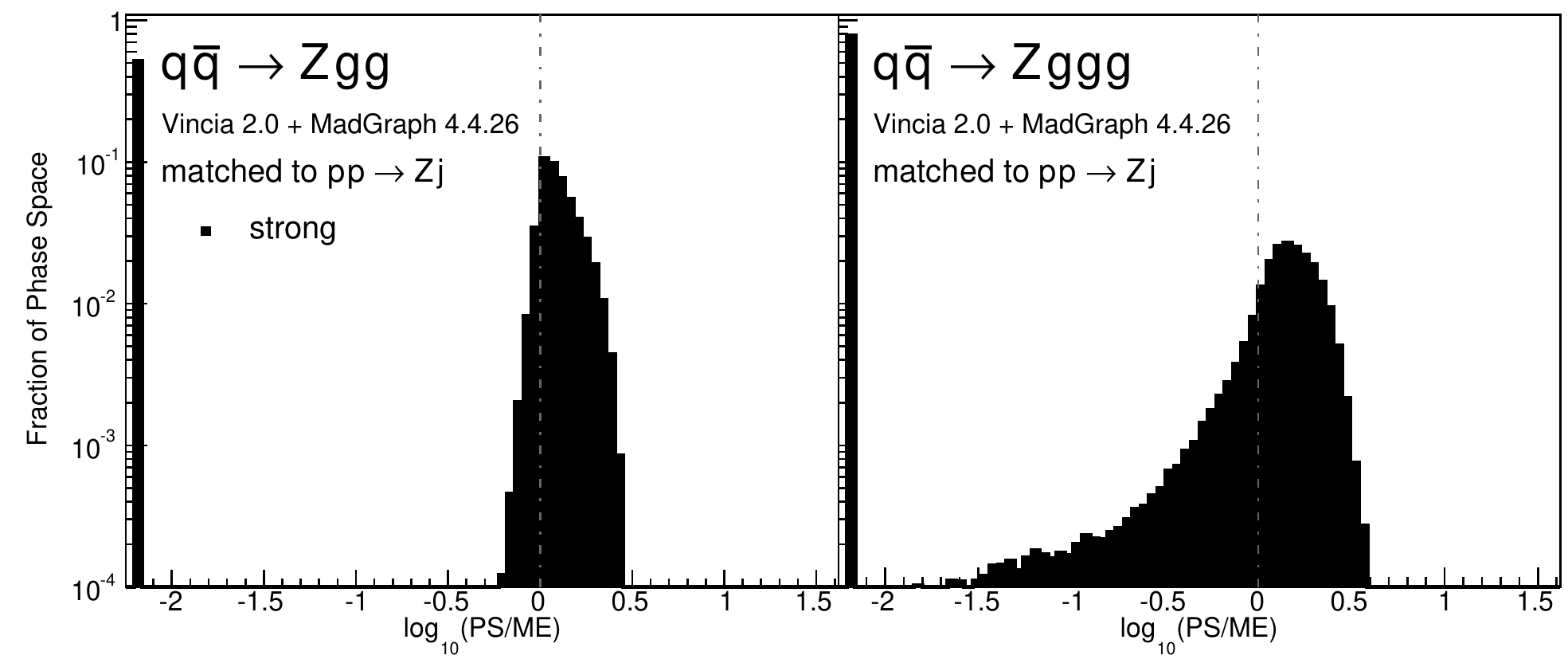}
\caption{\label{fig:PSMEstrong} Antenna shower, compared to matrix elements: distribution
of $\t{log}_{10}(\t{PS}/\t{ME})$ in a flat phase space scan of the full phase space.
Contents normalized to the number of generated points. Gluon emission only.}
\end{figure}

Histograms showing the logarithmic distribution of the PS-to-ME ratios for
$q\bar q\to Zgg$ and $q\bar q\to Zggg$, in a flat scan 
over the full phase space, comparing a strongly ordered shower with the LO amplitude 
squared, are shown in \figRef{fig:PSMEstrong}.
The spike on the very left of the histograms corresponds to the part of phase space 
where there are no ordered shower histories. Note that about $35\%$ of the 
whole phase space in a flat scan of $q\bar q\to Zgg$ does not have an ordered shower 
path, a significantly 
higher fraction than the roughly $2\%$ found for the final-state phase spaces 
in~\cite{Giele:2011cb}. We interpret this as due to the significantly larger size of 
the initial-state phase space, which is not limited by the original antenna invariant mass 
but only by the hadronic CM energy. 
The binning of the histogram is chosen such that the two bins around $0$ (marked
with a gray dashed line) correspond to the shower having less than $10\%$ deviation 
to the tree-level matrix element. For the shower with strong ordering about
$10\%$ of the total number of phase space points, corresponding to about
$15\%$ of the phase space with at least one ordered path, populate these two bins. 

\begin{figure}[tbp]
\centering
\includegraphics[width=.496\textwidth]{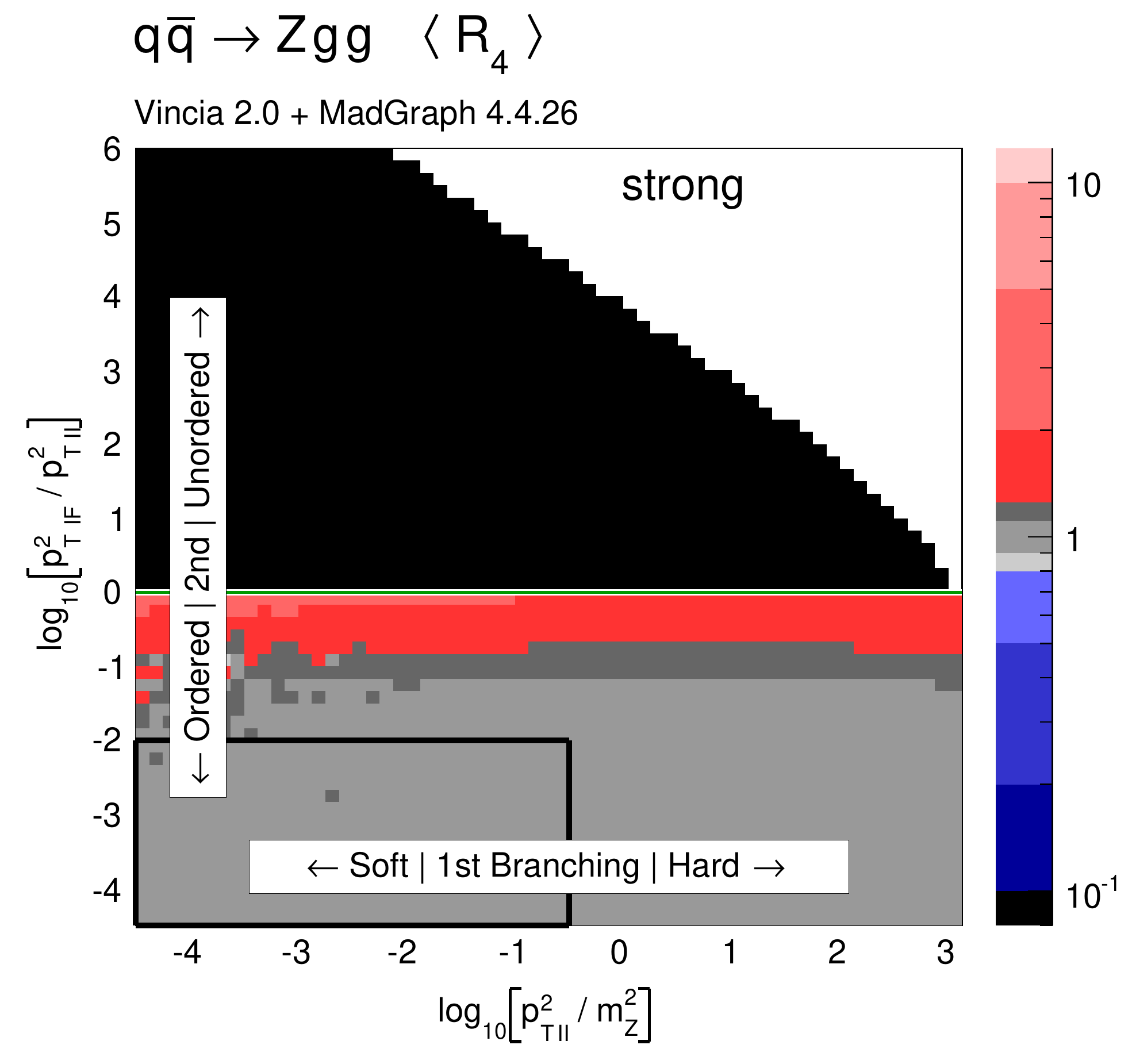}
\includegraphics[width=.496\textwidth]{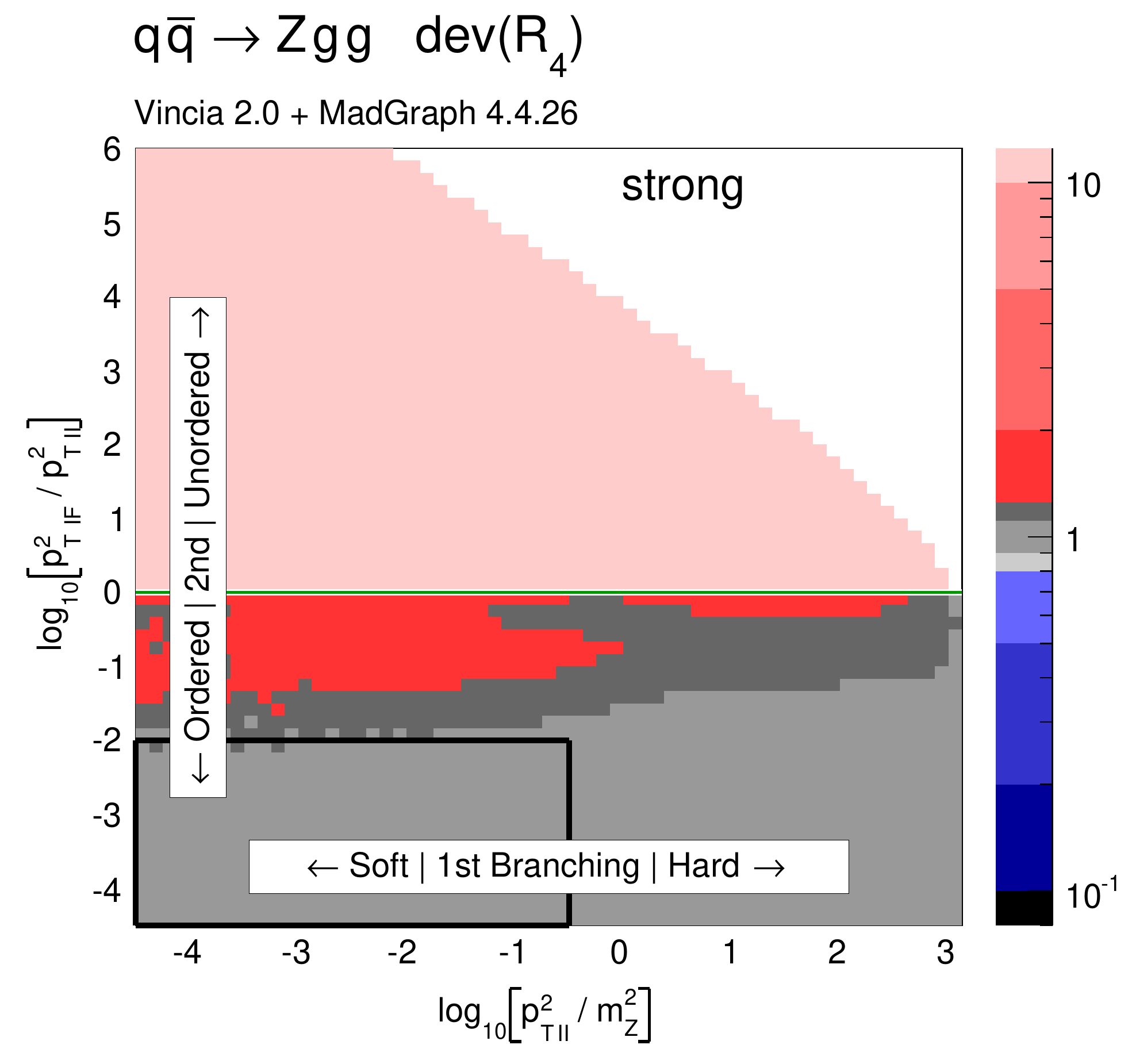}
\caption{\label{fig:PSME2Dstrong} The value of $\langle R_4 \rangle$ (left)
and $\t{dev}(R_4)$ (right), differentially over the 4-parton phase space, with 
$p_\perp^2$ ratios characterizing the first and second emissions on the $x$- and $y$ 
axis, respectively. Strong ordering in the shower, with gluon emission only.} \vspace*{5mm}
\includegraphics[width=.496\textwidth]{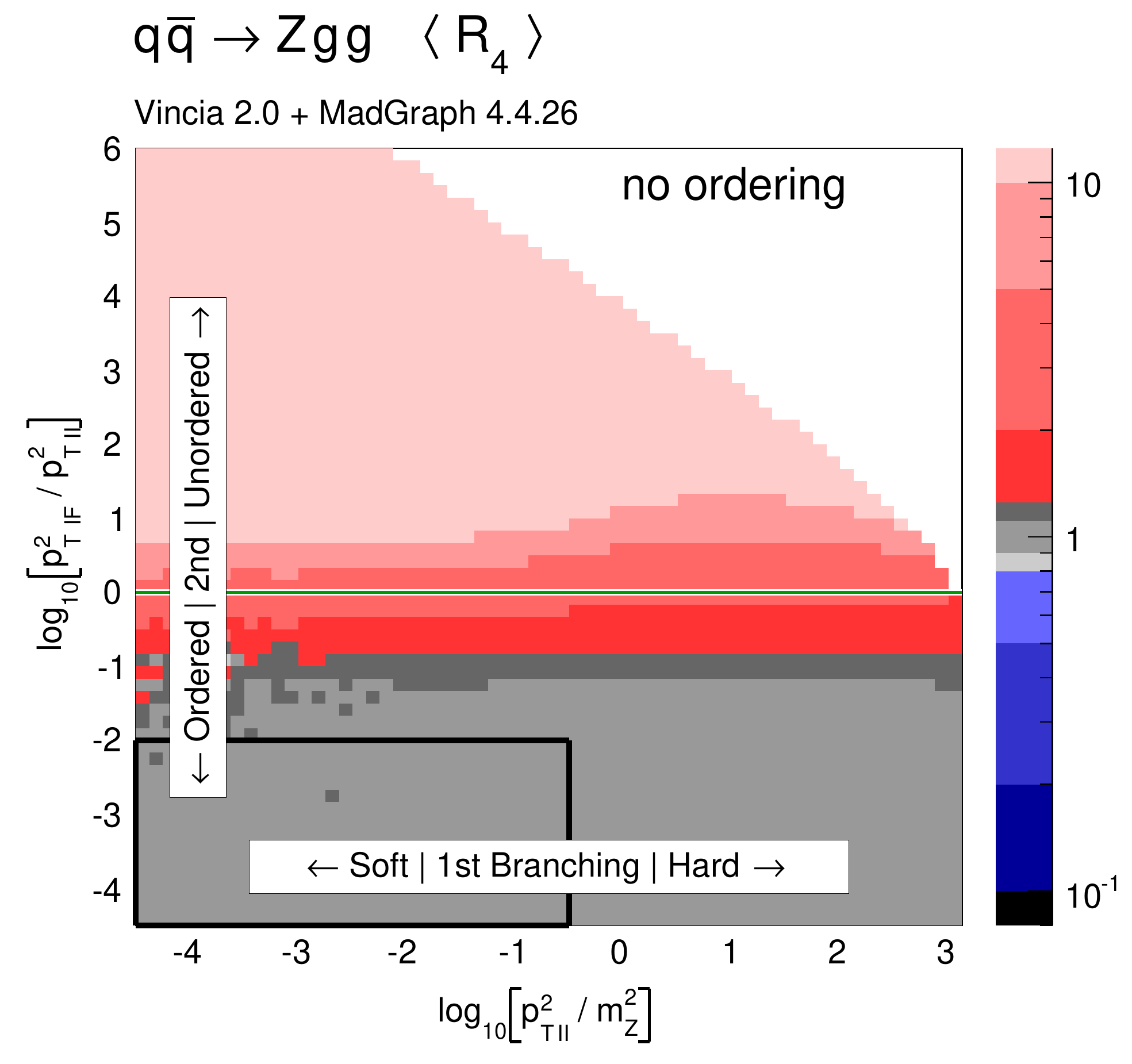}
\includegraphics[width=.496\textwidth]{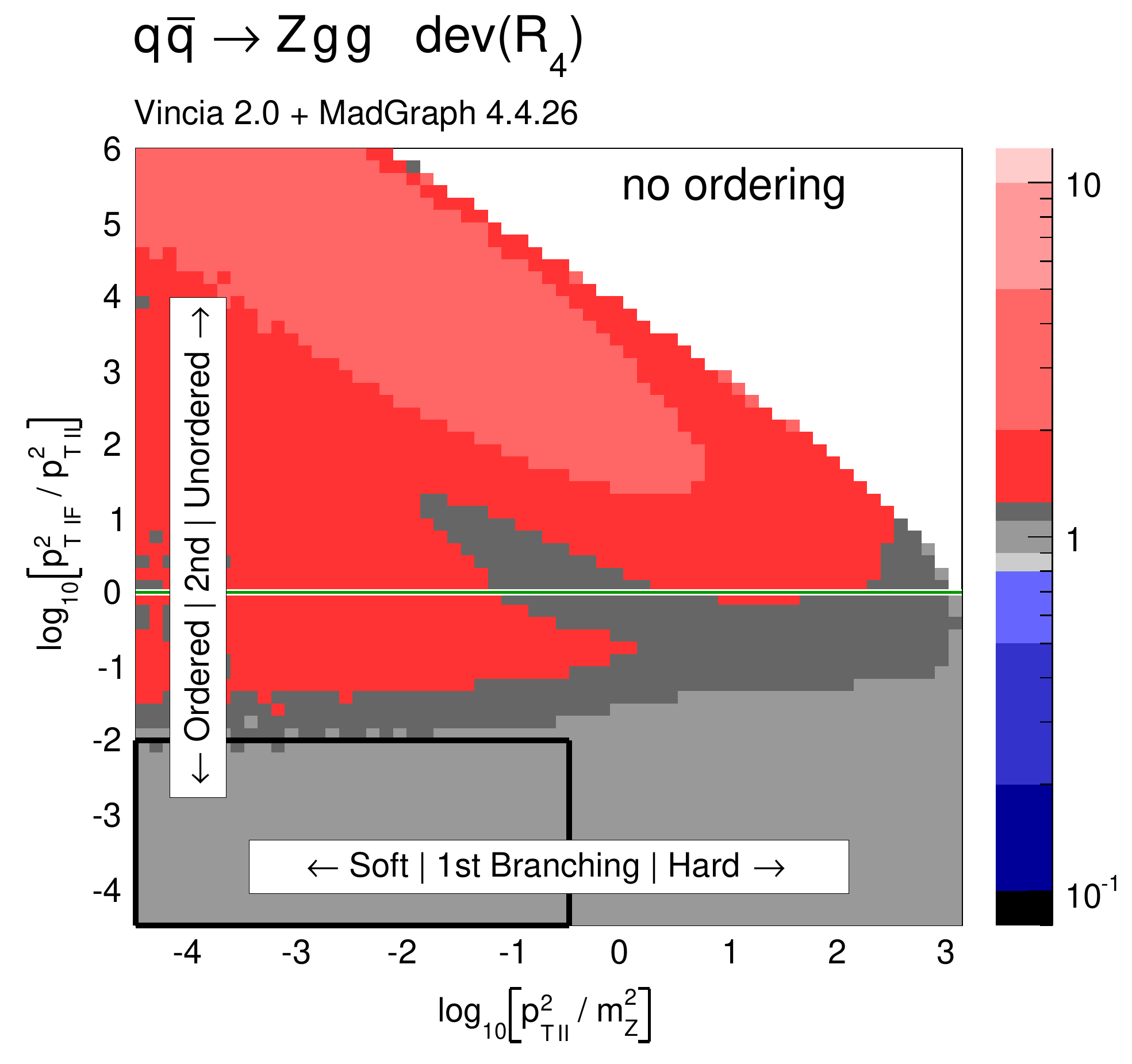}
\caption{\label{fig:PSME2DnoOrd} The value of $\langle R_4 \rangle$ (left)
and $\t{dev}(R_4)$ (right), differentially over the 4-parton phase space, with 
$p_\perp^2$ ratios characterizing the first and second emissions on the $x$- and $y$ 
axis, respectively. No ordering in the shower, with gluon emission only.}
\end{figure}

To gain an understanding of where in phase space significant deviations between the shower
approximation and the LO amplitudes squared occur, we consider the 2D distributions presented 
in \figsRef{fig:PSME2Dstrong} and \ref{fig:PSME2DnoOrd}. For all plots, the 
$x$ axes represent the degree of ordering of the first ($Z\to Zg$) emission, while the $y$ 
axis represents the degree of ordering of the second ($Zg \to  Zgg$) emission, defined more
precisely below. Note that, since the phase spaces have more than 2 dimensions, each bin still
represents an average of different phase-space points with the same $x$ and $y$ coordinates. 
Since the ratios on the axis are plotted logarithmically, zero denotes the border between 
ordered and unordered paths. The black-framed box in the lower left-hand corner of the 
plots highlights the strongly ordered region defined by 
$p_{\perp\,\t{IF}}^2 \ll p^2_{\perp\,\t{II}} \ll m^2_Z$, in which any (coherent) 
LL shower approximation is expected to give reasonable results.
In the left-hand panes, grey colours signify less than 20\% deviation 
from a ratio unity (with the middle shade corresponding to less than 10\% deviation, 
corresponding to near-perfect agreement). Red shades signify increasingly large deviations, 
with contours at $2$, $5$, and $10$. Blue contours extend to $1/2$, $1/5$, and $1/10$, 
while black indicates regions where the shower answer is less than one tenth of the 
matrix-element answer. In the right-hand panes, the same colour scale is used to show a 
measure of the width of the $R_4$ distribution in each bin, defined below. These plots
are intended to ensure that an average good agreement in the left-hand 
pane is not merely accidental, but also corresponds to a narrow distribution. 

In \figRef{fig:PSME2Dstrong}, the left-hand pane provides a clear illustration of the dead 
zone for the process $q\bar q\to Zgg$ 
in a strongly $p_\perp^2$-ordered antenna shower. Each bin of the two-dimensional 
histogram shows the average of the value of $R_4$ in eq.~\eqref{eq:R4} over all phase-space 
points populating that bin. For every phase-space point there are two possible
(not necessarily ordered) shower histories, with different scales for the first branching, 
$p_{\perp\,\t{II}}^2$, and second one, $p_{\perp\,\t{IF}}^2$. The combination of scales that
correspond to the path with the smaller scale of the second branching is used to characterize the
phase space point. 
The black region in \figRef{fig:PSME2Dstrong} for strong ordering corresponds to the spike in
\figRef{fig:PSMEstrong}. Since there are two shower histories, there is in principle the possibility
that the second history (which was not used to characterize the phase-space point) contributes
as an ordered history, but this does not appear to happen anywhere in the region classified as
unordered. 
The plot on the right shows the deviation within each bin, which we define to be
\begin{align}
  \t{dev}(R_4)=10^{\sqrt{\langle\t{log}_{10}^2(R_4)\rangle-
  \langle\t{log}_{10}(R_4)\rangle^2}}~,
\end{align}
since the distribution of $R_4$ is naturally a logarithmic one. We assign a deviation of
$10$ to the dead zone, since the log would otherwise not be defined. As mentioned above, the 
deviation is intended to illustrate whether an average value in the left plot is achieved by a 
broad or a narrow distribution.

One could force the dead zone to disappear by simply removing the ordering condition and 
starting the shower at the phase space maximum for each antenna.
However, as can be seen from \figRef{fig:PSME2DnoOrd}, this would highly overcount the matrix 
element in the unordered region, again parallelling the observations  for the equivalent 
case of final-state radiation in \cite{Giele:2011cb}. 
The strong-ordering condition is clearly a better approximation to QCD, even if it does not fill 
all of phase space. To improve the shower, we will therefore need to allow the shower to access 
the whole phase space while suppressing the overcounting in the unordered region.

\subsection{Smooth Ordering compared with Tree-Level Matrix Elements \label{sec:Impr24Branch}}

As we saw in the previous section, a strongly ordered shower has a significant dead zone for 
hard emissions, especially in the initial-state sector. We now want to focus on how to remove
them by generalising \vc's ``smooth ordering''~\cite{Giele:2011cb} to
initial-state phase spaces. Ref.~\cite{Giele:2011cb} shows that replacing 
the step function of an ordered shower with
a smooth suppression factor leads to a surprisingly good description of the unordered region in 
$Z$ decay. Based on this study, an improved version of the shower accept probability in
eq.~\eqref{eq:acceptProbShower}, which allows to take ``unordered'' branchings into account is 
\begin{align}
  \mc O(\hat t,t)\,P^\t{shower}=P_\t{imp}\,P^\t{shower}=
  \frac{\hat t}{\hat t + t}\,P^\t{shower}~,
  \label{eq:Pimp}
\end{align}
where $t$ is the scale of the trial branching at hand and $\hat t$ is the reference
scale.

\begin{figure}[t]
\centering
\begin{subfigure}[b]{0.5\textwidth}\includegraphics*[scale=0.436]{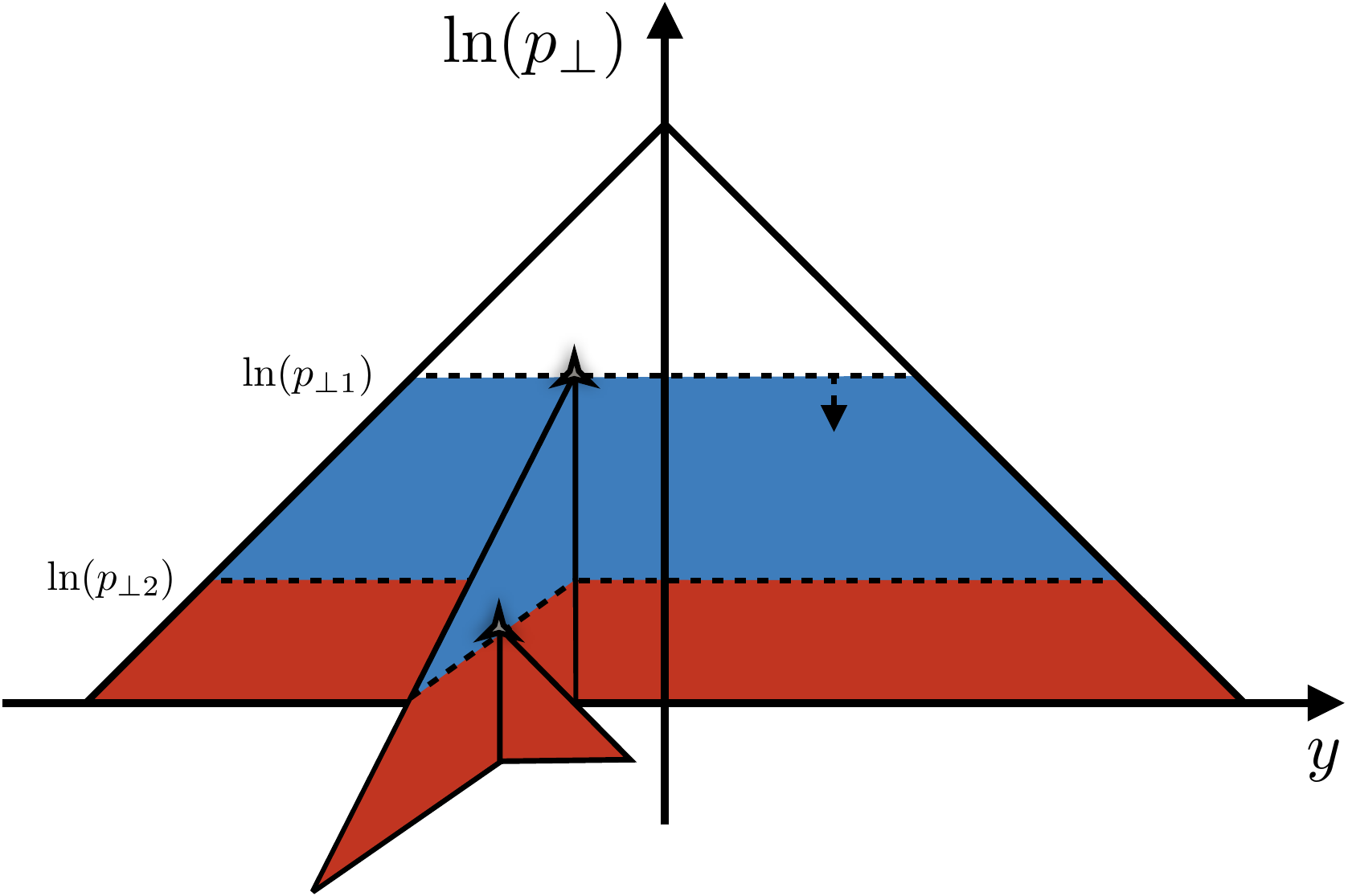}
\caption{Strong Ordering\label{fig:origamiStrong}}
\end{subfigure}%
\begin{subfigure}[b]{0.5\textwidth}\hspace*{1mm}\includegraphics*[scale=0.436]{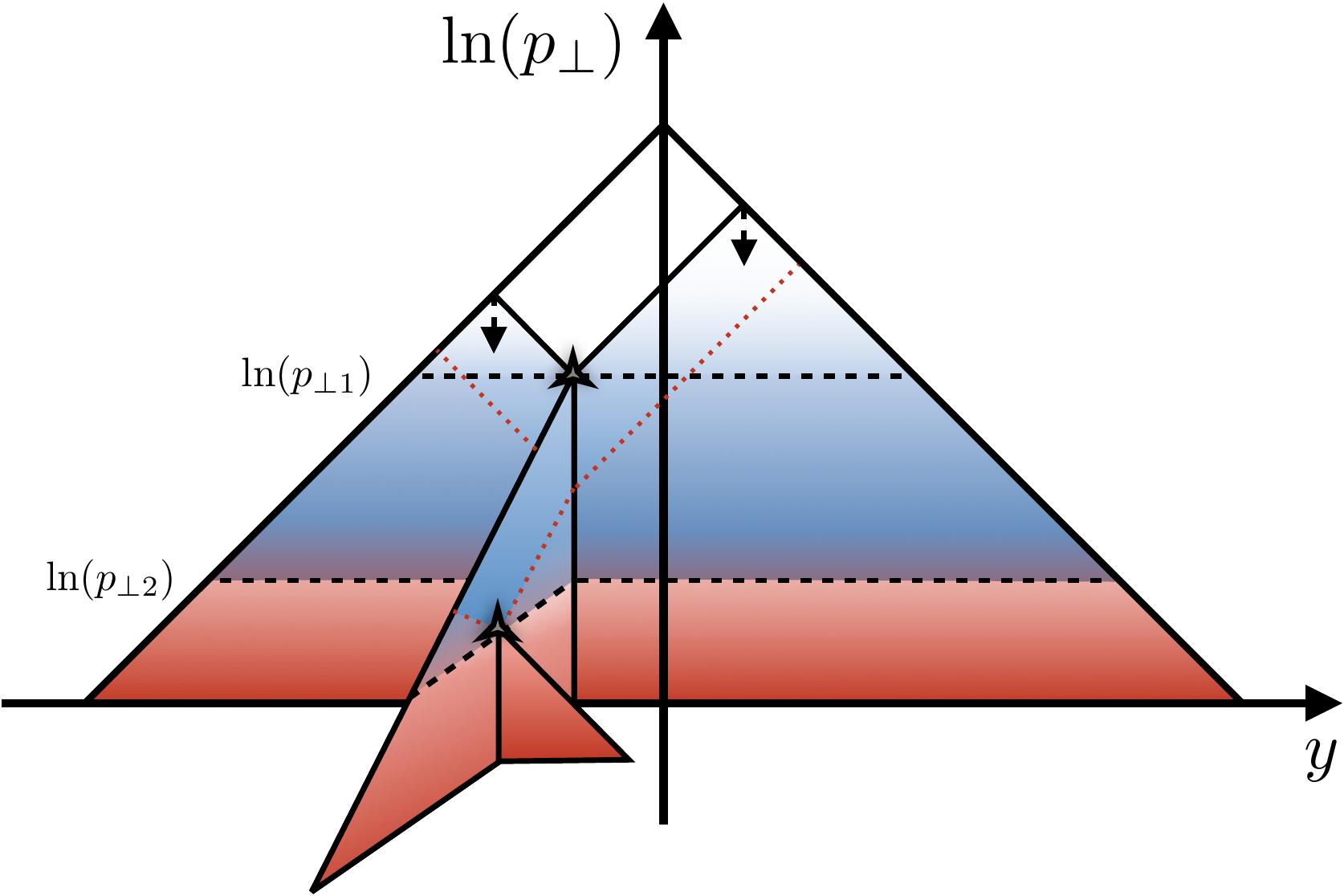}
\caption{Smooth Ordering\label{fig:origamiSmooth}}
\end{subfigure}
\caption{Illustration of the phase-space coverage of $p_\perp$-ordered dipole/antenna showers with (a)~strong and (b)~smooth ordering, in the ``origami'' plane of $\ln p_\perp$ vs rapidity.
\label{fig:origami}}
\end{figure}

The difference between conventional strong ordering and \vc's $P_\mrm{imp}$-suppressed smooth ordering can be illustrated by considering  so-called origami diagrams~\cite{Andersson:1990dp,Gustafson:1990qi,Andersson:1995jv}, in which the antenna (or, equivalently, dipole) phase space is depicted in terms of  $\ln (p_\perp^2)$ versus rapidity. Defining these by our gluon-emission evolution variable, $p_\perp^2 = m^2_{12} m^2_{23}/m^2$ and by $y = \frac12 \ln(m^2_{12}/m^2_{23})$ respectively, for an antenna with total invariant mass $m$ splitting into two smaller antennae with masses $m_{12}$ and $m_{23}$, the leading (double-logarithmic) contribution to the branching probability is transformed to just a constant over the antenna phase space,
\begin{equation}
\d P \sim \frac{\mc C \alpha_s}{2\pi} \, \mathrm{d}\ln p_\perp^2 \, \mathrm{d}y~,
\end{equation}
where $\mc C$ is the colour factor normalised so that $\mc C\to N_C$ in the leading-colour limit.
The phase-space boundary for gluon emissions with $p_\perp \ll m$ is determined by
$y_\mathrm{max} (p_\perp) = \frac12\ln(m^2/p_\perp^2)$, 
so that the rapidity range available for emissions at a given $p_\perp$ defines a triangular region,
\begin{equation}
\Delta y (p_\perp)~=~
\ln(m^2/p_\perp^2)
~=~
\ln(m^2) - \ln(p_\perp^2)~,
\label{eq:yRange}
\end{equation}
corresponding to the outer hulls of the diagrams shown in \figRef{fig:origami}. 
For an emission at any given value of $p_{\perp 1}^2 = m_{12}^2 m_{23}^2 / m^2$, the total rapidity range (at that $p_\perp$ value) is unchanged by the branching,
\begin{equation}
\Delta y(p_{\perp 1}) ~=~ \underbrace{\ln(m^2) - \ln(p_{\perp 1}^2)}_{\mbox{pre-branching}}
~=~ \underbrace{ \ln(m_{12}^2) - \ln(p_{\perp 1}^2) + \ln(m_{23}^2) - \ln(p_{\perp 1}^2)}_{\mbox{post-branching}}~,
\end{equation} 
cf.~the dashed line at $\ln(p_\perp) = \ln(p_{\perp 1})$ in the
figure. For soft emissions, however, say at a reference value of
$p_{\perp} = 1\,\mathrm{GeV}$, the post-branching configuration covers
a total rapidity range which is larger by,
\begin{equation}
\Delta y (1\,\mrm{GeV})_\mrm{post} - \Delta y (1\,\mrm{GeV})_\mrm{pre}~=~ 
\ln(m_{12}^2) + \ln(m_{23}^2) - \ln(m^2)~=~\ln(p_\perp^2)~.
\end{equation}
The additional phase space ``opened up'' by the branching can hence be
represented by adding a double-sided isoceles right triangle to the
origami diagram, with side lengths $\ln(p_{\perp 1})$, which --- for
lack of a better direction --- is drawn pointing out of the original
plane. Restricting the subsequent shower evolution to populate only
the region below the $p_{\perp 1}$ scale produces a strongly ordered
shower, illustrated in \figRef{fig:origamiStrong} with the blue and
red shaded regions representing the phase space accessible to a second
and third branching, respectively. The case of smooth ordering is
illustrated in \figRef{fig:origamiSmooth} for the same sequence of
branchings. In this case, each of the antennae produced by the first
branching are allowed to evolve over their full phase spaces, and
their respective full phase-space triangles are therefore now included
in the diagram, using solid black lines for the first branching and
red dotted lines for the phase-space limits after the second
branching. The suppression of the branching probability near and above
the branching scale is illustrated by reducing the amount of shading
of the corresponding regions. Comparing the figures, one can see that
we expect no change  in the total range or integrated rate of soft
emissions (at the bottom of the diagrams). The only effects occur near
and above the branching scale where the strongly ordered (LL) shower
formalism is anyway unpredictive. In \secRef{sec:SmoothVsStrong}
below, we show explicitly that the leading-logarithmic
structure of smoothly-ordered showers is identical to that of
strongly ordered ones, but for the remainder of this section we
constrain our attention to comparisons with fixed-order matrix
elements.  

A further point that must be addressed in the context of the ordering criterion is that our matrix-element-correction formalism, discussed below, requires a Markovian (history-independent) definition of the $\hat{t}$ variable in the $P_\mathrm{imp}$ factor in  \eqRef{eq:Pimp}. Rather than using the scale of the preceding branching directly (which depends on the shower path and hence would be history-dependent), we therefore compute this scale in a Markovian way as follows:
Given a $n$-parton state we determine the values of the evolution variable corresponding
to all branchings the shower could have performed to get from any $(n-1)$- to the given
$n$-parton state. The reference scale $\hat t$ is then taken as the minimum of those scales. 
The dead zone, equivalent to the unordered region, is now populated by allowing 
branchings of a restricted set of antennae to govern the full relevant phase space.
Such antennae are are called unordered, while other antennae are called ordered. It
is in principle permissible to treat all antennae in an event as unordered. To
mimic the structure of effective $2\to 4$ and higher branchings, we however only tag
those antennae which are connected to partons that partook in the branching that gave 
rise to the chosen value for $\hat t$ as unordered. Branchings of ordered antennae may
then contribute below the scale $\hat t$.

For example, consider the case of a gluon emission being associated with  
the smallest value of the evolution variable. In this case the gluon as well as the two 
partons playing the role of the parent antenna that emitted the gluon, are marked for
unordering and therefore all antennae in which these three partons participate are
allowed to restart the evolution at their phase space limits. This limited unordering 
reflects that no genuinely new region of phase space would be opened up by allowing 
partons/antennae completely unrelated to the ``last branching'' to be unordered, as 
these will already have explored their full accessible phase spaces during the prior 
evolution. 

We note that for the final state the available phase space reduces for each successive 
branching, limiting the effect of the smooth ordering. In~\cite{Hartgring:2013jma} it
is shown, that for final-state radiation, the damping factor in \eqRef{eq:Pimp}
does not modify the LL $1/t$ behaviour and only generates explicitly subleading 
$\hat{t}/t^2$ corrections in the strongly unordered limit, $t \gg \hat{t}$.
For the initial state, the phase space boundaries are governed by the hadronic 
centre-of-mass energy leading to possibly large unordered regions and therefore 
a rather large effect of the smooth ordering. As the main purpose of the smooth 
ordering is to fill the all available phase space for the MECs, we restrict it 
to the ME corrected branchings by default and keep all following shower emissions 
strongly ordered. In this case, all damping factors get replaced by the MEC weight,
see \secRef{sec:MG4MECs}, by virtue of the Sudakov veto algorithm.

\begin{figure}[tbp]
\centering
\includegraphics[width=.9\textwidth]{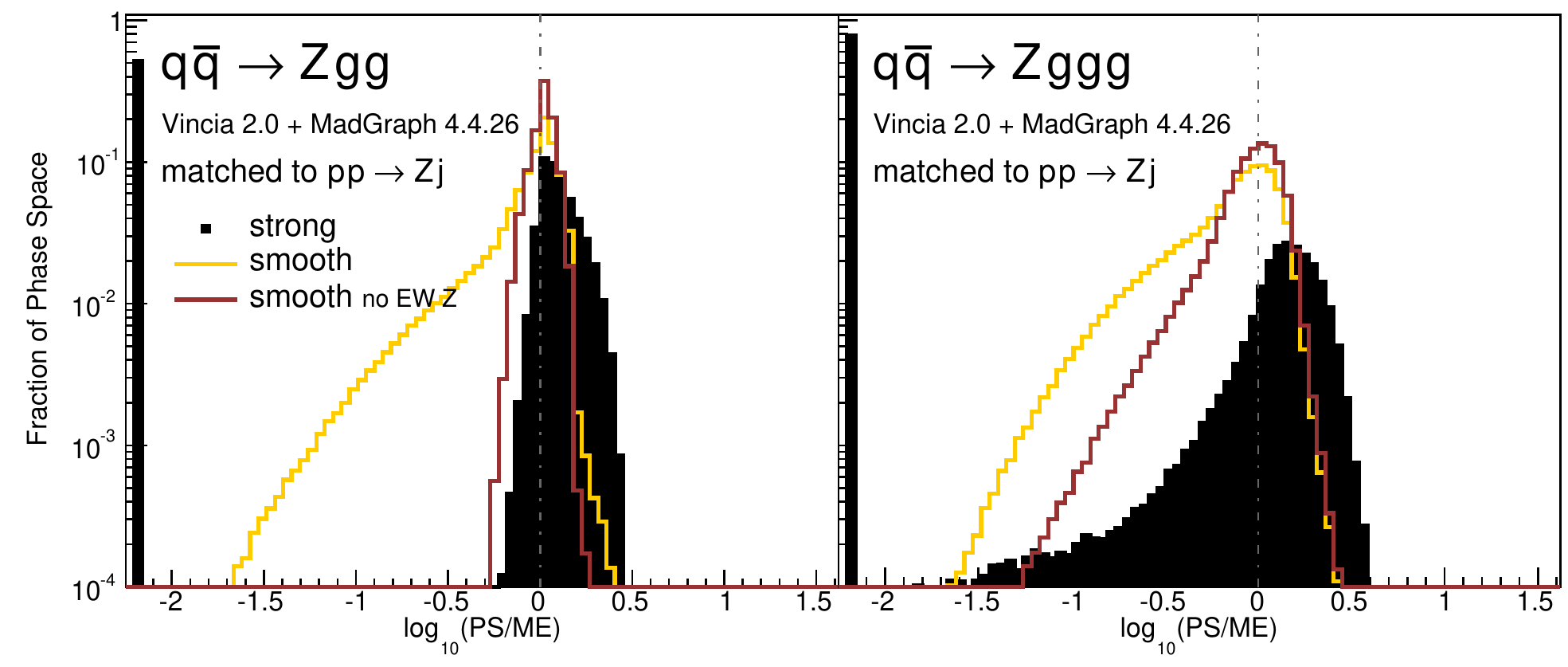}
\caption{\label{fig:PSME2} Antenna shower, compared to matrix elements: distribution
of $\t{log}_{10}(\t{PS}/\t{ME})$ in a flat phase space scan of the full phase space
with strong and smooth ordering and smooth ordering with a cut on $m_{\perp\,Z}^2$.
Contents normalized to the number of generated points. Gluon emission only.}
\end{figure}

We compare the logarithmic distributions of the ratio of the shower approximation to the
matrix element for $q\bar q\to Zgg$ and $q\bar q\to Zggg$ for both strong and smooth 
ordering in \figRef{fig:PSME2}.
When applying smooth ordering, the distribution 
gets narrower on the side where the shower overcounts the tree-level matrix element, and 
that the dead-zone spike is replaced by an extended tail towards low ratios on the other 
side. This tail is due to configurations that look like a hard-QCD process accompanied 
with a radiated $Z$. Such phase-space points should in principle be
populated by an electroweak shower, such as the one presented in \cite{Christiansen:2014kba}; 
not having developed the required formalism in the antenna context yet, however, we still 
allow our QCD shower to populate this region of phase space; it will in any case be corrected 
with matrix elements, see 
\secRef{sec:MECs}. To focus on the improvement in the QCD regions of phase
space we apply a cut on the transverse mass of the $Z$ boson and require it
to be larger than the branching scale of the path that has been chosen to
characterize the phase space point,
\begin{align}
  \label{eq:EWcut1}
  p_{\perp\,\t{IF}}^2 < m_{\perp\,Z}^2=k_{\perp\,Z}^2+m_Z^2~.
\end{align}
We define $k_{\perp\,Z}^2$ to be the minimum of all possible
\begin{align}
  \label{eq:EWcut2}
  k_{\perp\,Z\,q}^2= \t{min}(E_Z^2,E_q^2)(1-\cos\theta_{Zq})~.
\end{align}
The resulting distributions are shown in red in \figRef{fig:PSME2}.
Applying the cut leads to a removal of the part of phase space where the $Z$ should
have been generated as an emission rather than as part of the hard process. 
The distribution is now dominated by QCD and the smoothly ordered shower produces
a narrower as well as more symmetric distribution, compared to the strongly ordered
shower.

\begin{figure}[tp]
\centering
\includegraphics[width=.496\textwidth]{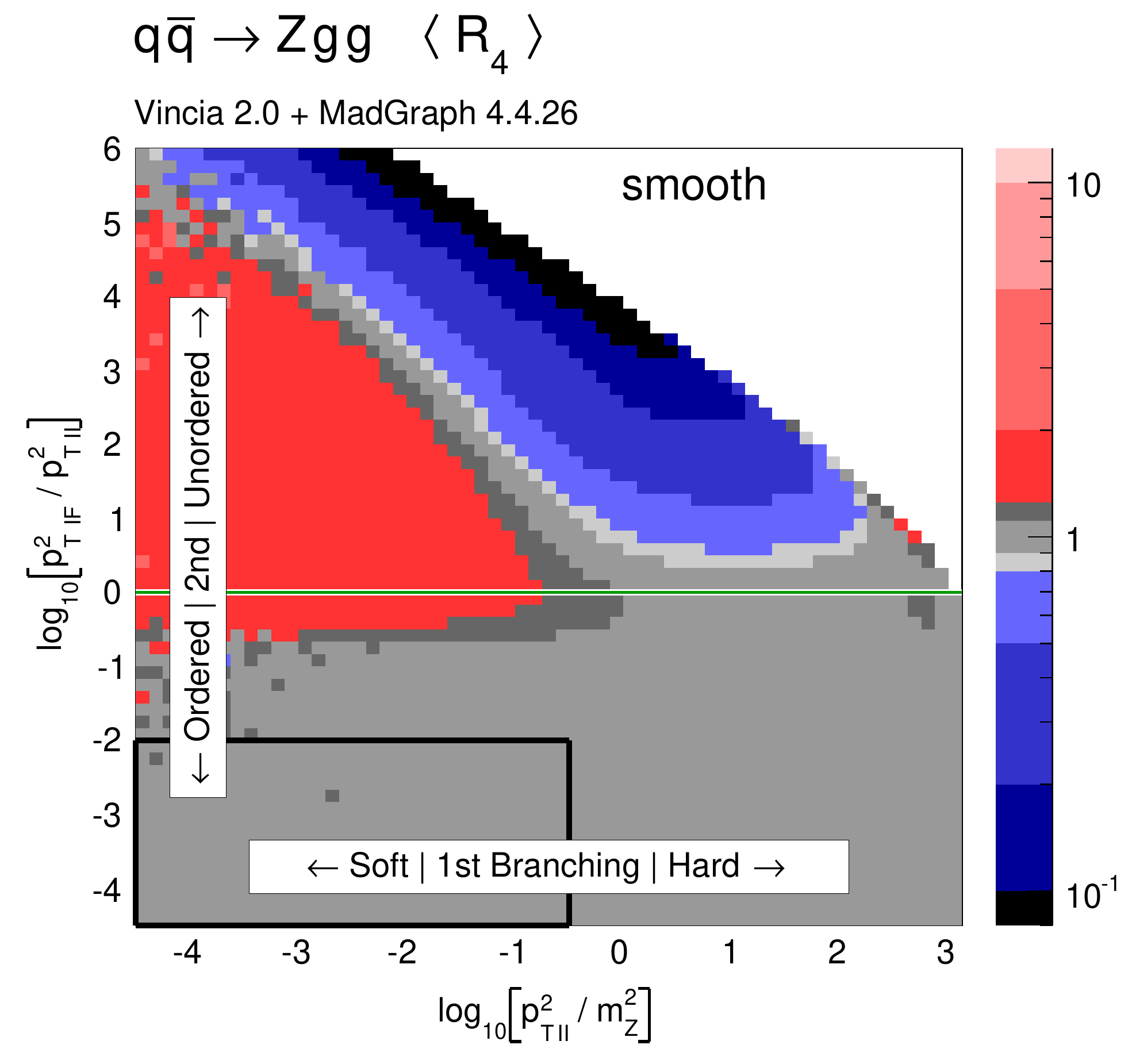}
\includegraphics[width=.496\textwidth]{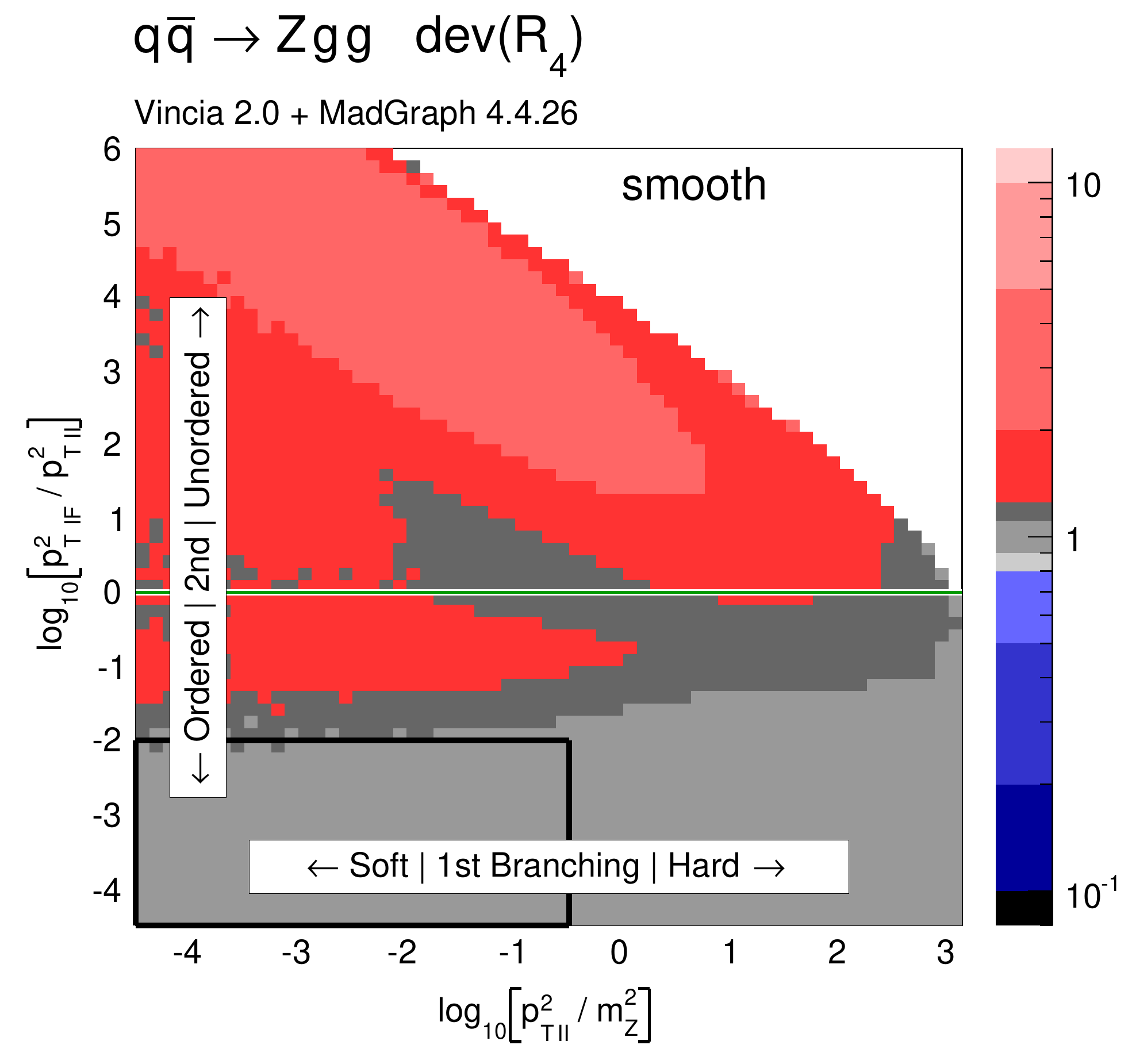}
\caption{\label{fig:PSME2Dsmooth} The value of $\langle R_4 \rangle$ (left)
and $\t{dev}(R_4)$ (right), differentially over the 4-parton phase space, with 
$p_\perp^2$ ratios characterising the first and second emissions on the $x$ and $y$ 
axis, respectively. Smooth ordering in the shower, with gluon emission only.} \vspace*{5mm}
\includegraphics[width=.496\textwidth]{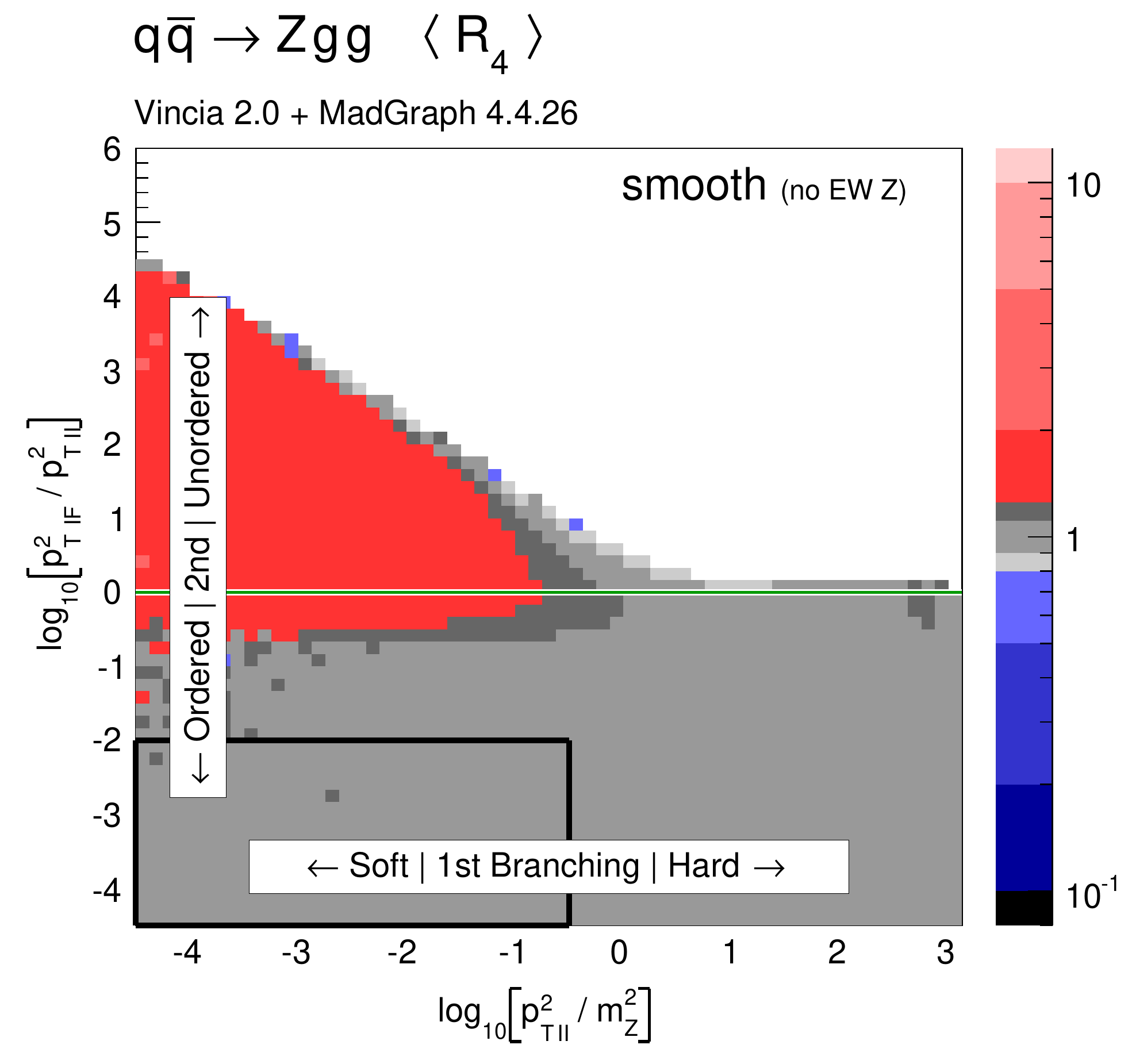}
\includegraphics[width=.496\textwidth]{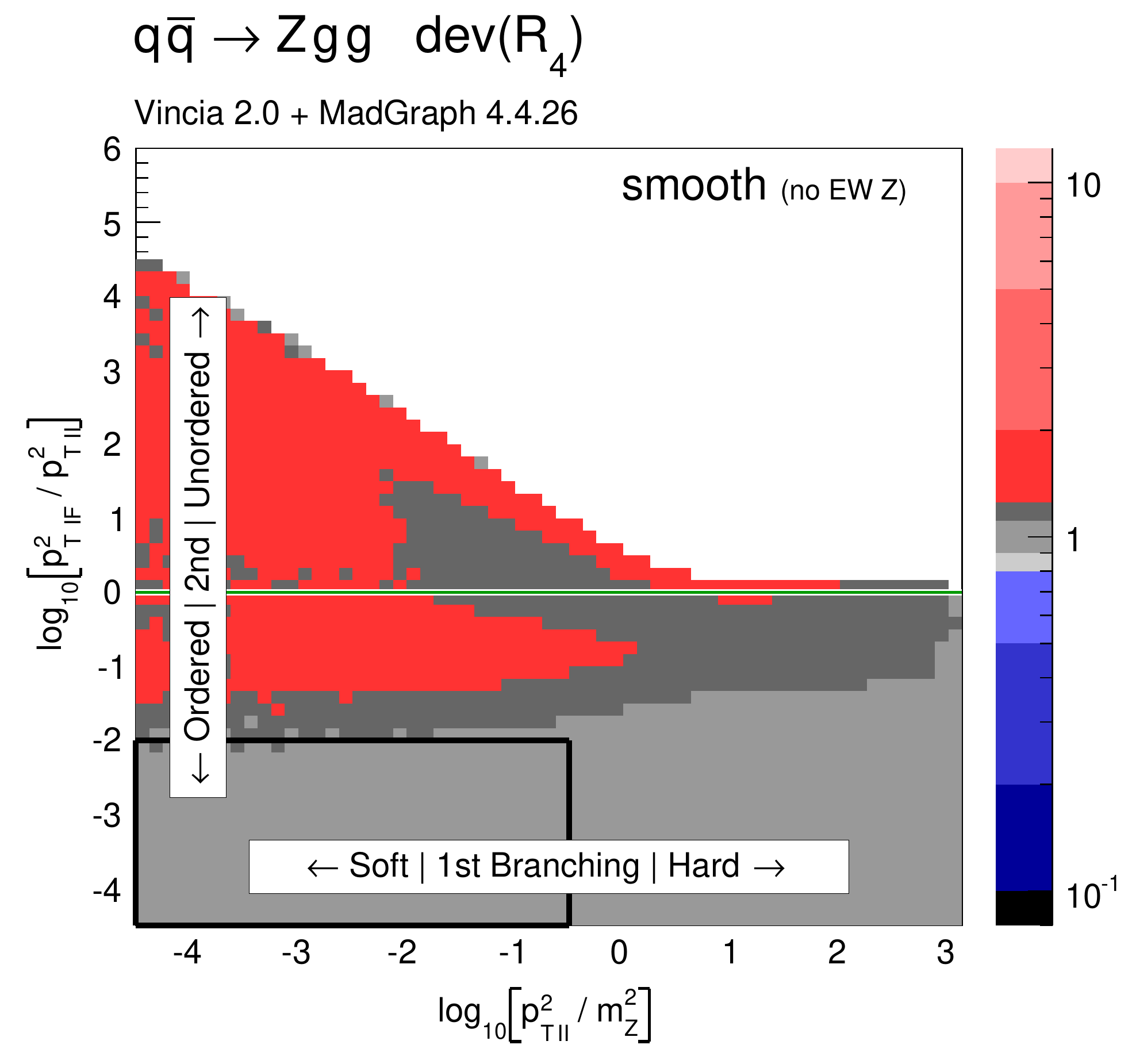}
\caption{\label{fig:PSME2DsmoothNoEWZ} The value of $\langle R_4 \rangle$ (left)
and $\t{dev}(R_4)$ (right), differentially over the 4-parton phase space, with 
$p_\perp^2$ ratios characterizing the first and second emissions on the $x$ and $y$ 
axis, respectively. Smooth ordering in the shower, with a cut on $m_{\perp\,Z}^2$ and
gluon emission only.}
\end{figure}

Similarly we repeat the two-dimensional histograms for the smoothly-ordered antenna 
shower in \figRef{fig:PSME2Dsmooth} without and in~\figRef{fig:PSME2DsmoothNoEWZ} 
with the cut on $m_{\perp\,Z}^2$. As expected, we obtain an improved description as compared 
to both the strong and unordered showers, \figsRef{fig:PSME2Dstrong} 
and~\ref{fig:PSME2DnoOrd} respectively.
Due to the form of the improvement factor in \eqRef{eq:Pimp} we get a factor
of $0.5$ at the green line, around where the scales of the two branchings coincide, 
leading to a better description already of this region. Once again these plots show that 
the shower undercounts the region where the $Z$ boson is very soft and should have been 
generated with a weak shower, representing a path that is not available in \vc\ yet. The 
strongly unordered region remains somewhat overcounted, though by less than a factor 2, 
far better and with narrower distributions than was the case for the fully unordered shower,
\figRef{fig:PSME2DnoOrd}.

An extended set of plots, including Higgs production processes, can be found
in \appRef{app:MEplots}.

\subsection{Smooth Ordering vs. Strong Ordering \label{sec:SmoothVsStrong}}

This section presents a comparison of strong and smooth ordering,
first in terms of their analytical leading-logarithmic structures, and
then using jet clustering 
scales, investigating the processes $e^+e^-\to$ jets as well as $pp\to Z+$jets. The 
analyses are adapted from the code used in~\cite{Hoche:2015sya}, originally written by
S.~H{\"o}che.
In order to focus on the shower properties we present parton-level
distributions, with MECs switched off, a fixed strong coupling with
$\alpha_s(m_Z)=0.13$, and a very low cutoff, $10^{-3}~\t{GeV}$ for
$e^+e^-\to$ jets and $10^{-2}~\t{GeV}$ for $pp\to Z+$jets. To furthermore put the magnitude of the differences between smooth and strong ordering into perspective, an $\alpha_s(m_Z)$-variation band for the strongly ordered result is included in figures \ref{fig:jetResZ} and \ref{fig:jetResZprod}. 

We emphasise that, even leaving the $\alpha_s$ and cutoff settings
aside, the distributions in this section are meant for validation only. The event generation modus used below does not make use of \vc's matrix-element correction features. When using MECs, the main purpose of the smooth ordering is to fill the available phase space with non-vanishing weight, which allows a reweighting to reproduce the correct LO matrix-element result. Keeping this disclaimer in mind, it is still useful to investigate how the phase space is filled before MECs are applied.

\subsubsection*{Leading Logarithms}
As discussed in the preceding section, the leading (double-pole) behaviour of the
gluon-emission antenna functions is just a constant over phase
space when expressed in terms of the origami variables $\ln(p_\perp)$
and $y$. We begin by considering a conventional strongly-ordered antenna shower, such
as that of \ari~\cite{Gustafson:1987rq,Lonnblad:1992tz} (or \vc\ with
strong ordering). 
The leading contribution to the Sudakov factor $\Delta (Q_\perp^2, p_\perp^2)$
representing the no-branching probability between two resolution
scales $Q_\perp^2 > p_\perp^2$ (e.g., following a preceding branching
which happened at the scale $Q_\perp$), is then, cf.~\eqRef{eq:yRange},
\begin{equation}
-\ln \Delta_\mathrm{strong}  ~\stackrel{\mathrm{LL}}{\sim}~ \frac{{\cal C} \alpha_s}{2\pi}
\int_{\ln p_\perp^2}^{\ln Q_\perp^2} \mathrm{d}\ln q_\perp^2 \int_{-\ln(m/q_\perp)}^{\ln(m/q_\perp)} \mathrm{d}y 
~ = ~ \frac{{\cal C} \alpha_s}{2\pi}
\int_{\ln p_\perp^2}^{\ln Q_\perp^2} \mathrm{d}\ln q_\perp^2  
\ln\left[\frac{m^2}{q_\perp^2}\right] 
\label{eq:lnDeltaStrong}
\end{equation}
\begin{equation}
 = ~ \frac{{\cal C} \alpha_s}{2\pi} \left( \frac12 \ln^2
      \left[\frac{Q_\perp^2}{p_\perp^2}\right] +
      \ln\left[\frac{Q_\perp^2}{p_\perp^2}\right]\ln\left[\frac{m^2}{Q_\perp^2}\right]\right)~, \label{eq:LLstrong}
\end{equation}
for a final-final antenna\footnote{For initial-initial antennae, replace $m$ in the phase-space limit on the rapidity integral in \eqRef{eq:lnDeltaStrong} by $\sqrt{s}=\sqrt{s_{AB}/(x_A x_B)}$, assuming $x_Ax_B \ll 1$. For initial-final antennae, replace it by $\sqrt{s_{AK}/x_A}$ assuming $x_A\ll 1$.} with invariant mass $m$ and assuming $p_\perp^2\ll 
m^2$. This agrees with the LL limit for dipole showers derived in
\cite{Hoche:2015sya}. 
We note that the second term is absent from 
\cite[eq.~(8)]{Bellm:2016rhh} due to a phase-space restriction 
placed in eq.~(2) of that paper, which we believe is appropriate to
remove double-counting of soft emissions in showers based  on DGLAP
kernels. In the context
of antenna showers however, the antenna functions already have the
correct (eikonal) soft limits, and the
imposition of this additional phase-space constraint would have the
(undesired) effect of removing the added rapidity range corresponding to the extra origami
fold discussed in \secRef{sec:Impr24Branch}, producing an 
``undercounting'' of soft emissions. We therefore regard the
expression above, \eqRef{eq:LLstrong}, as the reference expression
which an LL-correct antenna shower should reproduce.

A counter-example, illustrating an incorrect LL behaviour, can be furnished
by considering a so-called ``power shower''~\cite{Plehn:2005cq} in
which the upper boundary of the integral above is replaced by $m^2$ rather
than $Q_\perp^2$ (e.g., letting newly created antennae evolve over
their full phase spaces, irrespective of the ordering scale, and
without any suppression). This produces an extra logarithm which
is not present in the strongly ordered case:
\begin{equation}
-\ln \Delta_{\mathrm{pwr}}  \stackrel{\mathrm{LL}}{\sim}
 \frac{{\cal C} \alpha_s}{2\pi} \left(
  \frac12 \ln^2 \left[\frac{Q_\perp^2}{p_\perp^2}\right] 
+ \ln\left[\frac{Q_\perp^2}{p_\perp^2}\right]\ln\left[\frac{m^2}{Q_\perp^2}\right]
 + \textcolor{red}{\frac12 \ln^2 \left[\frac{m^2}{Q_\perp^2}\right]}\right)~, \label{eq:LLpower}
\end{equation}
where we have rewritten the $\frac12\ln^2(m^2/p_\perp^2)$ result to make the two
first terms identical to the ones produced in the strongly ordered
case, so that the third term, highlighted in red,  represents the
difference. 

For smooth ordering, with the 
$P_\mathrm{imp}$ suppression factor defined in \eqRef{eq:Pimp}, the relevant integral
is:
\begin{equation}
\int_{p_\perp^2}^{m^2} \frac{1}{1+\frac{q_\perp^2}{Q_\perp^2}}
  \frac{\mathrm{d}q_\perp^2}{q_\perp^2} \ln \left[\frac{ m^2}{q_\perp^2}\right]~,
\end{equation}
which after a bit of algebra can be cast in the following form:
\begin{equation}
\frac12 \ln^2\left[\frac{Q_\perp^2}{p_\perp^2}\right]  +  \ln
\left[\frac{Q_\perp^2}{p_\perp^2}\right] \ln \left[\frac{m^2}{Q_\perp^2}\right] 
+ \ln\left[1+\frac{p_\perp^2}{Q_\perp^2}\right]\ln\left[\frac{m^2}{p_\perp^2}\right] - \mrm{Li}_2\left[
\frac{-Q_\perp^2}{m^2}\right] - \mrm{Li}_2\left[\frac{-p_\perp^2}{Q_\perp^2}\right] - \frac{\pi^2}{6},
\end{equation}
where the two first terms are again identical to those of
\eqRef{eq:LLstrong}. In the third term, $\ln(1+p_\perp^2/Q_\perp^2)
\to 0$ for $p_\perp^2/Q_\perp^2 \to 0$, and the fourth and fifth
terms are bounded by $-\pi^2/12<\mrm{Li}_2(-x)<0$ (with $0$
corresponding to the limit $x \to 0$ and $-\pi^2/12$
for $x \to 1$). We  thus conclude that the LL properties of the 
antenna shower are not spoiled by changing from strong to smooth
ordering.

\subsubsection*{Hadronic $Z$ Decays}

\begin{figure}[tbp]
\centering
\begin{minipage}[t]{0.49\textwidth}
  \graphNoSpace{width=0.99\textwidth}{LL_JetRates_Zdecay/log10_y_mm1.pdf}
  \graphNoSpace{width=0.99\textwidth}{LL_JetRates_Zdecay/log10_y_mm1-ref1.pdf}
  \graphNoSpace{width=0.99\textwidth}{LL_JetRates_Zdecay/log10_y_mm1-ref2.pdf}
  \graphNoSpace{width=0.99\textwidth}{LL_JetRates_Zdecay/log10_y_mm1-ref3.pdf}
  \includegraphics[width=0.99\textwidth]{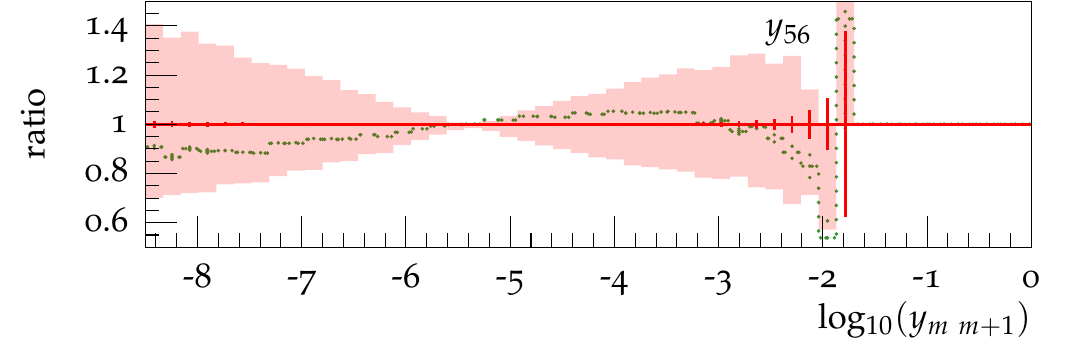}
\end{minipage}
\begin{minipage}[t]{0.49\textwidth}
  \graphNoSpace{width=0.99\textwidth}{LL_JetRates_Zdecay/log10_y_mm1_y_m1m.pdf}
  \graphNoSpace{width=0.99\textwidth}{LL_JetRates_Zdecay/log10_y_mm1_y_m1m-ref1.pdf}
  \graphNoSpace{width=0.99\textwidth}{LL_JetRates_Zdecay/log10_y_mm1_y_m1m-ref2.pdf}
  \includegraphics[width=0.99\textwidth]{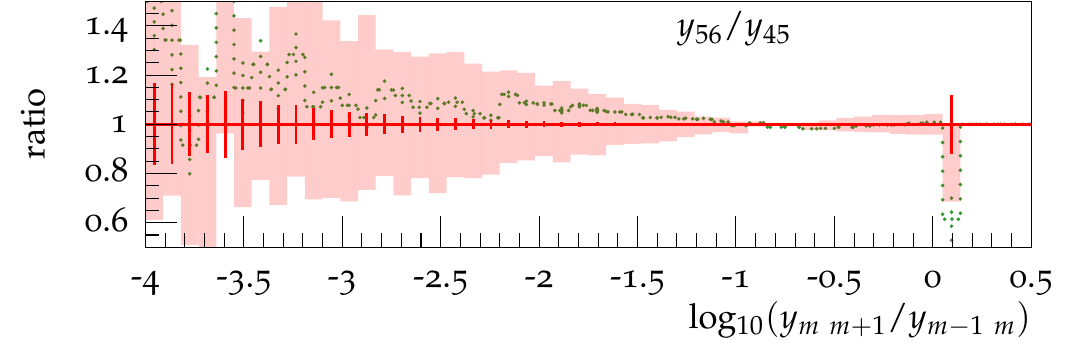}
\end{minipage}
\caption{\label{fig:jetResZ} 
Logarithmic distributions of differential jet resolutions and their ratios for heavy
$Z$ decays ($m_Z=1000~\t{GeV}$). Predictions of \vc~2.0 with strong (smooth) ordering
are shown in solid red (dotted green) lines. The red band shows an $\alpha_s$
variation with $\alpha_s(m_Z)=0.12$ and $\alpha_s(m_Z)=0.14$.}
\end{figure}

To increase the available phase space we used a heavy $Z$ with
$m_Z=1000~\t{GeV}$ which decays hadronically.
In \figRef{fig:jetResZ} we present the parton-level result
for four successive jet resolution measures, $y_{m\,m+1}$ (with 
$m\in\{2,3,4,5\}$), and their ratios $y_{m\,m+1}/y_{m-1\,m}$,
using the Durham jet algorithm.
Jet resolution scales exhibit a Sudakov
suppression for low values, and exhibit fixed-order 
behaviour for large values. We note that 
in realistic calculations (and in experimental data), low-scale values are typically strongly affected by hadronisation corrections, which are absent here since we are at parton level, with a fixed $\alpha_s$. We also exclude values of $y_{m\,m+1}$ corresponding to scales below the shower cutoff. 
Small values of the ratios $y_{m\,m+1}/y_{m-1\,m}$
highlight the modelling in the region of large scale separation, i.e.\ where 
effects of resummation become relevant. Large values of $y_{m\,m+1}/y_{m-1\,m}$ 
are associated with the region of validity of fixed-order calculations.

In the distributions of the jet resolution scales themselves we observe
moderate differences between the different ordering modes, up to 
$\mc O(20\%)$. Smooth ordering generates more events with larger $y_{m\,m+1}$
separation and, consequently, fewer events with small separation, compared
to strong ordering. 

While the prediction with smooth ordering lies below the strongly ordered
one for small values of the $y_{34}/y_{23}$ ratio, it eventually slightly exceeds
the strong ordering in the $y_{56}/y_{45}$ ratio.
This behaviour is a combination of two effects: Smooth ordering allows more 
phase-space coverage, while at the same time, the Markovian restart scale means 
that emissions from ``ordered" antennae have more stringent phase-space restrictions
than in the strongly ordered case. Thus, if more ordered antennae are present, which
is only the case after several branchings, the Markovian restarting scale may lead 
to a softer multi-emission pattern than in the strongly ordered case.
However, recall that MECs are an essential ingredient in the evolution, and that 
for emissions beyond the highest ME multiplicity, no smooth ordering is applied. 
This means that for lower multiplicities, the effect of smooth ordering is 
effectively removed and replaced by the full fixed-order result. For higher
multiplicities, the shape change due to the Markovian restart scale is also absent,
since smooth ordering is not applied. This suggests that smooth ordering of the 
entire cascade, and without MECs, exhibits some undesirable features. However, 
it is worth noting that the differences are largest in the soft region, where
non-perturbative physics and tuning are expected to have large impact, as 
e.g.\ exemplified by a large dependence on the value of $\alpha_s(m_Z)$.
Finally we note that the prediction with smooth ordering lie well within
the $\alpha_s(m_Z)$-variation band of the strong ordering.

\subsubsection*{Drell-Yan}

\begin{figure}[tbp]
\centering
\begin{minipage}[t]{0.49\textwidth}
  \graphNoSpace{width=0.99\textwidth}{PP_JetRates_Zprod/log10_d_mm1.pdf}
  \graphNoSpace{width=0.99\textwidth}{PP_JetRates_Zprod/log10_d_mm1-ref1.pdf}
  \graphNoSpace{width=0.99\textwidth}{PP_JetRates_Zprod/log10_d_mm1-ref2.pdf}
  \graphNoSpace{width=0.99\textwidth}{PP_JetRates_Zprod/log10_d_mm1-ref3.pdf}
  \includegraphics[width=0.99\textwidth]{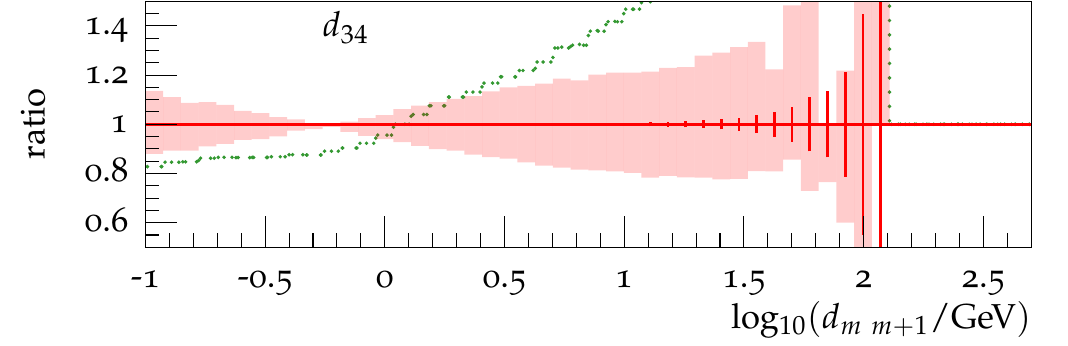}
\end{minipage}
\begin{minipage}[t]{0.49\textwidth}
  \graphNoSpace{width=0.99\textwidth}{PP_JetRates_Zprod/log10_d_mm1_d_m1m.pdf}
  \graphNoSpace{width=0.99\textwidth}{PP_JetRates_Zprod/log10_d_mm1_d_m1m-ref1.pdf}
  \graphNoSpace{width=0.99\textwidth}{PP_JetRates_Zprod/log10_d_mm1_d_m1m-ref2.pdf}
  \includegraphics[width=0.99\textwidth]{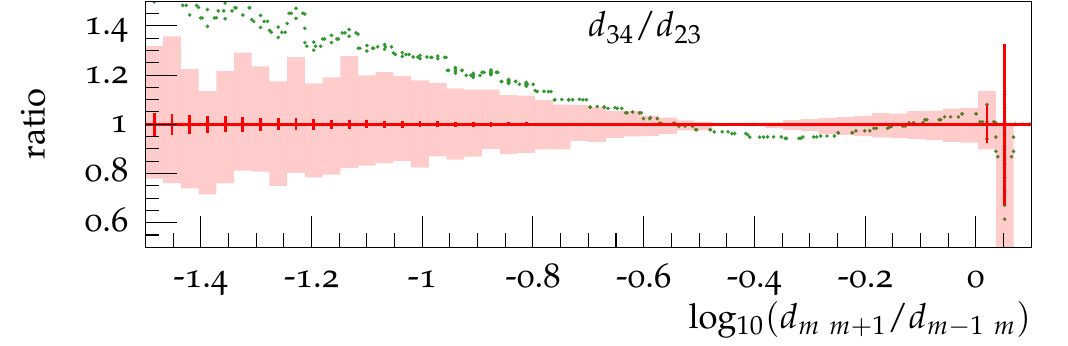}
\end{minipage}
\caption{\label{fig:jetResZprod} 
Logarithmic distributions of differential jet resolutions and their ratios for
$Z+$jets events. Predictions of \vc~2.0 with strong (smooth) ordering
are shown in solid red (dotted green) lines. The red band shows an $\alpha_s$
variation with $\alpha_s(m_Z)=0.12$ and $\alpha_s(m_Z)=0.14$.}
\end{figure}

The parton-level results for $Z+$jets events are presented in 
\figRef{fig:jetResZprod}:
four successive jet resolution measures, $d_{m\,m+1}$ (with $m\in\{0,3\}$), 
and their ratios $d_{m\,m+1}/d_{m-1\,m}$, using the longitudinally 
invariant $k_\perp$ jet algorithm with $R=0.4$.
As before, jet resolution scales show a fixed-order behaviour for large values, a
Sudakov suppression and potentially large non-perturbative corrections for low 
values. The ratios $y_{m\,m+1}/y_{m-1\,m}$ are used to more clearly reveal the successive scale hierarchies. 

The observations for both, the jet resolution scales, and their rations, are
qualitatively similar to the $e^+e^-\to$~jets case, though quantitatively the effects here are larger. 
We notice the same turn-over when going from $d_{12}/d_{01}$ to $d_{34}/d_{23}$
we saw for $Z$ decays, with the explanation being very similar as before. 
Smooth ordering will allow to fill a additional phase space regions with harder 
emission (cf.\ figure \ref{fig:origami}). Due to the unitarity of the parton 
shower algorithm, this naively means that fewer soft emissions occur. This is
counter-acted by the Markovian restart scale, which means that the smoothly 
ordered shower yields softer emissions from ``ordered" antennae. At low multiplicity, 
the former dominates, as all antennae are allowed to fill their available phase 
space, while at higher multiplicity, the latter drives the differences. \FigRef{fig:jetResZprod} shows trends in $d_{01}$ and 
$d_{12}$ similar to the ones visible in \figRefCite{10} and \figRefCite{20} of
\cite{Bellm:2016rhh}. Note again that the additional, compensating effect of the
Markovian restarting scale starts playing an important role for higher multiplicities.

\subsection{Hard Jets in QCD processes \label{sec:hardJetsQCD}}

\begin{figure}[tbp]
\centering
\includegraphics[width=.9\textwidth]{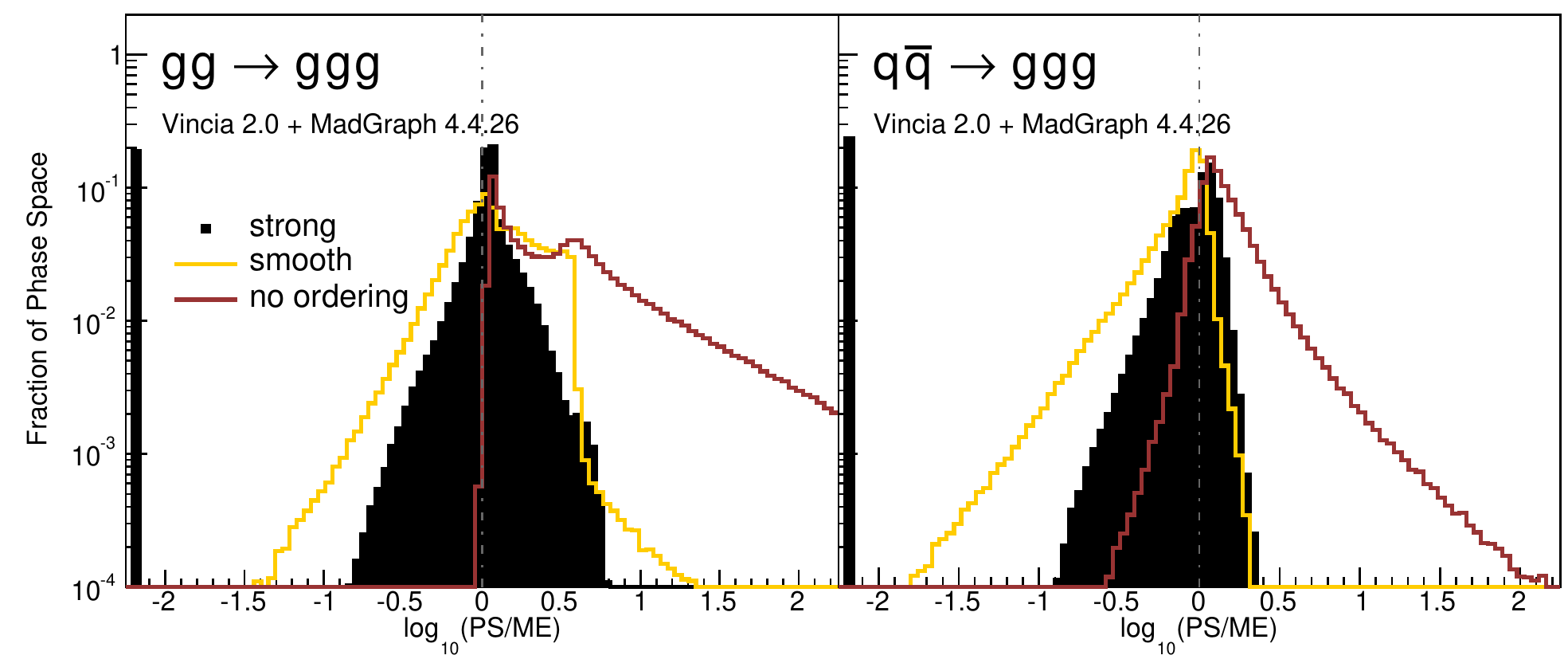}
\caption{\label{fig:PSMEQCD} Antenna shower, compared to matrix elements: distribution
of $\t{log}_{10}(\t{PS}/\t{ME})$ in a flat phase space scan of the full phase space
with strong, smooth, and no ordering with respect to the factorization scale of the
Born process.
Contents normalized to the number of generated points. Gluon emission only.}
\end{figure}

We already discussed our strategy to include hard branchings in non-QCD processes in
\secRef{sec:hardJets}. For processes with QCD jets in the final state we apply
a different formalism, as the Born process already comes with a QCD scale.
The first branching is allowed to populate all of phase space; however, the region
with scales above the factorization scale, $t>t_\t{fac}$, is treated with smooth
ordering, as described in \secRef{sec:Impr24Branch}. In \figRef{fig:PSMEQCD} we show
the PS-to-ME ratios for $gg\to ggg$ and $q\bar q\to ggg$ where the factorization
scale is chosen to be the transverse momentum of the final state partons
in the Born $2\to2$ process. We show a comparison of strong ordering, i.e. not 
including $t>t_\t{fac}$, smooth ordering with $\hat t=t_\t{fac}$ in the 
$P_\t{imp}$ factor, and no ordering, which corresponds to 
adding an event sample with $t>t_\t{fac}$. The plots indicate that the smooth
ordering is preferred over adding hard jets as a separate event sample.
Note that the asymmetric distribution of the PS-to-ME ratio for $gg\to ggg$
is the result of combining the distributions of different colour flows.

One could imagine applying the same treatment to non-QCD processes as well. 
However, this is not done in \vc\ as the factorization scale in these processes is 
not a QCD scale and therefore not suited to enter the $P_\t{imp}$ factor.

\subsection{Matrix-Element Corrections with \mg~4 \label{sec:MG4MECs}}

In this section we review the GKS procedure for iterative matrix-element corrections
(MECs)~\cite{Giele:2011cb}. To first order, the formalism is equivalent to that by Bengtsson and
Sj{\"o}strand in refs.~\cite{Bengtsson:1986et,Bengtsson:1986hr}, and to the approach used for real
corrections in \textsc{Powheg}~\cite{Frixione:2007vw,Nason:2004rx}. 
In the context of final-state showers, the approach 
was generalised to multiple emissions in \cite{Giele:2011cb} where it was successfully used to 
include MECs through $\mc O(\alpha_s^4)$ for hadronic $Z$ decays. A generalisation at the one-loop 
level has also been developed~\cite{Hartgring:2013jma}, though so far limited to 
$\mc O(\alpha_s^2)$. Here, we focus on tree-level corrections only. 

Matrix-element corrections take the all-orders approximation of the shower as their starting point, 
and apply ME-based corrections to this structure order by order in perturbation theory. At 
tree level, the following multiplicative correction factor is applied to each antenna function for matching to leading-colour matrix elements,
\begin{align}
  \label{eq:acceptProbMECs}
  \mc C_i\,\bar a_i~~\rightarrow~~\mc C_i\,\bar a_i\,P_n^\t{ME}\qquad\t{with}
  \quad P_n^\t{ME}=\frac{|\mc M_n|^2}{\sum_j \mc C_j\,\bar a_j\,|\mc M_{n-1}|^2}~,
\end{align}
with the $n$-particle matrix element squared $|\mc M_n|^2$, see \eqsRef{eq:LCME}
and \eqref{eq:fullCol} for more details on colour ordering.
Given \vc's invertible kinematics maps and the explicit forms of the physical antenna functions 
defined in \secRef{sec:antShowers}, the denominator is exactly calculable (taking the 
smooth-ordering $P_\t{imp}$ factors defined in the previous section into account). The numerator 
is obtained by using amplitudes derived from \mg~4~\cite{Alwall:2007st}, stored in \vc's
\texttt{interfaces/MG4} subdirectory. Minor extensions were required to 
include processes with initial-state coloured partons, and several new matrix-element routines were added in the context of this work. The F77 syntax for calling a \vc-modified MG4 matrix element is (using the specific 
example of a $b\bar{b}\to Hggg$ matrix element):
\begin{verbatim}
SUBROUTINE Sbbx2gggh(MCMODE,ICOL,P1,HEL1,ANS)
\end{verbatim}
where
\begin{itemize}
  \item \texttt{INTEGER MCMODE} selects between Leading Colour (0), \vc\ Colour (1), and 
    Full Colour (2), as defined below, 
  \item \texttt{INTEGER ICOL} selects which colour ordering is desired for 
    \texttt{MCMODE=0,1},
  \item \texttt{DOUBLE PRECISION P1(0:3,NEXTERNAL)} the momenta of the particles (in this 
    example \texttt{NEXTERNAL=6}),
  \item \texttt{INTEGER HEL1(4)} holding up to 4 helicity configurations to be summed over, 
    sufficient to average over an unpolarised initial 2-parton state or decaying vector boson, 
    with specified final-state helicities. The enumeration of helicity configurations follows 
    \mg's normal helicity-counting convention. 
  \item The requested matrix element squared is saved in the double-precision \texttt{ANS} 
    variable, which in \vc\ always has only a single element. 
\end{itemize}
\begin{emergency}{.8\textwidth}
From within \vc\, these matrix elements are accessed via C++ wrappers accessible via the
\texttt{VinciaPlugin::mgInterface.ME2()} methods, with definitions contained in the 
\texttt{MG4interface.h} and \texttt{MG4interface.cc} files. 
The input is a number of particles with partons being colour ordered, i.e. 
ordered in colour chains such as $q-g-g-\bar q$, where initial partons are crossed
into the final state. The diagonal entry in \mg's colour matrix, $\mc C^\t{MG}_{ii}$,
associated with the given colour order, is chosen with \texttt{ICOL}. 
Using the more recent convention of \mg~5~\cite{Alwall:2014hca}\footnote{In \mg~4 the
colour matrix for amplitudes with multiple quark pairs is more complicated and
required a decomposition by hand to separate the leading- from the subleading-colour parts, as is now done automatically by \mg~5.} 
we define the leading-colour matrix element as
\begin{align}
  \label{eq:LCME}
  \left|\mc M_n\right|^2 = \mc C^\t{MG}_{ii}~\left|\mc J_n^{(i)}\right|^2~,
\end{align}
with the colour-stripped $n$-particle amplitude $\mc J_n^{(i)}$ corresponding directly to a \texttt{JAMP} in \mg's nomenclature.
\end{emergency}

\paragraph{Full-Colour Matrix-Element Corrections:}
The full matrix element contains contributions that cannot be associated with a 
single colour ordering, i.e.\ the off-diagonal entries of the colour matrix, representing interferences between different colour orderings.
To include those subleading
colour contributions while remaining within a formalism that provides strictly positive-definite correction factors, we use the following prescription~\cite{Giele:2011cb}
(\vc\ colour),
\begin{align}
  \left|\mc M_n\right|^2 = \mc C^\t{MG}_{ii}~\left|\mc J_n^{(i)}\right|^2
  ~~\rightarrow~~\mc C^\t{MG}_{ii}~\left|\mc J_n^{(i)}\right|^2~
  \frac{\sum_{j,k}\mc C^\t{MG}_{jk}~\mc J_n^{(j)}~\mc J_n^{(k)\,*}}
  {\sum_j \mc C^\t{MG}_{jj}~\left|\mc J_n^{(j)}\right|^2}~.
  \label{eq:fullCol}
\end{align}
The matrix element for each colour structure gets a correction 
from the subleading colour part of the full matrix element that is proportional to the 
relative weight of that colour structure such that the sum over all colour flows reproduces the full colour-summed matrix element norm squared.

Note that, though we show all matrix-element comparisons with leading colour, the 
conclusions do not change when replacing leading with full colour.

\paragraph{Interference between different Born-level processes:}
In previous versions of \vc\ the interference contributions from different 
Born-level processes were ignored; e.g., the interference between 
$Z\to d\bar{d}(g\to u\bar{u})$ and $Z\to u\bar{u}(g\to d\bar{d})$ contributing to
$Z\to d\bar{d}u\bar{u}$ was not included. 
As those interferences can become fairly large and are already present for
the first branching, e.g., $qg\to qgg$ can arise from $gg\to gg$ or $qg\to qg$ 
Born-level processes, we developed a more general formalism capable of handling 
these cases.
Yet more interesting and illustrative are the interferences between $gg\to H$ and 
$Q\bar{Q}\to H$ Born processes, which both contribute to $Qg \to QH$ (with $Q$ a 
heavy quark) but involve completely different types and orders of couplings.
For this special case of Higgs production and decay we provide an option to
allow/disallow such interferences.

\begin{figure}[tbp]
\centering
\includegraphics[width=0.45\textwidth]{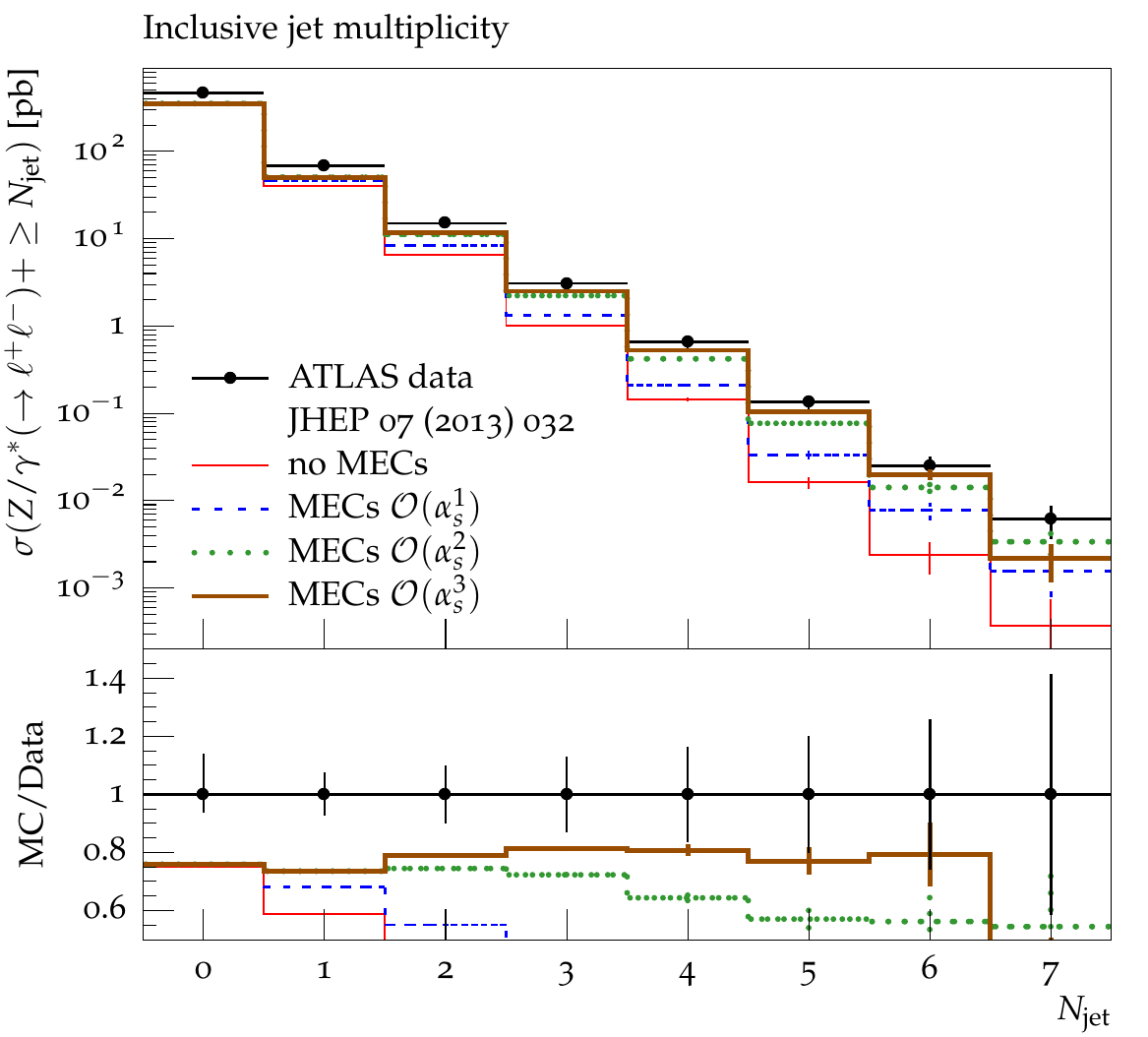}
\hspace*{5mm}
\includegraphics[width=0.45\textwidth]{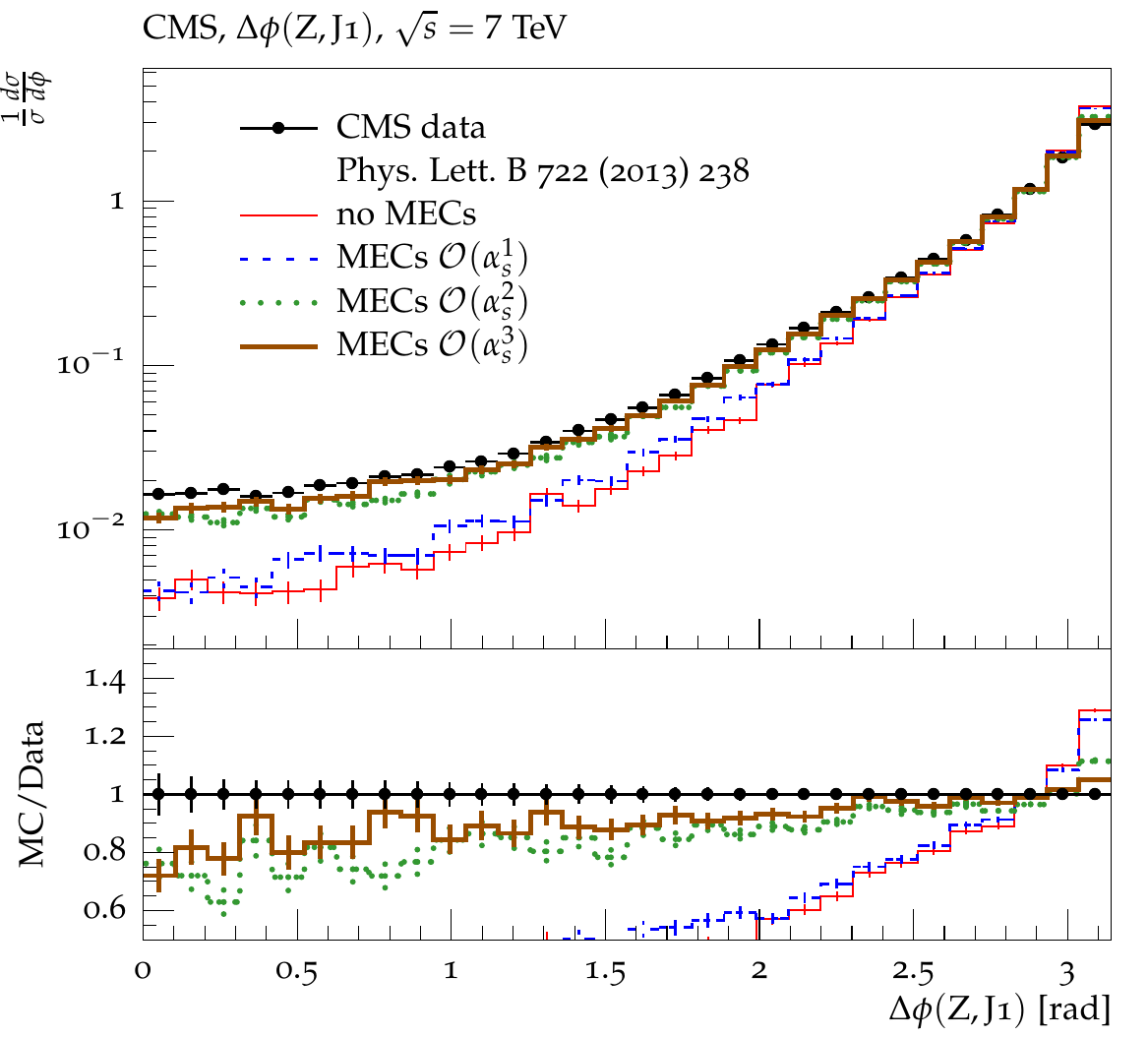}
\vspace*{3mm} \\
\includegraphics[width=0.45\textwidth]{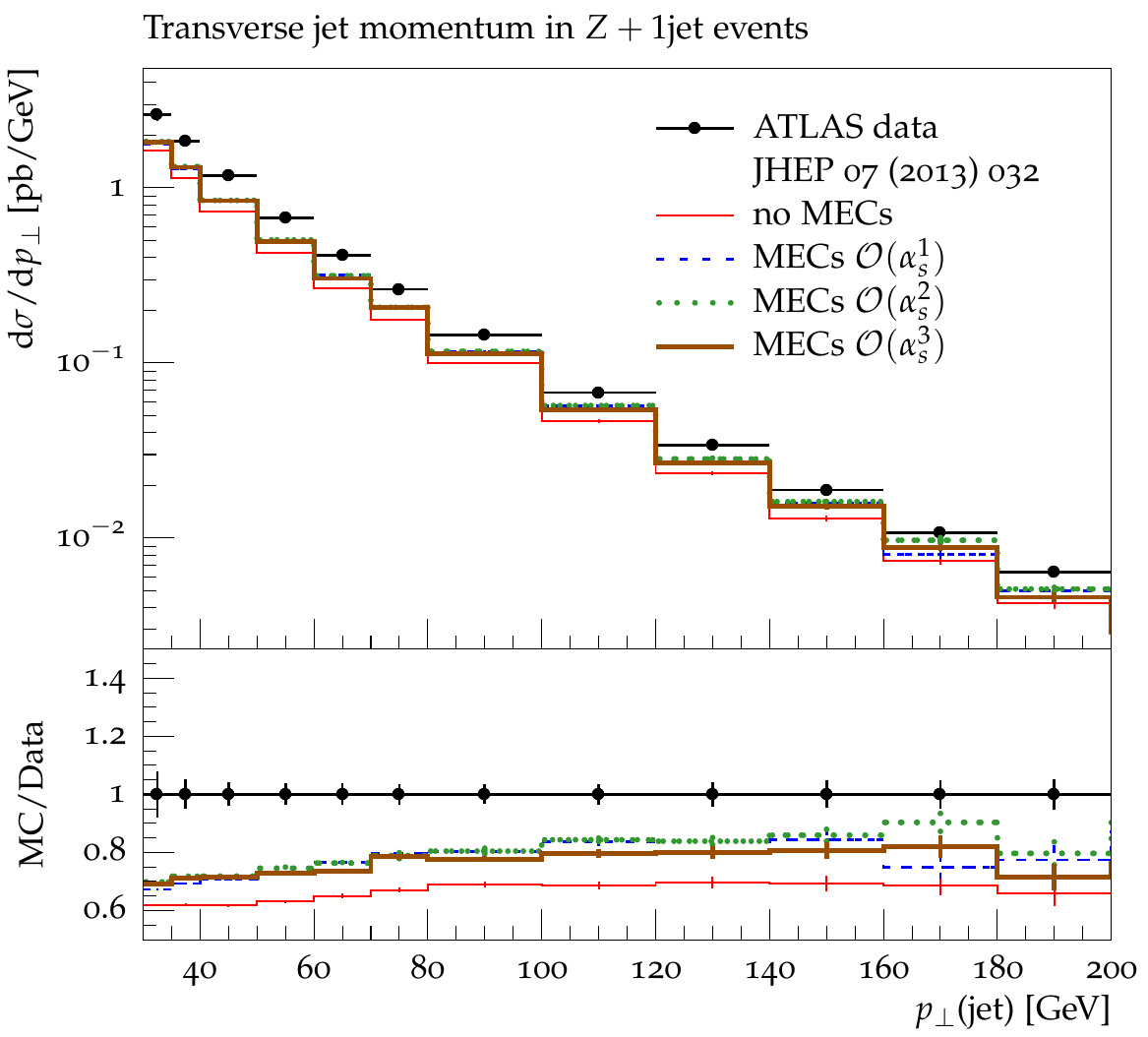}
\caption{\label{fig:dataCompMECs}
Inclusive cross section for the Drell-Yan lepton pair plus $\ge N$ jets
({\sl top left}), distribution of the azimuthal angle between the $Z$ boson 
and the hardest jet ({\sl top right}), and jet $p_\perp$ in $Z+1\,\t{jet}$
events ({\sl bottom}). Parton-level predictions of \vc~2.0 for 
increasing order of MECs included, compared to ATLAS data from 
\cite{Aad:2013ysa} and CMS data from \cite{Chatrchyan:2013tna}.}
\end{figure}

\paragraph{Impact of Matrix-Element Corrections:}
In \figRef{fig:dataCompMECs} we show parton-level predictions of \vc\ in $Z$ 
production events, i.e. multi-parton-interactions and hadronisation turned off, 
to focus solely on the shower properties and the impact of successive MECs.
Comparisons to data including multi-parton-interactions and hadronisation will
be presented in the section \secRef{sec:results}. We compare \vc\ with increasing orders 
of MECs included to ATLAS~\cite{Aad:2013ysa} and CMS~\cite{Chatrchyan:2013tna} data.
The inclusive cross section and the azimuthal angle between the reconstructed
$Z$ boson and the hardest jet (shown in the upper panel of \figRef{fig:dataCompMECs})
clearly highlight that MECs improve the description of data sensitive to multiple
hard emissions. The progressive improvements that are introduced through iterated 
MECs is particularly obvious in the inclusive jet multiplicity.
It is worthwhile mentioning that jet multiplicities beyond the third jet are only 
described by the approximate shower result. However, the combination of MECs up to 
third order seems to yield a good starting point for the shower, such that also high 
jet multiplicities are well-described.
Note that correcting only the hardest emission leads only to a modest improvement, 
since \vc's antenna functions already provide a good approximation of the $Z+\t{jet}$ 
matrix element.
The lower panel of \figRef{fig:dataCompMECs} shows the jet transverse momentum in 
exclusive $Z+\t{jet}$ events. This observable should be dominated by the MEC of the 
hardest emission. Indeed, the description improves over plain showering, and is very 
stable upon iteratively including MECs to higher multiplicities. This showcases that 
MECs to higher-multiplicity states do not degrade the quality of the description of
lower-multiplicity observables.

\section{Preliminary Results and Tuning \label{sec:results}}

\subsection{The Strong Coupling} 

All components of \vc\ (i.e., both matrix-element corrections and showers) use a single reference value for strong coupling constant, with the default value
$\alpha_s^{\overline{\mrm{MS}}}(M_Z) = 0.118$, in agreement with the current world
average~\cite{Agashe:2014kda,d'Enterria:2015toz}. By default, we use 2-loop running expressions, with the number of active flavours changing at each quark-mass threshold (including at $m_t$), though options for 1-loop running or even  fixed $\alpha_s$ values are provided as well. The  inclusion of 3-loop running effects is not relevant at the present (LO+LL) level of precision of the shower. In the infrared, the behaviour of $\alpha_s$ is regulated by allowing to evaluate it at a slightly displaced scale, $\alpha_s(\mu) \to
\alpha_s(\mu + \mu_0)$ and by imposing an upper bound $\alpha_s <
\alpha_s^\mrm{max}$. The set of default parameter values are:
{\small \begin{verbatim}
Vincia:alphaSvalue    = 0.118 ! Default alphaS(mZ) MSbar 
Vincia:alphaSorder    =     2 ! Default is 2-loop running 
Vincia:alphaSmuFreeze =   0.4 ! mu0 scale in alphaS argument, in GeV
Vincia:alphaSmax      =   1.2 ! max numerical value of alphaS
\end{verbatim} }

Within the context of an LO+LL calculation, however, the value $\alpha_s(M_Z)=0.118$  produces a  poor agreement with collider measurements; direct ``tunings'' at the LO+LL level typically find effective values closer to $\alpha_s(M_Z)=0.140$, see e.g.~\cite{Giele:2011cb,Skands:2014pea}. To permit analogous tunings of \vc, a user-specifiable prefactor is applied to the renormalisation-scale argument for each branching type,
\begin{eqnarray}
    \mbox{Gluon Emission : } \alpha_s(p_\perp) & \to & \alpha_s(k_\mu \ p_\perp)~,\label{eq:alphaSkmu}\\
    \mbox{Gluon Splitting : } \alpha_s(m_{qq}) & \to & \alpha_s(k_\mu^{\mrm{split}} \ m_{qq})~,
\end{eqnarray}
with equivalent parameters for splittings involving initial-state partons. 
The $k_\mu$ and $k_\mu^{\mrm{split}}$ parameters provide the same range of tuning possibilities for the effective coupling constant as in other parton-shower models, while they are simultaneously straightforward to interpret e.g.\ in the context of NLO matrix-element merging schemes.

The \vc\ shower algorithms do nonetheless incorporate a translation (on by default) between the $\overline{\mrm{MS}}$ value given above and the so-called CMW (or MC) scheme which is appropriate for soft-gluon emission in coherent parton showers~\cite{Catani:1990rr}. Since this translation is only rigorously defined in the limit of vanishing gluon energy, there is an ambiguity as to precisely how it should be applied to finite gluon energies. We address this by applying the CMW translation only to the coupling constant accompanying the eikonal (double-pole) term of the gluon-emission antenna functions, 
\begin{eqnarray}
    \alpha_s^{\overline{\mrm{MS}}} a_\mrm{Emit} & = & \alpha_s^{\overline{\mrm{MS}}} \left(a_\mrm{eik} + a_\mrm{coll} + a_\mrm{hard}\right) \\ & \to &  
\alpha_s^{\mrm{CMW}} a_\mrm{eik} \ + \ 
    \alpha_s^{\overline{\mrm{MS}}} \left(a_\mrm{coll} + a_\mrm{hard}\right) ~,
\end{eqnarray}
with a few different options provided for how the eikonal term should be extrapolated to finite gluon energies. In a future study we shall aim to bring these ambiguities under better control by systematic application of one-loop corrected antenna functions, but this is still (far) beyond the scope of the present work.

\subsection{\vc~2.0 Default Tune}

Two main tools were used to perform the analyses: \vc's own ROOT-based analysis tool,
\textsc{VinciaRoot}~\cite{Giele:2011cb}, and \riv~\cite{Buckley:2010ar}. For the 
hadron-collider distributions, we compare \vc~2.0 with \py~8.2. 
For the $e^+e^-\to\mrm{hadrons}$ analyses, we also include \vc~1.2,
since this version  
included NLO corrections to $e^+e^-\to 3\ \mrm{jets}$ which have not
yet been migrated to \vc~2.0. Note, however, that even without the NLO
corrections the two \vc\ versions are not exactly identical due to
a slightly revised definition of the smooth-ordering criterion, to
make it truly Markovian.  

We note that these tunings were done manually (by ``eye''), rather than by automated minimisation of $\chi^2$ or equivalent measures. The latter is not as straightforward as it may sound, due to correlations between measurements and the influence of regions of low theoretical accuracy. These issues can be at least partially addressed by combining global knowledge and experience to (subjectively) choose binwise weighting factors. Nevertheless, manual and automated approaches may be considered complementary, with the former certainly competitive for the purpose of determining a set of ``reasonable default values'', which is our principal aim here. 

\subsubsection*{Hadronic $Z$ Decays}

The final-state showering and hadronisation parameters are constrained
using hadronic $Z$ decays, mainly from the LEP experiments. 
In the context of \vc~2.0, the rates of perturbative
final-state branchings depend on the effective renormalisation scheme and
scale choice, cf.~\eqRef{eq:alphaSkmu}, for which we have 
chosen the default values: 
{\small \begin{verbatim}
Vincia:CMWtypeFF       =    2  ! CMW rescaling for FF antennae
Vincia:alphaSkMuF      =  0.6  ! muR prefactor for gluon emissions
Vincia:alphaSkMuSplitF =  0.5  ! muR prefactor for gluon splittings 
                               ! (g -> qqbar)
\end{verbatim} }
\begin{figure}[t]
\centering%
\includegraphics*[width=0.333\textwidth]{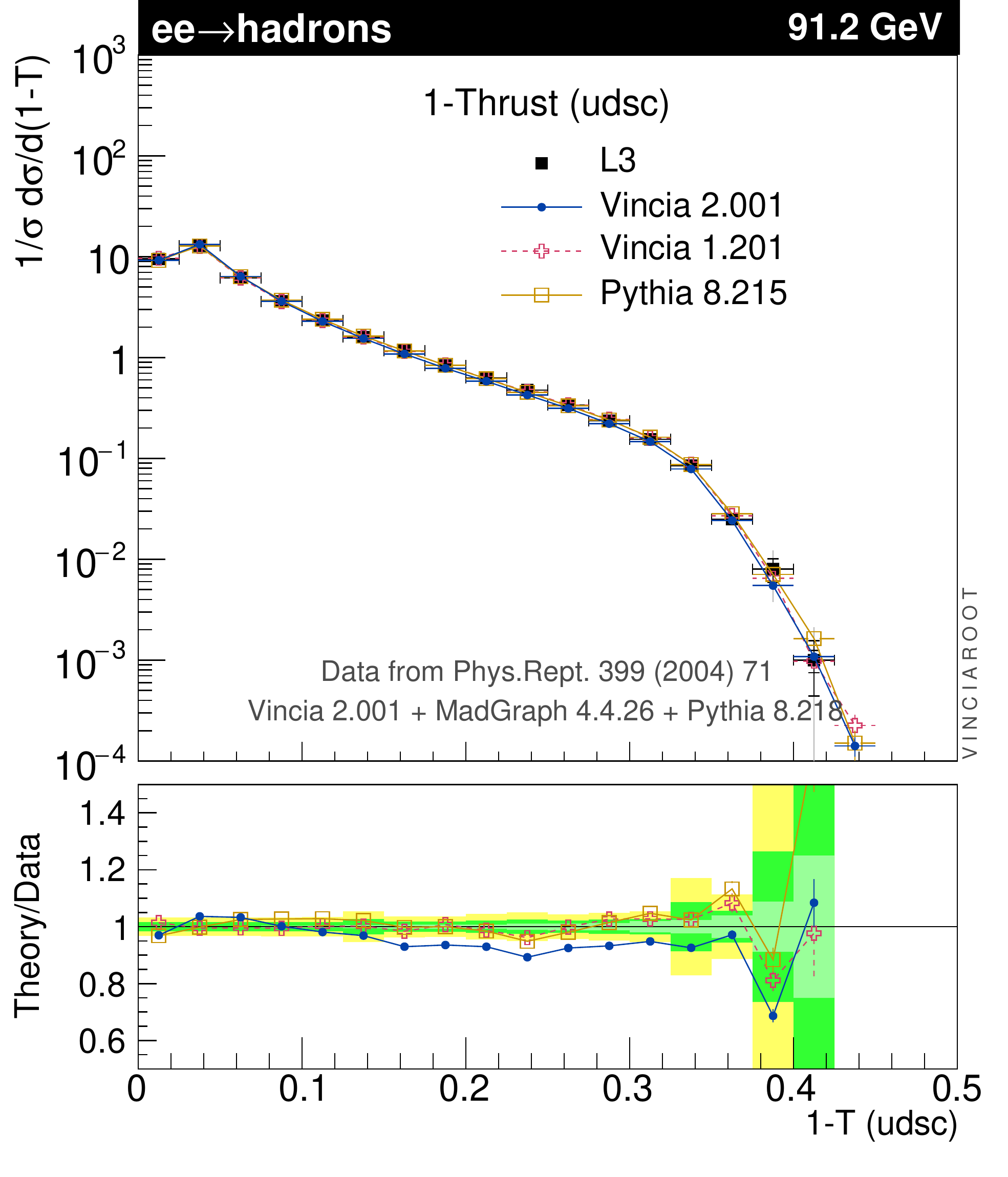}%
\includegraphics*[width=0.333\textwidth]{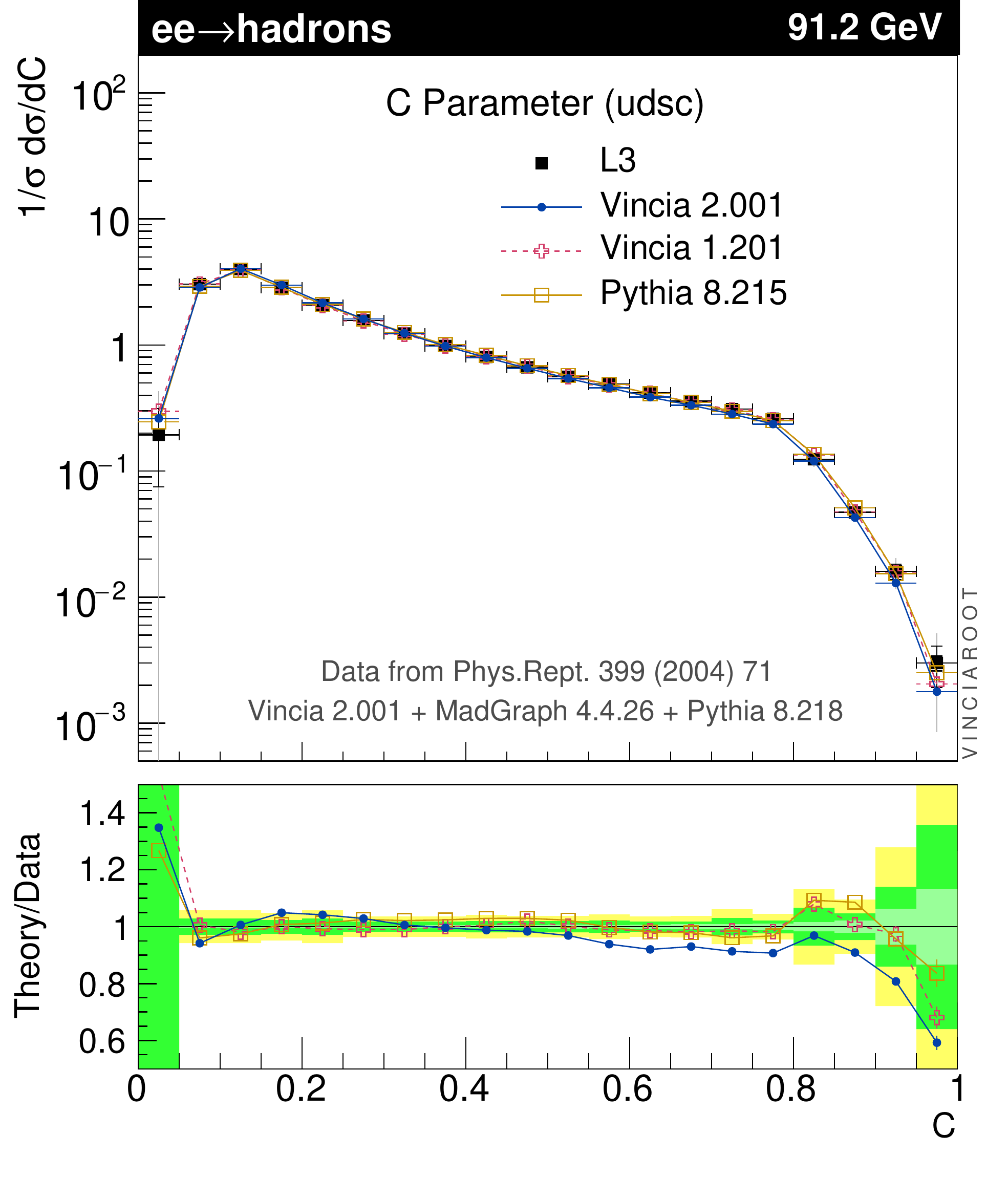}%
\includegraphics*[width=0.333\textwidth]{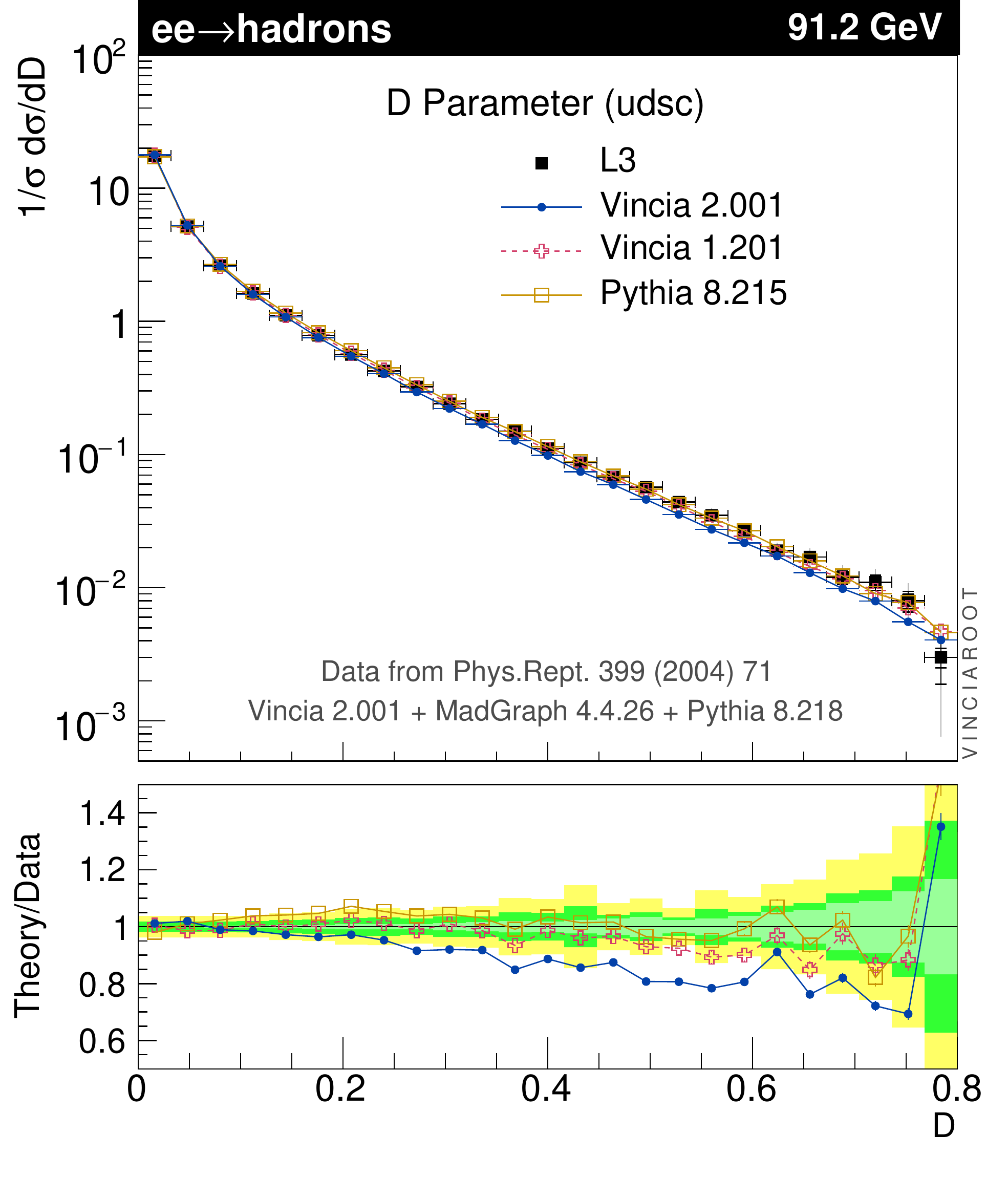}\\
\includegraphics*[width=0.333\textwidth]{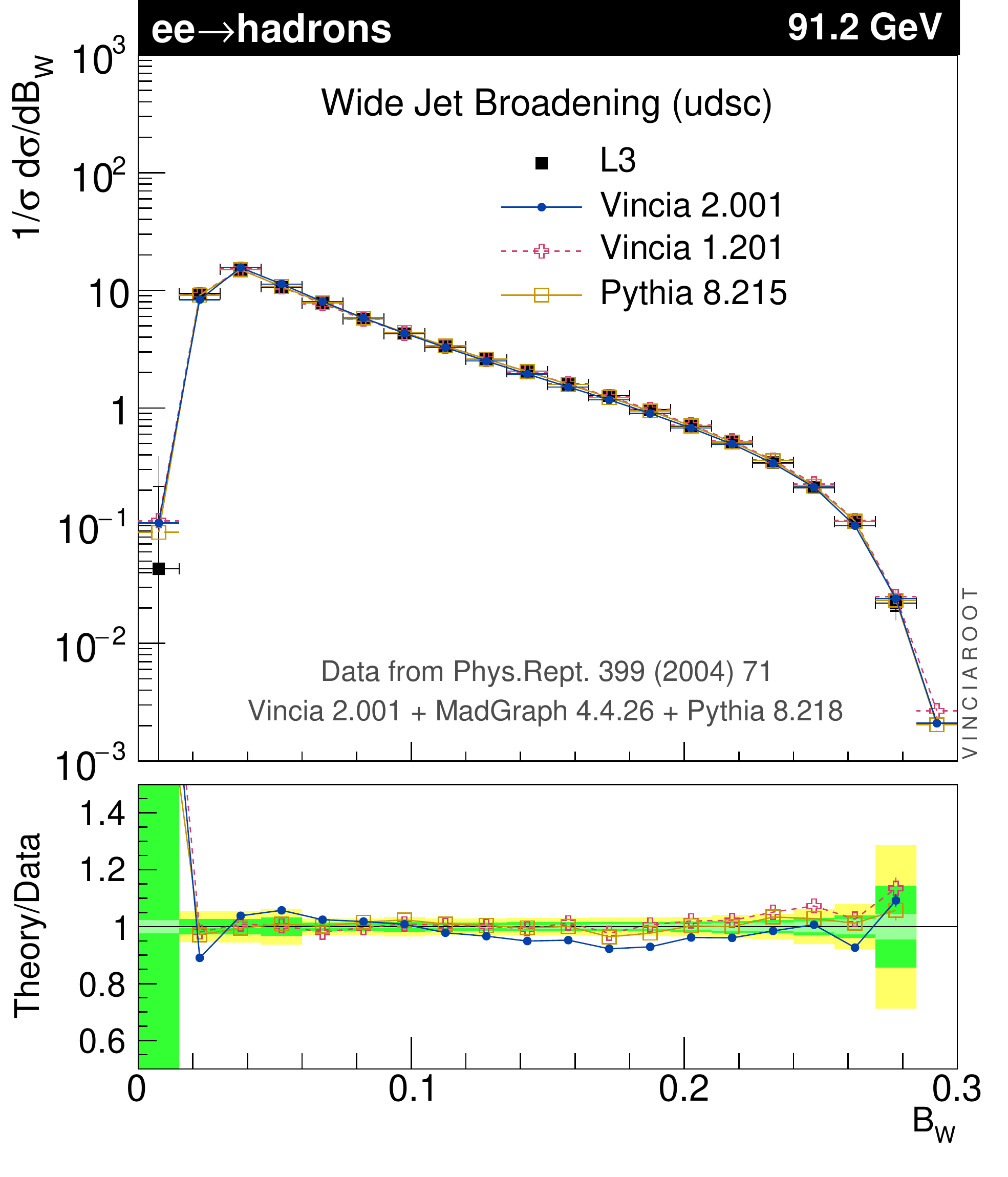}%
\includegraphics*[width=0.333\textwidth]{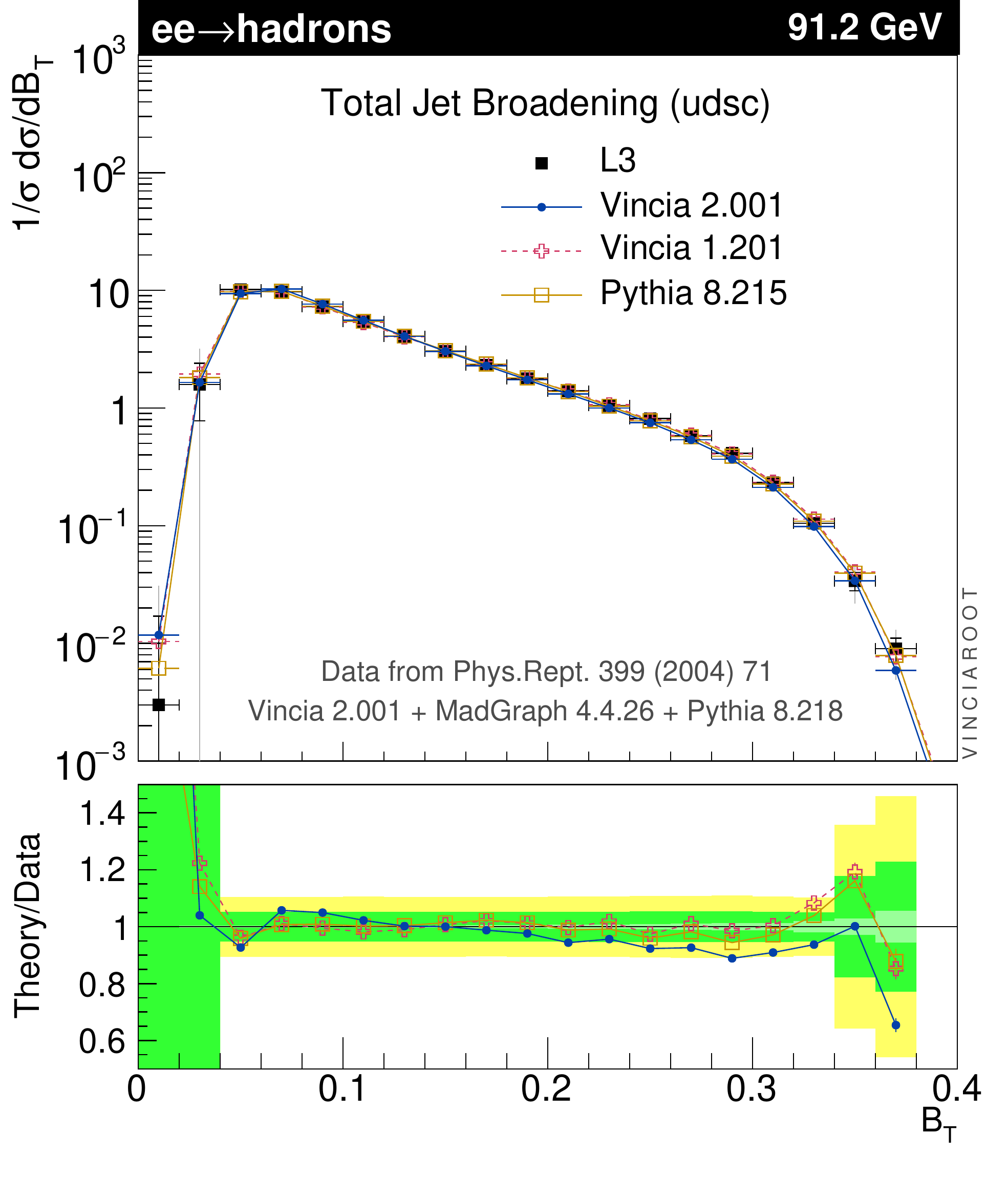}%
\caption{Event-shape variables compared with measurements performed by
  the L3 experiment. 
\label{fig:tuningFSR}}
\end{figure}
\FigRef{fig:tuningFSR} shows the event-shape observables\footnote{For
definitions, see e.g.~\cite{Achard:2004sv}.} that were used as the
primary tuning constraints, compared with
light-flavour tagged data from the L3
experiment~\cite{Achard:2004sv}. 
In the main (top) plot panes, experimental data
  is represented by black square symbols, with 1-$\sigma$ and 2-$\sigma$
  uncertainties represented by black vertical error bars and
  light-grey extensions, respectively. In the ratio panes, the inner (green) bands
  indicate the 1-$\sigma$ uncertainties on the data; outer (yellow)
  bands represent 2 $\sigma$.

Note that, since \vc~2.0 does not incorporate the NLO corrections to
$Z \to 3$ jets internally (unlike \vc~1.2~\cite{Hartgring:2013jma}), we have chosen to allow the default tune 
to undershoot the reference data slightly in regions
dominated by hard, resolved 3-jet events. This hopefully produces 
a more universal global tuning which should also be
appropriate for use with the NLO merging strategies that are available
within \py, notably UNLOPS~\cite{Lonnblad:2012ix}. 

\begin{figure}[t]\centering
\includegraphics*[width=0.33\textwidth]{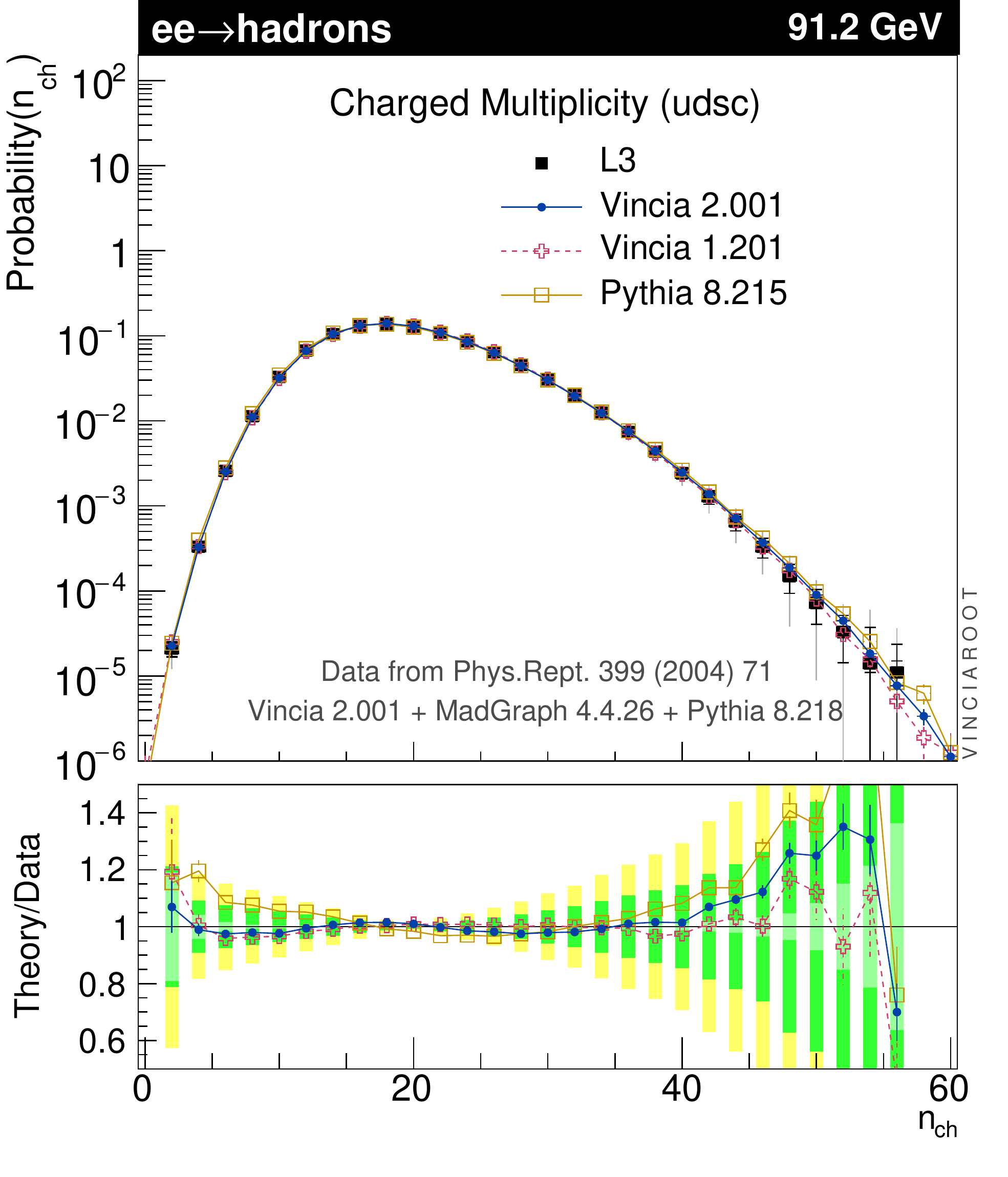}
\includegraphics*[width=0.33\textwidth]{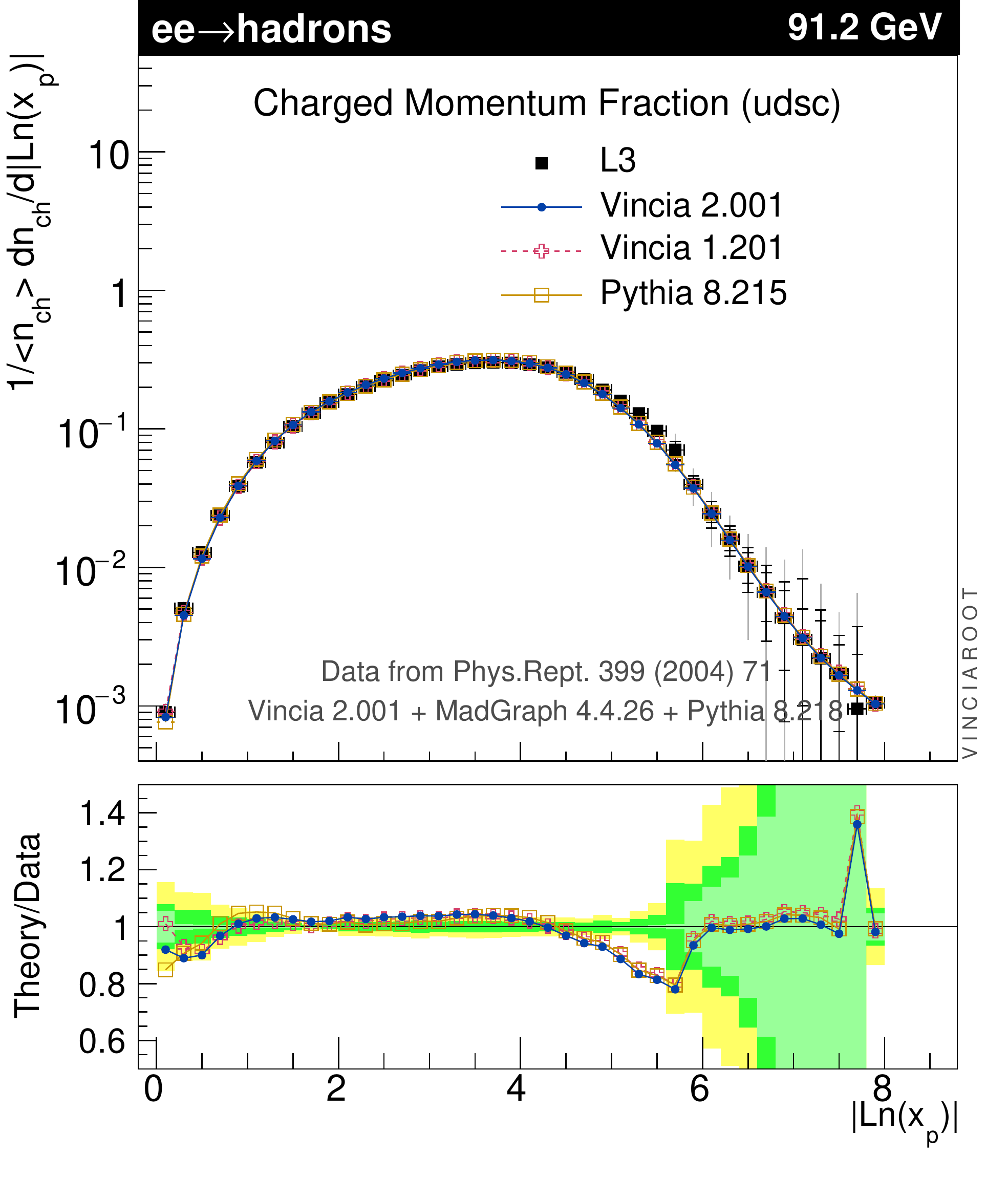}%
\caption{
Charged-track multiplicity and momentum spectra, 
compared with measurements performed by
  the L3 experiment.\label{fig:nChX}.}
\end{figure}

\begin{figure}[t]
\centering
\includegraphics*[width=0.334\textwidth]{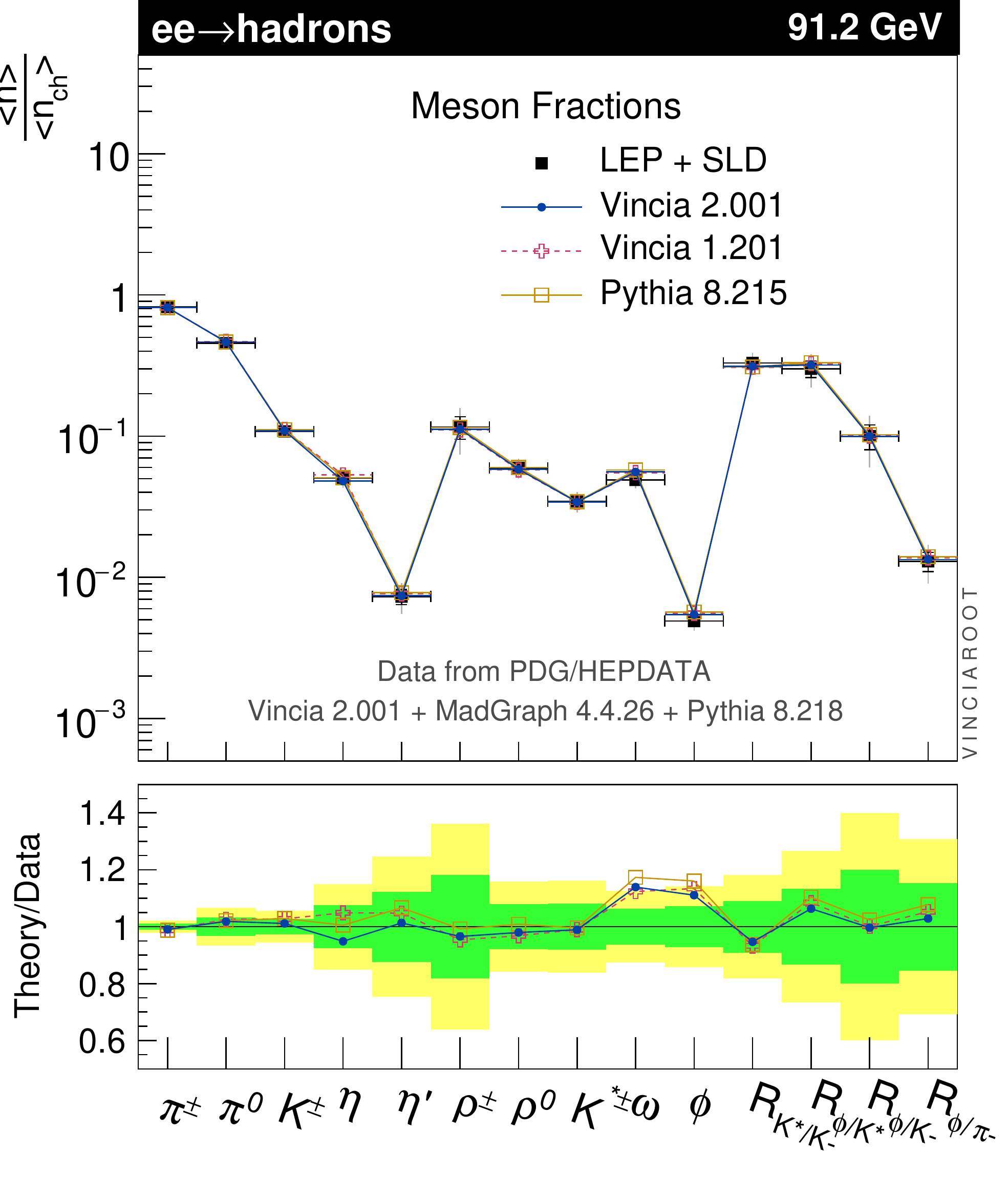}
\includegraphics*[width=0.334\textwidth]{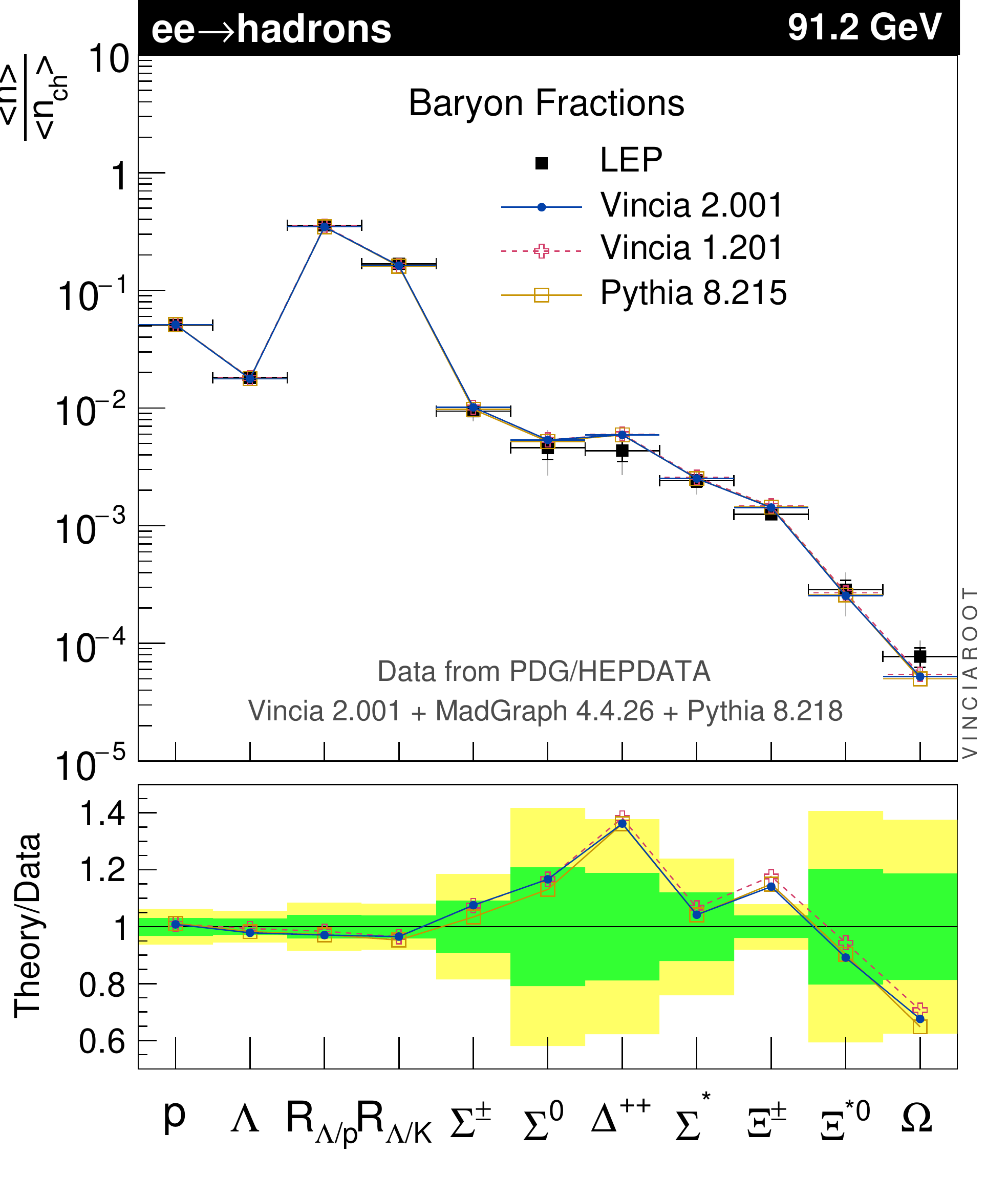}%
\caption{Identified-particle rates (expressed as fractions of the
  charged-particle multiplicity, or as indicated by $R$ symbols),
  compared with the Monash 2013 reference values.\label{fig:IDparticle}}
\end{figure}

\begin{figure}[t]
\centering
\includegraphics*[width=0.333\textwidth]{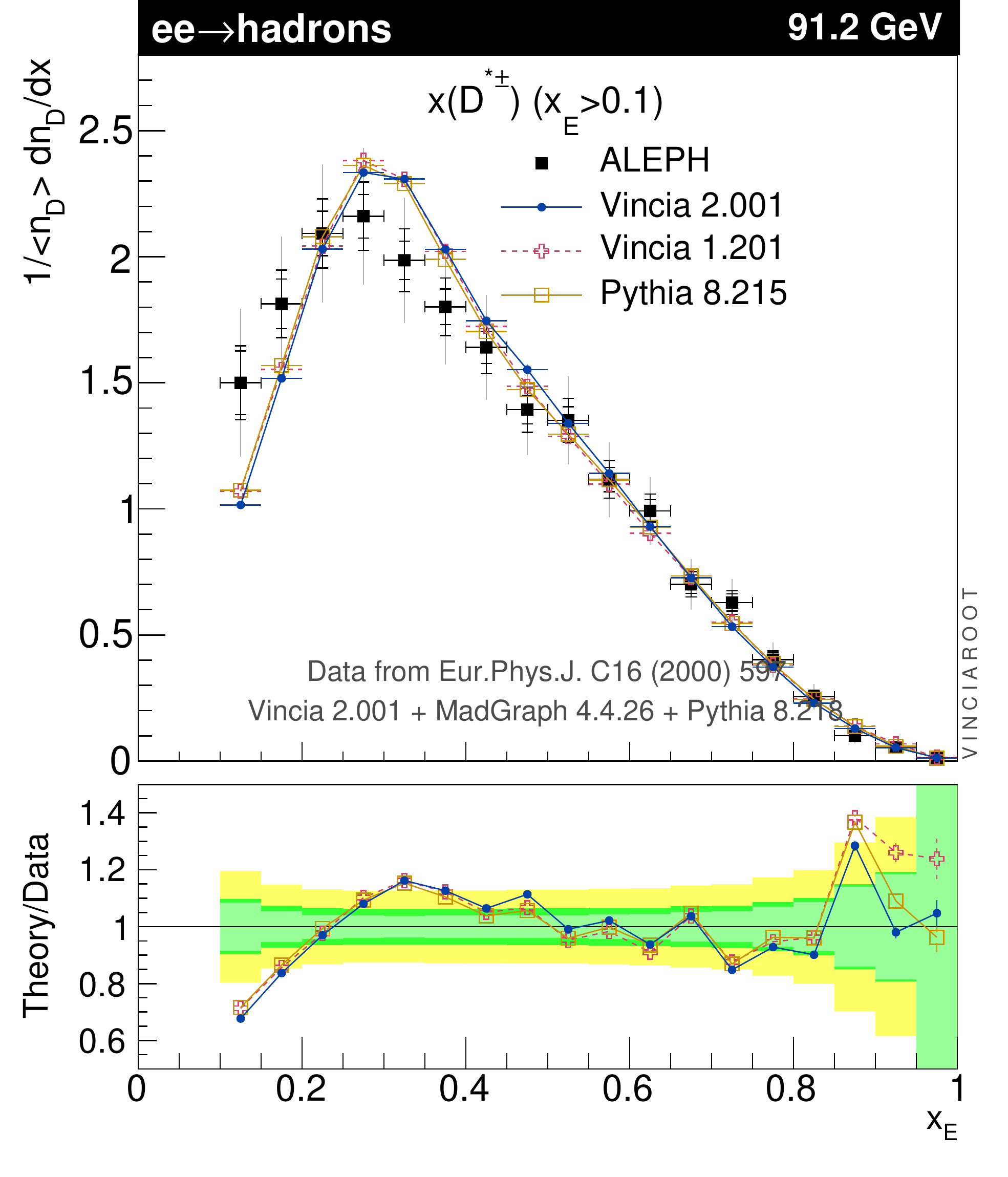}%
\includegraphics*[width=0.333\textwidth]{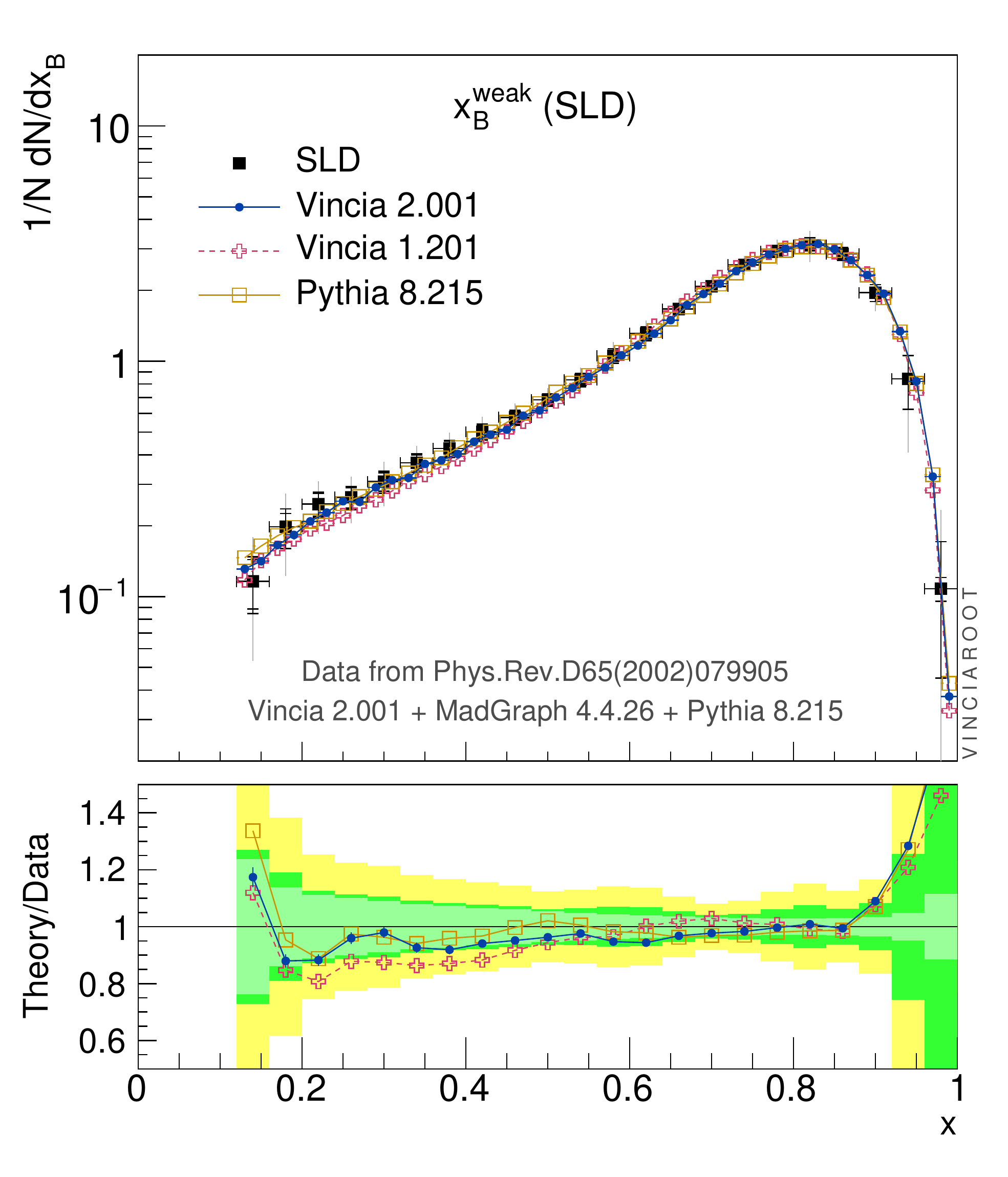}%
\includegraphics*[width=0.333\textwidth]{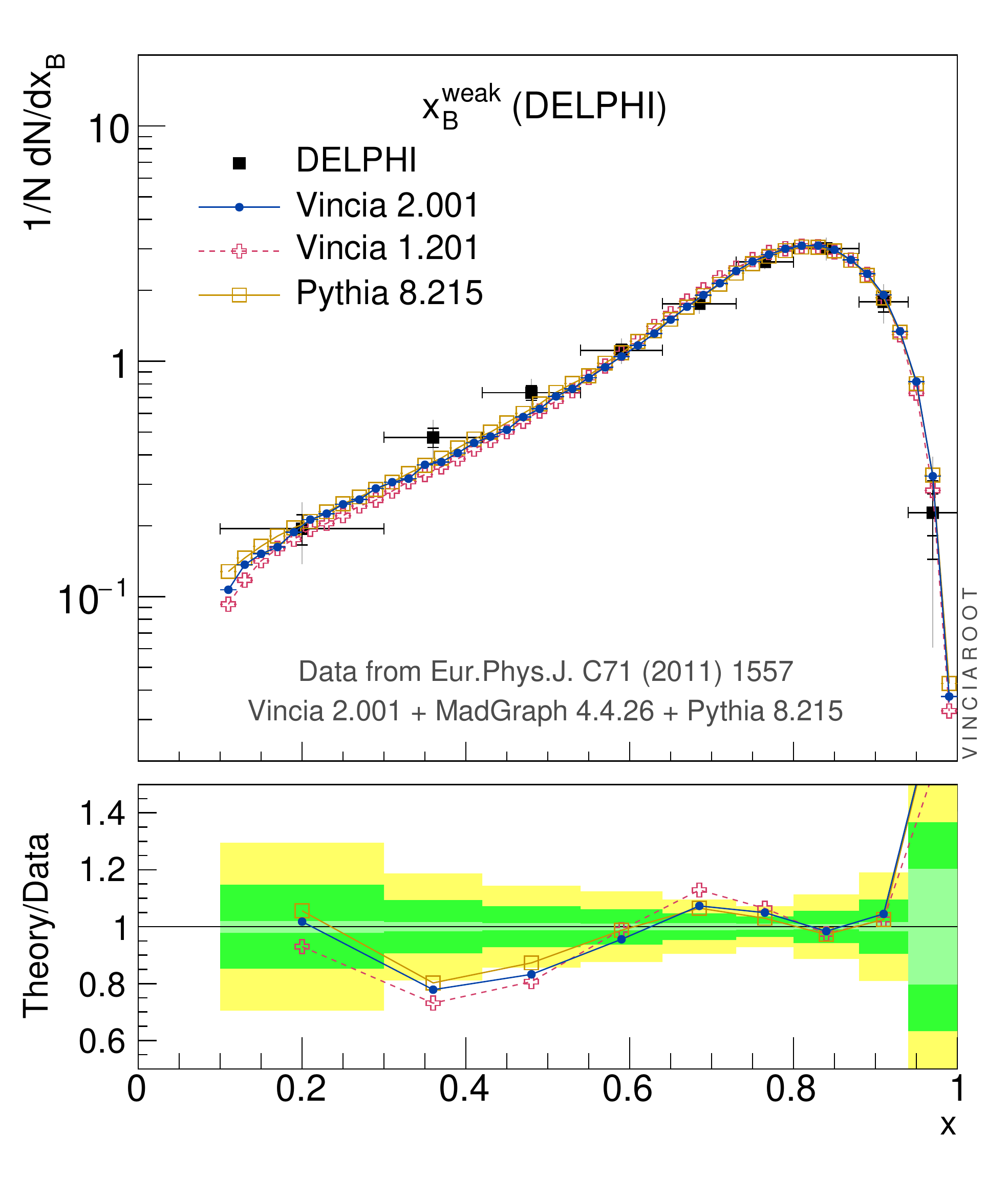}
\caption{Distributions sensitive to heavy-quark fragmentation. {\sl Left:} the energy-fraction spectrum of charged $D^*$ mesons compared with ALEPH data~\cite{Barate:1999bg}. {\sl Center and Right:} the momentum-fraction spectrum of weakly decaying $B$ hadrons compared to measurements by SLD~\cite{Abe:2002iq} and DELPHI~\cite{DELPHI:2011aa}, respectively. 
\label{fig:massive}}
\end{figure}

The Lund string
model~\cite{Andersson:1983jt,Andersson:1983ia,Andersson:1998tv} is
used for hadronisation, with 
parameters (re)optimised for use with \vc's shower model. 
The main parameters are the shower IR cutoff, the Lund
fragmentation-function $a$ and $b$ parameters --- which are defined by 
\begin{equation}
f(z) \propto \frac{(1-z)^a}{z}\exp\left( \frac{-bm_\perp^2}{z} \right)~,
\end{equation}
with $z = E_\mrm{hadron}/E_\mrm{parton}$ and $m_\perp^2 = m^2 +
p_\perp^2$ --- and the transverse-momentum
broadening in string breaks, expressed as a Gaussian with width
$\sigma_\perp\sim {\cal O}(\Lambda_\mrm{QCD})$. The default \vc~2.001 hadronisation-parameter values are,
{\small \begin{verbatim}
Vincia:cutoffScaleFF = 0.9   ! Cutoff value in GeV for FF antennae
StringZ:aLund        = 0.5   ! Lund a parameter
StringZ:bLund        = 1.15  ! Lund b parameter
StringZ:aExtraDiquark= 1.12  ! (extra for diquarks)
StringPT:sigma       = 0.295 ! Soft pT in string breaks
\end{verbatim} }
The inclusive charged-particle multiplicity distribution and momentum
($x_p= 2|p|/E_\mrm{cm}$) spectrum is shown in \figRef{fig:nChX},
again compared with light-flavour tagged L3 data from~\cite{Achard:2004sv}.
Finally, we show the rates for identified light-flavour mesons and
baryons in \figRef{fig:IDparticle}; these hardly change between the
\py, \vc~1, and \vc~2 defaults. Note that we  here compare to the 
reference measurement values derived for the Monash tune~\cite{Skands:2014pea} of
\py~8, which are not identical to the corresponding PDG values in
particular for some of the baryon rates, see~\cite{Skands:2014pea}. 
The corresponding full set of default parameter values are:
{\small \begin{verbatim}
! * String breakup flavour parameters
StringFlav:probStoUD      = 0.21    ! Strangeness-to-UD ratio
StringFlav:mesonUDvector  = 0.45    ! Light-flavor vector suppression
StringFlav:mesonSvector   = 0.555   ! Strange vector-meson suppression
StringFlav:mesonCvector   = 1.03    ! Charm vector-meson suppression
StringFlav:mesonBvector   = 2.2     ! Bottom vector-meson suppression
StringFlav:probQQtoQ      = 0.077   ! Diquark rate (for baryon production)
StringFlav:probSQtoQQ     = 1.0     ! Optional Strange diquark suppression
StringFlav:probQQ1toQQ0   = 0.027   ! Vector diquark suppression
StringFlav:etaSup         = 0.53    ! Eta suppression
StringFlav:etaPrimeSup    = 0.105   ! Eta' suppression
StringFlav:decupletSup    = 1.0     ! Optional Spin-3/2 Baryon Suppression
StringFlav:popcornSpair   = 0.9     ! Popcorn
StringFlav:popcornSmeson  = 0.5     ! Popcorn
StringZ:rFactC            = 1.60    ! Bowler parameter for c quarks
StringZ:rFactB            = 1.1     ! Bowler parameter for b quarks
StringZ:useNonstandardB   = true    ! Special treatment for b quarks
StringZ:aNonstandardB     = 0.82    ! a parameter for b quarks
StringZ:bNonstandardB     = 1.4     ! b parameter for b quarks
\end{verbatim} }
Note that the last 6 parameters govern $c$- and particularly $b$-quark
fragmentation. Since massive-quark effects are not explicitly addressed
 in this version of \vc, these parameters have been chosen merely on a
``best-effort'' basis. We plan to return to this in a future
update. A minimal set of checks on the
level of agreement with 
heavy-quark spectra can be carried out 
using the \texttt{vincia03-root} and \texttt{vincia05-root} example
programs included with the code. The former includes cross checks on the $g\to c\bar{c}$ and $g\to b\bar{b}$ rates as well as a $D^*$
spectrum, sensitive to $c$-quark fragmentation,  while the latter
focuses on constraints from $b$-tagged events. For completeness, the $D^*$ and $B$-hadron spectra produced by these example programs are reproduced in \figRef{fig:massive}.
For the SLD $x_B$ spectrum, be advised that the current distribution of \vc~(version 2.001) contains the spectrum obtained from the HepData archive~\cite{Buckley:2010jn} at the time of writing. However, the corrections contained in an erratum subsequently published by SLD~\cite{Abe:2002iq} were missing from this table. The figure we show here contains the updated values (from the erratum). The updated table will be included in the next public release of \vc, with corresponding updates expected in the HepData archive in due course. 

\subsubsection*{Drell-Yan}

\begin{figure}[tbp]
\centering\vspace*{-4mm}
\includegraphics[width=0.45\textwidth]{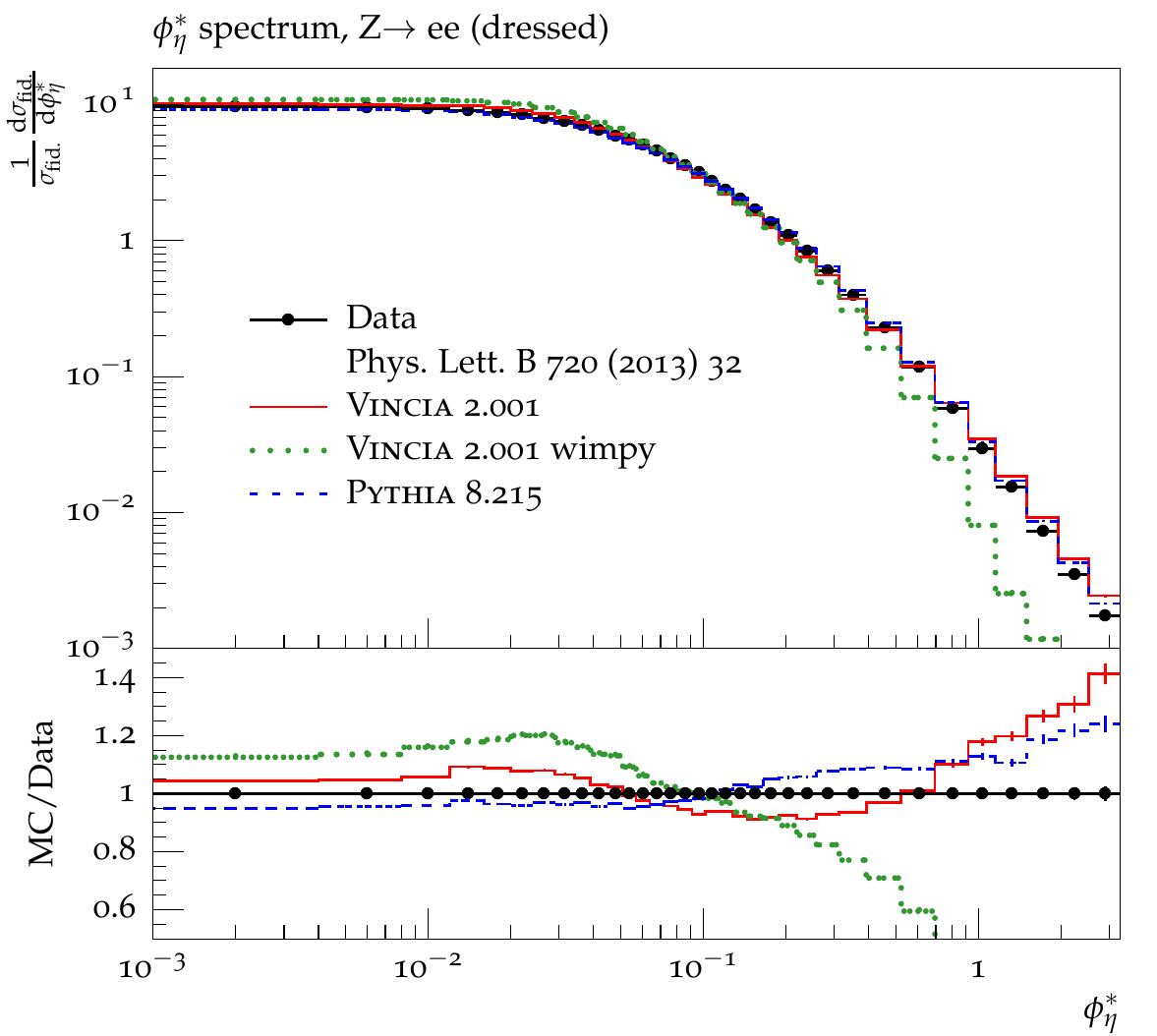}
\hspace*{5mm}
\includegraphics[width=0.45\textwidth]{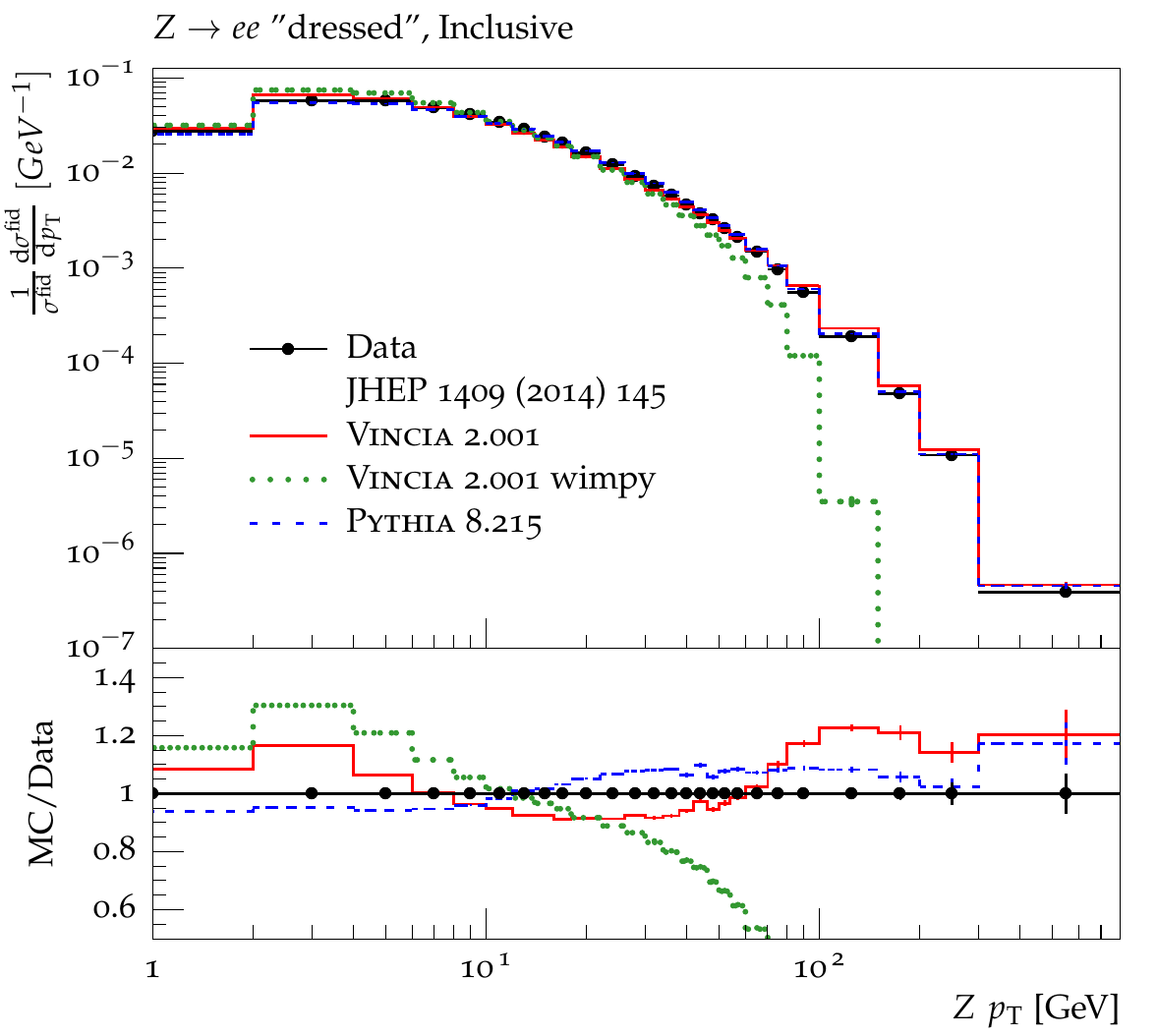}
\caption{\label{fig:ZpTphiStar} 
Angular correlations ({\sl left}) and the transverse momentum spectrum ({\sl right})
of the Drell-Yan lepton pair. Predictions of default \vc~2.0 in red, \vc~2.0 wimpy
in green, and \py~8.2 in blue, compared to ATLAS data from \cite{Aad:2012wfa} and 
\cite{Aad:2014xaa}.}
\vskip6mm
\includegraphics[width=0.45\textwidth]{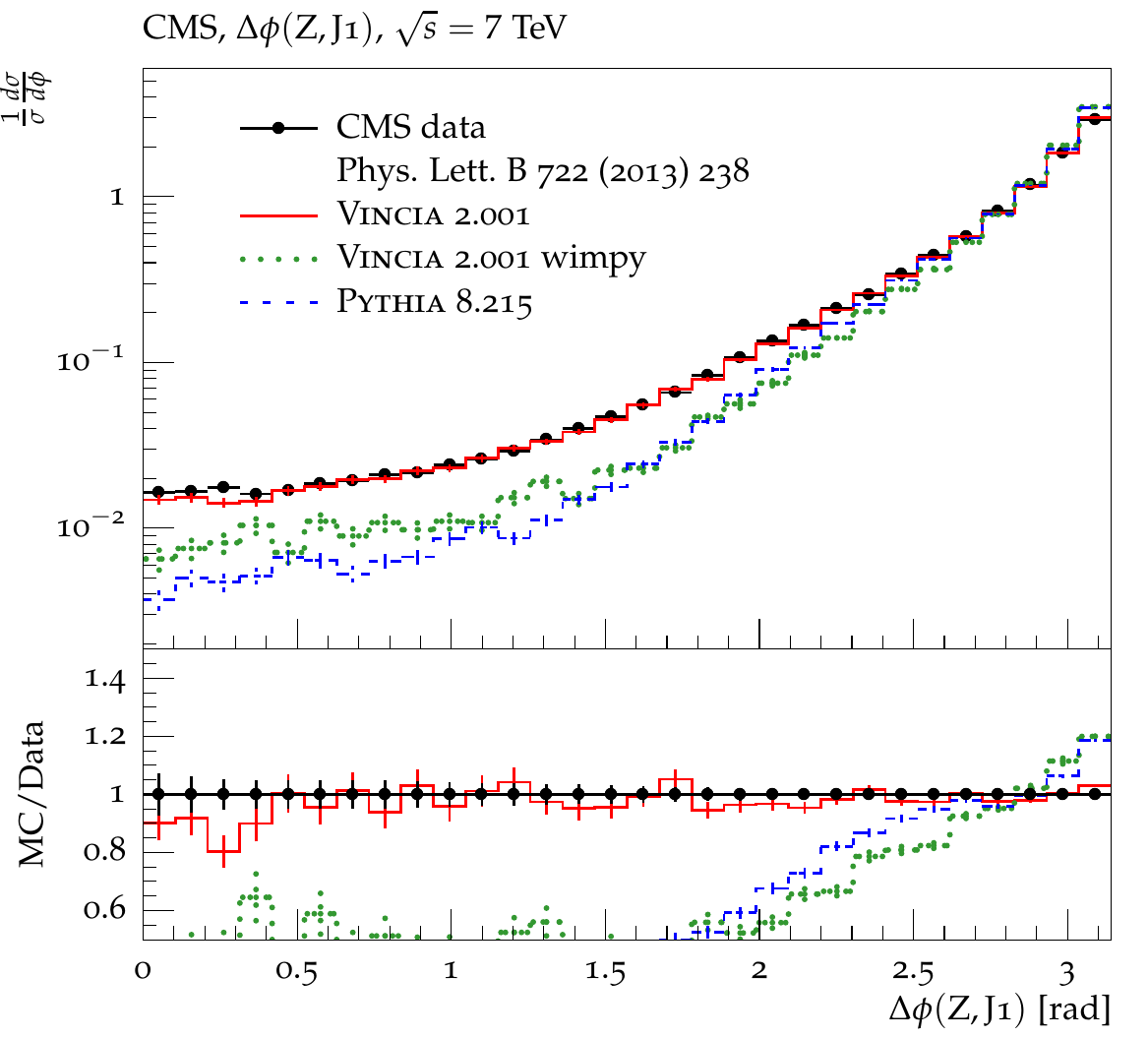}
\hspace*{5mm}
\includegraphics[width=0.45\textwidth]{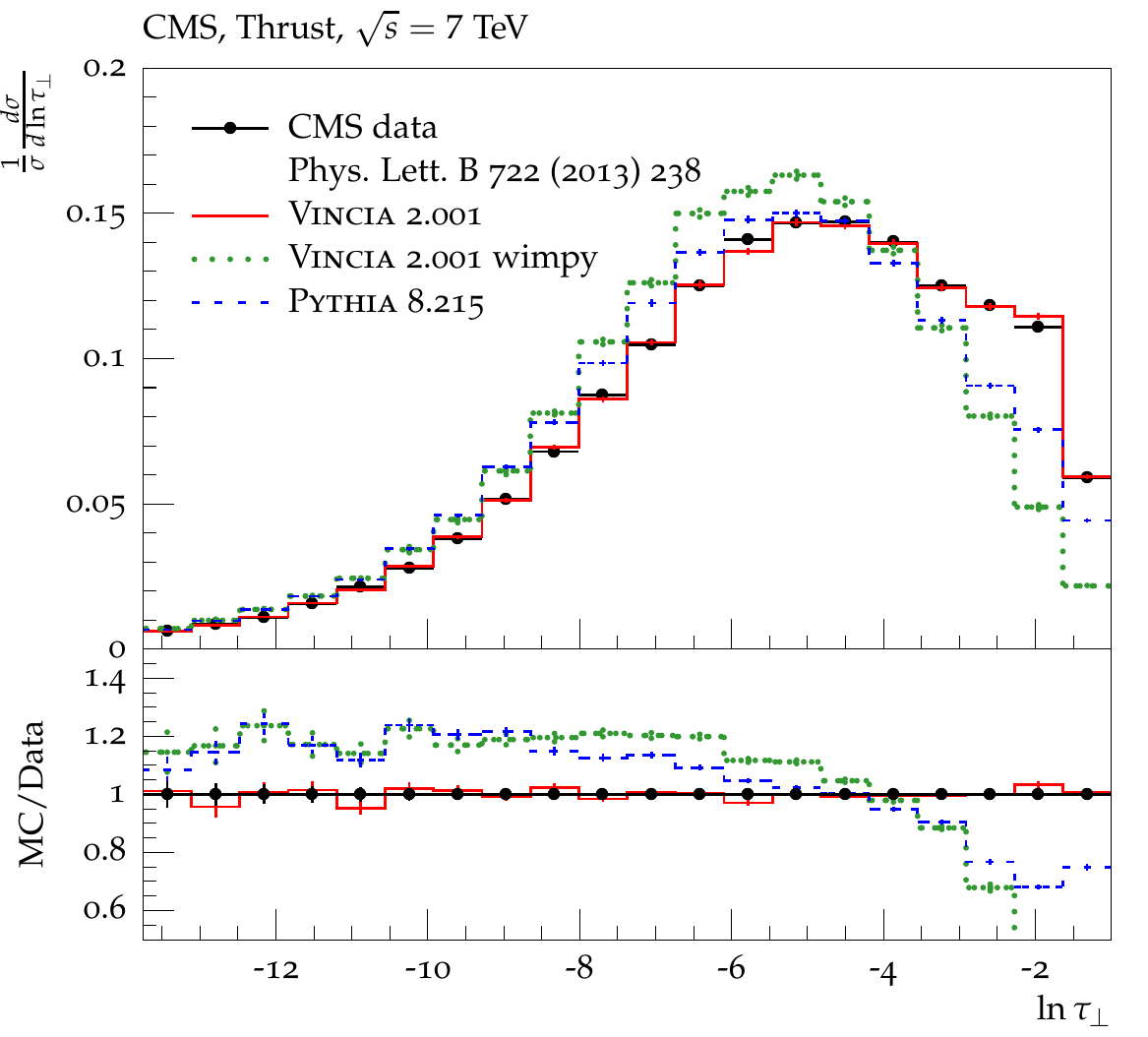}
\caption{\label{fig:Zshapes}
Distribution of the azimuthal angle between the $Z$ boson and the hardest jet ({\sl left}) 
and thrust ({\sl right}). Predictions of default \vc~2.0 in red, \vc~2.0 wimpy
in green, and \py~8.2 in blue, compared to CMS data from \cite{Chatrchyan:2013tna}.}
\end{figure}

\begin{figure}[tp]
\centering\vspace*{-4mm}
\includegraphics[width=0.45\textwidth]{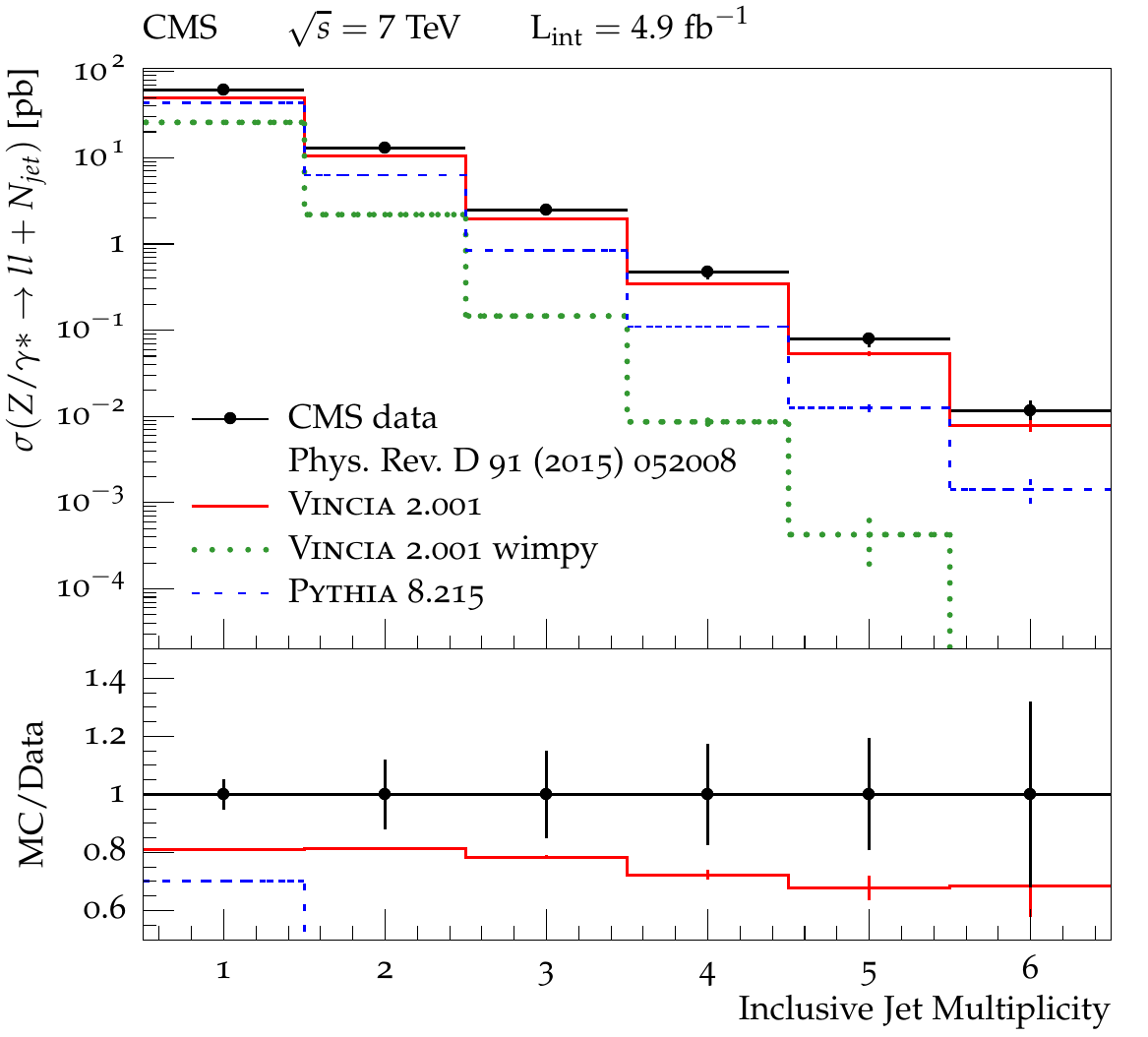}
\hspace*{5mm}
\includegraphics[width=0.45\textwidth]{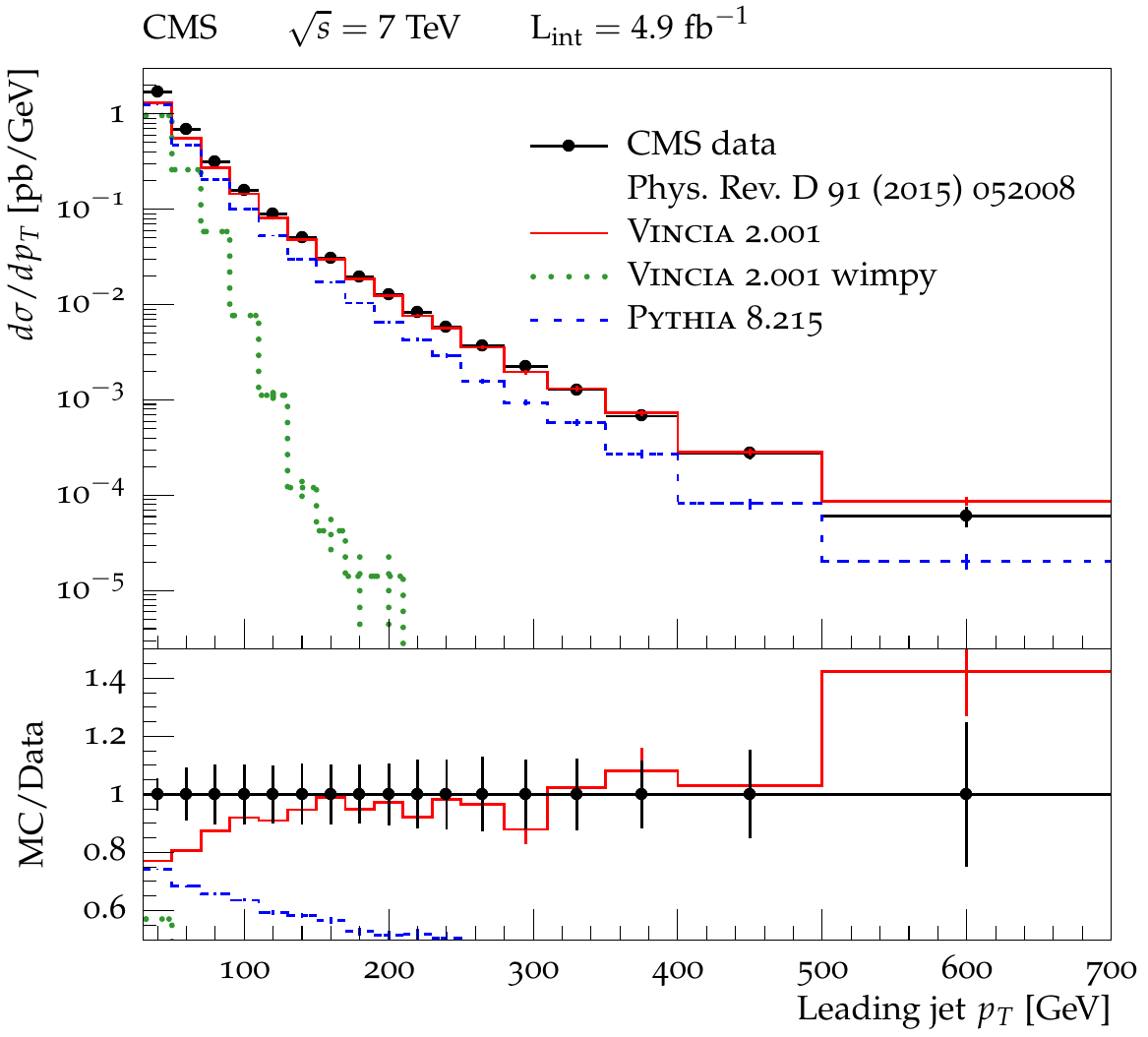}
\vspace*{3mm} \\
\includegraphics[width=0.45\textwidth]{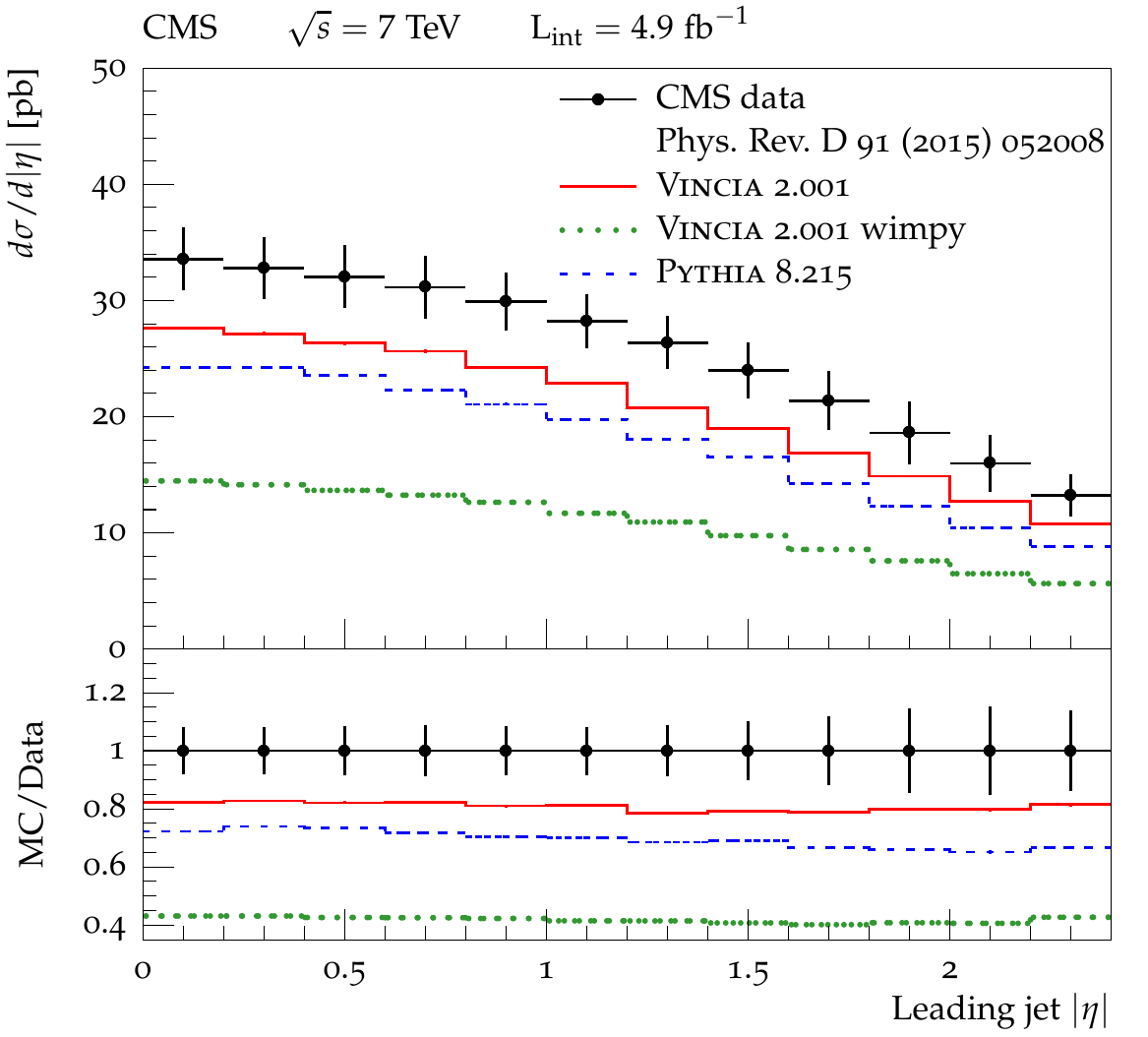}
\caption{\label{fig:Zjets} 
Inclusive cross section for the Drell-Yan lepton pair plus $\ge N$ jets
({\sl top left}), the transverse momentum ({\sl top right}) and the pseudorapidity 
spectrum of the leading jet ({\sl bottom}). Predictions of default \vc~2.0 in red, 
\vc~2.0 wimpy in green, and \py~8.2 in blue, compared to CMS data from 
\cite{Khachatryan:2014zya}.}
\end{figure}

In \figsRef{fig:ZpTphiStar}-\ref{fig:Zjets} we show a set of observables in Drell-Yan 
events with ATLAS data from \cite{Aad:2012wfa} and \cite{Aad:2014xaa} and CMS data from 
\cite{Chatrchyan:2013tna} and \cite{Khachatryan:2014zya}. We show predictions of default 
\vc~2.0 in red, \vc~2.0 wimpy (representing an ordinary shower, starting at the 
factorization scale, i.e.~no hard jets, no MECs, and strong ordering) in green, and 
\py~8.2 in blue. The \vc~2.001 results correspond to the following
default parameter choices:
{\small \begin{verbatim}
# Perturbative shower parameters
Vincia:CMWtypeII       = 2    ! CMW rescaling of Lambda for II antennae
Vincia:CMWtypeIF       = 2    ! CMW rescaling of Lambda for IF antennae
Vincia:alphaSkMuI      = 0.75 ! Renormalization-scale prefactor for ISR 
                              ! emissions
Vincia:alphaSkMuSplitI = 0.7  !   -"- for g->qq splittings
Vincia:alphaSkMuConv   = 0.7  !   -"- for ISR conversions
# Shower IR cutoff and primordial kT
Vincia:cutoffScaleII   = 1.0  ! Cutoff value (in GeV) for II antennae
Vincia:cutoffScaleIF   = 0.9  ! Cutoff value (in GeV) for IF antennae
BeamRemnants:primordialKThard = 1.05 ! Primordial kT for hard interactions
BeamRemnants:primordialKTsoft = 0.7  ! Primordial kT for soft interactions
\end{verbatim} }

\FigRef{fig:ZpTphiStar} shows angular correlations and the transverse momentum spectrum
of the Drell-Yan lepton pair. As one would expect the spectrum of \vc~2.0 wimpy dies
out at the $Z$ mass. The prediction of default \vc~2.0 shows too much activity in the
hard tail of the spectrum which is caused by the reweighting of the event sample that 
includes high-$p_\perp$ jets, see \secRef{sec:hardJets}. The tuning of
the renormalisation-scale prefactors was chosen to produce as good a
compromise as possible between the regions above and below 
$p_\perp \sim m_Z/2$.

\FigRef{fig:Zshapes} shows the improved predictions when MECs are included.
The left plot shows the relative azimuthal angle between the $Z$ boson and the 
hardest jet, $\Delta\phi(\t Z,\t J_1)$, where multiple shower emissions are required to 
obtain values below $\pi$. This plots shows that although \py's power 
shower is matrix-element corrected for the first emission and results in a very good 
description of the $Z$ transverse momentum, its prediction for $\Delta\phi(\t Z,\t J_1)$ 
is worse than that of \vc~2.0 wimpy. For this observable as well as for the thrust in 
the right plot in \figRef{fig:Zshapes} default \vc~2.0 agrees well with 
the data.

\FigRef{fig:Zjets} shows the inclusive cross section for the Drell-Yan lepton pair
plus $\ge N$ jets, the transverse momentum and the pseudorapidity spectrum of the
leading jet. For all observables we find default \vc~2.0 to produce a fairly good
description of the data.
As expected, \vc~2.0 wimpy is not able to produce enough jets and can not populate 
the full spectrum of the transverse momentum of the hardest jet.

\subsection*{Underlying Event}

Although soft-inclusive QCD physics is not the main focus of this
version of \vc, it is nonetheless relevant to verify that a reasonable description
of the underlying event (UE) is obtained. We rely on the basic
multi-parton-interaction (MPI) modelling of \py~8~\cite{Sjostrand:1987su,Sjostrand:2004ef,Sjostrand:2014zea} 
including its default colour-reconnection (CR) model, with parameters
reoptimised for use with \vc's initial- and final-state showers. 

The MPI and CR parameter choices for the default \vc~2.001 tune are as
follows:
{\small \begin{verbatim}
! UE/MPI tuning parameters
SigmaProcess:alphaSvalue            = 0.118
SigmaProcess:alphaSorder            = 2
MultiPartonInteractions:alphaSvalue = 0.119
MultiPartonInteractions:alphaSorder = 2
MultiPartonInteractions:pT0ref      = 2.00
MultiPartonInteractions:expPow      = 1.75
MultiPartonInteractions:ecmPow      = 0.21
! Parameters for PYTHIA 8's baseline CR model
ColourReconnection:reconnect = on
ColourReconnection:range     = 1.75
! VINCIA is not compatible with perturbative diffraction
Diffraction:mMinPert = 1000000.0
\end{verbatim} }
Note that we choose 2-loop running for $\alpha_s$,
analogously to the rest of \vc, whereas the default \py~8.2 Monash
tune~\cite{Skands:2014pea} uses 1-loop running. We also set the
$\alpha_s(M_Z)$ reference value for 
hard processes (\texttt{SigmaProcess:alphaSvalue}) to the same value
(0.118) as used for the showers, and use a similar value (0.119) for
MPI, whereas the default \py~tune employ larger values $\sim$
0.13. The remaining MPI parameters were optimised using the
7-TeV charged-track summed-$p_\perp$ and number densities from~\cite{Aad:2010fh},
as well as their 900-GeV equivalents to constrain the energy-scaling
parameter. The colour-reconnection strength was determined  using the high-multiplicity region of the  $\left<p_\perp\right>(N_\mrm{ch})$ distribution measured by ATLAS~\cite{Aad:2010fh} in minimum-bias events. It should be noted however that \vc\ is not suitable for (low-multiplicity) minimum-bias physics in its present form. This is partly related to the last parameter, which is included to switch off \py's perturbative treatment of hard diffraction, with which \vc\ is not yet compatible. 

\begin{figure}[tbp]
\centering
\includegraphics[width=0.45\textwidth]{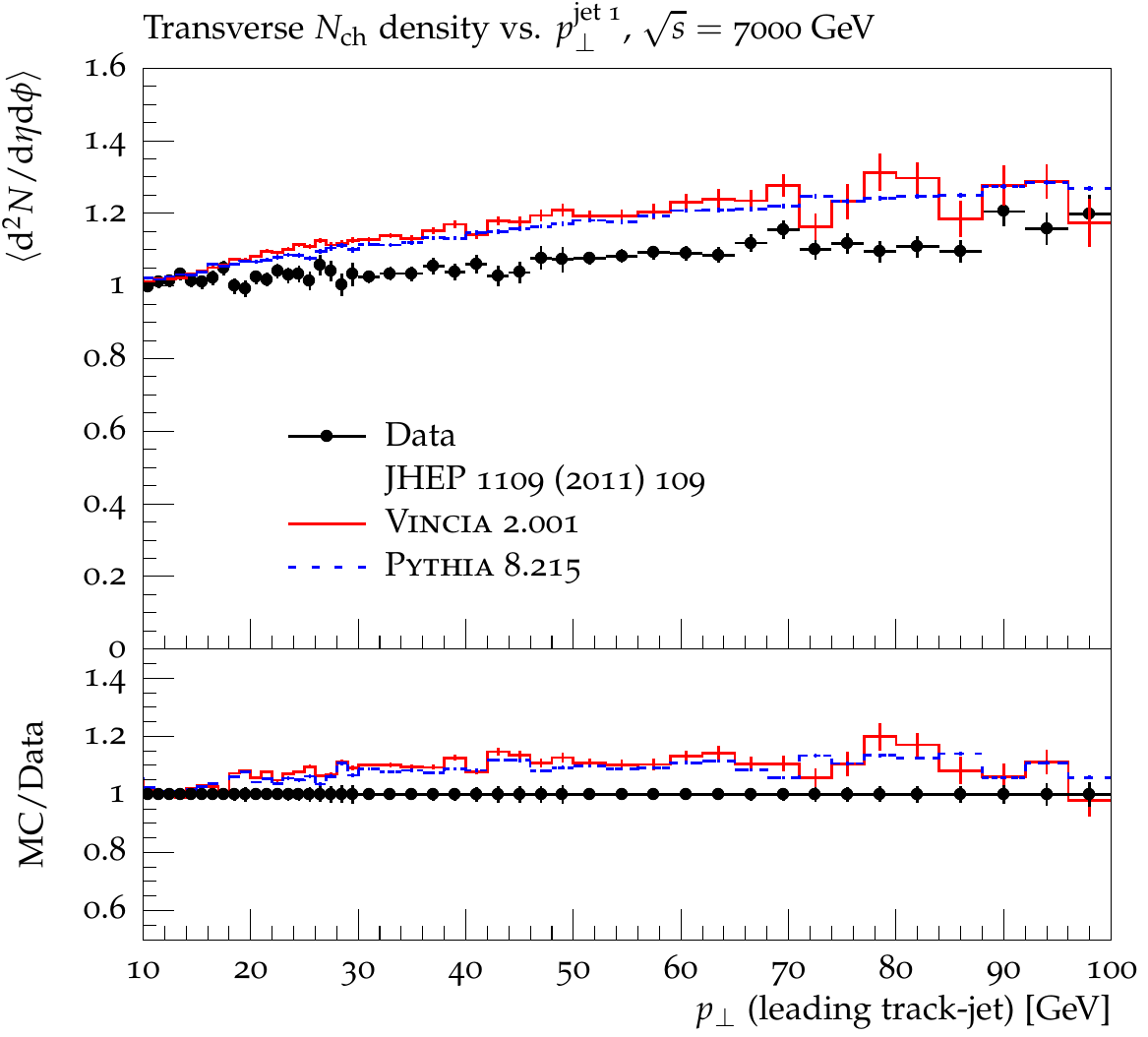}
\hspace*{5mm}
\includegraphics[width=0.45\textwidth]{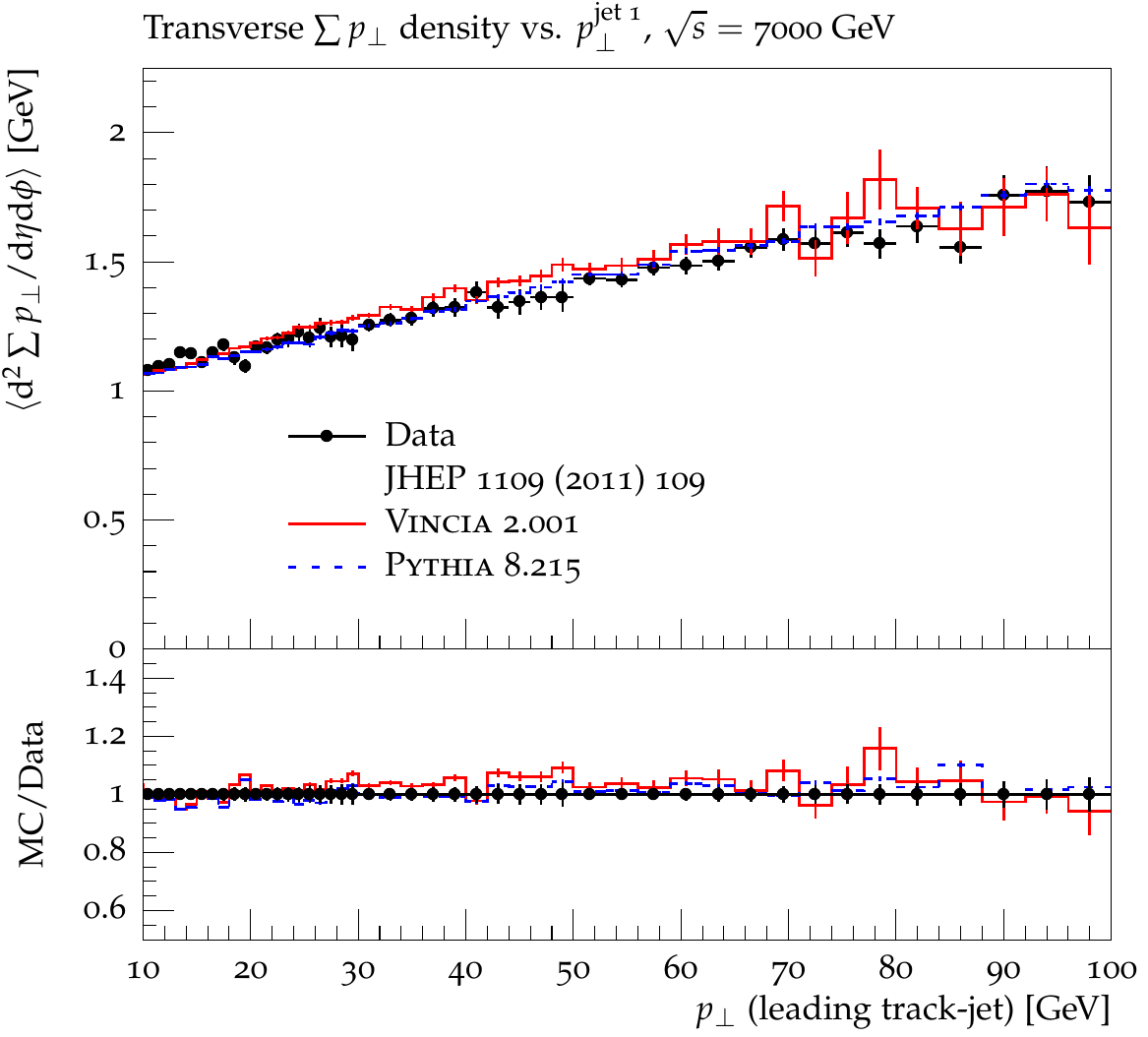}
\vspace*{3mm} \\
\includegraphics[width=0.45\textwidth]{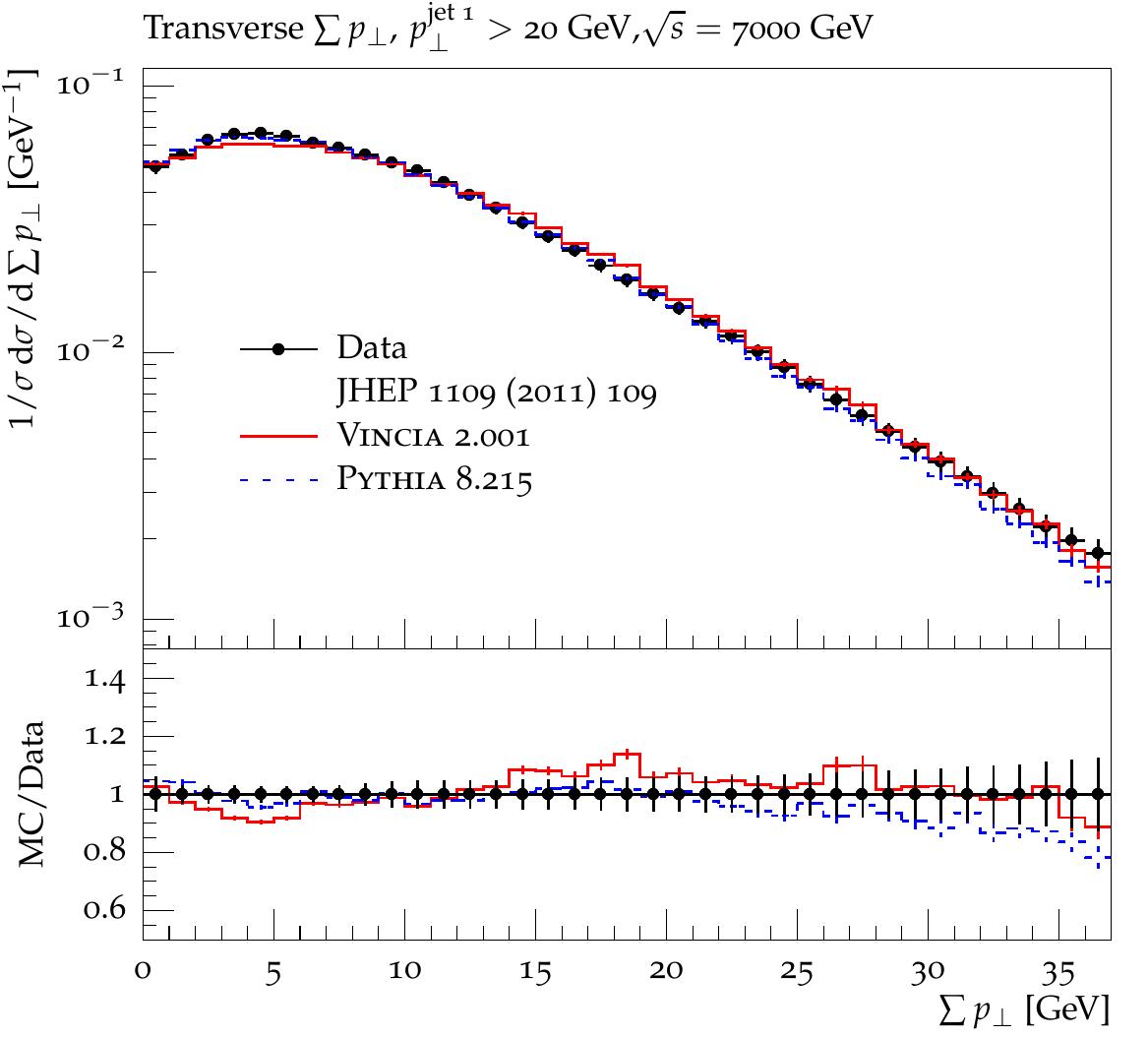}
\caption{\label{fig:UE}
The underlying event in $pp$ collisions at 7 TeV:
Measurement of charged particles with $p_\perp > 0.5~\mrm{GeV}$ and
$|\eta|<2$ in the transverse region; average multiplicity ({\sl top left}) and 
average scalar $\sum p_\perp$ ({\sl top right}) as a function of the transverse 
momentum of the leading track-jet, and normalized scalar $\sum p_\perp$ distribution
for leading track-jets with $p_\perp > 20~\mrm{GeV}$ ({\sl bottom}).
Predictions of default \vc~2.0 in red and \py~8.2 in blue,
compared to CMS data from \cite{Chatrchyan:2011id}.
Note that we use a cut of $p_\perp > 15~\mrm{GeV}$ in the hard process for the 
MC predictions and are therefore not showing the region of 
$1~\mrm{GeV}<p_\perp~\t{(leading track-jet)}<10~\mrm{GeV}$ for the top histograms.}
\end{figure}

In \figRef{fig:UE}, we compare default \vc~2.0 with default \py~8.2, to three 
basic observables measuring the level of activity in the region transverse to the 
leading (hardest) charged-particle jet in the central pseudorapidity region, $|\eta|<2$, 
for LHC collisions at 7 TeV. 
We use the conventional definition of the transverse region, spanning
$60^\circ < \Delta\phi < 120^\circ$ in azimuth with respect to the
leading charged-particle jet, and compare to CMS data~\cite{Chatrchyan:2011id}. 
These comparisons satisfy us that at least the global properties of the UE are in acceptable agreement with the measurements, in particular in regards to the average $p_\perp$ density (top right-hand plot) and its event-to-event fluctuations (bottom right-hand plot). The charged-track multiplicity (top left-hand plot) is a more difficult observable to predict since it is less IR safe and hence more dependent on details of the hadronisation modelling; we presume that the small (${\cal O}(10\%)$) discrepancies observed for both \py\ and \vc\ in this observables may be due to imperfections in \py's still rather crude modelling of colour reconnections. 

\subsubsection*{QCD Jets \label{sec:QCDjets}}

As our final set of validation checks, we consider the following observables in hard-QCD events: azimuthal dijet decorrelations, jet cross sections, and jet shapes. A technical aspect is that, due to the steeply falling nature of the jet $p_\perp$ spectrum, we use weighted events for all MC results in this section.  
The basic $2\to2$ QCD process at the scale $\hat p_\perp$ is oversampled by an amount of
$(\hat p_\perp/10)^4$, while the compensating event weight is $(10/\hat p_\perp)^4$.
This allows to fill the low-cross-section tails of the distributions with a reasonable amount of events. Note however that for observables that are not identical to the biasing variable (which are all observables since no one-to-one measurement of the partonic $\hat{p}_\perp$ is possible), rare events with large weights can then produce ``spurious'' peaks or dips in distributions, accompanied by large error bars. Such features are to be expected in some of the distributions we show below; removing them would require generating substantially more events.
While these features appear in the predictions of \vc, they are not present in \py's
distributions. The reason is as follows: The aforementioned event weight becomes large
for small values of $\hat p_\perp$. As this value serves as the starting scale in \py's shower,
the event will not produce any high-$p_\perp$ jets. In \vc, however, the full phase space
for the first emission is explored with the suppression factor $P_\t{imp}$ which is necessary for
the application of MECs. In the rare cases, where \vc\ produces a jet with $p_{\perp\,j}\gg
\hat p_\perp$, the large event weight becomes visible in distributions which require
high-$p_\perp$ jets.

A second technical aspect is that, as shown in \figRef{fig:PSMEQCD}, the PS-to-ME ratios for QCD processes result in rather 
broad distributions already for the first order correction with gluon emission only. This 
complicates including MECs for QCD processes, as violations in the Sudakov veto algorithm 
for generating emission and no-emission probabilities in the shower become more likely. By 
default, we neglect such violations. It is however possible for the user to check the effect 
of taking the violations into account properly via the procedure outlined in 
Ref.~\cite{Hoeche:2009xc}, which has been included in \vc.

\begin{figure}[b!]
\centering
\begin{minipage}[b]{0.49\textwidth}
  \includegraphics[width=0.99\textwidth]{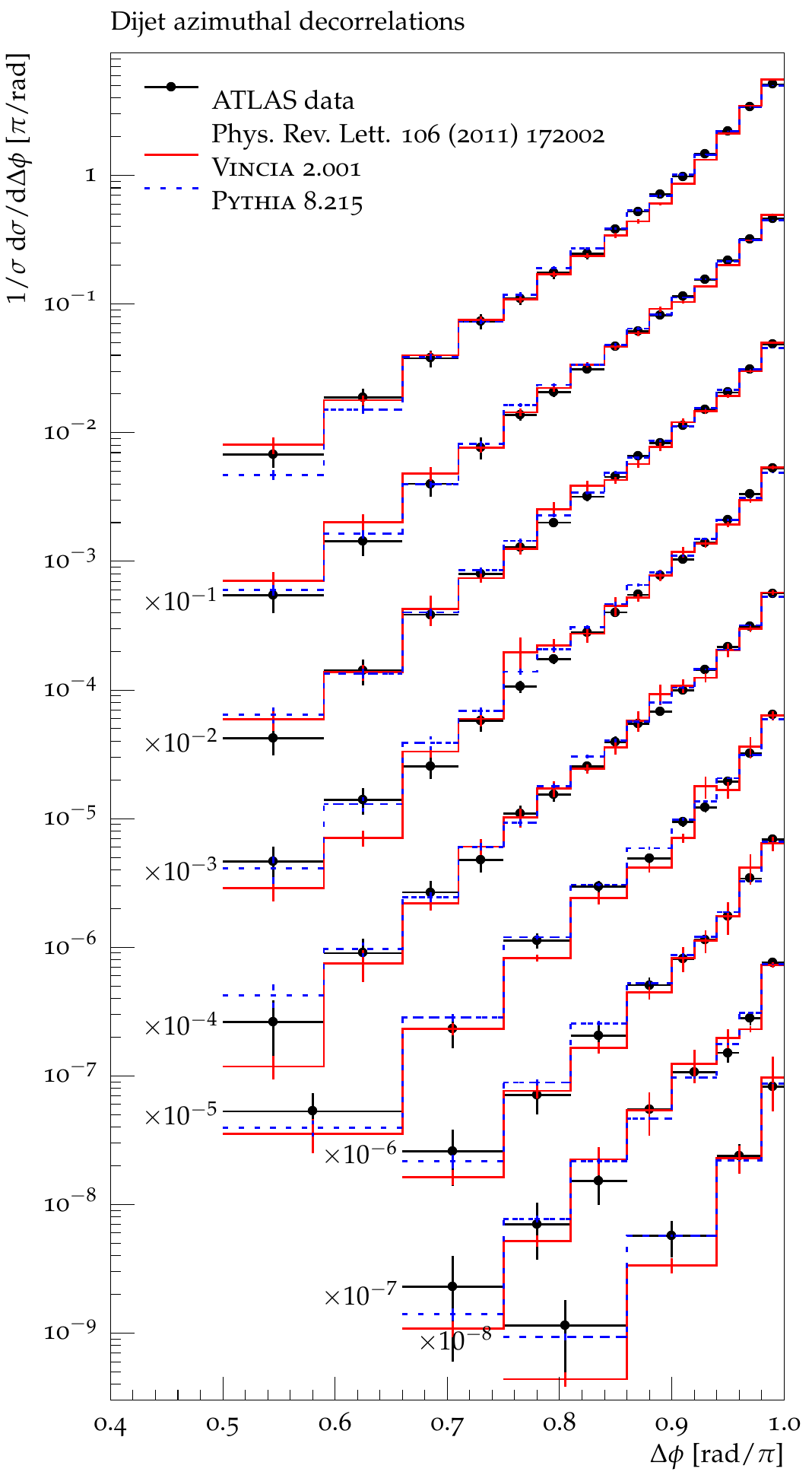}
\end{minipage}
\begin{minipage}[b]{0.49\textwidth}
  \graphNoSpace{width=0.99\textwidth}{ATLAS_2011_S8971293/d01-x01-y01-ref1.pdf}
  \graphNoSpace{width=0.99\textwidth}{ATLAS_2011_S8971293/d01-x01-y01-ref2.pdf}
  \graphNoSpace{width=0.99\textwidth}{ATLAS_2011_S8971293/d01-x01-y01-ref3.pdf}
  \graphNoSpace{width=0.99\textwidth}{ATLAS_2011_S8971293/d01-x01-y01-ref4.pdf}
  \graphNoSpace{width=0.99\textwidth}{ATLAS_2011_S8971293/d01-x01-y01-ref5.pdf}
  \graphNoSpace{width=0.99\textwidth}{ATLAS_2011_S8971293/d01-x01-y01-ref6.pdf}
  \graphNoSpace{width=0.99\textwidth}{ATLAS_2011_S8971293/d01-x01-y01-ref7.pdf}
  \graphNoSpace{width=0.99\textwidth}{ATLAS_2011_S8971293/d01-x01-y01-ref8.pdf}
  \includegraphics[width=0.99\textwidth]{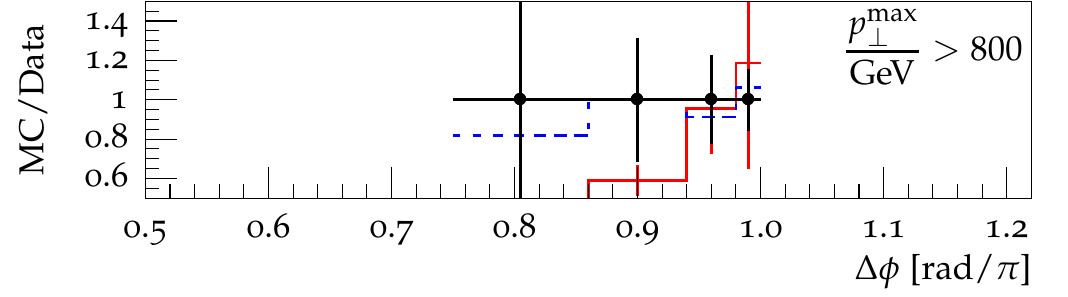}
\end{minipage}
\caption{\label{fig:DijetAzimCorr} 
Distribution of dijet azimuthal decorrelations; predictions of \vc~2.0 in red and 
\py~8.2 in blue, compared to ATLAS data from \cite{daCosta:2011ni}.}
\end{figure}

In \figRef{fig:DijetAzimCorr} we show the predictions of \vc~2.0 and \py~8.2 
for dijet azimuthal decorrelations for different ranges of the jet transverse 
momentum and compare to ATLAS data from \cite{daCosta:2011ni}.
While we observe no glaring discrepancies with the data --- the general trends of 
the distributions are well reproduced by both \vc\ and \py --- 
there still appears to be some room for improvement, in particular with \vc\ 
undershooting the precisely measured data points around $\Delta\phi \sim 0.9$ 
in the lower two $p_\perp^\t{max}$ bins by about 10--20\%.

\afterpage{
\begin{figure}[p!]
\centering
\begin{minipage}[b]{0.49\textwidth}
  \includegraphics[width=0.99\textwidth]{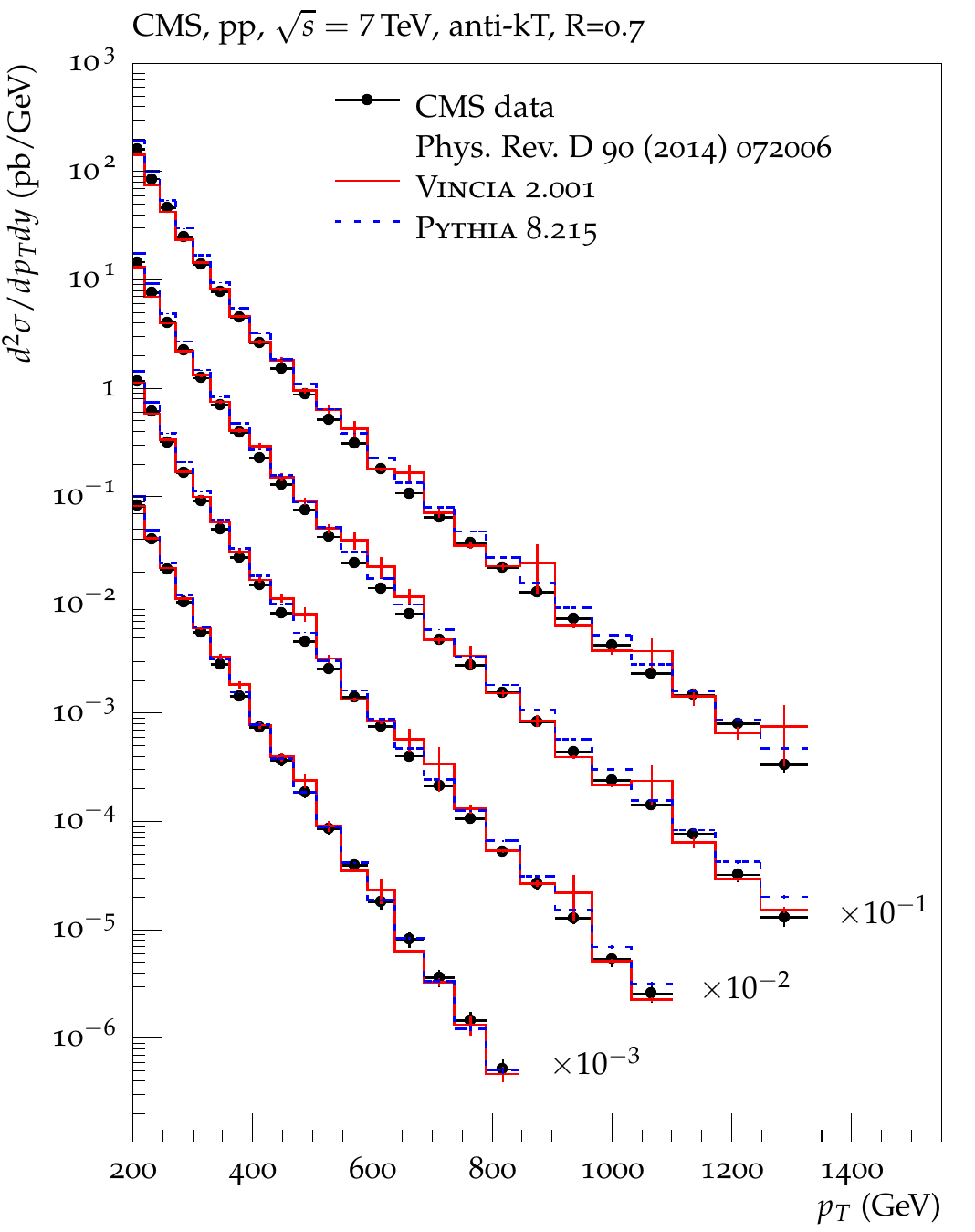}
\end{minipage}
\begin{minipage}[b]{0.49\textwidth}
  \graphNoSpace{width=0.99\textwidth}{CMS_2014_I1298810/d07-x01-y01-ref1.pdf}
  \graphNoSpace{width=0.99\textwidth}{CMS_2014_I1298810/d07-x01-y01-ref2.pdf}
  \graphNoSpace{width=0.99\textwidth}{CMS_2014_I1298810/d07-x01-y01-ref3.pdf}
  \includegraphics[width=0.99\textwidth]{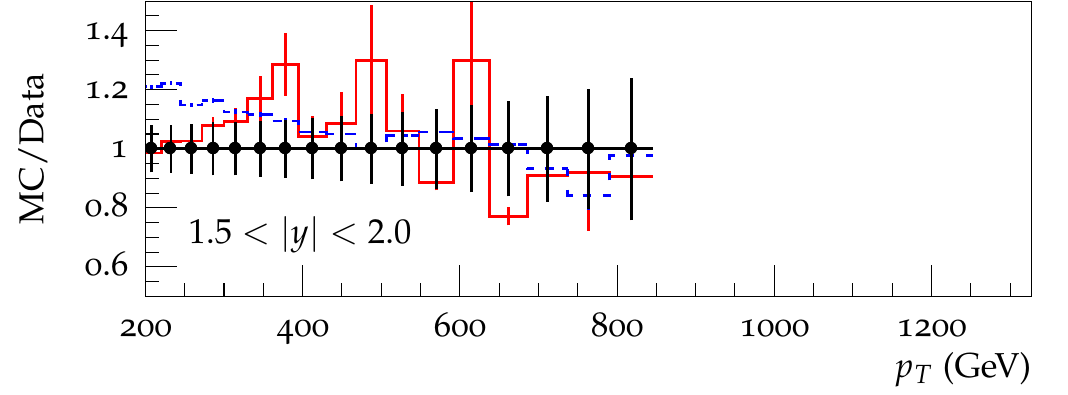}
\end{minipage}\vspace*{-2mm}
\caption{\label{fig:QCDpT} 
Inclusive jet cross section for 4 different rapidity bins as a function
of the jet $p_\perp$. Predictions of \vc~2.0 in red and \py~8.2 in blue. Data from CMS~\cite{Chatrchyan:2014gia}.}
\vskip6mm
\centering
\begin{minipage}[b]{0.49\textwidth}
  \includegraphics[width=0.99\textwidth]{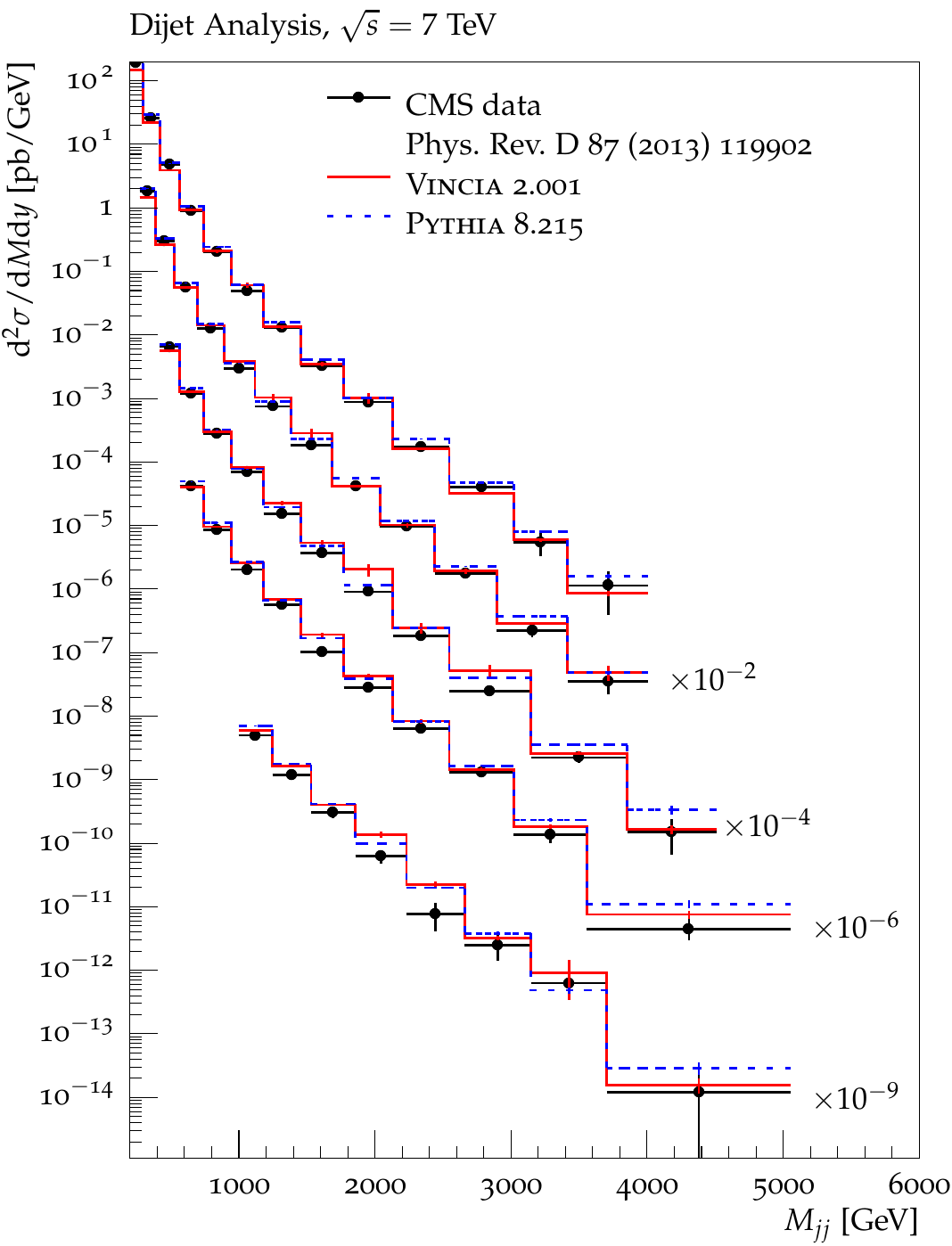}
\end{minipage}
\begin{minipage}[b]{0.49\textwidth}
  \graphNoSpace{width=0.99\textwidth}{CMS_2013_I1208923/d02-x01-y01-ref1.pdf}
  \graphNoSpace{width=0.99\textwidth}{CMS_2013_I1208923/d02-x01-y01-ref2.pdf}
  \graphNoSpace{width=0.99\textwidth}{CMS_2013_I1208923/d02-x01-y01-ref3.pdf}
  \graphNoSpace{width=0.99\textwidth}{CMS_2013_I1208923/d02-x01-y01-ref4.pdf}
  \includegraphics[width=0.99\textwidth]{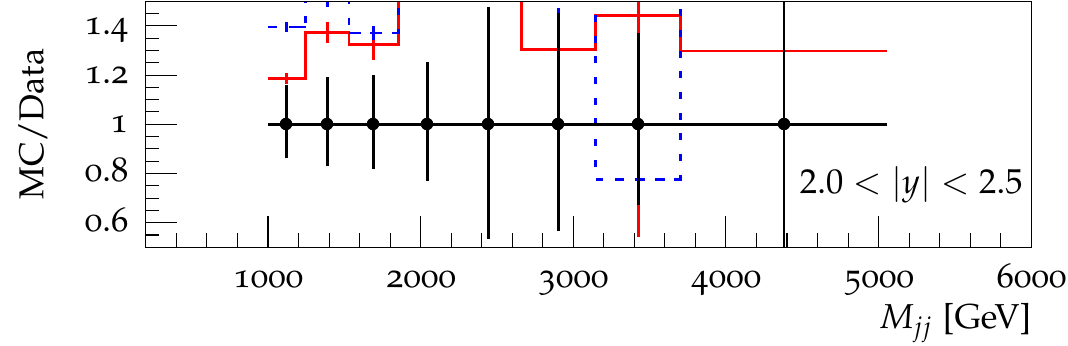}
\end{minipage}\vspace*{-2mm}
\caption{\label{fig:QCDmjj} 
Inclusive dijet cross sections for 5 different rapidity bins as a function
of the dijet mass. Predictions of \vc~2.0 in red and \py~8.2 in blue. Data from CMS~\cite{Chatrchyan:2012bja}.}
\end{figure}
\clearpage}

\afterpage{
\begin{figure}[p]
\centering
\begin{minipage}[b]{0.49\textwidth}
  \includegraphics[width=0.99\textwidth]{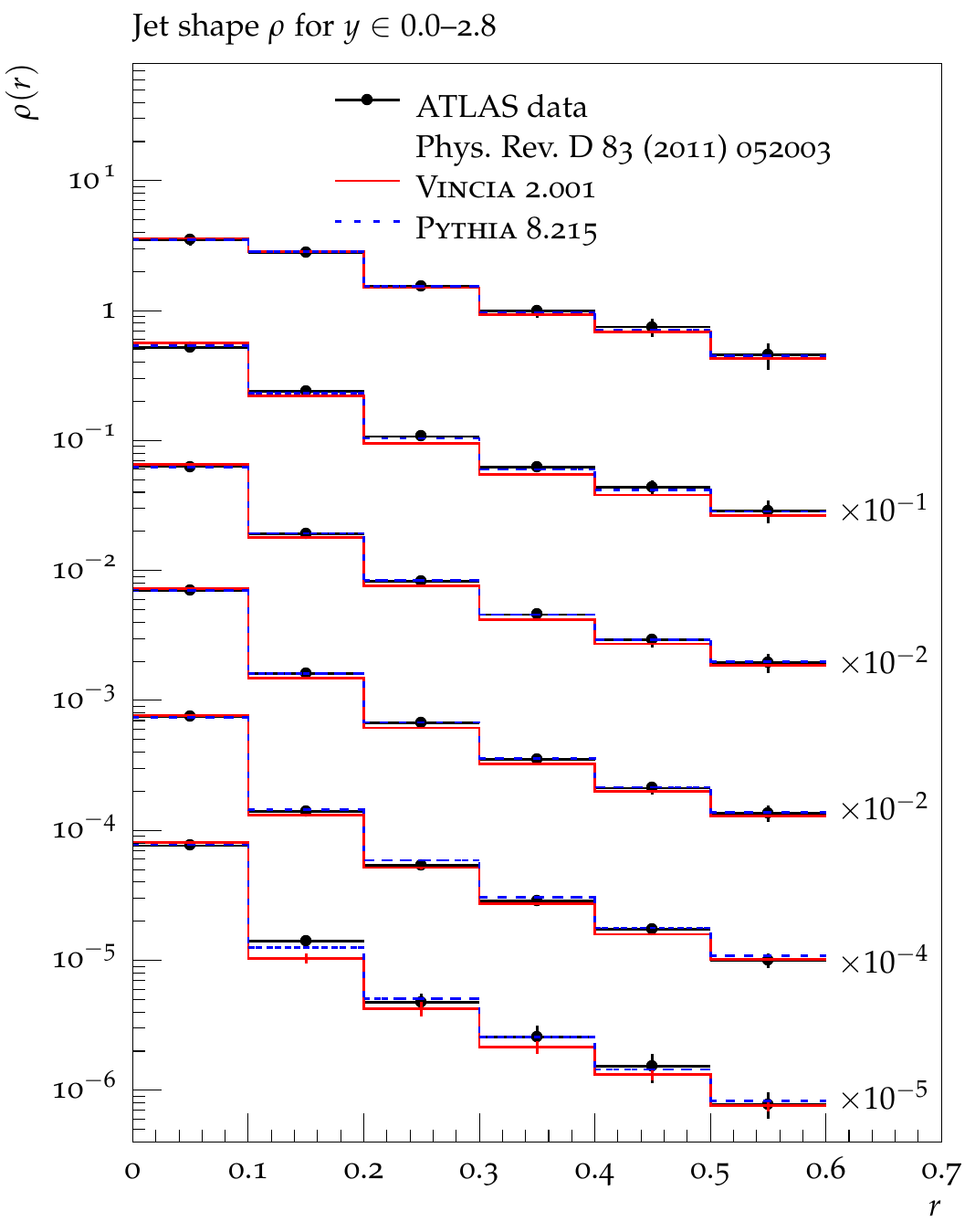}
\end{minipage}
\begin{minipage}[b]{0.49\textwidth}
  \graphNoSpace{width=0.99\textwidth}{ATLAS_2011_S8924791/d01-x06-y01-ref1.pdf}
  \graphNoSpace{width=0.99\textwidth}{ATLAS_2011_S8924791/d01-x06-y01-ref2.pdf}
  \graphNoSpace{width=0.99\textwidth}{ATLAS_2011_S8924791/d01-x06-y01-ref3.pdf}
  \graphNoSpace{width=0.99\textwidth}{ATLAS_2011_S8924791/d01-x06-y01-ref4.pdf}
  \graphNoSpace{width=0.99\textwidth}{ATLAS_2011_S8924791/d01-x06-y01-ref5.pdf}
  \includegraphics[width=0.99\textwidth]{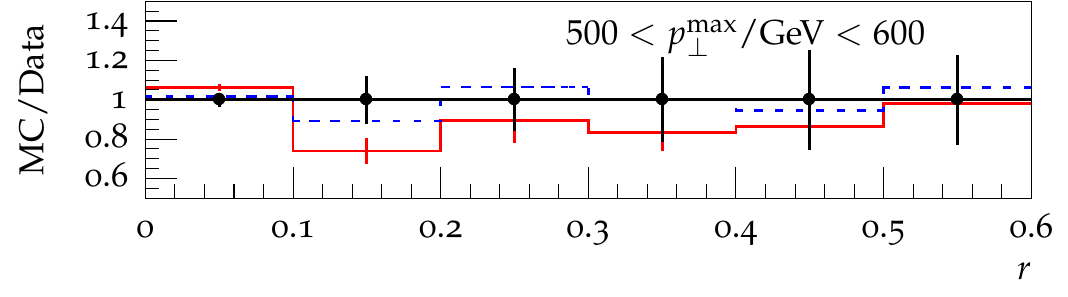}
\end{minipage} \\
\begin{minipage}[b]{0.49\textwidth}
  \includegraphics[width=0.99\textwidth]{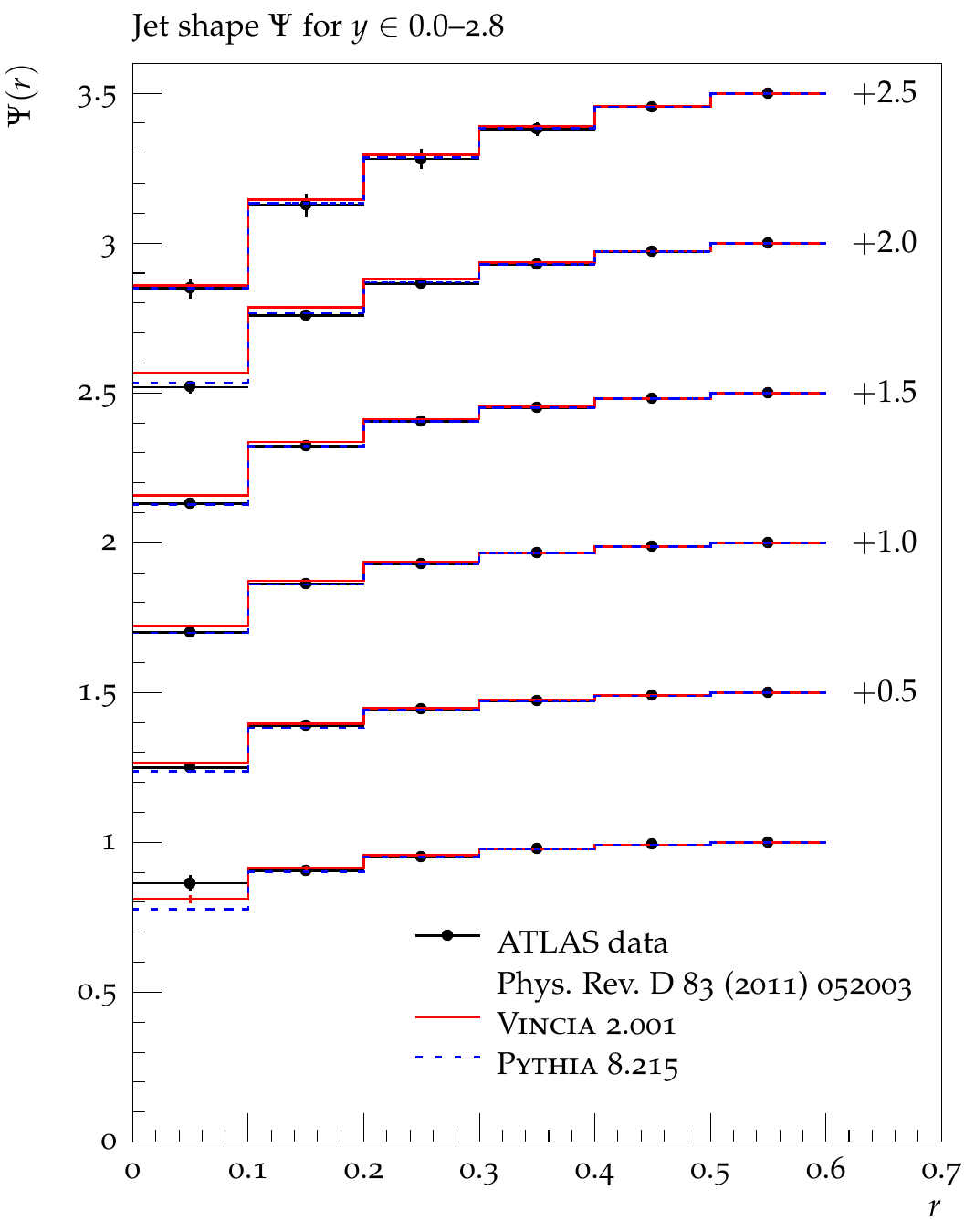}
\end{minipage}
\begin{minipage}[b]{0.49\textwidth}
  \graphNoSpace{width=0.99\textwidth}{ATLAS_2011_S8924791/d01-x06-y02-ref1.pdf}
  \graphNoSpace{width=0.99\textwidth}{ATLAS_2011_S8924791/d01-x06-y02-ref2.pdf}
  \graphNoSpace{width=0.99\textwidth}{ATLAS_2011_S8924791/d01-x06-y02-ref3.pdf}
  \graphNoSpace{width=0.99\textwidth}{ATLAS_2011_S8924791/d01-x06-y02-ref4.pdf}
  \graphNoSpace{width=0.99\textwidth}{ATLAS_2011_S8924791/d01-x06-y02-ref5.pdf}
  \includegraphics[width=0.99\textwidth]{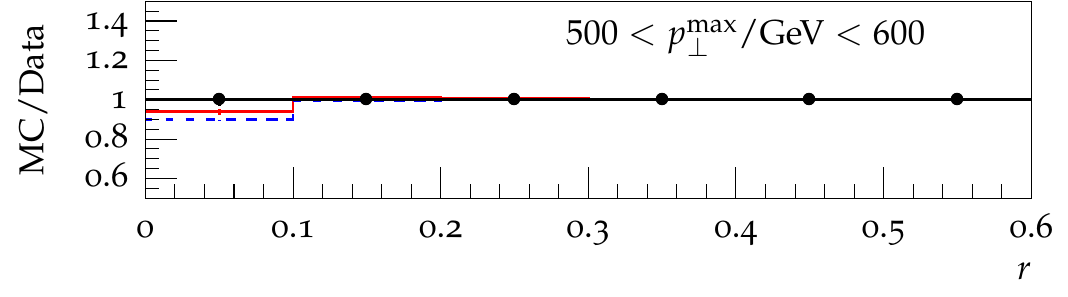}
\end{minipage}
\caption{\label{fig:RhoPsi} 
Distributions of the jet shape variables $\rho(r)$ ({\sl top}) and $\Psi(r)$ 
({\sl bottom}) for different ranges of the jet transverse momentum. Predictions 
of \vc~2.0 in red and \py~8.2 in blue, compared to ATLAS data from \cite{Aad:2011kq}.}
\end{figure}
\clearpage}

\FigsRef{fig:QCDpT} and \ref{fig:QCDmjj} show the transverse momentum and jet mass
spectra for different ranges of the jet rapidity and compare the MC predictions
to CMS data from \cite{Chatrchyan:2014gia} and \cite{Chatrchyan:2012bja}
respectively. We note that, whereas \py\ lies systematically above the data here, the lower default 
$\alpha_s$ value chosen in \vc\ causes the \vc\ normalisations to be substantially lower, even to the point of undershooting the measurements. 
This is not surprising given that the inclusive-jet cross section in \py/\vc\ is calculated at LO. The tails of the distributions unfortunately suffer from rather large weight-fluctuation effects, as was discussed above; nonetheless we note that the bins for which a reasonable statistical precision is obtained are generally closer to the data than the \py\ reference comparison. 

Finally, in \figRef{fig:RhoPsi} we show the differential jet shape variable $\rho(r)$ and 
its cumulative integral $\Psi(r)$ for different ranges of the jet transverse momentum, compared with ATLAS data from \cite{Aad:2011kq}. This validates that the FSR broadening of QCD jets is in reasonable agreement with the experimental measurements, though we note that \vc's distributions may be slightly too narrow, which we again regard as being consistent with the LL nature of \vc's antenna functions and analogous to the slightly too narrow Thrust distribution we allowed in the $e^+e^-$ event shapes. As far as a first default set of parameters goes, we are satisfied with this level of tuning, with future directions being informed both by lessons from combinations with external matrix-element matching and merging schemes and by attempts to integrate NLO antenna-function corrections into the shower itself, e.g.\ in the spirit of~\cite{Hartgring:2013jma}.

\section{Summary and Conclusions \label{sec:conclusions}}
We presented the first publicly available antenna shower for initial and final state in
\vc~2.0, with focus on antenna functions and kinematic maps for initial state radiation.
\vc~2.0 includes two different methods to explore the full phase space for the first 
emission, depending on the hard process at hand, without the disadvantages of a ``power 
shower''. The full phase space of subsequent emissions is populated in a Markovian way.
We compare explicitly to tree-level matrix elements for $pp\to Z/H\,jj(j)$ and $pp\to jjj$
to check the validity of our approximations.

We extended the iterative MEC approach to the initial state and include MECs for
QCD up to $\mathcal{O}(\alpha_s^4)$ (4 jets), and for Drell-Yan and Higgs production 
up to $\mathcal{O}(\alpha_s^3)$ ($V/H$ + 3 jets). This is the first time MECs beyond one leg have been applied to hadron collisions. 
However, this implementation was not without its complications; the large phase space available
for initial-state branchings implies that ``unordered'' emissions account for a larger fraction of the full 
phase space than was the case for FSR, and the MEC factors are less well behaved and therefore more difficult / less efficient to implement, compared  to pure final-state MECs. We also saw in \secRef{sec:QCDjets} that biased event samples result in larger weight fluctuations for \vc\ than in the case of pure \py, presumably due to unordered emissions in \vc\ allowing a larger range of corrections to each event. In the context of future developments of \vc, these aspects will therefore merit further consideration. 

We presented first validation results with \vc~2.0 for the main benchmark processes for FSR and ISR, including hadronic $Z$ decays, Drell-Yan, and QCD jets. We observe good agreement with experimental data from the LEP/SLD and LHC experiments. 

The development of a more highly automated interface to \mg~5 is among the main 
development targets for the near future. The feasibility of an interface to
\textsc{Njet2}~\cite{Badger:2013vpa} is also being explored.

\subsection*{Acknowledgements}
We thank the HepForge project, \url{www.hepforge.org} \cite{Buckley:2006nm} 
for providing the hosting and repository services used for developing and 
maintaining the \vc\ code. HepForge is funded via the UK STFC and hosted at 
the Durham IPPP. We thank Johannes Bellm for useful discussions on
\cite{Bellm:2016rhh}.
This work was supported 
in part by the ARC Centre of Excellence for Particle Physics at the Terascale.
SP is supported by the US Department of Energy under contract DE-AC02-76SF00515.
The work of MR was supported by the European Research Council under Advanced
Investigator Grant ERC--AdG--228301 and ERC Advanced Grant no.~320651, 
``HEPGAME''. PS is the recipient of an Australian Research Council Future 
Fellowship, FT130100744: ``Virtual Colliders: high-accuracy models for high 
energy physics''. 

\appendix

\section{Details of the Shower Algorithm \label{app:showerDetails}}

In this section we present some details of the shower algorithm, starting
with the construction of the kinematics after the branching. Thereafter we
will give a brief overview on how the antenna functions correctly reproduce
the DGLAP functions in the collinear limit.

\subsection{Construction of the Post-Branching Momenta}
The antenna picture does not distinguish between the emitter / splitter 
and recoiler. Therefore, we have one mapping for each configuration, initial-initial,
initia-final, and final-final. We express the momenta in 
terms of branching invariants only which leads to very simple expressions.

\subsubsection*{Initial-Initial Antennae \label{app:IIkinematics}}
For a branching of type $AB\to abj$, the invariant mass and rapidity of the recoiler,
$R\to r$, are not changed. The kinematics are constructed in the lab frame, where the 
post-branching momenta read as follows,
\begin{align}
  p_a^\mu &= \sqrt{\frac\sab\sAB\frac{\sAB+\sjb}{\sAB+\saj}}\,p_A^\mu~, \\
  p_b^\mu &= \sqrt{\frac\sab\sAB\frac{\sAB+\saj}{\sAB+\sjb}}\,p_B^\mu~, \\
  p_j^\mu &= \sqrt{\frac{\sjb^2}{\sab\sAB}\frac{\sAB+\sjb}{\sAB+\saj}}\,p_A^\mu +
             \sqrt{\frac{\saj^2}{\sab\sAB}\frac{\sAB+\saj}{\sAB+\sjb}}\,p_B^\mu +
             \sqrt{\frac{\saj\sjb}\sab}\,p_\perp^\mu~, \\
  p_r^\mu &= p_a^\mu + p_b^\mu - p_j^\mu~,
\end{align}
with $p_\perp=(0,\cos\phi,\sin\phi,0)$, where $\phi$ is chosen uniformly in $[0,2\pi]$.

\subsubsection*{Initial-Final Antennae \label{app:IFkinematics}}
For a branching of type $AK\to akj$ the kinematics are constructed in the centre-of-mass
frame of the parent antenna, which we define to be the rest frame of $p_A+p_K$ here, rotated so they are aligned with the $z$ axis. (The inverse of the corresponding Lorentz transformation is applied afterwards to bring the system back to the lab frame.) 
The post-branching momenta read as follows,
\begin{align}
  p_a^\mu &= \frac{\sAK+\sjk}\sAK\,p_A^\mu~, \\
  p_k^\mu &= \frac{\sjk\saj}{\sAK(\sAK+\sjk)}\,p_A^\mu +
              \frac\sak{\sAK+\sjk}\,p_K^\mu -
             \frac{\sqrt{\sjk\sak\saj}}{\sAK+\sjk}\,p_\perp^\mu~, \\
  p_j^\mu &= \frac{\sjk\sak}{\sAK(\sAK+\sjk)}\,p_A^\mu +
             \frac\saj{\sAK+\sjk}\,p_K^\mu +
             \frac{\sqrt{\sjk\sak\saj}}{\sAK+\sjk}\,p_\perp^\mu~,
\end{align}
with $p_\perp$ defined as in the previous paragraph.

\subsubsection*{Final-Final Antennae \label{app:FFkinematics}}
For a branching of type $IK\to ijk$ the kinematics are constructed in the
centre-of-mass frame of the parent antenna, with the direction of parton $I$ defining the positive $z$ axis. (The inverse of the corresponding Lorentz transformation is applied afterwards to bring the system back to the lab frame.)  
A first set of post-branching momenta is constructed with parton $i$ aligned with the $z$ axis and using the $xz$ plane to represent the branching plane,   
\begin{align}
  p_i^\mu &= E_i(1,0,0,1)~, \\
  p_k^\mu &= E_k(1,\sin\theta_{ik},0,\cos\theta_{ik}), \\
  p_j^\mu &= E_j(1,-\sin\theta_{ij},0,\cos\theta_{ij}),
\end{align}
with the energies
\begin{align}
  E_i = \frac{\sIK-\sjk}{2\,\sqrt{\sIK}}~,\quad
  E_k = \frac{\sIK-\sij}{2\,\sqrt{\sIK}}~,\quad
  E_j = \frac{\sIK-\sik}{2\,\sqrt{\sIK}}~,
\end{align}
and angles between the partons
\begin{align}
  \cos\theta_{ik} = \frac{2E_iE_k - \sik}{2E_iE_k}~,\quad
  \cos\theta_{ij} = \frac{2E_iE_j - \sij}{2E_iE_j}~.
\end{align}
The azimuth angle of the emitted gluon in the $xy$ plane (defining the orientation of the branching plane) is generated by rotating the above momenta around the $z$ axis by a uniformly chosen random angle $\phi$.

Finally, there remains one more global orientation angle, which can be cast as the angle between parton $i$ and the original parton $I$, $\psi_{Ii}$, around an axis perpendicular to the branching plane (still in the centre-of-mass frame), i.e., specifying the degree to which $p_i$ is not aligned with the $z$ axis after the branching. 
Different choices are implemented in \vc (see \cite{Lonnblad:1992tz,Giele:2007di}), with the default being
\begin{align}
  \cos\psi_{Ii} = 1+\dfrac{2y_{ii}}{\sIK-\sjk}\qquad\t{with}\qquad
  y_{ii} &= -\frac{(1-\rho)\sik/\sIK+2f\sij\sjk/\sIK^2}{2(1-\sij/\sIK)}~,\nonumber\\
  f &= \frac{\sjk}{\sij+\sjk}~,\qquad
  \rho = \sqrt{1+4f(1-f)\sij\sjk/\sik\sIK}~.
  \nonumber
\end{align}
The final post-branching momenta are constructed by rotating the $ijk$ system by the angle  $\psi_{Ii}$ around 
the axis perpendicular to the branching plane, and then finally 
performing the inverse Lorentz transform to bring the post-branching partons
back to the lab frame. 

\subsection{Collinear Limits of the Antenna Functions.}
\label{app:collLimit}

In this paragraph, we collect, for the convenience of the reader, the collinear 
limits of the antenna functions used in \vc. We also relate the antenna functions 
in this limit to corresponding DGLAP splitting kernels,
which we will denote by $P(x\to yz)$. Note that the apparent difference in colour
factors for DGLAP splitting kernels and antenna functions is due to the 
phase-space and coupling factor, which, for antennae, is
\begin{align}
\frac{\alpha_s\,\mc C_\t{ant}}{4\pi}\quad\t{with}\quad
\mc C_\t{ant} \in [C_A,2C_F,T_R]~,
\end{align}
whereas DGLAP kernels are conventionally defined with
\begin{align}
\frac{\alpha_s\,\mc C_\t{DGLAP}}{2\pi}\quad\t{with}\quad
\mc C_\t{DGLAP} \in [C_A,C_F,T_R/2]~.
\end{align}
We note that antenna functions with gluons as parent partons only reproduce
half of a DGLAP kernel, as gluons take part in two antennae.
In addition a factor of $1/z$ will multiply the DGLAP kernels in case of
initial state radiation.

The collinear limits of the antenna functions below agree with the limits
found in 
Refs.~\cite{GehrmannDeRidder:2005hi,GehrmannDeRidder:2005aw,GehrmannDeRidder:2005cm,Daleo:2006xa}.

\subsubsection*{Initial-Initial Antennae \label{app:IIdglapLimit}}
In the case of initial-initial antenna functions the energy-sharing variable
is $z=\sAB/\sab$ and we arbitrarily pick the invariant mass of one of the parton 
pairs, $Q^2=\saj$, and its scaled version, $y=Q^2/\sAB$.
For an easy comparison with the DGLAP kernels we rewrite the antenna
functions in terms of these variables,
\begin{align}
  \bar a_{q\bar q\,g}^\t{II} &=~
    \frac1{Q^2}\,\frac1z\,\left(2\frac{z}{1-z-zy}+
    \left(1-z-zy\right)\right)+\mc O(Q^2)~, \\
  \bar a_{gg\,g}^\t{II} &=~
    2\,\frac1{Q^2}\,\frac1z\,\left(\frac{z}{1-z-zy}+\frac1z-1-y+
    \left(1-z-zy\right)\frac{z}{1+zy}\right)+\mc O(Q^2)~, \\
  \bar a_{qx\,q}^\t{II} &=~
    \frac12\,\frac1{Q^2}\,\frac1z\,
    \left(\frac{\left(1-z-zy\right)^2+1}z\right)~, \\
  \bar a_{gx\,\bar q}^\t{II} &=~
    \frac1{Q^2}\,\frac1z\,\left(1-2z+2\frac{z^2}{1-zy}\right)~.
\end{align}
Note that we made use of $\sjb=\sAB(1-z-zy)/z$ and wrote the terms
in the gluon emission antennae, that do not contain $1/Q^2$ singularities
as $\mc O(Q^2)$, as they will vanish in the unresolved limit anyway.

Given this new form of the antenna functions, the collinear limits,
$y\to 0$, are simple to read off,
\begin{alignat}{3}
  \bar a_{q\bar q\,g}^\t{II} &\to~
    \frac1{Q^2}\,\frac1z\,\left(\frac{1+z^2}{1-z}\right)
    ~&=&~ \frac1{Q^2}\,\frac1z\,P(q\to qg)~,\\
  \bar a_{gg\,g}^\t{II} &\to~
    2\,\frac1{Q^2}\,\frac1z\,\left(\frac{z}{1-z}+\frac{1-z}z+z(1-z)\right)
    ~&=&~ \frac1{Q^2}\,\frac1z\,P(g\to gg)\, \\
  \bar a_{qx\,q}^\t{II} &\to~ 
    \frac12\,\frac1{Q^2}\,\frac1z\,\left(\frac{(1-z)^2+1}z\right)
    ~&=&~ \frac12\,\frac1{Q^2}\,\frac1z\,P(q\to gq)~, \\
  \bar a_{gx\,\bar q}^\t{II} &\to~ 
    \frac1{Q^2}\,\frac1z\,\left(z^2+(1-z)^2\right)
    ~&=&~ \frac1{Q^2}\,\frac1z\,P(g\to q\bar q)~.
\end{alignat}
Note in particular that the second and fourth antenna functions include the full 
DGLAP kernels for $g\to gg$ and $g\to q\bar{q}$, respectively. This is different 
from their final-state counterparts (see below) in which two neighbouring antenna 
functions must be summed over to recover the full DGLAP kernels. This difference 
arises from the fact that there is no ``emission into the initial state'' --- the 
initial-state gluon only occurs as a hard leg, not as the emitted parton.

\subsubsection*{Initial-Final Antennae \label{app:IFdglapLimit}}
We start with the collinear limit of the initial state side, $Q^2=\saj$
and $y=Q^2/\sAK$, and energy-sharing variable $z=\sAK/(\sAK+\sjk)$ and 
rewrite the antenna functions,
\begin{align}
  \bar a_{qq\,g}^\t{IF} &=~
    \frac1{Q^2}\,\frac1z\,\left(\frac{1+z^2-2zy}{1-z}\right)+\mc O(Q^2)~,\\
  \bar a_{gg\,g}^\t{IF} &=~
    2\,\frac1{Q^2}\,\frac1z\,\left(\frac{z(1-zy)}{1-z}+\frac{(1-z)(1-zy)}z+
    z(1-z)\right)+\mc O(Q^2)~, \\
  \bar a_{qx\,q}^\t{IF} &=~
    \frac12\,\frac1{Q^2}\,\frac1z\,\left(\frac{(1-z)^2+(1-zy)^2}z\right)~, \\
  \bar a_{gx\,\bar q}^\t{IF} &=~
    \frac1{Q^2}\,\frac1z\,\left(z(1-2z)y+z^2+(1-z)^2\right)~.
\end{align}
Note that we made use of $\sjk=\sAK(1-z)/z$ and $\sak=\sAK(1-yz)/z$
and, as before, wrote the terms
in the gluon emission antennae, that do not contain $1/Q^2$ singularities
as $\mc O(Q^2)$.

Given this new form of the antenna functions, the collinear limits,
$y\to 0$, are simple to read off,
\begin{alignat}{3}
  \bar a_{qq\,g}^\t{IF} &\to~
    \frac1{Q^2}\,\left(\frac{1+z^2}{1-z}\right)
    ~&=&~ \frac1{Q^2}\,\frac1z\,P(q\to qg)~,\\
  \bar a_{gg\,g}^\t{IF} &\to~
    2\,\frac1{Q^2}\,\frac1z\,\left(\frac{z}{1-z}+\frac{1-z}z+z(1-z)\right)
    ~&=&~ \frac1{Q^2}\,\frac1z\,P(g\to gg)\, \\
  \bar a_{qx\,q}^\t{IF} &\to~
    \frac12\,\frac1{Q^2}\,\frac1z\,\left(\frac{(1-z)^2+1}z\right)
    ~&=&~ \frac12\,\frac1{Q^2}\,\frac1z\,P(q\to gq)~, \\
  \bar a_{gx\,\bar q}^\t{IF} &\to~
    \frac1{Q^2}\,\frac1z\,\left(z^2+(1-z)^2\right)
    ~&=&~ \frac1{Q^2}\,\frac1z\,P(g\to q\bar q)~.
\end{alignat}

Now we continue with the collinear limit of the final state side,
$Q^2=\sjk$ and $y=Q^2/\sAK$, and energy-sharing variable $z=\sak/\sAK$.
The antenna functions, rewritten in terms of the new variables
and using $\saj=\sAK(1-z+y)$, read
\begin{align}
  \bar a_{qq\,g}^\t{IF} &=~
    \frac1{Q^2}\,\frac1z\,\left(\frac{2z+(1-z+y)^2}{1-z+y}\right)+\mc O(Q^2)~,\\
  \bar a_{gg\,g}^\t{IF} &=~
    \frac1{Q^2}\left(2\,\frac{z}{1-z+y}+z(1-z)\right)+\mc O(Q^2)~, \\
  \bar a_{xq\,\bar q}^\t{IF} &=~
    \frac12\,\frac1{Q^2}\left((1-z+y)^2+z^2\right)~.
\end{align}

Given this new form of the antenna functions, the collinear limits,
$y\to 0$, are simple to read off,
\begin{alignat}{3}
  \bar a_{qq\,g}^\t{IF} &\to~
    \frac1{Q^2}\,\left(\frac{1+z^2}{1-z}\right)
    ~&=&~ \frac1{Q^2}\,P(q\to qg)~,\\
  \bar a_{gg\,g}^\t{IF} + \bar a_{gg\,g}^\t{IF} [z\leftrightarrow 1-z] &\to~
    2\,\frac1{Q^2}\,\left(\frac{z}{1-z}+\frac{1-z}z+z(1-z)\right)
    ~&=&~ \frac1{Q^2}\,P(g\to gg)\, \\
  \bar a_{xq\,\bar q}^\t{IF} &\to~
    \frac12\,\frac1{Q^2}\,\left((1-z)^2+z^2\right)
    ~&=&~ \frac12\,\frac1{Q^2}\,P(g\to q\bar q)~.
\end{alignat}

\subsubsection*{Final-Final Antennae \label{app:FFdglapLimit}}
In the case of final-final antenna functions the energy-sharing variable
is $z=\sik/\sIK$ and we arbitrarily pick the invariant mass of one of the parton 
pairs, $Q^2=\sjk$, and its scaled version, $y=Q^2/\sIK$.
For an easy comparison with the DGLAP kernels we rewrite the antenna
functions in terms of these variables (leaving out the finite parts as
their choice is arbitrary), 
\begin{align}
  \bar a_{q\bar q\,g}^\t{FF} &=~
    \frac1{Q^2}\,\left(\frac{2z+(1-z-y)^2}{1-z-y}\right)+\mc O(Q^2)~,\\
  \bar a_{gg\,g}^\t{FF} &=~
    \frac1{Q^2}\,\left(2\,\frac{z}{1-z-y}+(1-z-y)(z+y)\right)+\mc O(Q^2)~,\\
  \bar a_{xq\,\bar q}^\t{FF} &=~
    \frac12\,\frac1{Q^2}\left((1-z-y)^2+z^2\right)~.
\end{align}
Note that we made use of $\sij=\sIK(1-z-y)$ and, as before, wrote the terms
in the gluon emission antennae, that do not contain $1/Q^2$ singularities
as $\mc O(Q^2)$.

Given this new form of the antenna functions, the collinear limits,
$y\to 0$, are simple to read off,
\begin{alignat}{3}
  \bar a_{q\bar q\,g}^\t{FF} &\to~
    \frac1{Q^2}\,\left(\frac{1+z^2}{1-z}\right)
    ~&=&~ \frac1{Q^2}\,P(q\to qg)~,\\
  \bar a_{gg\,g}^\t{FF} + \bar a_{gg\,g}^\t{FF} [z\leftrightarrow 1-z] &\to~
    2\,\frac1{Q^2}\,\left(\frac{z}{1-z}+\frac{1-z}z+z(1-z)\right)
    ~&=&~ \frac1{Q^2}\,P(g\to gg)\, \\
  \bar a_{xq\,\bar q}^\t{FF} &\to~
    \frac12\,\frac1{Q^2}\,\left((1-z)^2+z^2\right)
    ~&=&~ \frac12\,\frac1{Q^2}\,P(g\to q\bar q)~.
\end{alignat}

\subsection{Phase-Space Variables and Limits \label{app:PSdetails}}

In \tabRef{tab:tZetaDefsLimits} we give an overview on combinations of the evolution 
variable $t$ and complementary phase-space variable $\zeta$ that are used
in the shower.

\begin{table}[tbp]
\caption{Definitions of the evolution variable $t$ and the
complementary phase-space variable $\zeta$ for II, IF and FF 
configurations, with the $\zeta$ boundaries in the last two
columns. \label{tab:tZetaDefsLimits}}
\begin{flushleft}
\begin{tabular}{lllll}
  \toprule
   & Evolution & Definition & 
  \multicolumn{2}{l}{$\zeta$ Boundaries} 
  \\
   & Variable $t$ & of $\zeta$ & $\zeta_\t{min}$ & $\zeta_\t{max}$ 
  \\ \midrule
  II & $\frac{\saj\sjb}\sab$ & $\frac\saj\sab$ &
  $\frac{s-\sAB}{s}-\sqrt{\frac{t_\t{max}-t}{s}}$ &
  $\frac{s-\sAB}{s}+\sqrt{\frac{t_\t{max}-t}{s}}$
  \\ [2.5mm]
   & & $\frac\saj\sAB$ &
  $\frac{s-\sAB}{2\sAB}-\frac1\sAB\sqrt{s(t_\t{max}-t)}$ &
  $\frac{s-\sAB}{2\sAB}+\frac1\sAB\sqrt{s(t_\t{max}-t)}$
  \\ [2.5mm]
   & & $\frac\sjb\sAB$ &
  $\frac{s-\sAB}{2\sAB}-\frac1\sAB\sqrt{s(t_\t{max}-t)}$ &
  $\frac{s-\sAB}{2\sAB}+\frac1\sAB\sqrt{s(t_\t{max}-t)}$
  \\ [2.5mm]
   & $\saj$ & $\frac\sab\sAB$ & $\frac{\sAB+t}\sAB$ & $\frac{s}\sAB$ 
  \\ [2.5mm]
   & $\sjb$ & $\frac\sab\sAB$ & $\frac{\sAB+t}\sAB$ & $\frac{s}\sAB$ 
  \\ \midrule
  IF & $\frac{\saj\sjk}{\sAK+\sjk}$ & $\frac{\sjk+\sAK}\sAK$ & 
  $\frac{\sAK+t}\sAK$ & $\frac1{x_A}$ 
  \\ [2.5mm]
   & & $\frac\saj{\sAK+\sjk}$ & $\frac{t~x_A}{\sAK(1-x_A)}$ & $1$ 
  \\ [2.5mm]
   & & $\frac\saj\sjk$ & $\frac{t~x_A}{\sAK(1-x_A)^2}$ & $\frac{\sAK+t}t$ 
  \\ [2.5mm]
   & $\saj$ & $\frac{\sjk+\sAK}\sAK$ & $\t{max}\left(1,\frac{t}\sAK\right)$ & $\frac1{x_A}$ 
  \\ [2.5mm]
   & $\sjk$ & $\frac\saj{\sAK+\sjk}$ & $0$ & $1$ 
  \\ \midrule
  FF & $4\,\frac{\sij\sjk}{\sIK}$ & $\frac\sij{\sij+\sjk}$ &
  $\frac12\left(1-\sqrt{1-\frac t\sIK}\right)$ & $\frac12\left(1+\sqrt{1-\frac t\sIK}\right)$
  \\ [2.5mm]
   & $\sij$ & $\frac\sjk\sIK$ & $0$ & $\frac{\sIK-t}\sIK$
  \\ [2.5mm]
   & $\sjk$ & $\frac\sij\sIK$ & $0$ & $\frac{\sIK-t}\sIK$
  \\ \bottomrule
\end{tabular}
\end{flushleft}
\end{table}

\section{Comparison with Matrix Elements \label{app:MEplots}}

\begin{figure}[b]
\centering
\includegraphics[width=.84\textwidth]{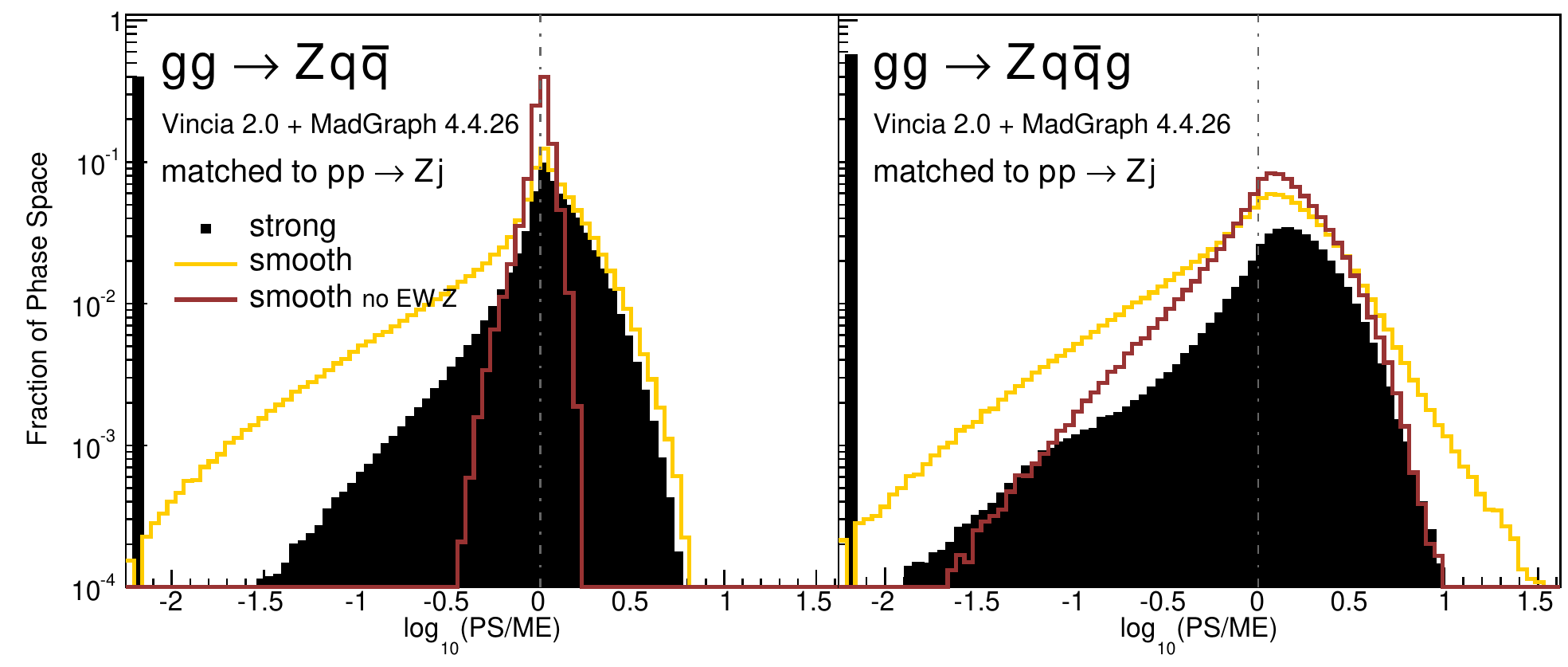}
\caption{\label{fig:PSMEggZ} Antenna shower, compared to matrix elements: distribution
of $\t{log}_{10}(\t{PS}/\t{ME})$ in a flat phase space scan of the full phase space.
Contents normalized to the number of generated points.} \vspace*{5mm}
\includegraphics[width=.42\textwidth]{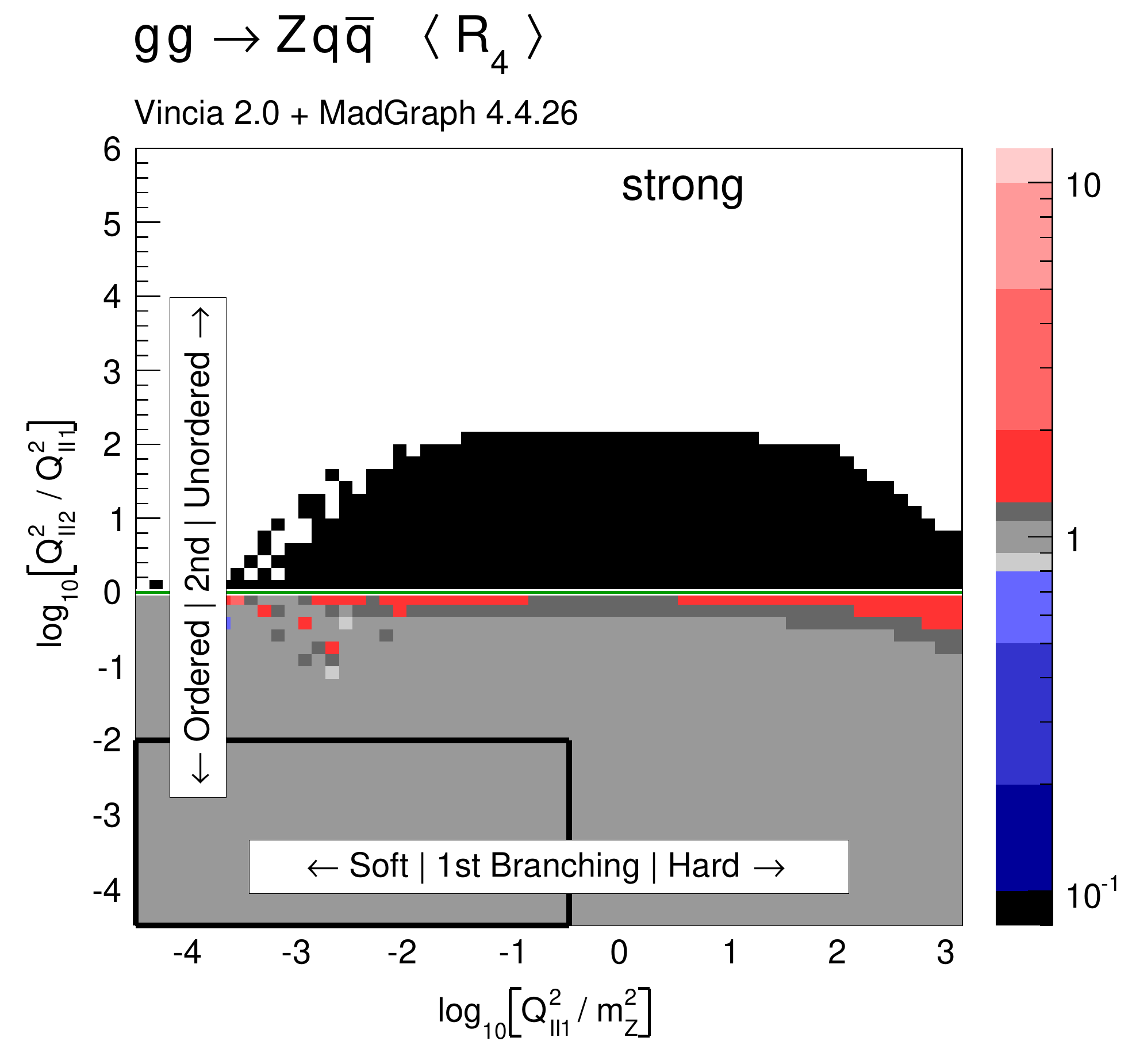}
\includegraphics[width=.42\textwidth]{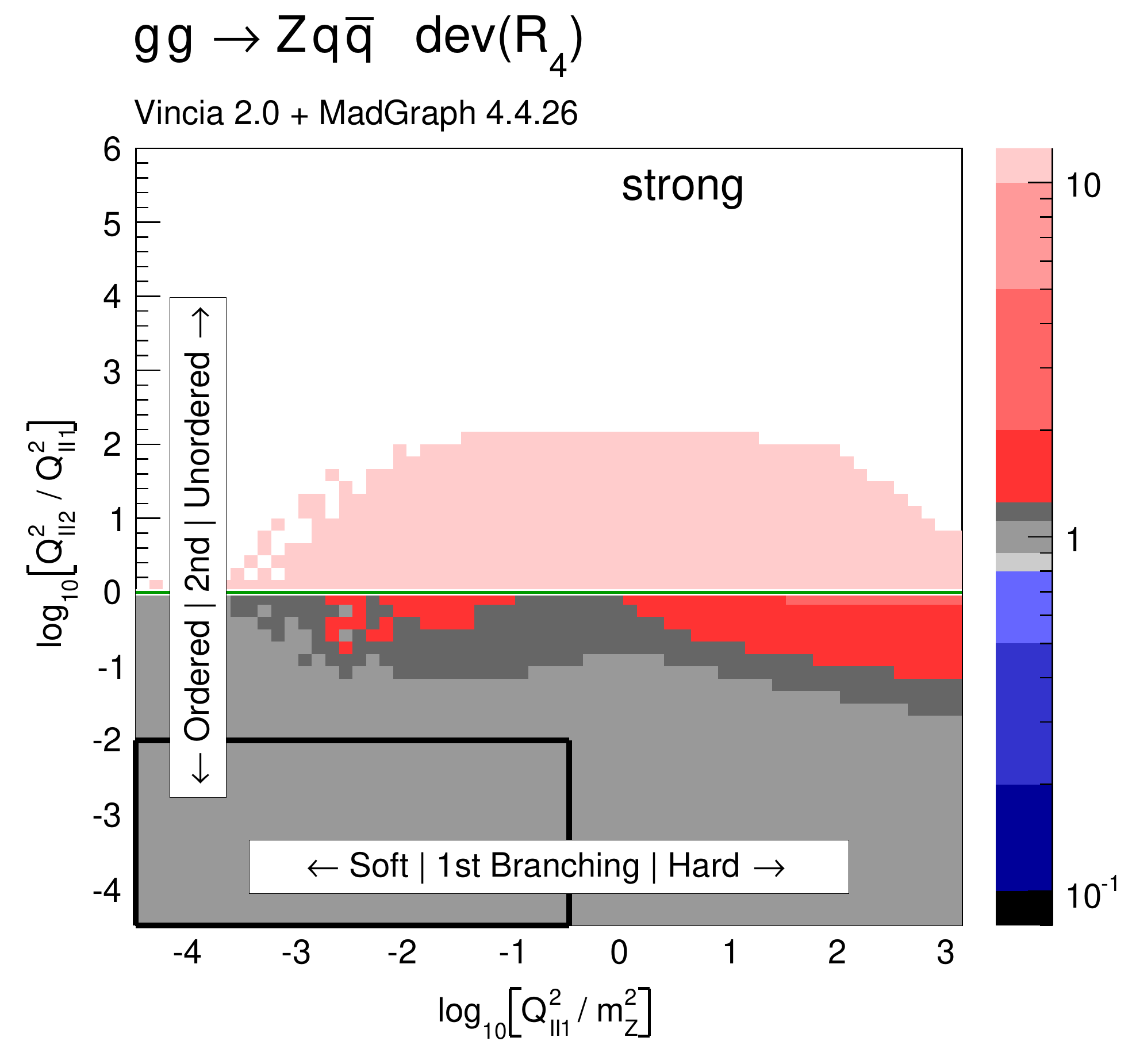}
\includegraphics[width=.42\textwidth]{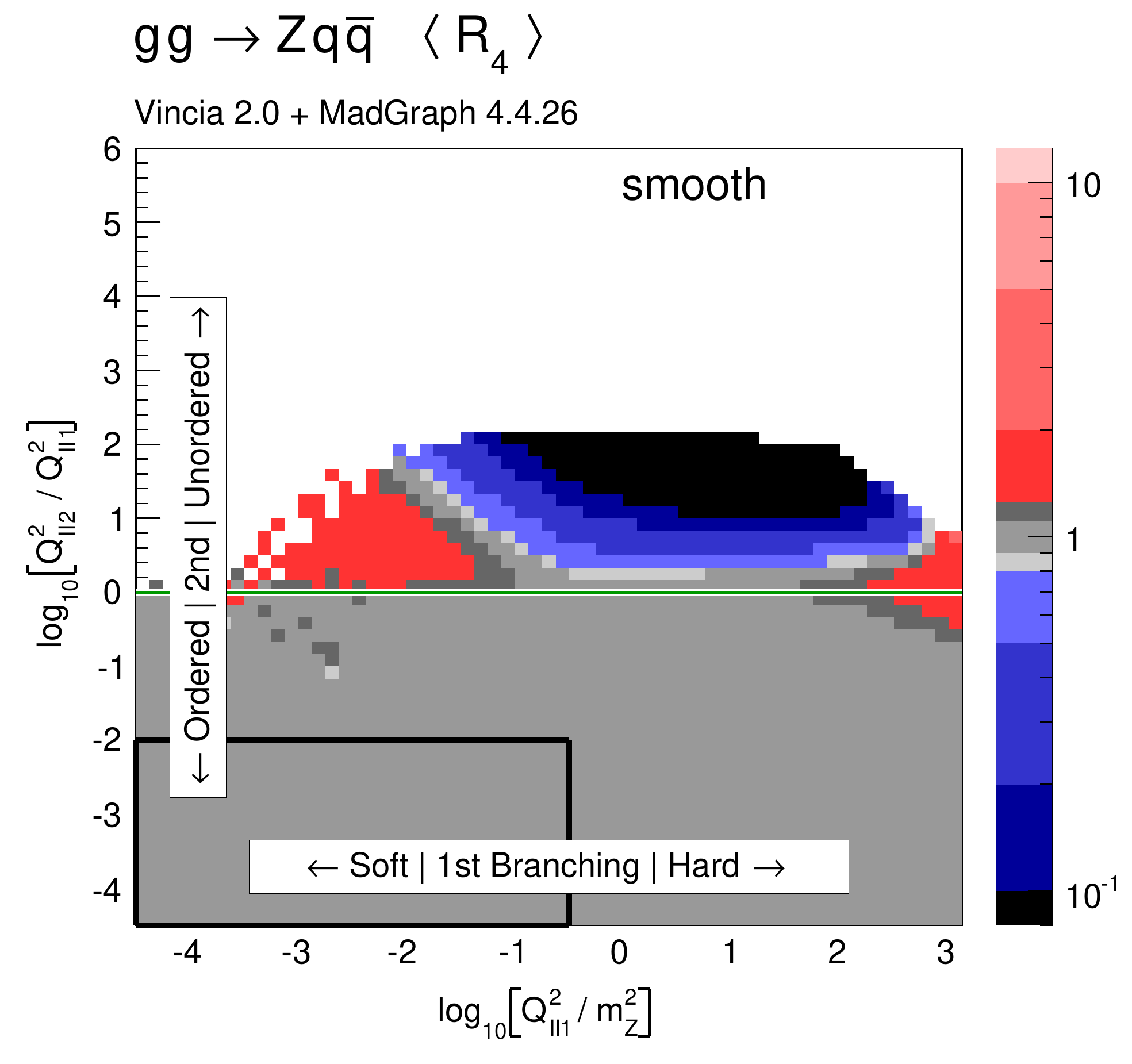}
\includegraphics[width=.42\textwidth]{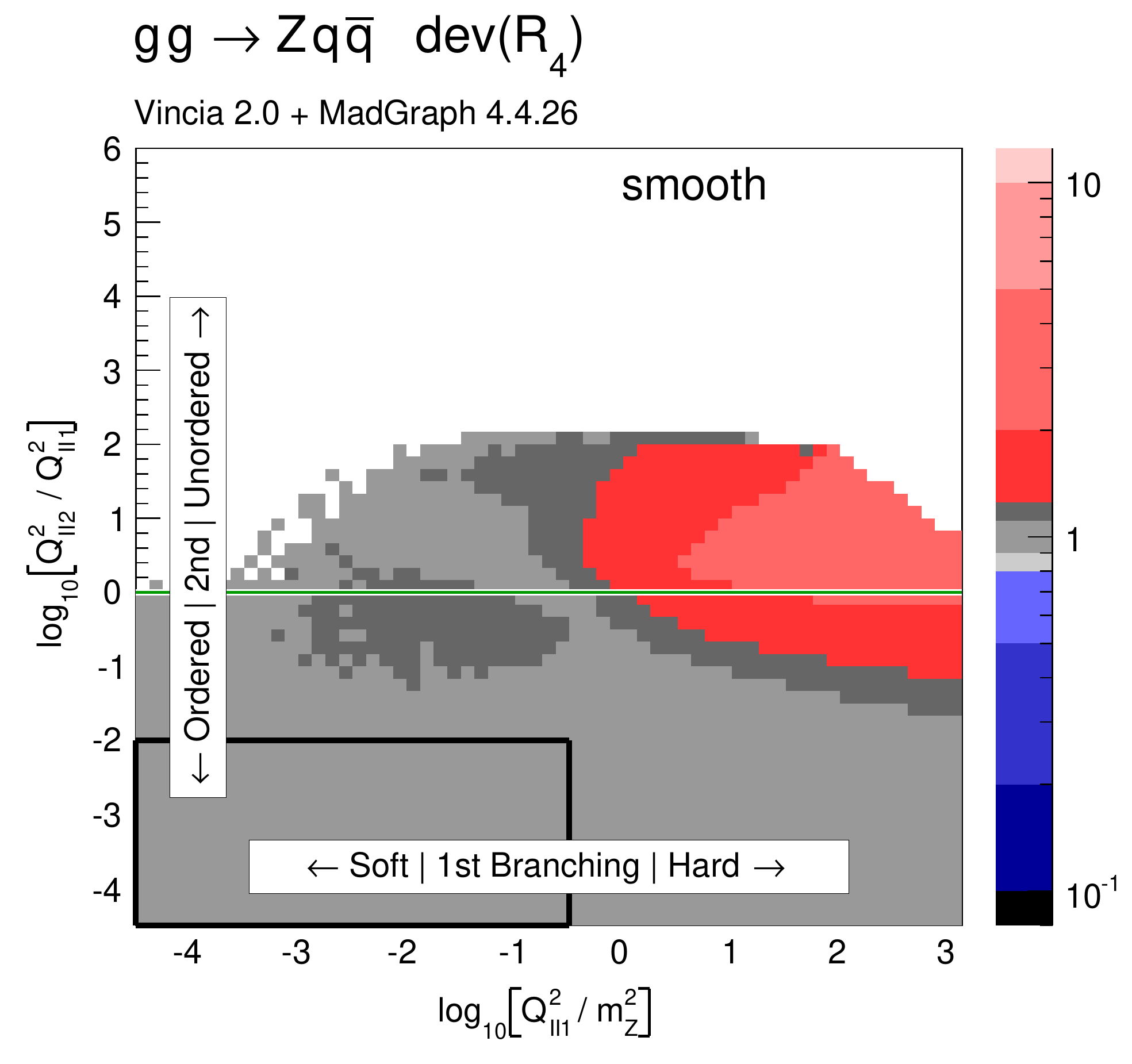}
\caption{\label{fig:PSMEggZ2D} The value of $\langle R_4 \rangle$ (left)
and $\t{dev}(R_4)$ (right), differentially over the 4-parton phase space, with 
$Q^2$ ratios characterising the first and second emissions on the $x$ and $y$ 
axis, respectively. Strong (top) and smooth (bottom) ordering in the shower.}\vspace*{-5mm}
\end{figure}

\begin{figure}[p]
\centering
\includegraphics[width=.84\textwidth]{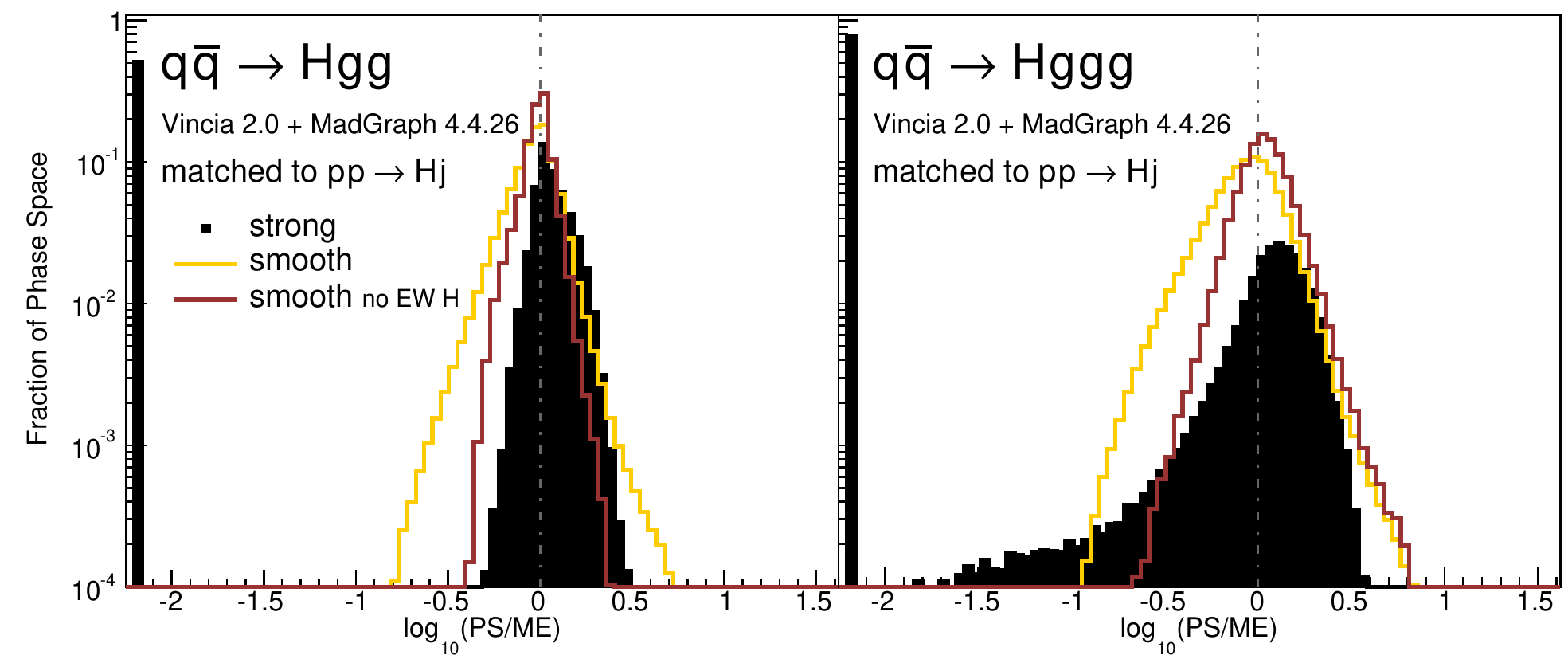}
\caption{\label{fig:PSMEqqH} Antenna shower, compared to matrix elements: distribution
of $\t{log}_{10}(\t{PS}/\t{ME})$ in a flat phase space scan of the full phase space.
Contents normalized to the number of generated points.
Gluon emission only.} \vspace*{5mm}
\includegraphics[width=.42\textwidth]{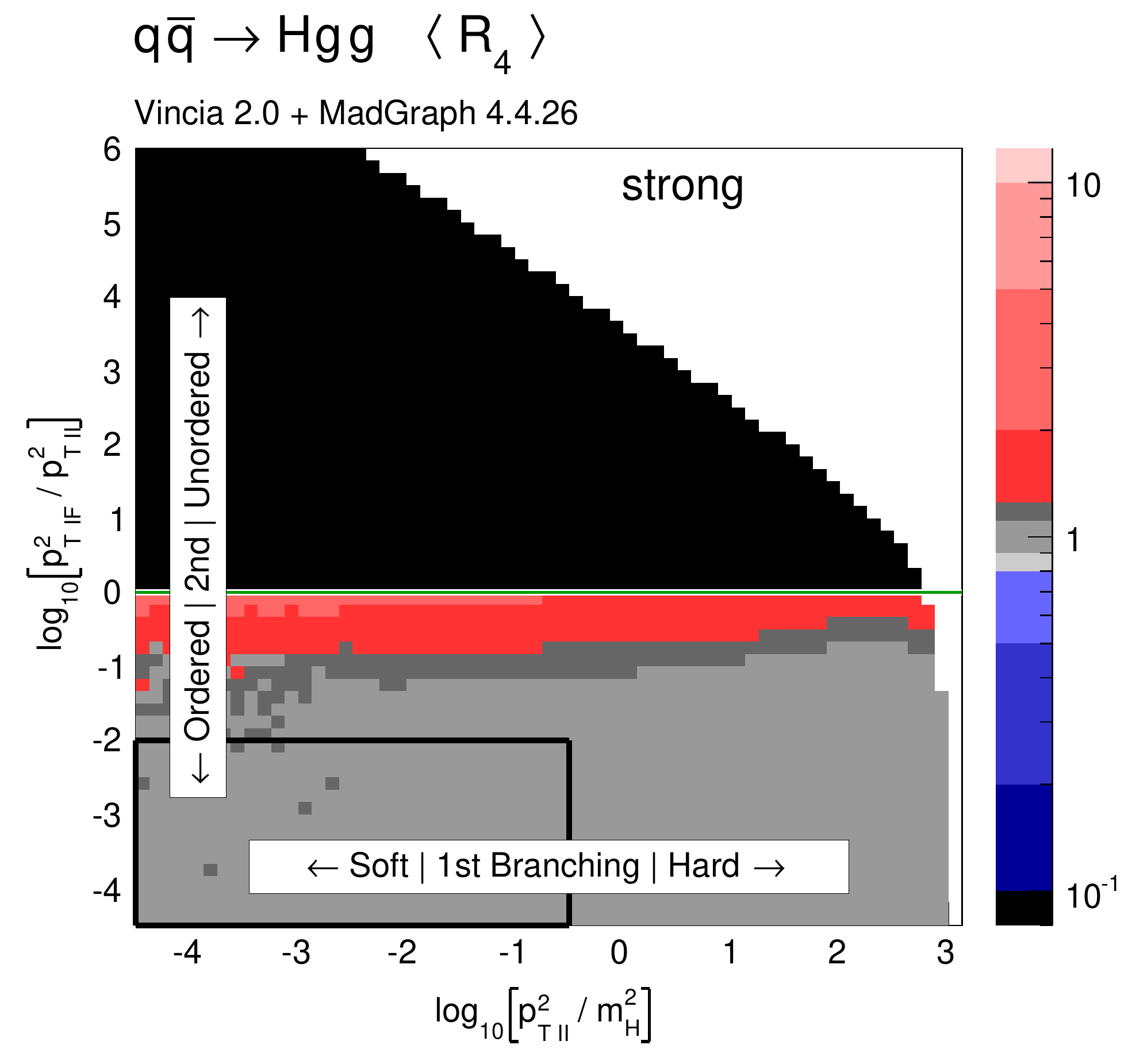}
\includegraphics[width=.42\textwidth]{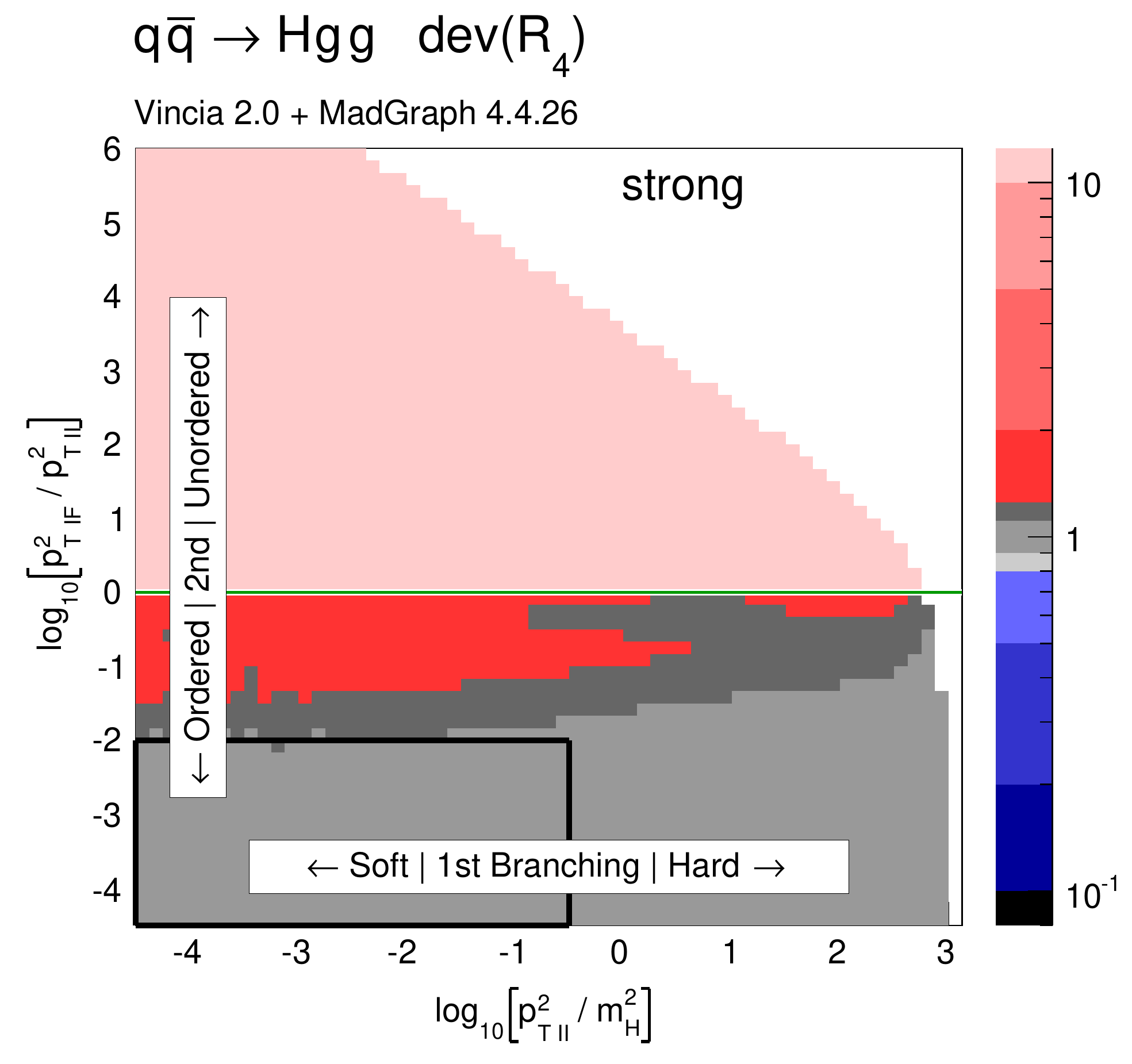}
\includegraphics[width=.42\textwidth]{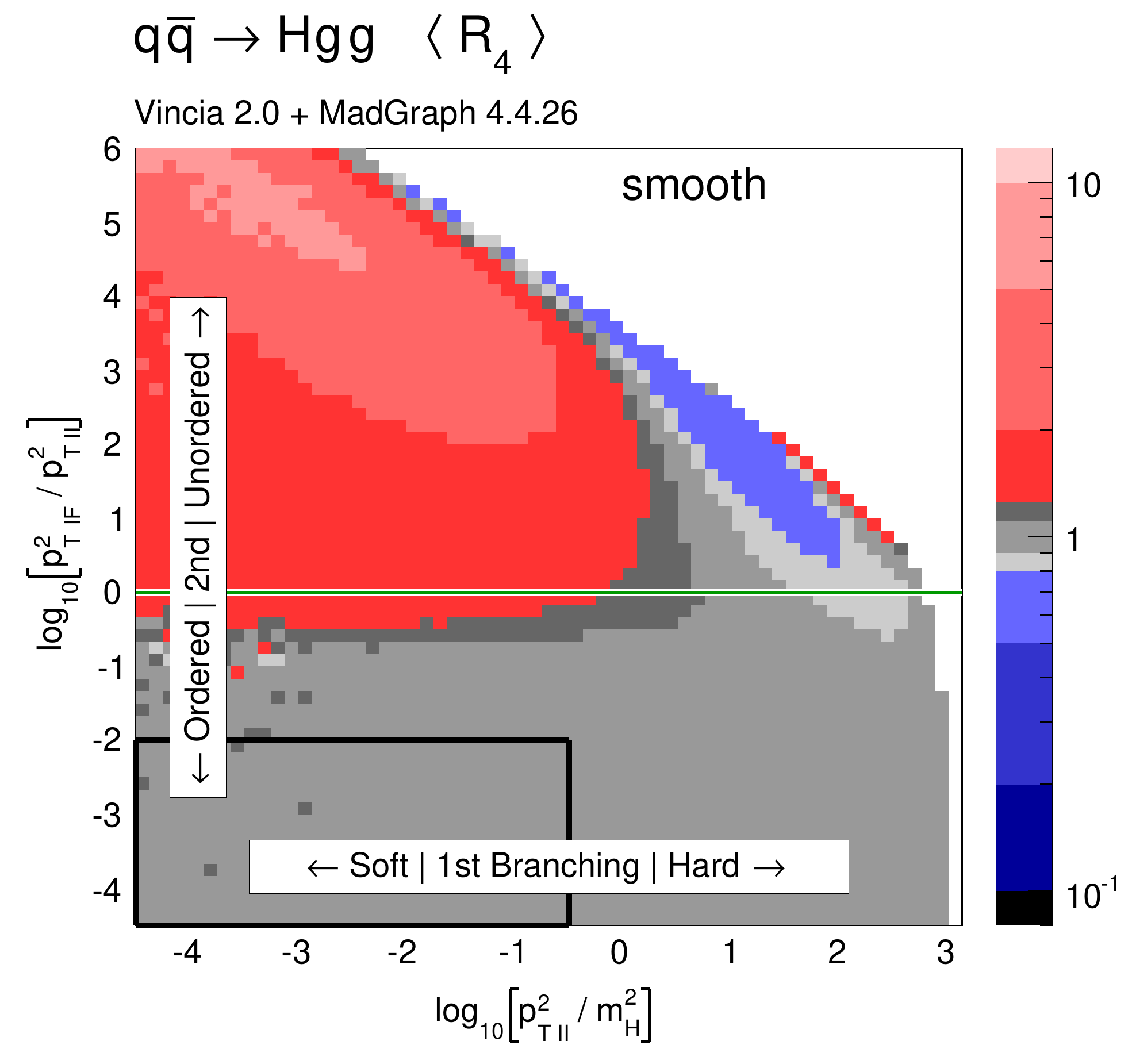}
\includegraphics[width=.42\textwidth]{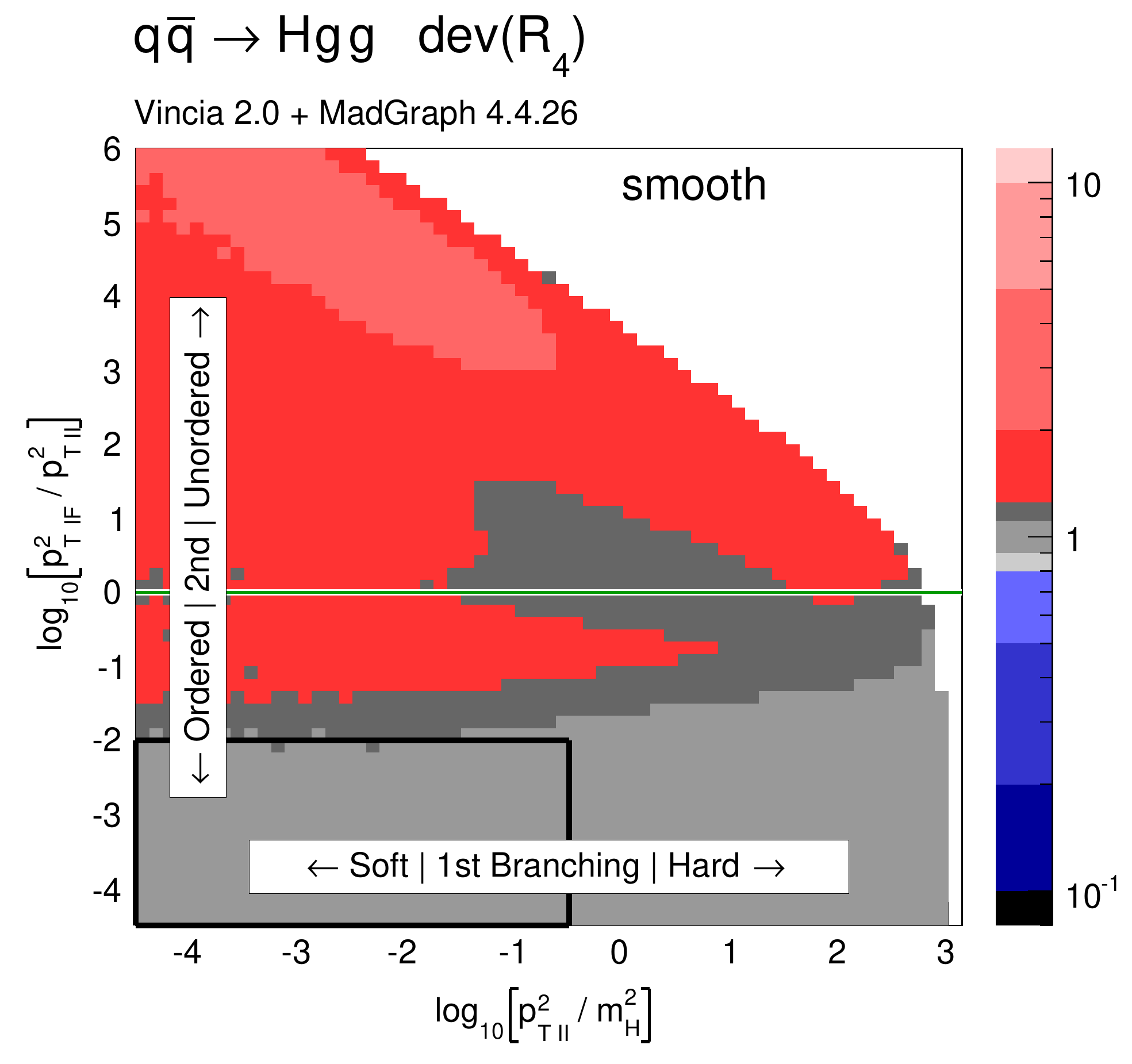}
\caption{\label{fig:PSMEqqH2D} The value of $\langle R_4 \rangle$ (left)
and $\t{dev}(R_4)$ (right), differentially over the 4-parton phase space, with 
$p_\perp^2$ ratios characterising the first and second emissions on the $x$ and $y$ 
axis, respectively. Strong (top) and smooth (bottom) ordering in the shower, 
with gluon emission only.}
\end{figure}

\begin{figure}[p]
\centering
\includegraphics[width=.84\textwidth]{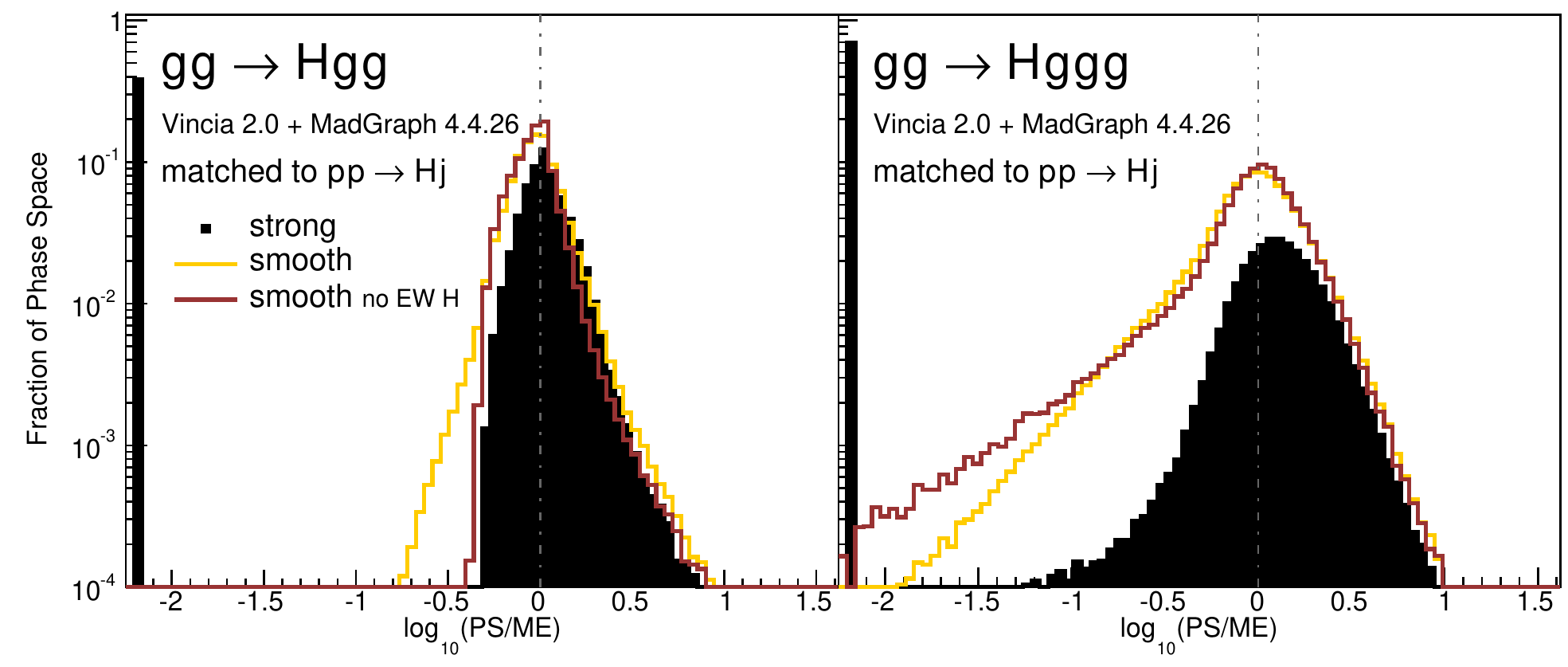}
\caption{\label{fig:PSMEggH} Antenna shower, compared to matrix elements: distribution
of $\t{log}_{10}(\t{PS}/\t{ME})$ in a flat phase space scan of the full phase space.
Contents normalized to the number of generated points.
Gluon emission only.} \vspace*{5mm}
\includegraphics[width=.42\textwidth]{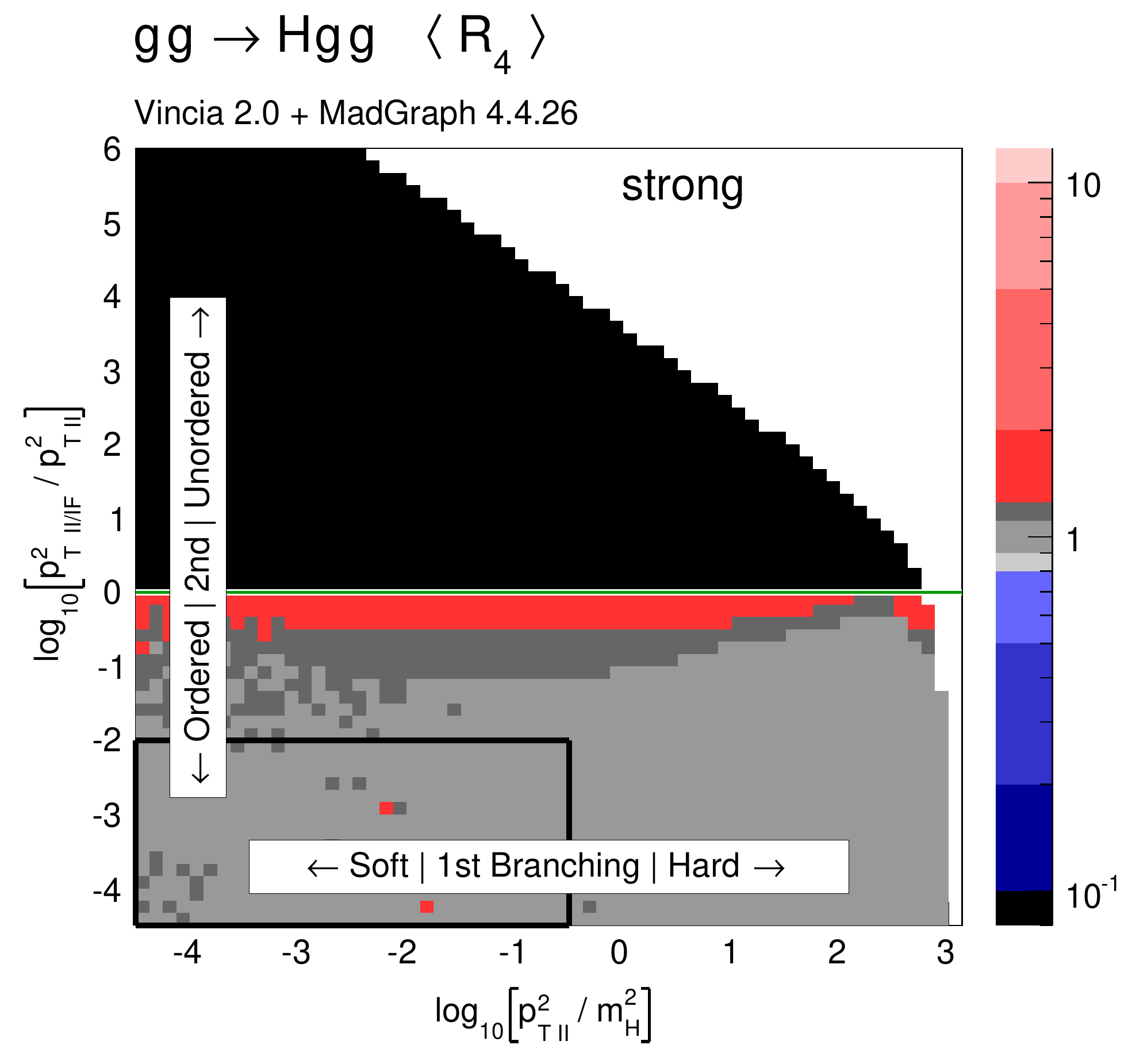}
\includegraphics[width=.42\textwidth]{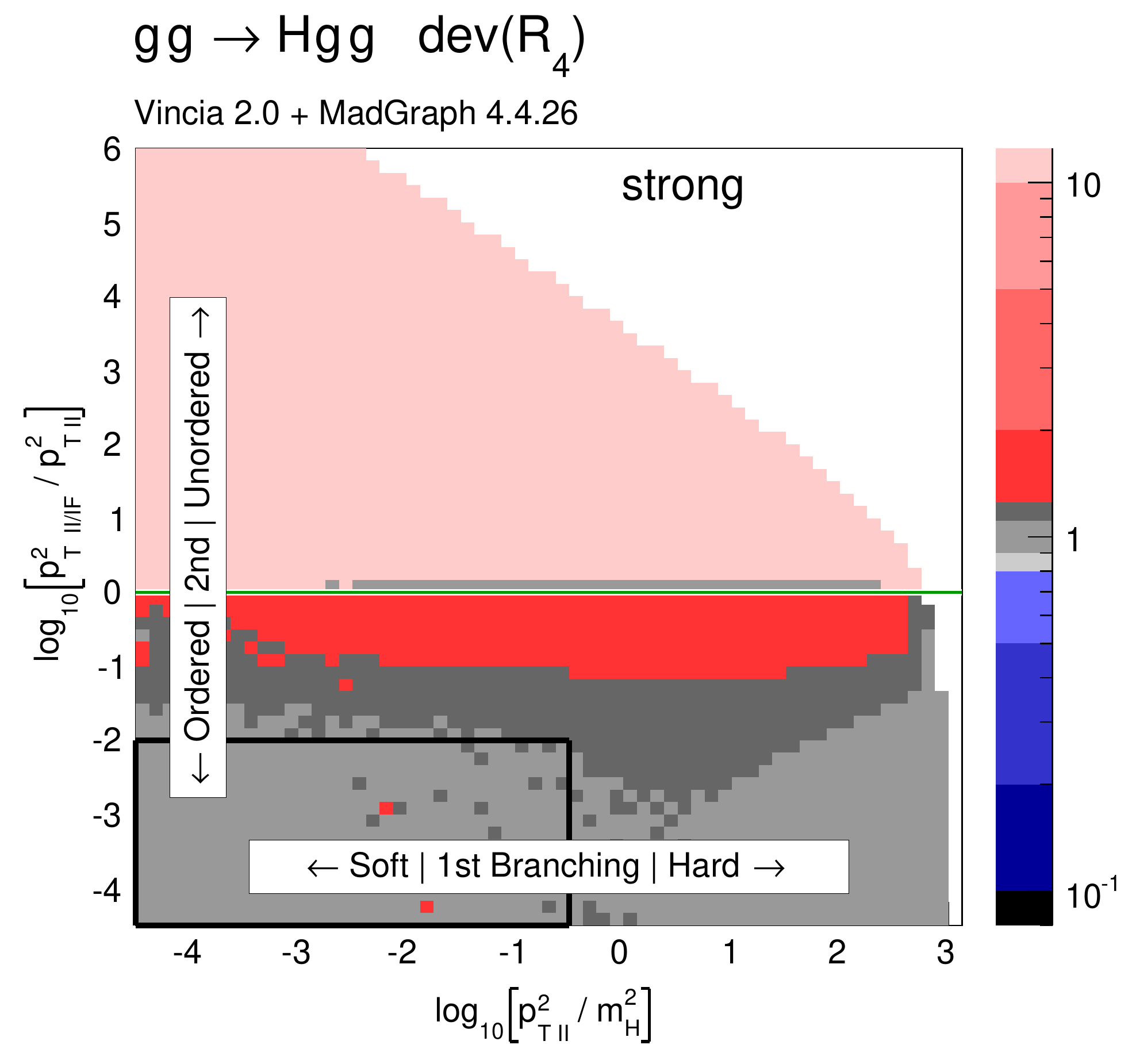}
\includegraphics[width=.42\textwidth]{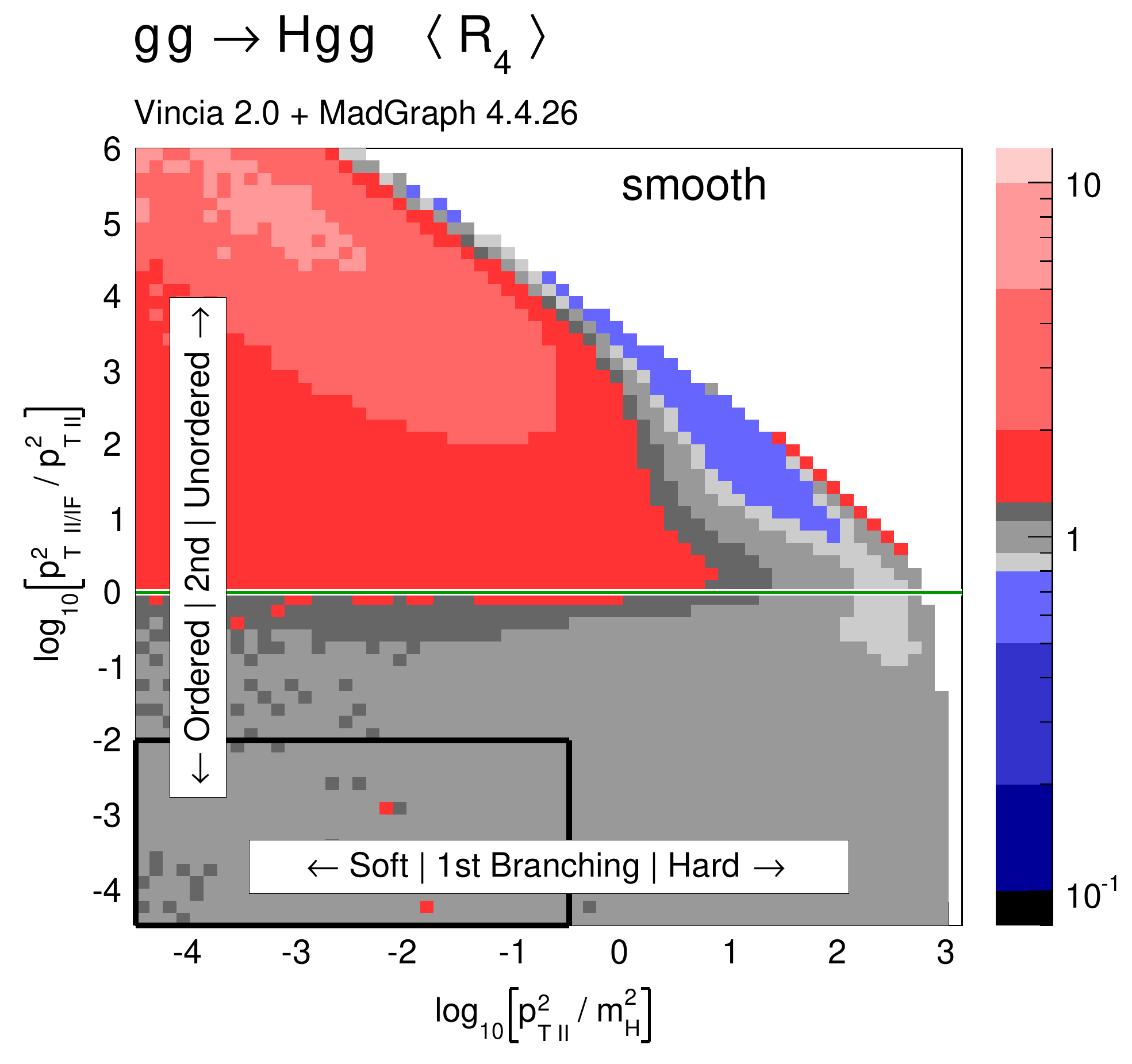}
\includegraphics[width=.42\textwidth]{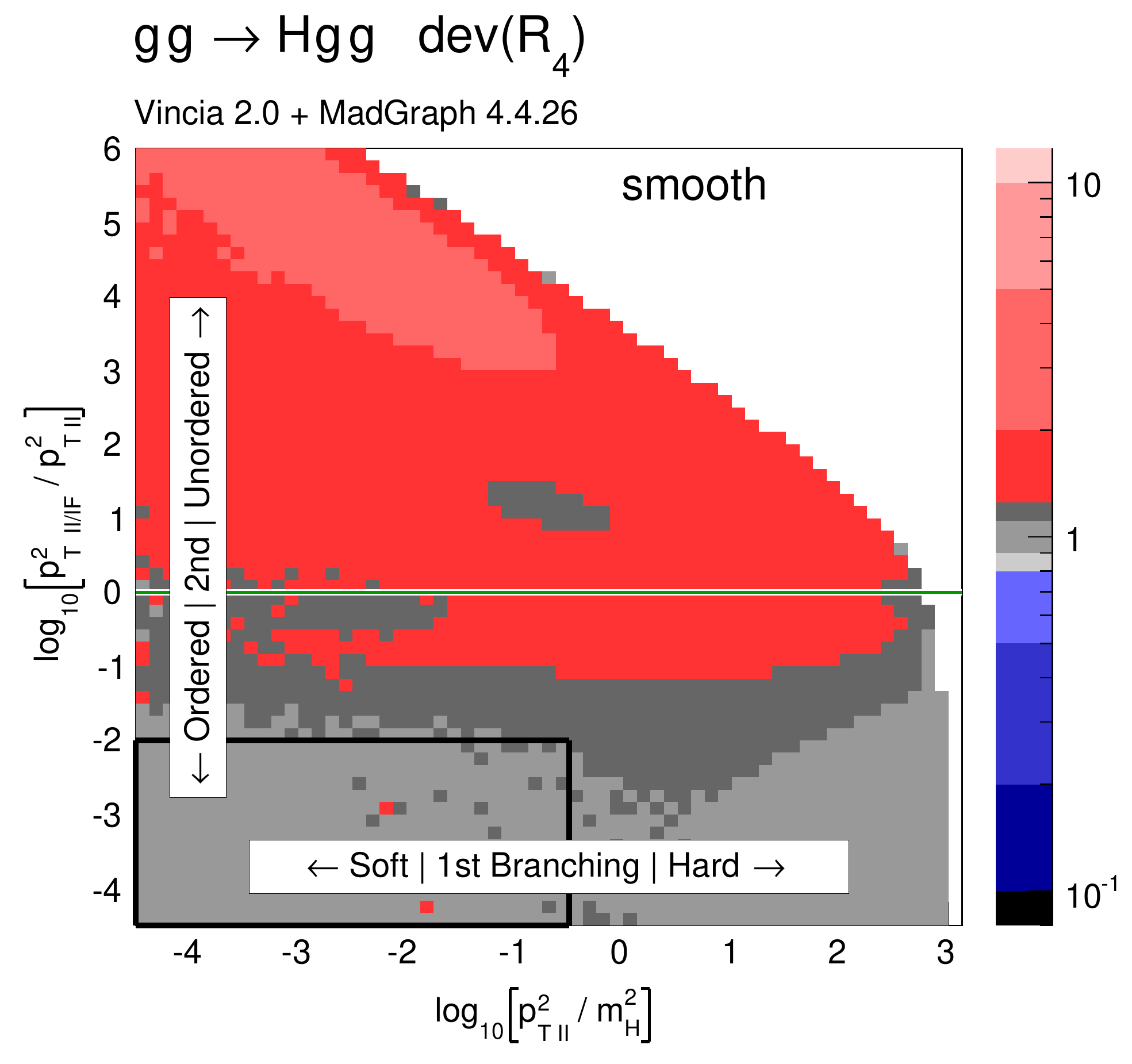}
\caption{\label{fig:PSMEggH2D} The value of $\langle R_4 \rangle$ (left)
and $\t{dev}(R_4)$ (right), differentially over the 4-parton phase space, with 
$p_\perp^2$ ratios characterising the first and second emissions on the $x$ and $y$ 
axis, respectively. Strong (top) and smooth (bottom) ordering in the shower, 
with gluon emission only.}
\end{figure}

In this section we show an extended set of plots where we compare the shower 
approximation to leading-order matrix elements; see \secsRef{sec:validations}
and \ref{sec:Impr24Branch} for a description of the observables. 
We show the one- and two-dimensional distributions of the PS-to-ME ratios
for $gg\to Zq\bar q(g)$ in \figsRef{fig:PSMEggZ} and \ref{fig:PSMEggZ2D},
for $q\bar q\to Hgg(g)$ in \figsRef{fig:PSMEqqH} and \ref{fig:PSMEqqH2D}, and
for $gg\to Hgg(g)$ in \figsRef{fig:PSMEggH} and \ref{fig:PSMEggH2D}. As before,
see \eqsRef{eq:EWcut1} and \eqref{eq:EWcut2},
we include distributions with a cut on the transverse mass of the boson,
$m_{\perp\,Z}^2$ (labelled ``no EW Z'') and $m_{\perp\,H}^2$ (labelled ``no EW H'')
respectively.

\afterpage{

\bibliographystyle{ants}
\bibliography{ants.bbl}

\clearpage

}

\end{document}